\documentclass[PAPER, atlasdraft=false, UKenglish,cernpreprint, texmf, orcidlogo]{atlasdoc}
 
\usepackage{atlaspackage}
 
\usepackage[acronym,shortcuts]{glossaries}
\glsdisablehyper
\usepackage{atlasbiblatex}
 
\usepackage{atlasphysics}

\graphicspath{{logos/}{figures/}}
 
\usepackage{ANA-TRIG-2022-01-PAPER-defs}
 
\usepackage{diagbox}
\usepackage{adjustbox}
\usepackage{multirow}

\AtlasTitle{The ATLAS Trigger System for LHC Run 3 and Trigger performance in 2022}

\AtlasAbstract{
The ATLAS trigger system is a crucial component of the ATLAS experiment at the LHC.
It is responsible for selecting events in line with the ATLAS physics programme. This paper presents
an overview of the changes to the trigger and data acquisition system during the second long
shutdown of the LHC,
and shows the performance of the trigger system and its components in the proton--proton collisions
during the 2022 commissioning period as well as its expected performance in
proton--proton and heavy--ion collisions for the remainder of the third LHC data-taking period (2022--2025).
}
 
\AtlasRefCode{TRIG-2022-01}
 
\PreprintIdNumber{CERN-EP-2023-299}

\AtlasJournalRef{JINST 19 (2024) P06029}
\AtlasDOI{10.1088/1748-0221/19/06/P06029}

\hypersetup{pdftitle={ATLAS document},pdfauthor={The ATLAS Collaboration}}
 
\begin{document}
 
\maketitle
 
\tableofcontents
 
\newpage
\section{Introduction}
\label{sec:intro}

The trigger system~\cite{PERF-2011-02, TRIG-2016-01} is an essential component of the ATLAS experiment~\cite{atlas-det-run3}
as it is responsible for deciding whether or not to permanently record data from a given bunch-crossing interaction
(also referred to as an \textit{event}) for further study.
It aims to minimise the rate of the recorded events while maintaining an excellent and unbiased efficiency for physics processes of interest.
During the \runi (2009 -- 2013) and \runii (2015 -- 2018) operational periods at the Large Hadron Collider (LHC), the
ATLAS trigger system collected data from proton--proton (\pp) collisions at a centre-of-mass energy
up to $\sqrt{s}=13\,$TeV, a peak instantaneous luminosity of \lumiruntwopeak and up to a mean of 60
interactions per bunch crossing (pile-up). As part of the LHC heavy--ion (HI) programme, lead-lead nuclei collisions (Pb+Pb) at a
centre-of-mass energy up to $\sqrt{s}=5\,$TeV per nucleon and a peak instantaneous luminosity of \lumiHIruntwo
were achieved, with other ion species and collision configurations used in addition.
 
During the second LHC Long Shutdown (LS2, 2019 -- 2021), the ATLAS trigger system underwent a major upgrade.
The purpose of this 
upgrade was to enhance the physics reach of the experiment
for the ongoing operation in \runiii (2022 -- 2025). In \runiii the centre-of-mass energy of the \pp collisions is $\sqrt{s}=13.6\,$TeV
with an instantaneous luminosity planned to be kept constant at its peak value, approximately 
$2.4\times\lumi{e34}$
for a duration as long as ten hours per LHC fill and 60--70 interactions per bunch crossing.
 
The ATLAS Run-3 trigger system,
its performance in the \pp\ collisions during the 2022 commissioning period, including rates and efficiencies, as well as expected
trigger algorithm modifications for HI data taking
are described in this paper.
After a brief introduction to the ATLAS detector in Section~\ref{sec:ATLASdetector}, Section~\ref{sec:TDAQsystemChanges}
summarises the changes to the trigger and data acquisition during LS2. Section~\ref{sec:menu} gives overviews
of trigger selections implemented for \runiii followed by an introduction to the reconstruction algorithms
used at the \ac{HLT} in Section~\ref{sec:HLTreco}. The implementation of different triggers and their performance
are discussed in Sections~\ref{sec:sigPerf} and~\ref{sec:special}. The trigger software performance is
presented in Section~\ref{sec:softwarePerf}.
 
The results presented in this paper are based on the \pp\ collision data recorded by the ATLAS experiment and the corresponding simulated \ac{MC} events.
The rates are quoted for an instantaneous luminosity value of $1.8\times\lumi{e34}$.
The integrated luminosity used varies over the figures and is measured to be at most 31~\ifb\ for the 2022 data set~\cite{ATL-DAPR-PUB-2023-001}.
Due to the absence of HI physics data taking in 2022, there are no HI results to show in this paper.
 
Trigger efficiencies are calculated with respect to either reconstructed or true physics objects
from \ac{MC} simulation. Typical methods used to determine trigger efficiency in data~\cite{TRIG-2018-05}
are either the tag-and-probe or bootstrap methods. The tag-and-probe method is based on a sample of events with a known resonance, such as
$Z$ boson or $J/\psi$, decaying to two charged leptons. The bootstrap method involves
successive measurements of trigger efficiency with respect to unbiased or lower-threshold triggers.
\ac{MC} generator information about the physics objects can be used to determine trigger efficiency in the \ac{MC} simulation.


\section{The Run-3 ATLAS detector}
\label{sec:ATLASdetector}
 
\newcommand{\AtlasCoordFootnote}{
ATLAS uses a right-handed coordinate system with its origin at the nominal interaction point (IP)
in the centre of the detector and the \(z\)-axis along the beam pipe.
The \(x\)-axis points from the IP to the centre of the LHC ring,
and the \(y\)-axis points upwards.
Polar coordinates \((r,\phi)\) are used in the transverse plane,
\(\phi\) being the azimuthal angle around the \(z\)-axis.
The pseudorapidity is defined in terms of the polar angle \(\theta\) as \(\eta = -\ln \tan(\theta/2)\).
Angular distance is measured in units of \(\Delta R \equiv \sqrt{(\Delta\eta)^{2} + (\Delta\phi)^{2}}\).}
 
The ATLAS detector at the LHC has undergone a substantial upgrade with improvements to various detector subsystems and their electronics, in order to
enable the broad physics programme planned for the Run-3 data taking. The original configuration of the detector (as it was built for the start of \runi of the LHC) is described in Ref.~\cite{PERF-2007-01}.
 
The ATLAS detector covers nearly the entire solid angle around the collision point\footnote{\AtlasCoordFootnote}. It consists of an inner tracking detector surrounded by a superconducting solenoid, electromagnetic and hadron calorimeters, and a muon spectrometer incorporating three large superconducting air-core toroidal magnets.
 
The \ac{ID} system is immersed in a \SI{2}{\tesla} axial magnetic field and provides charged-particle tracking in the range \(|\eta| < 2.5\).
The high-granularity silicon pixel detector (PIX) covers the vertex region and typically provides four measurements per track,
the first hit normally being in the insertable B-layer (IBL) installed before \runii~\cite{PIX-2018-001}.
It is followed by the silicon microstrip tracker (SCT), which usually provides eight measurements per track.
These silicon detectors are complemented by the outermost of the three tracking subsystems, the transition radiation tracker (TRT),
which enables radially extended track reconstruction up to \(|\eta| = 2.0\).
The TRT also provides electron identification information
based on the fraction of hits (typically 30 in total) above a higher energy-deposit threshold corresponding to transition radiation.
The TRT gas configuration has a significant impact on the particle identification.
Due to a number of leaks in flexible active gas exhaust pipes that developed during Run 1 and Run 2,
it became too costly to operate the entire detector with the baseline xenon-based gas mixture.
For Run 3 an argon-based gas mixture is used in the entire barrel and within a few endcap wheels on one side of the detector~\cite{atlas-det-run3}.
Due to poor absorption of transition radiation photons by the argon gas, the particle identification function is significantly reduced in the barrel region.
However, in combination with \dedx measurements, it still contributes to the ATLAS electron identification, particularly at particle energies below 10\,\GeV.
The particle identification performance of the endcaps is largely preserved.
 
ATLAS uses two sampling calorimeter technologies covering the pseudorapidity range \(|\eta| < 4.9\).
Within the region \(|\eta|< 3.2\), electromagnetic (EM) calorimetry is provided by barrel and endcap
high-granularity lead/liquid-argon (LAr) calorimeters, consisting of three layers with varying granularities. In addition, a LAr presampler covering \(|\eta| < 1.8\) corrects for energy loss in material upstream of the calorimeters.
Hadron calorimetry is provided by the steel/scintillator-tile calorimeter (Tile calorimeter),
segmented into three barrel structures within \(|\eta| < 1.7\), and two copper/LAr hadron endcap calorimeters.
The solid angle coverage is completed with forward copper/LAr and tungsten/LAr calorimeter modules
optimised for electromagnetic and hadronic energy measurements respectively.
During LS2, the LAr calorimeter electronics was augmented with a new digital trigger path~\cite{Aad:2022noz}
providing finer granularity inputs to the upgraded trigger system discussed in Section~\ref{sec:TDAQsystemChanges}.
 
The muon spectrometer (MS) comprises separate trigger and high-precision tracking chambers
measuring the deflection of muons in a magnetic field generated by the superconducting
air-core toroidal magnets.
The field integral of the toroids ranges between \num{2.0} and \SI{6.0}{\tesla\metre}
across most of the detector.
Three stations of precision chambers, each consisting of layers of monitored drift tubes (MDTs), cover the region \(|\eta| < 2.7\).
In the innermost station of the endcaps, \(|\eta| > 1.3\),
the detectors used in Runs~1 and 2, Small Wheels, have been replaced by the New Small Wheels (NSWs)~\cite{atlas-det-run3}.
The NSWs use two technologies: small-strip thin gap chamber (sTGC) and Micromegas (MM) detectors, both with high-rate tolerance and improved resolution.
The muon trigger system covers the range \(|\eta| < 2.4\) with resistive plate chambers (RPCs)
in the barrel ($|\eta|<1.05$), and thin gap chambers (TGCs) in the endcap regions ($1.05 < |\eta| < 2.4$).
In response to the increasing number of gas leaks due to cracks in the gas inlets of the RPC
system that developed over time during \runii, significant work was undertaken during LS2 to reinforce
the RPC gas inlets and recover a large number of channels that had become inactive.
Inlets were repaired and no-return valves installed.
Such maintenance work is expected to continue throughout \runiii in periods with no data taking.
The resulting impact on the muon trigger efficiency is described in Section~\ref{sec:TDAQsystemChanges}.
 
\subsection{Overview of ATLAS Trigger and Data Acquisition System}
\label{sec:TDAQsystem}
 
The selection and recording of events is handled by the Trigger and Data Acquisition (TDAQ) system shown in Figure~\ref{fig:TDAQsystem}.
 
\begin{figure}
\centering
\includegraphics[trim={0 4cm 0 0},clip,width=0.9\textwidth]{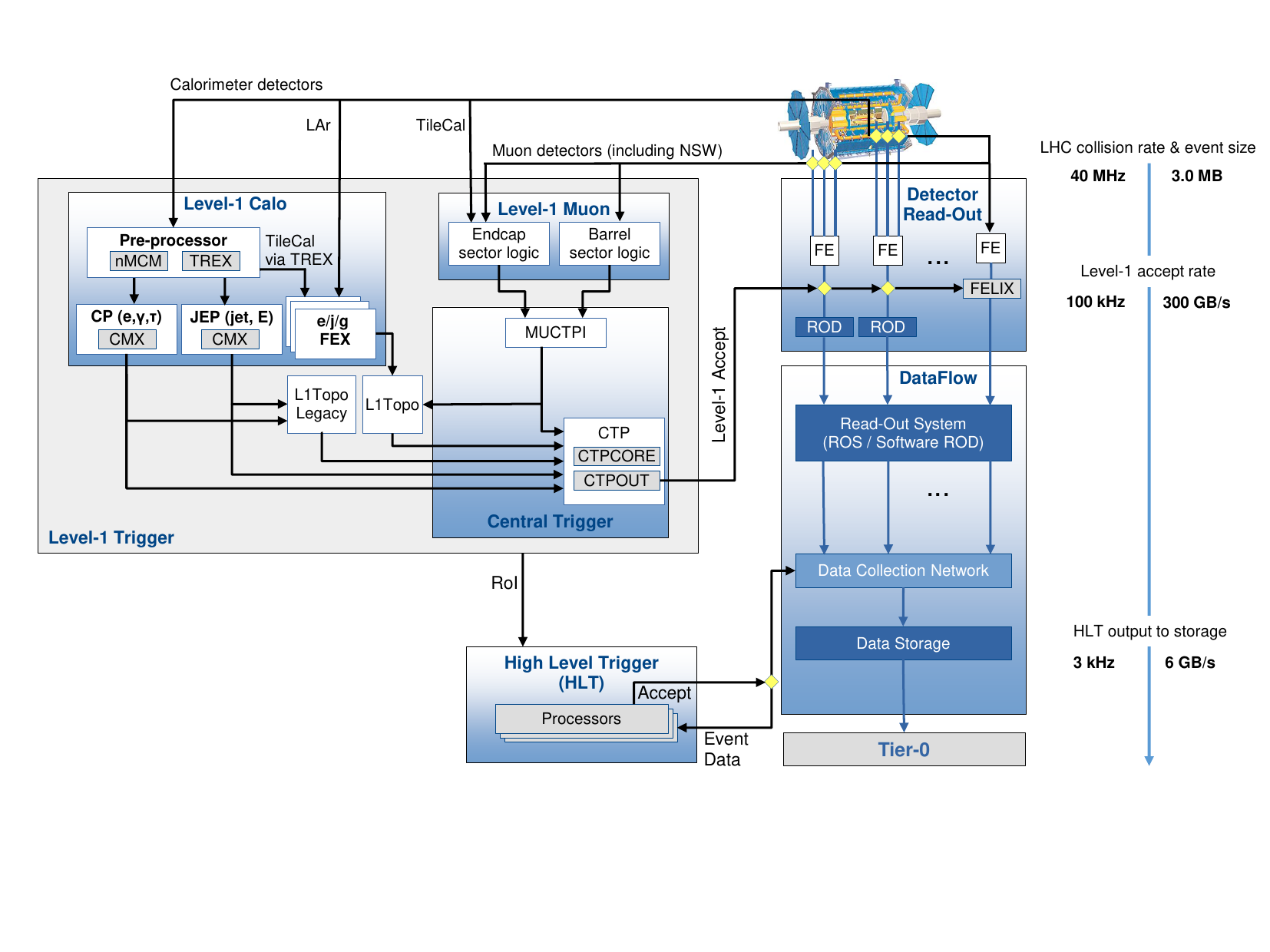}
\caption{The ATLAS TDAQ system in \runiii with emphasis on the components relevant for triggering as well as the detector read-out and data flow.
Level-1 Calo, Level-1 Muon and the Central Trigger all send data to the Read-Out System (or FELIX/Software ROD), described in Section~\ref{sec:DAQ},
primarily for the purposes of offline validation and error checking. This is not shown in the diagram for simplicity. Abbreviations used are defined in Sections~\ref{sec:TDAQsystem} and \ref{sec:TDAQsystemChanges}.}
\label{fig:TDAQsystem}
\end{figure}
 
The first-level (Level-1, L1) trigger is mainly based on two independent systems which use custom electronics to trigger on reduced-granularity information from either the calorimeters (L1Calo) or the muon detectors (L1Muon).
The L1 topological processor (L1Topo) system uses kinematic information from objects reconstructed in the L1Calo and L1Muon systems and applies topological selections.
Changes to these systems for Run 3 are described in detail in Section~\ref{sec:TDAQsystemChanges}.
 
The L1 trigger decision is formed by the \ac{CTP},  based on inputs received from the L1Calo trigger system, the L1Muon trigger system through the Muon-to-Central Trigger Processor Interface (MUCTPI)~\cite{MUCTPI}, the L1Topo system as well as several other subsystems. These subsystems are the Minimum Bias Trigger Scintillators (MBTS)~\cite{MBTS},
the ATLAS Forward Proton (AFP)~\cite{AFP1} detector and ALFA detector~\cite{ALFA} discussed further in Section~\ref{sec:minbias};
the LUCID-2 Cherenkov Counter~\cite{LUCID}, used for the primary luminosity measurements~\cite{DAPR-2021-01} complemented by measurements using the inner detector and calorimeters;
and the Zero Degree Calorimeter (ZDC)~\cite{ZDC} which is installed for heavy ion data taking (see Section~\ref{sec:menuHI}).
The CTP is also responsible for applying \textit{dead time}, a mechanism to limit the number of close-by L1 accepts~\cite{Bertelsen:2063035}.
 
As detailed in Section~\ref{sec:menu}, events that satisfy the \textit{trigger menu} requirements based on object type,
threshold and multiplicity are accepted at a rate up to the maximum detector read-out rate of 100 kHz
(down from the bunch crossing rate of about 40 MHz) at a fixed latency (detector read-out time window) below 2.5 $\mu s$. Up to 512 distinct L1 trigger items may be configured in the CTP.
 
If accepted by the L1 trigger, events are then sent to a software-based \ac{HLT}. Here \emph{online} algorithms reconstruct the event at progressively higher levels of detail than at L1, either in the full detector volume or in restricted Regions-of-Interest (RoIs), which are detector regions in which candidate trigger objects have been identified by the L1 trigger.
The HLT software is incorporated in the same software framework as is used \textit{offline} to reconstruct recorded events. 
For \runiii, this software framework was redesigned to support multi-threaded execution~\cite{AthenaMT} as detailed in Section~\ref{sec:AthenaMT}.
The physics output rate of the HLT during an ATLAS data-taking run 
is expected to be 3~kHz on average (see Section~\ref{sec:menu} for more details).
 
The Data Acquisition (DAQ) system~\cite{ATLASTDAQ:2016pov}, described in Section~\ref{sec:DAQ},
transports data from custom subdetector electronics to offline processing, according to the decisions made by the trigger.
Data are compressed prior to processing from the original event size of about 3\,MB to below 2\,MB.
An extensive software suite~\cite{ATL-SOFT-PUB-2021-001} is used in the reconstruction
and analysis of real and simulated data, in detector operations, and in the trigger and data acquisition systems of the experiment.
 
\subsection{The ATLAS run structure}
\label{sec:ATLASrun}
An ATLAS run is a period of data acquisition with a stable detector configuration. In the case of collecting data for physics analyses (\textit{physics data taking}), it usually coincides with an LHC fill, which can last many hours.
A unique number is assigned to every run at its beginning by the DAQ system.
A run is divided into Luminosity Blocks (LB), defined as intervals of approximately constant instantaneous luminosity and stable detector conditions (including the trigger system and its configuration), with a nominal length of one minute.
To define a data sample appropriate for physics studies, quality criteria are applied to select LBs where conditions are acceptable for a particular analysis. The instantaneous luminosity in a given LB is multiplied by the LB duration to obtain the integrated luminosity delivered in that LB.
From a data quality point of view, the LB represents the smallest quantity of data that can be declared good or bad for physics analysis. 
Further details on the LHC fill cycle, fill patterns and ATLAS run structure can be found in Ref.~\cite{TRIG-2019-04}.


\section{Changes to the Trigger/DAQ system for Run 3}
\label{sec:TDAQsystemChanges}

This section describes changes in the L1Calo, L1Muon and L1Topo systems as well as the HLT software
and DAQ system implemented for the \runiii, including some initial performance data where available.
More details about the hardware upgrades can be found in Ref.~\cite{atlas-det-run3}.


 
\subsection{Level-1 calorimeter trigger}
\label{sec:L1Calo}
 
\begin{figure}[htbp]
\centering
\includegraphics[width=0.49\textwidth]{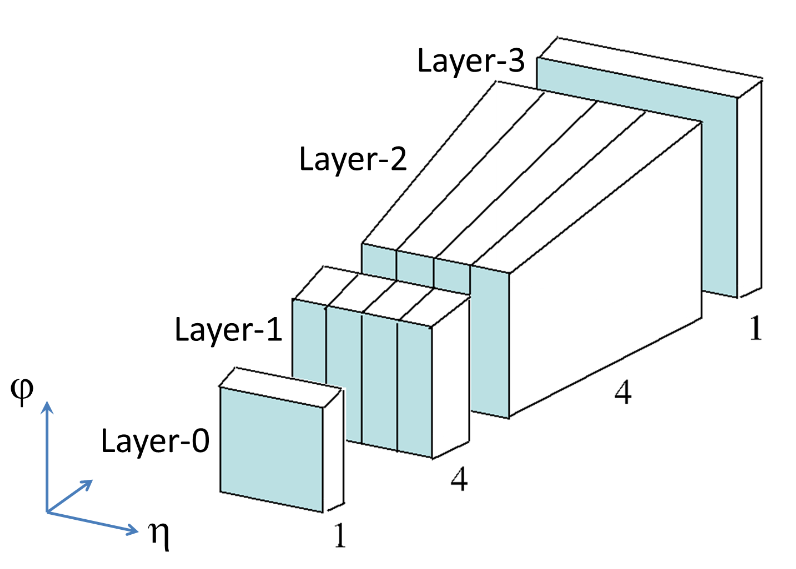}
\caption{
The trigger granularity from each $0.1\times 0.1$ trigger tower after the upgrade of the LAr
Calorimeter electronics. Ten \et\ values are provided from ``1-4-4-1'' longitudinal/transverse samples,
each forming a SuperCell.
}
\label{fig:sc}
\end{figure}
 
During LS2, the L1Calo trigger was upgraded to perform on-detector digitisation of transverse energies from the LAr calorimeters.
With this change in \runiii, L1Calo can receive finer-granularity input data from the LAr calorimeter.
While in Runs~1 and 2, these inputs consisted of \textit{trigger towers} spanning $0.1 \times 0.1$ in $\eta$ and $\phi$, the electromagnetic calorimeter information is now provided in the form of \textit{SuperCells} containing sums of four or eight calorimeter cells. Each trigger tower contains ten SuperCells as shown in Figure~\ref{fig:sc}. These trigger towers are distributed to Feature Extraction (FEX) processors.
In this system, the digitisation and calibration of the LAr calorimeter data are now performed in the LAr calorimeter electronics.
The Tile calorimeter data are still received in analog format, digitised by the pre-processors as shown in Figure~\ref{fig:TDAQsystem} and detailed below,
and transmitted by the new Tile Rear EXtension (TREX) modules to both the FEXs and the
Run-2 L1Calo system (hereafter referred to as the legacy system).
 
The upgraded L1Calo system includes new electromagnetic (eFEX) and jet (jFEX) feature extractors as well as a global feature extractor (gFEX).
The full \textit{SuperCell} granularity is available at the eFEX to reconstruct EM objects and hadronically decaying $\tau$ leptons,
as well as shower-shape variables used for their identification~\cite{atlas-det-run3}. Three programmable threshold parameters (commonly referred to as loose, medium and tight) are available for three shower-shape variables for the EM triggers and one shower-shape variable for tau triggers.
The resulting threshold pass bits for each shower shape variable are sent to the Run-3 L1Topo system and allow for the configuration of different levels of
background rejection for these triggers.
In the region with $|\eta|<2.5$ jFEX receives towers with a granularity of $0.1\times 0.1$ in $\eta - \phi$ space, which is a factor of two better in both dimensions than the inputs used for the Run-2 L1-jet-trigger system.   In the region with $|\eta| > 2.5$ the granularity is similar to \runii with some improvements in the far forward region.
The jFEX is used to reconstruct small-radius (small-$R$) jets with $R=0.4$. Its performance is expected to be similar to that of the legacy L1Calo system for the single jet triggers,
while improving the reconstruction for the nearby jets in the multi-jet triggers. In addition, jFEX brings new capability to reconstruct
large-radius (large-$R$) jets with $R=0.8$, hadronically decaying $\tau$ leptons in the range of $|\eta|\le 2.5$, and electromagnetic objects in the forward region of $2.3\le|\eta|\le 4.9$.
The latter two triggers have an optional isolation requirement.
Missing transverse momentum (\MET) is computed by summing transverse energy values in slices of constant $\phi$ and then weighted by $\mathrm{cos}(\phi)$ and $\mathrm{sin}(\phi)$ in order to determine the $x$- and $y$-components, respectively.
The gFEX has been designed with a coarser granularity than jFEX (similar to the Run-2 system) so that the data from the entire calorimeter can be processed on a single module, facilitating the identification of boosted objects and global observables.
The gFEX identifies large-$R$ jets within a ``circular'' $1.8 \times 1.8$ area ($R < 0.9$)
and provides local pile-up density and substructure information.
The gFEX computes \MET by separating the transverse energy (\ET) sums into ``hard'' and ``soft'' where the hard term consists of the \ET sum of towers satisfying $\ET > 25$~GeV and the soft term consists of the \ET sum of the remaining towers. The \MET is then computed as a linear combination of the hard and soft terms.
 
Both jFEX and gFEX have energy sum algorithms based on vector and scalar sums of $E_T$.
The pile-up subtraction algorithms introduced in jFEX and gFEX improve L1 jet and energy sum triggers efficiency, rates and purity.
More details on all these algorithms can be found in Ref.~\cite{atlas-det-run3}. 
Complementary information on the same types of objects (e.g. hadronically decaying $\tau$ leptons from eFEX and jFEX or
\met from jFEX and gFEX) can be used in the Run-3 L1Topo selection to further improve their L1 trigger performance.
 
The legacy L1Calo system~\cite{TRIG-2016-01,TDAQ-2019-01} can be operated in parallel to the upgraded L1Calo system in Run 3.
In the legacy system, analogue detector signals are digitised and calibrated in the multi-chip modules (nMCM) in the pre-processor system and sent in parallel to the Cluster Processors (CP) and the Jet/Energy-sum Processors (JEP). The CP system identifies electron, photon, and $\tau$-lepton candidates above a programmable threshold, and the JEP system identifies jet candidates and produces global sums of total and missing transverse energy.
 
During the commissioning and validation of the upgraded L1Calo system in 2022,
calorimeter-based physics triggers were provided by the legacy L1Calo system. In parallel,
the electromagnetic objects for the new Run-3 L1Calo trigger were commissioned to be used from
the start of 2023 data taking. After final tuning, they allow for a significant reduction in
the L1 trigger rate of up to 10\,kHz and an improved efficiency for electron triggers requiring
isolation. Figure~\ref{fig:l1calo} shows the 2022 performance of the legacy trigger in comparison to
the new one for the barrel region.\footnote{For \runiii,
the L1 threshold nomenclature is changed so that the threshold value corresponds to a 50\% efficiency
point, not the efficiency plateau as was the case in \runii.
Thus L1 
Run-3 triggers with thresholds of 14\,\gev\ for a single muon and 8\,\gev\ for a pair of muons
have the same performance, respectively, as a 20\,\gev\ single muon and a 10\,\gev\ di-muon L1 triggers 
in \runii.}
At the time of submission of this manuscript,
the final optimisation of the Run-3 L1Calo trigger system was still ongoing.
 
\begin{figure}[htbp]
\centering
\includegraphics[width=0.49\textwidth]{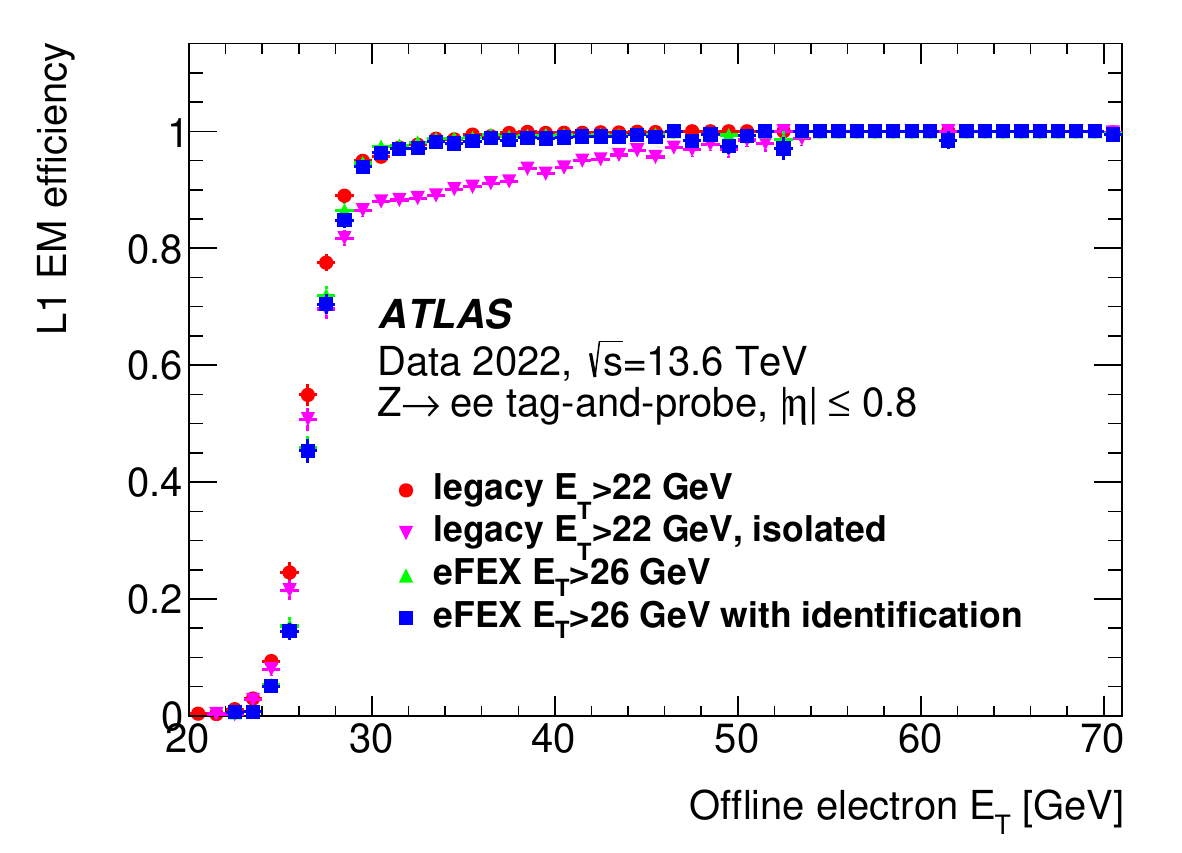}
\includegraphics[width=0.49\textwidth]{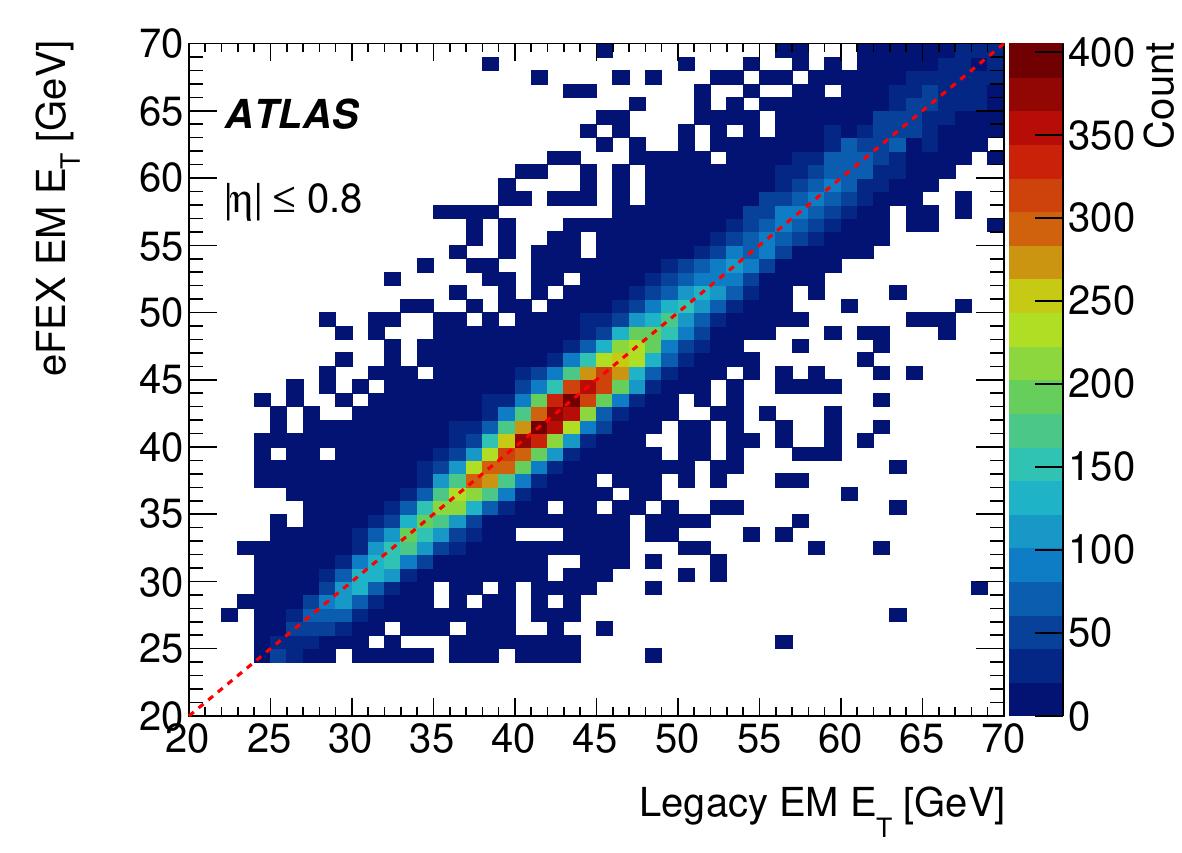}
\caption{(left) Efficiency of the L1 single-object electromagnetic (EM) trigger and
(right) the correlation of the L1Calo transverse energies measured by the legacy
Cluster Processor system and the Run-3 eFEX.
Both measurements are done in the inner EM barrel within $|\eta|<0.8$.
Isolated legacy trigger~\cite{TRIG-2018-05} and eFEX trigger with identification~\cite{atlas-det-run3} 
have the same rate. Only statistical uncertainties are shown.
}
\label{fig:l1calo}
\end{figure}


\subsection{Level-1 muon trigger}
\label{sec:L1Muon}
 
The L1Muon trigger uses hits from the RPCs (in the barrel) and the TGCs (in the endcaps) to determine the deviation of the hit pattern from that of a muon with infinite momentum~\cite{TRIG-2018-01}.
For Run 3 the L1Muon transverse momentum (\pT) thresholds have been redefined to improve the performance of both the high-\pT single muon and the low-\pT multi-muon triggers.
 
To reduce the rate of the low-momentum charged particles 
in the endcap regions, 
the L1Muon trigger in \runii applied coincidence requirements between the outer TGC station and either the inner TGC stations or the tile calorimeter.
In \runiii, the replacement of the Small Wheels by the NSWs allows for a further rate reduction.
Additional RPC modules have also been deployed in the inner barrel station for \runiii.
These new Barrel Inner Small (BIS78) chambers are located in the transition $1.0<\eta<1.3$
region between the barrel and endcap on one side of the detector only and will be utilised to reduce the trigger rate of the endcap region which is not covered by inner endcap TGC
stations. These new chambers are considered to be a pilot project for
the inner barrel upgrade in preparation for Run 4 (2029--2032) when the second side of BIS78 will be installed.
 
\begin{figure}[htbp]
\centering
\includegraphics[width=0.49\textwidth]{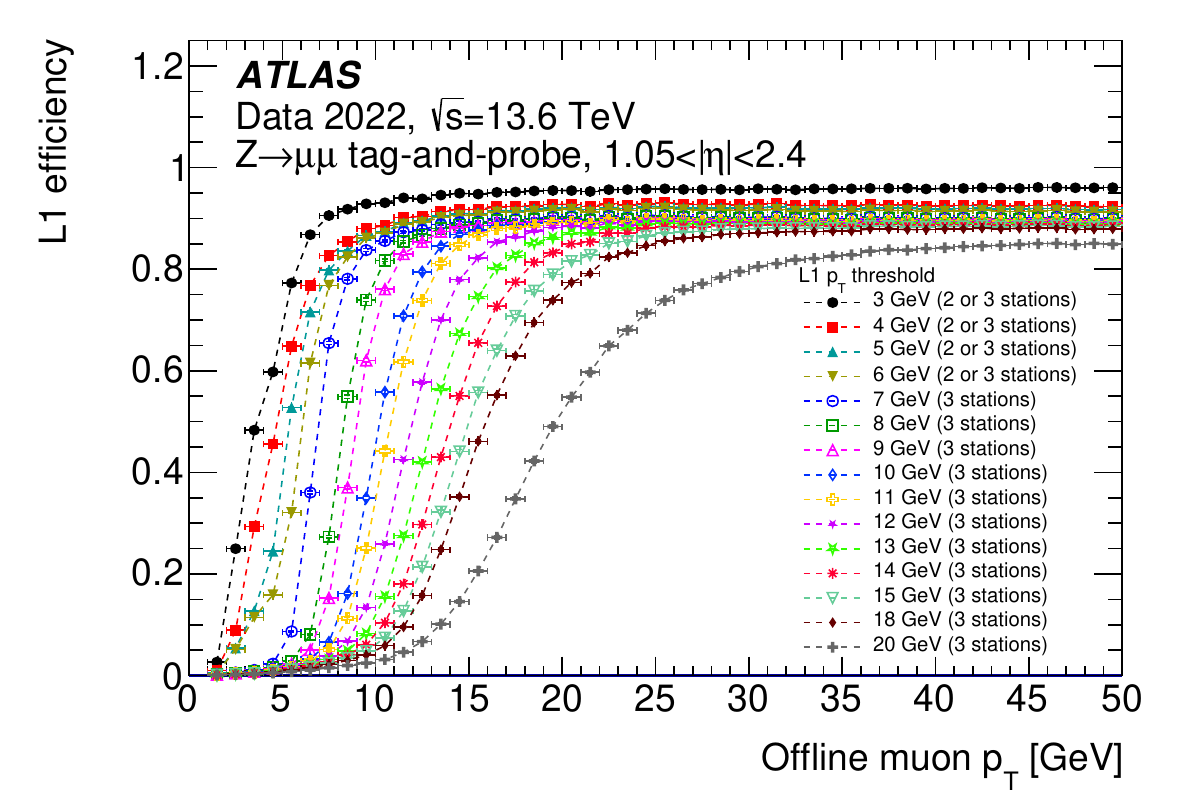}
\includegraphics[width=0.49\textwidth]{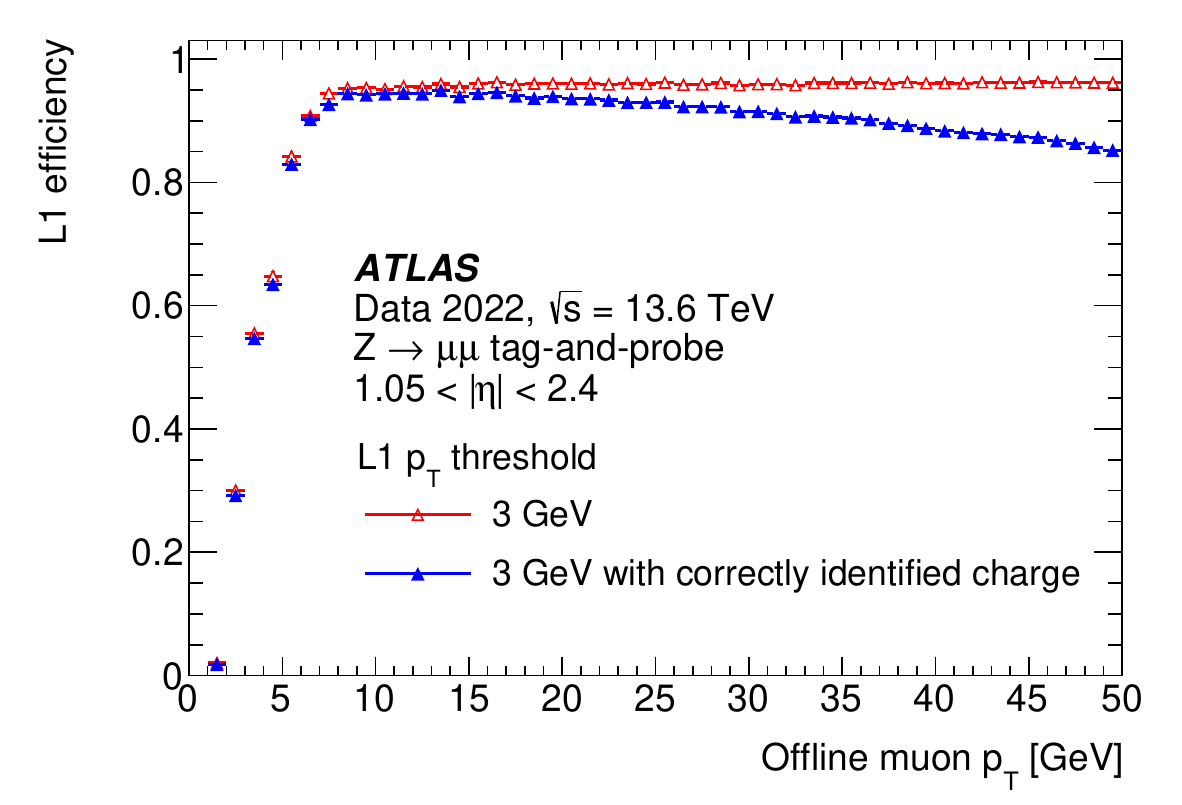}
\caption{(left) Efficiency of L1 muon triggers in the endcap region for various \pT thresholds.
(right) The efficiency of the charge identification for L1 muons with \pt$>3\,$\GeV.
Only statistical uncertainties are shown.
}
\label{fig:l1muon_tgc}
\end{figure}
 
The upgrade of the endcap trigger processor board, called Endcap Sector Logic, increases the number of TGC thresholds from 6 to 15 and supports the attaching of up to
four flags to muon candidates.
Figure~\ref{fig:l1muon_tgc}~(left) shows the efficiency of single-muon triggers for the endcap region.
The finer granularity of low-\pT TGC thresholds results in a better resolution of the invariant mass of two muons in the Run-3 L1Topo system,
crucial for $B$-physics and light states triggers discussed in Section~\ref{sec:bls}.
The rate of these low-\pT multi-muon triggers can be further reduced by using information on
the charge of the muon candidate, which is provided by one of the new flags.
As can be seen from Figure~\ref{fig:l1muon_tgc}~(right),
the accuracy of muon charge identification is close to 100\% at low \pT,
although it gets worse with increasing \pT, due to the reduction of curvature of the muon path in the magnetic field.
The other three flags indicate the following conditions: the muon candidate satisfies the coincidence in all three outer stations of the TGC (the Full station flag);
the Inner Coincidence flag described in the previous paragraph (in 2022 only the inner endcap TGC stations in the region of \(1.05 < |\eta| < 1.3\) were used to satisfy it, outside this region the flag was always set);
the muon candidate does not pass through regions of weak magnetic field, where the momentum resolution of such candidates is poor (the Hot-RoIs flag).
All three flags were required by default for the 2022 single L1 muon triggers.
The di-muon triggers use only the Full station flag.
 
\begin{figure}[htbp]
\centering
\includegraphics[width=0.47\textwidth]{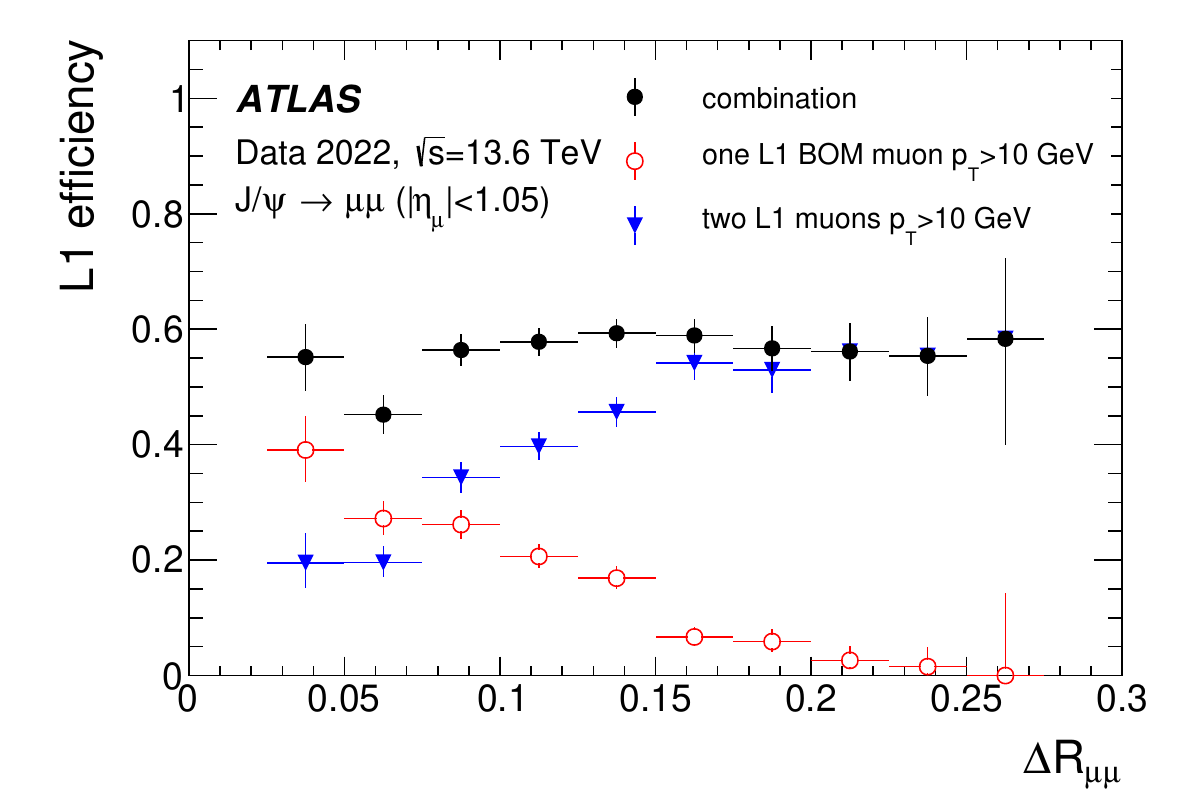}
\includegraphics[width=0.47\textwidth]{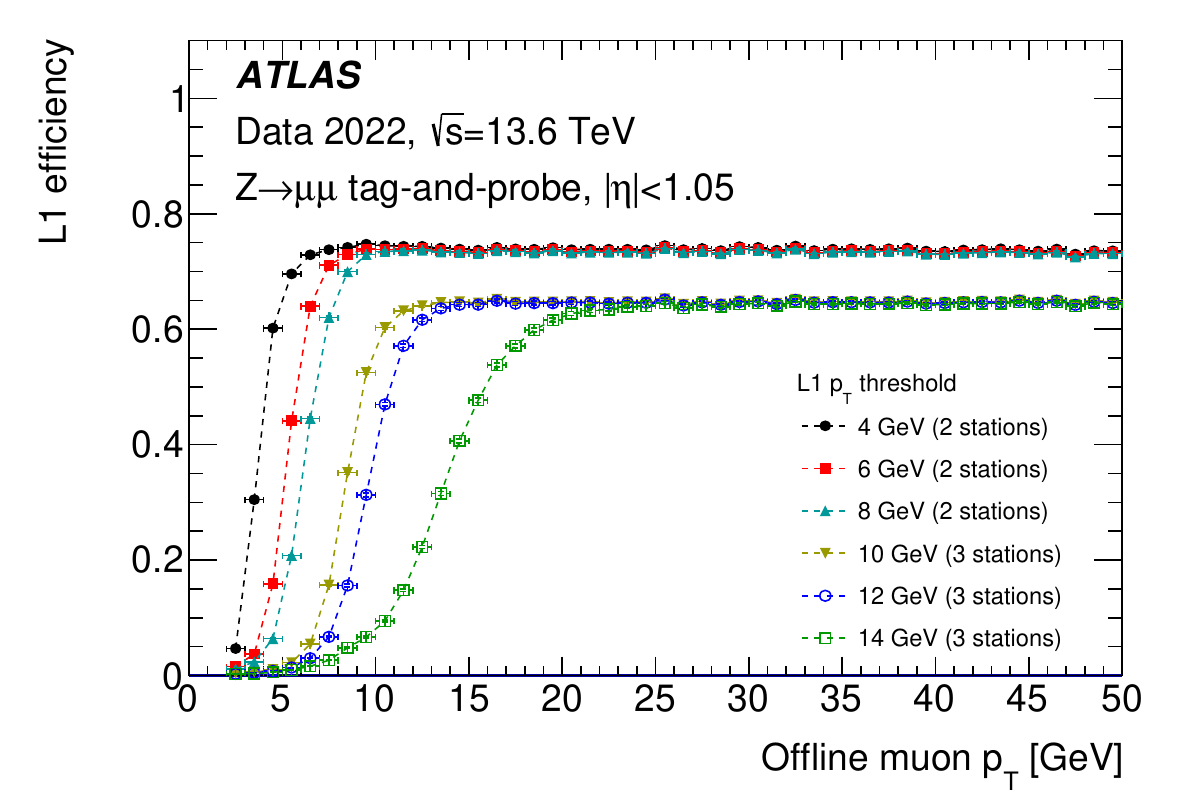}
\caption{(left) The efficiency of trigger reconstruction with respect to offline muon selection
as a function of distance ($\Delta R_{\mu\mu}$) between two close-by muons from $J/\psi$ decay in the barrel region
for a L1 trigger with two 
muon candidates of \pt$>10\,$\GeV\
and a L1 trigger with a single Barrel-Only (\trig{BO}) muon candidate of \pt$>10\,$\GeV\ and close-by flag (\trig{M}).
(right) Efficiency of L1 muon triggers for the barrel region for various \pT thresholds.
Only statistical uncertainties are shown.
}
\label{fig:l1muon_rpc}
\end{figure}
 
Muon triggers in the barrel region 
are provided by three concentric layers of RPC doublets, with the trigger decision relying on coincidence logic.
For the low-\pT thresholds, the coincidence of three out of four layers in the two inner doublets of the RPC, which was required in \runii, was
relaxed during 2022 by requiring at least one hit in each of the two doublets. This increased the trigger efficiency with a minimal impact on the rates.
The high-\pT trigger thresholds require the low-\pT trigger logic to be satisfied, as well as an additional hit in one of two layers in the outer barrel station.
The number of RPC thresholds remains the same as in \runii, with a total of six thresholds: three low-\pT (2-station) and three high-\pT (3-station) thresholds.
The firmware of the RPC Pad and of the Barrel Sector Logic boards is upgraded
to flag the possibility of two muons in a single tower\footnote{A tower is a region of approximately $0.2\times 0.2$ in $\eta - \phi$ space
comprising four possible RoI positions which are read out by a Pad Logic board.}, which would otherwise be identified as only one candidate.
By propagating this information to the HLT, the identification of close-by muons (e.g. from boosted $J/\psi$ decays) is enhanced, as shown in Figure~\ref{fig:l1muon_rpc}~(left).
Figure~\ref{fig:l1muon_rpc}~(right) shows the L1Muon efficiency of various single muon triggers in the barrel region with respect to offline muon selection.
Their maximum efficiencies are about 5\% lower than at the end of \runii,
due to inefficiencies in the RPC detectors mainly caused by leaks in the gas distribution system.
The lower L1 efficiency of the 3-station triggers is due to the additional hit requirements at the outer station and the detector coverage.
 
The MUCTPI is upgraded to provide full-granularity muon RoI information to the L1Topo system
and to be able to interface with the new Endcap Sector Logic.
A special SL2MuCTPI board~\cite{atlas-det-run3} is introduced to interface the Barrel Sector Logic with the new MUCTPI.
The MUCTPI receives muon candidate information from the barrel and endcap Sector Logic and calculates the overall muon candidate multiplicity
for each \pT threshold, taking into account the possible overlap between trigger sectors in order to avoid double counting of muon candidates.
Additional logic in the MUCTPI allows for the exclusion of candidates from the RPC feet areas, or to flag Barrel-Only 
and Endcap-Only muon candidates.
The MUCTPI sends to the CTP information for up to 32 combinations of the \pT thresholds and flags, which is an increase from the six available in \runii.
Combinations with different thresholds in the barrel and endcap regions can be also implemented.
 
The commissioning of the L1Muon trigger proceeded systematically in several steps. It started with the commissioning of
the upgraded endcap trigger system using test pulses injected into the on-detector electronics. This allowed for
the detection of hardware problems and incorrect fibre connections, as well as the adjustment of the trigger timing and the validation of the trigger logic.
Subsequently, the use of cosmic rays provided a full integration test with the barrel and endcap trigger systems with the new MUCTPI and the CTP.
The commissioning of the upgraded L1Muon system (without the NSW) was finalised in 2022, while the commissioning of the NSW and the BIS78 RPC
chambers is still ongoing at the time of writing.


\subsection{Level-1 topological trigger}
\label{sec:L1Topo}
 
The addition of the L1Topo trigger in 2016 allowed for both the reduction of energy thresholds without an increase in trigger rates as well as
a higher complexity of the algorithms available at L1 through topological selections. This resulted in a significant improvement of the background rejection and enhanced
acceptance of physics signal events, despite the increase of luminosity during \runii.
A technical description of the Run-2 L1Topo system can be found in Ref.~\cite{TRIG-2019-02}.
 
The principle of operation of the L1Topo system remains the same between \runii and \runiii.
The L1Topo system receives Trigger OBjects (TOBs) containing kinematic (e.g. \et, position) and further qualifying information (e.g. flags)
from the L1Calo and L1Muon systems and applies topological selections.
The Run-3 upgrades of both the L1Calo and the L1Muon systems have an impact on the L1Topo,
since they provide the input data directly for the topological selections.
The outputs from the L1Calo and the MUCTPI are remapped in the Topo-Fiber Optic eXchange (TopoFOX) module before arriving at the L1Topo,
contrary to the Run-2 system where the Common Merger Module (CMX) and MUCTPI2Topo~\cite{ATLAS-TDR-23} were needed for e.g. TOB sorting.
In \runiii, the L1Topo hardware is upgraded with three new modules (TOPO1, TOPO2 and TOPO3) designed to deal with the new input format.
Each module has two Field Programmable Gate Array (FPGA) processors on which to run the algorithms.
 
\begin{figure}[htbp]
\centering
\includegraphics[width=0.49\textwidth]{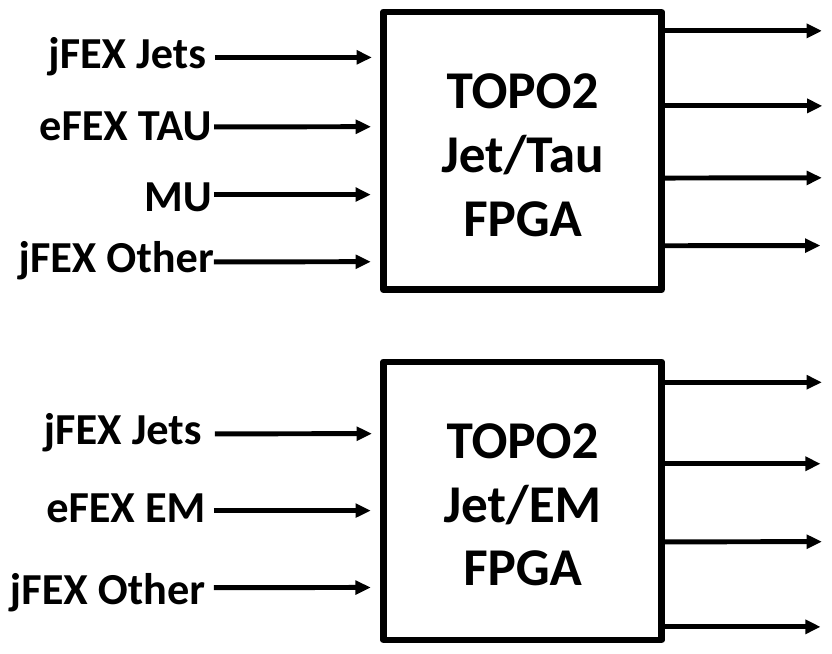}
\includegraphics[width=0.49\textwidth]{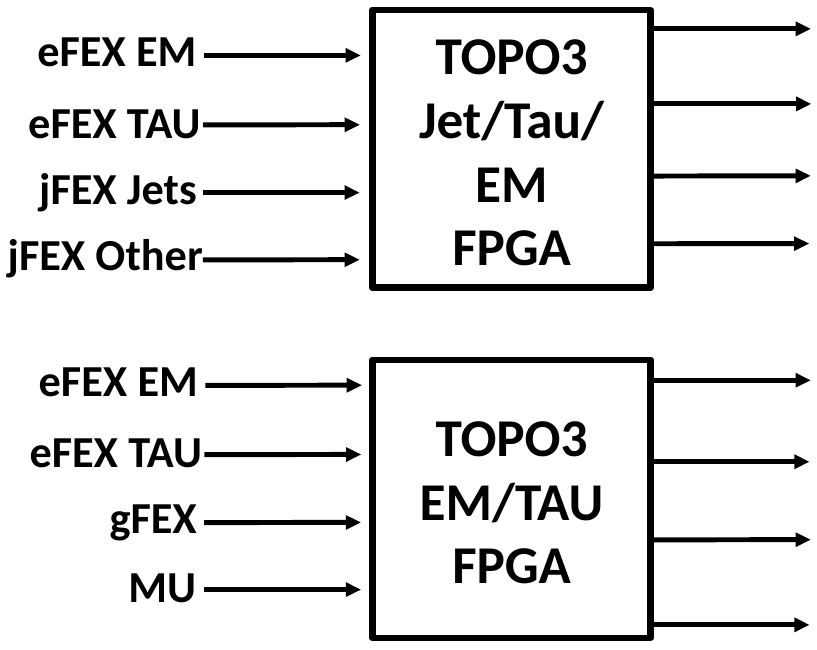}
\caption{Distribution of L1 TOB inputs for TOPO2 and TOPO3 modules.}
\label{fig:l1topo}
\end{figure}
 
In \runiii, the selections based on the multiplicities of the L1Calo objects passing pre-defined energy thresholds are performed in the TOPO1 module.
These algorithms count the number of objects passing a given threshold, e.g. \ET,
or located within a given region in $\eta$.
The output multiplicity bits are transmitted to the CTP.
The number of available thresholds and multiplicity bits per L1Calo object type is configurable in L1Topo
firmware and could be modified in the course of \runiii according to the trigger menu
requirements discussed in Section~\ref{sec:menu}.
 
The TOPO2 and TOPO3 modules are used for the topological selections as in the Run-2 L1Topo system~\cite{TRIG-2019-02}.
Available inputs for these two L1Topo models, shown in Figure~\ref{fig:l1topo},
impose constraints on the possible combinations of L1Calo and L1Muon TOBs in topological algorithms.
The L1Topo algorithms are distributed across the FPGAs as evenly as possible to achieve optimal resource usage.
The configuration of algorithms with their parameters is stored in the trigger menu.
The L1Topo algorithm decisions are transmitted to the CTP
and are limited to 32 bits per FPGA.
 
The L1Topo legacy system consists of two Run-2 L1Topo modules with their original Run-2 firmware configuration,
but with inputs only from the legacy L1Calo system and not from L1Muon.
The decisions of only 22 out of 113 topological algorithms that were implemented in \runii~\cite{TRIG-2019-02} are sent to the CTP in \runiii.
The L1Topo legacy system was used to collect physics events in 2022, while the upgraded system underwent commissioning.
When the three new L1Topo modules are fully validated, the two L1Topo legacy modules will be disconnected.


\subsection{HLT software}
\label{sec:AthenaMT}
 
The ATLAS software framework Athena~\cite{Athena} was used in Runs~1 and 2 at all stages of the event data processing path, from detector simulation to event reconstruction and physics analysis (referred to as ``offline''), as well as for the HLT selections (referred to as ``online''). Athena is based on the inter-experiment framework Gaudi~\cite{Gaudi}.
In \runii, memory limitations of the simulation and reconstruction workflows led to the development of a multi-process event processing model
(AthenaMP), which was adopted as an intermediate solution, also at the HLT. In this approach, the main process is forked after initialisation into a  number of worker processes equal to the number of events which should be processed in parallel. Each worker processes events independently using a single thread, sharing read-only memory with other workers via the copy-on-write mechanism.
Thanks to this mechanism, a reduction of the overall memory requirements was possible, with respect to the instantiation of totally independent processes.
 
A typical online reconstruction sequence makes use of dedicated fast trigger algorithms to provide early background rejection, followed by more precise and CPU-intensive algorithms that are similar or identical to those used for offline reconstruction to make the final selection.
Reconstruction algorithms process detector data to extract features and hypothesis algorithms test the selection hypothesis for all active trigger chains,
where a chain consists of a L1 trigger item and a series of
HLT algorithms organised into distinct steps that reconstruct physics objects and apply kinematic selections to them.
 
For \runiii, the ATLAS software framework was adapted to support multi-threaded execution (AthenaMT), based on the concurrent-processing version of Gaudi, named \textit{GaudiHive}~\cite{GaudiHive} which is itself based on Intel's Threading Building Blocks (TBB) library~\cite{TBB}. HLT requirements were taken into account during this transition~\cite{ATL-SOFT-PUB-2016-001}. AthenaMT allows greater memory sharing across compute cores than is possible with AthenaMP and, consequently, greater flexibility and efficiency when running on hardware with limited memory per core. It is also a prerequisite for the eventual use of compute accelerators such as GPUs in online and offline data processing. All data processing steps (except event generation) will use AthenaMT in \runiv.
 
Three types of parallelism are included: inter-event parallelism (multiple events are processed in parallel),
intra-event parallelism (multiple algorithms can run in parallel for a single event) and in-algorithm parallelism (algorithms can
internally utilise multi-threading and vectorisation).
The execution order of the job's algorithms is determined by the input data required, and output data produced, by each algorithm.
These data dependencies must be exposed by the developer to the AthenaMT scheduler via ``data handles'', which are used by the scheduler to appropriately order the algorithms for execution.
The execution depends on the configured number of threads (made available to execute algorithms) and event slots (made available to concurrently process events).
Asynchronous or time-varying data (conditions data) whose lifetime can be longer than an event are handled by the framework with special conditions algorithms, which are scheduled appropriately by the framework so that the required conditions data are available when needed. 
 
To take advantage of the major changes in AthenaMT, the ATLAS HLT framework was to a large extent redesigned and rewritten. While in the Run-2 HLT framework all trigger algorithms were implemented using an HLT-specific interface (with offline algorithms requiring a wrapper), the Run-3 framework requires no HLT-specific interfaces and takes full advantage of the AthenaMT scheduler and other AthenaMT components to control event execution in the HLT.
These modifications eased online integration of the offline reconstruction developments due to the seamless integration between the offline and HLT frameworks.
 
The development of the AthenaMT framework was a common project of the offline and HLT groups from the start~\cite{ATL-SOFT-PUB-2016-001}, incorporating the key principles of the HLT event selection in ATLAS that remain unchanged from Runs~1 and 2:
\begin{enumerate}
\item Where possible, trigger selections make use of partial-event data which are processed step-wise (from low to high granularities) using regional reconstruction inside RoIs corresponding to a part (or the whole) detector.
\item Event processing is terminated as soon as the event is known to have failed all active trigger selections.
\end{enumerate}
 
These two key principles require some extensions to AthenaMT, namely the concept of \textit{EventViews} to accommodate the regional reconstruction as well as the introduction of the control flow to allow for early rejection.
EventViews allow for an algorithm to be executed multiple times in a single event on either partial or full event data, making use of already existing framework functionality to execute the reconstruction steps.
Their execution is prepared by Input Maker algorithms that are scheduled to run before the reconstruction algorithms.
Subsequent combinatorial hypothesis algorithms allow for topological selections to be defined between sets of candidate physics objects.
HLT-specific filter algorithms reside between the steps of the step-wise HLT processing. They provide the possibility of early rejection by gating which steps are allowed to be executed on any given event.
 
The scheduling of the execution of algorithms in the HLT is augmented by a control-flow logic determined when the HLT configuration is prepared, applying an additional layer of steering logic to the event processing. This layer is defined on nested lists of algorithms and is in addition to the basic execution steering based on each algorithm's input and output data. Figure~\ref{fig:menu_control_flow} shows an example of a data flow graph. The control flow in conjunction with the data dependencies ensure that for a given step, its components are executed in a fixed order: Filter $\xrightarrow{}$ Input Maker $\xrightarrow{}$ reconstruction $\xrightarrow{}$ hypothesis algorithms. The additional control logic is required because each processing step can have multiple Filters: for example, the step 1 electron and muon paths
in Figure~\ref{fig:menu_control_flow} each have two feeding Filters (electron, electron+muon) and (muon, electron+muon), respectively. These paths may activate in a given event due to one or both of these Filters passing. The control flow creates a graph representation of all possible execution paths at configuration time. The graph is built from a list of all physics selections configured to be executed and does not change during run time. However, each trigger chain corresponds to one path through the graph, and these chains can be individually enabled or disabled during run time or executed on only a fraction of events.
 
\begin{figure}
\centering
\includegraphics[width=0.9\textwidth]{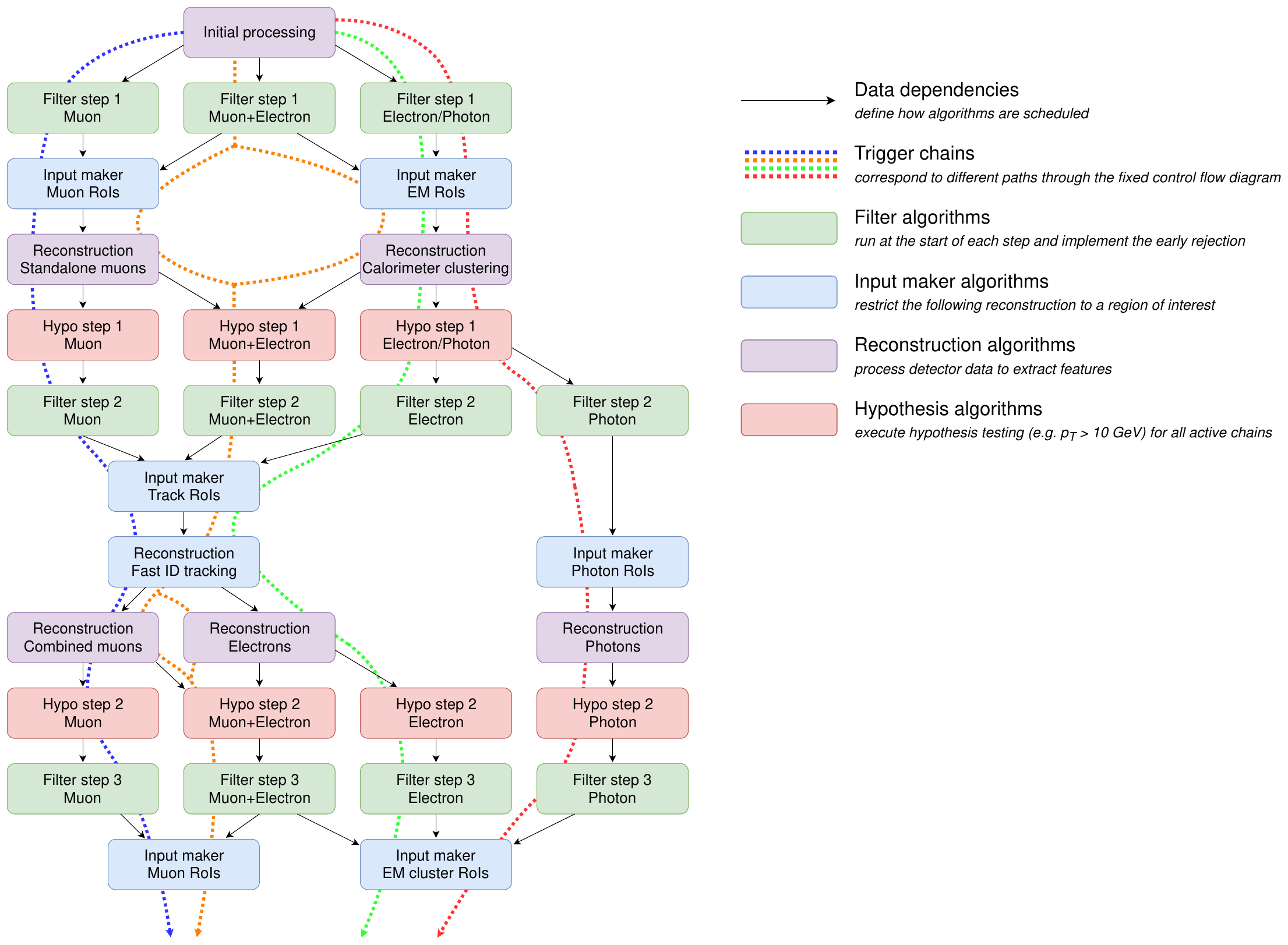}
\caption{Example of control flow for menu processing. The control flow graph is created during initialisation and the steps are executed based on the available data. If a filter passes, processing continues through the next steps until all filters fail where the processing of the following steps is stopped. If the last step is reached with a chain passing all of its steps the event will be accepted.}
\label{fig:menu_control_flow}
\end{figure}
 
The HLT software configuration is stored in the trigger-specific Oracle database in JSON~\cite{JSON}
format with blobs of entire JSON job configurations being uploaded to the database. This allows for a considerable simplification of the database schema compared to Run 2~\cite{TRIG-2019-04}
allowing further improvements to associated tools and metadata. Several tools allow developers easy inspection, modification or downloading of JSON files of database configuration data for debugging of problems and reproducing setups. The trigger configuration service in AthenaMT can be either run from the trigger database, from JSON files or from standard ATLAS python-based configurations.
 
In 2022 the HLT software was used in a mixed multi-processing / multi-threaded configuration with two event slots per worker during commissioning and low-luminosity data-taking periods. For the \ac{MC} production the configuration was pure MT with typically eight event slots per grid job~\cite{grid}.
The HLT software performance in 2022 is further detailed in Section~\ref{sec:softwarePerf}.
 
\subsection{DAQ system}
\label{sec:DAQ}
 
The DAQ system, as shown in Figure~\ref{fig:TDAQsystem}, supports the operation of the two trigger levels, beginning with detector-specific on and off-detector electronics which perform a variety of data processing and monitoring features before passing events either to the L1 trigger or to the downstream systems.
In Runs~1 and 2 the off-detector stage was performed in detector-specific custom hardware modules called Read-Out Drivers (RODs), typically sitting in VME crates. On receipt of a L1 accept, these RODs read out their data to the first common stage of the DAQ system, the Read-Out System (ROS). The ROS consists of a set of commodity server machines hosting custom-built I/O cards. These receive data from detector front-end electronics and store them in internal memory buffers, called Read-Out Buffers (ROBs). The Data Collection Manager (DCM) orchestrates the data flow from the ROS through the HLT Multi-Process Processing Units (HLTMPPUs). During processing the HLT requests data from the ROS as needed before either accepting an event for permanent storage or requesting its deletion from the ROS buffers.
In 2022 a dedicated HLT computing farm of around 56k (112k with hyperthreading) cores was running up to 90k HLTMPPU selection applications. Every processing node hosts one DCM and one or a few HLTMPPU selection applications, depending on the chosen multi-process and multi-threading configuration. Individual events are assigned to processing nodes in the computing farm by a HLT SuperVisor (HLTSV) application according to available free event nodes on each machine.
 
New in \runiii is the introduction of the Front-End Link eXchange (FELIX) read-out system and software, running on commodity servers (SW ROD). These new DAQ systems are integrated into the read-out path for those detector systems with new or upgraded front-end electronics (LAr digital trigger, L1Calo, L1Topo, NSW). The interface to the HLT is identical between ROS and SW ROD, with event data routed on-demand to HLT processing nodes in both cases.
While the ATLAS data flow architecture remains unchanged for \runiii, the change to a multi-threaded event processing environment required changes in the interface between the HLTMPPU and the DCM, and in the event assignment from the HLTSV.
Once HLT processing has been completed, accepted events are sent to a dedicated cluster of servers (known for historical reasons as Sub-Farm Outputs (SFOs)) for packing, compression, and transfer to offline storage. To accommodate the increased average rate of physics data to be written out, the Run-3 physics bandwidth to permanent storage has been doubled to allow for up to 8 GB/s.


\subsection{Validation strategy}
\label{sec:validation}
The ATLAS software is being continuously developed to adapt to changing requirements and conditions, fix defects, improve resource usage and in the particular case of the software used for recording data, to further optimise the performance of the selection algorithms.
As the software is evolving, its validation both in terms of the signature reconstruction performance (see Sections~\ref{sec:HLTreco} and~\ref{sec:sigPerf}) and the trigger menu functionality and execution speed (see Section~\ref{sec:menu}) is of great importance to ensure the reliability and predictability of its performance.
 
Validation jobs are run nightly using the ATLAS Release Tester (ART) system~\cite{Donszelmann:2020ldy} to monitor the status of the trigger software. To facilitate the monitoring and tracking of the ART test results, an automated system has been developed.
This system analyzes any log files produced by the tests, looking for errors and warnings. It is able to catch any newly introduced changes on a daily basis, including changes to the effectiveness of the software system with respect to time constraints and allocation of resources. Any newly introduced failures are flagged and followed up with the corresponding developer.
By tracking the history of the number of events accepted by a given trigger, it is possible to also monitor the evolution of the trigger algorithm performance and detect unexpected degradation or changes immediately.
 
In addition to the daily validation, the trigger software is periodically validated using high statistics simulated samples (e.g. $t\bar{t}$ as well as other more specific physics signal processes) and collision data. For both types of validation, sets of histograms representing the performance of the HLT and signature reconstruction are produced and checked against a reference.
The main purpose of validating with simulated samples is to check the reconstruction performance, including new developments and to validate new trigger chains that are foreseen to be used online in terms of selection efficiency.
Before a new software version is deployed on the HLT farm, it is used to process an enhanced-bias data set~\cite{ATL-DAQ-PUB-2016-002}. The enhanced-bias data set consists of roughly one million events, and
is collected with filters that ensure an even representation of events with different topologies and rarities.
This data reprocessing is, due to its large statistics, executed on the LHC computing grid using the ATLAS production system~\cite{Borodin:2015kfq}. Not only is the performance checked in terms of reconstruction efficiency and rates of trigger chains, but also the performance in terms of resource usage (such as run time, memory consumption, etc.) is checked to ensure smooth data taking. This is important in order to achieve an overall high data quality, and efficient data taking. Further information about the assessment of the data quality can be found in Ref.~\cite{TRIG-2019-04}.


\section{Trigger menu}
\label{sec:menu}

Events are selected to be recorded if they satisfy the conditions of one or more trigger chains.
The list of trigger chains used
for data taking is known as a trigger menu, which also includes prescales for each trigger chain\footnote{
To control the rate of accepted events and to manage CPU consumption at the HLT, a prescale value, or simply prescale, can be applied. For a
prescale value of $n$, a trigger chain has a probability of 1/$n$ to be activated in the event.
By default they are randomly generated for every individual trigger. However, a mechanism of coherent prescale sets
exists for defining groups of triggers whose prescales are correlated. Individual prescale factors can be
given to each chain at L1 or at the HLT, and can be any value greater than or equal to one.
The value -1 is used to disable triggers.}.
The trigger menu consists of \emph{Physics} triggers, detailed in Section~\ref{sec:sigPerf},  and \emph{Auxiliary}
triggers, detailed in Section~\ref{sec:special}. Triggers which use information from more than one
object type are called \emph{Combined}.
 
Physics triggers are used for physics analyses and can be subdivided into the following categories:
\begin{itemize}
\item \emph{Primary} triggers, which cover all signatures relevant to the ATLAS physics programme and are typically unprescaled.
\item \emph{Support} triggers, which are used for efficiency and performance measurements, background estimates, or for monitoring. These are typically operated at a small rate (of the order of 0.5-1.0 Hz each) using prescale factors. About 15\% of the HLT bandwidth in \runii was dedicated to support triggers.
\item \emph{Alternative} triggers, using alternative (sometimes experimental or new) reconstruction algorithms complementary to the primary or support selections, and often heavily overlapping with the primary triggers. They are often used, for example, as part of the commissioning process for future primary triggers.
\item \emph{Backup} triggers, with tighter selections and lower processing or output rate. They can replace the relevant primary triggers if their CPU usage or output rate becomes too high.
These triggers require almost no additional computing resources or output rate as they select a subset of the primary triggered events.
\end{itemize}
 
Auxiliary triggers can in turn be subdivided into the following categories:
\begin{itemize}
\item \emph{Calibration} triggers, used for detector calibrations.
\item \emph{Cosmic ray} triggers.
\item \emph{Beam-induced background} triggers, which are recorded in LHC bunches with single beam or no beam present.
\item \emph{Noise} triggers, which are collected by a \emph{random} trigger at L1\footnote{
The random trigger item at L1 corresponds to the read-out from the detector of events chosen at random. They are always prescaled.
L1 random triggers on filled bunches can be used to seed specific triggers to overcome potential
inefficiencies at L1, while L1 random triggers on the unfilled LHC bunches are typically used for noise and background studies.}.
\item \emph{Other} dedicated triggers and algorithms. 
\end{itemize}
 
To facilitate further processing and analysis, accepted events are recorded into different data sets, called \emph{streams}, which are designed
to have minimal overlap. The trigger menu defines the streams to which an event is written, depending on the trigger chains
that accepted the event. The five different types of data streams considered in the recording rate budget available at the
HLT during nominal \pp\ data taking are:
\begin{itemize}
\item \emph{Physics} streams: collision events of interest for physics studies. The events
contain full detector information and dominate in terms of processing, bandwidth and storage
requirements. There are three physics streams for the \pp\ data taking:
the Main stream, the $B$-physics and light states (BLS) stream and the Hadronic stream.
Events in the Main stream are promptly reconstructed after completion of the first-pass calibration and data quality assessment,
as described in Ref.~\cite{DAPR-2018-01}, while events
in the other two, \textit{Delayed}, streams are reconstructed when resources allow it.
\item \emph{Express} stream: a very small subset of the physics stream events
for prompt monitoring, detector calibration, and first-pass data quality checks.
It is fully reconstructed offline within a day of having been recorded.
\item \emph{Background} streams: background events of interest for physics and detector performance studies.
\item \emph{Debug} streams: events for which no trigger decision could be made are written to this stream.
Typical reasons are crashes, timeouts in the HLT processing, and HLT data payloads exceeding set thresholds.
These events need to be analysed and recovered separately to identify and fix possible problems
in the TDAQ system.
\item \emph{Calibration} streams: events containing only partial detector information for calibration of specific subdetectors.
\item \emph{Trigger-Level Analysis (TLA)} streams: events sent to this stream contain only specific physics objects reconstructed by the HLT,
and optionally only partial detector information. These data are used directly in the corresponding physics analysis (e.g. Ref.~\cite{EXOT-2016-20}).
The average TLA event size is 4.5 kB in 2022, which is approximately 0.3\% of the size of a full ATLAS event.
\item \emph{Monitoring} streams: events to be sent to dedicated monitoring nodes for online analysis for,
e.g., detector monitoring, but not recorded.
\end{itemize}
 
For special data-taking configurations, it is possible to introduce additional streams, such as, for example,
the \emph{Enhanced bias} physics data stream, which is used to record events for trigger rate predictions~\cite{TRIG-2019-04,ATL-DAQ-PUB-2016-002}.
With the exception of the debug streams, the streaming model is inclusive, which means that an event can be written
to multiple streams.
 
\begin{figure}[htbp]
\centering
\includegraphics[width=0.47\textwidth]{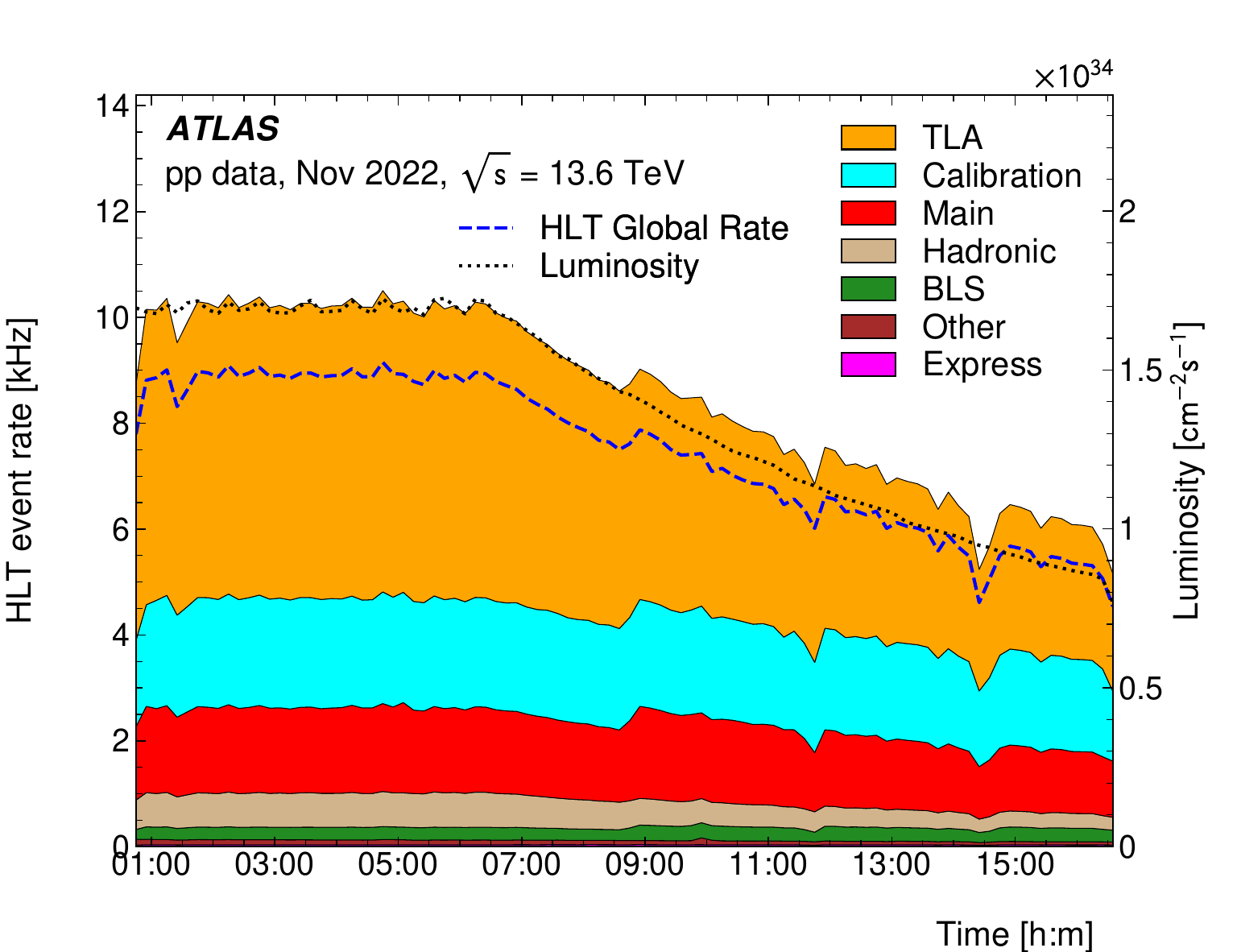}
\qquad
\includegraphics[width=0.47\textwidth]{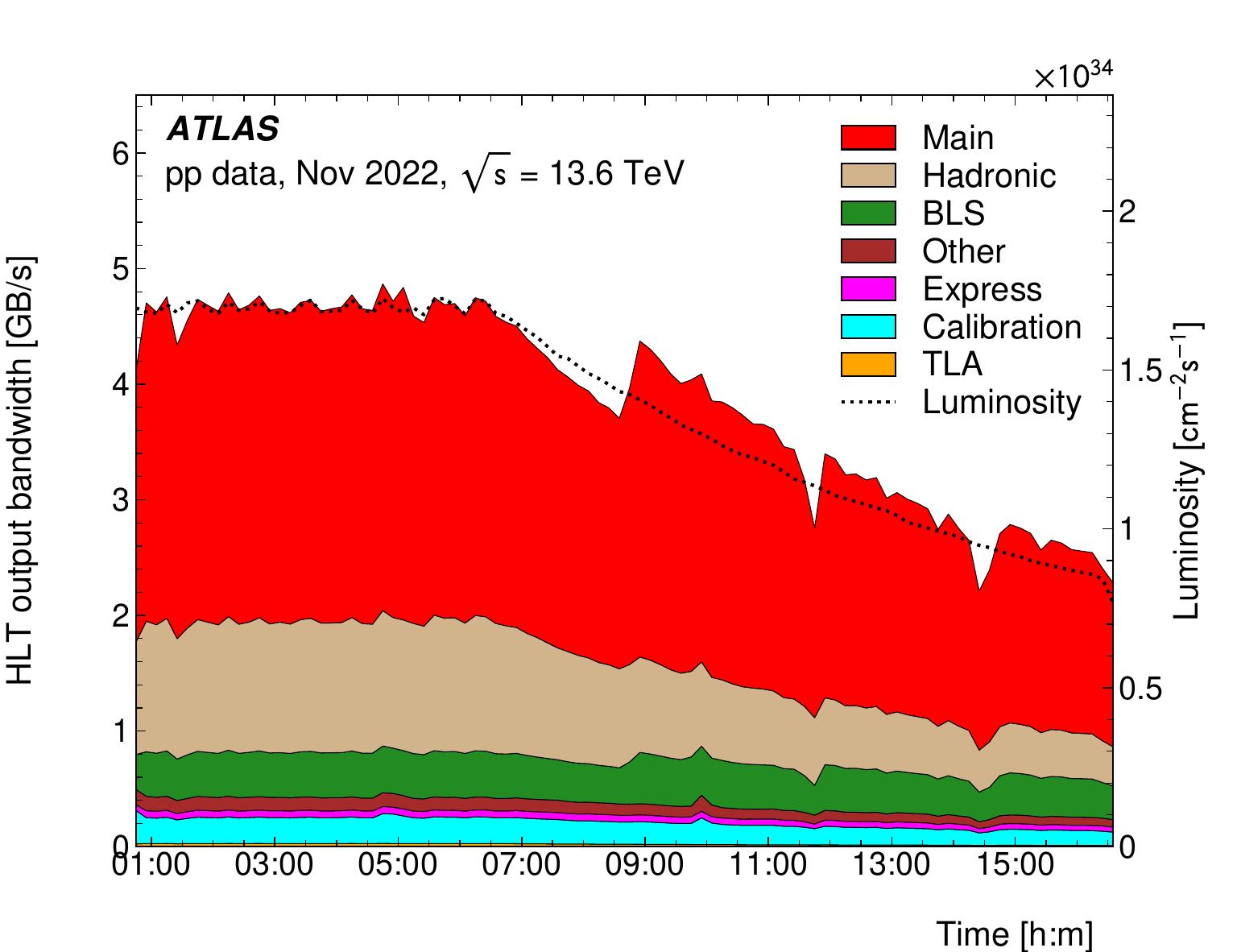}
\caption{An example of (left) the rate and (right) bandwidth  output by the HLT streams in a typical 2022 \pp\ run. The total HLT event rate is lower than the sum of the stream rates, because the same events may be written to multiple streams.
The peak bandwidth of the TLA stream is approximately 25\,MB/s, and consequently barely visible in the right hand figure.
The event size per stream can vary up to a factor of 300.
}
\label{fig:stream_rate_BW}
\end{figure}
 
The trigger menu composition and trigger thresholds are optimised for several luminosity ranges in order
to maximise the physics output of the experiment and to fit within the rate and bandwidth constraints of the ATLAS detector,
TDAQ system and offline computing.
The effect of such optimisation on the HLT stream output for a typical run is shown in Figure~\ref{fig:stream_rate_BW}.
For \runiii, the most relevant constraints are the maximum L1 rate of 100 kHz
(unchanged with respect to \runii) as defined by the ATLAS detector read-out capability, a target average HLT physics
output rate of 3 kHz (1.2 kHz in \runii) and CPU resources of the HLT farm, as detailed in Section~\ref{sec:softwarePerf}.
Substantial increases in the HLT output rate were enabled by the expanded SFO capacity, providing improved coverage for physics,
while also allowing space for sustained operation with very long periods (up to ten hours) of constant instantaneous luminosity
close to its peak value,
which can be seen in Figure~\ref{fig:stream_rate_BW} before 6:30. Such periods make the average Run-3
HLT rate almost equal to the peak rate, in contrast to Run 2 for which the average was about 2/3 of the peak.
To ensure an optimal trigger menu
within the rate constraints for a given LHC luminosity, prescale factors can be applied to L1 and HLT triggers and
these are changed during data taking in such a way that triggers may be disabled entirely or only executed for a
certain fraction of events. Supporting triggers usually run at a constant rate.
The small event size of triggers in the TLA stream allows it to be recorded at rates of the order of 1-10$\,$kHz
while taking up a minor fraction of the total HLT bandwidth.
This strategy is effective in avoiding high prescales at the HLT for low \pt\ TLA triggers.
Some triggers are enabled only later in an LHC fill when the luminosity and pile-up have reduced and the required HLT farm resources are available.
The effects of prescale changes towards the end of the run on the HLT output rate and bandwidth can be seen in Figure~\ref{fig:stream_rate_BW} around
8:30, 11:00 and 14:30. Further flexibility is provided by bunch groups, which allow triggers to include specific requirements on the
proton bunches in the LHC. These requirements include paired (colliding) bunch-crossings for
physics triggers, empty or unpaired crossings for background studies or searches for long-lived particle decays,
and dedicated bunch groups for detector calibration.
 
\subsection{Baseline physics trigger menu for proton--proton collisions}
\label{sec:menuBase}
 
The primary \pp\ menu triggers cover all signatures relevant to the ATLAS physics programme including electrons,
photons, muons, taus, jets and \met which are used for \ac{SM} precision
measurements including decays of the Higgs, $W$ and $Z$ bosons, and searches for physics beyond the SM
such as heavy particles, supersymmetry or exotic particles. A set of low-\pt di-muon and di-electron
triggers is used to collect $B$-meson decays, which are essential for the $B$-physics programme of ATLAS.
 
The Run-3 trigger menu aims to maximise the physics impact of the Run-3 data set by exploiting the newly implemented detector features, more performant HLT hardware, and algorithmic advancements, while simultaneously maintaining a level of consistency with the Run-2 trigger menu to allow for combined analyses on both data sets. Trigger thresholds at L1 and HLT were generally kept the same as during \runii, benefiting from improvements to reduce trigger rate.
The trigger menu strategy remains focused on assigning the majority of the rate to inclusive triggers rather than analysis-specific triggers.
In particular, the Run-3 trigger menu maintains the unprescaled isolated single-electron and single-muon trigger \pt\ thresholds around 25~\GeV,
as described in Sections~\ref{subsec:egammamenu} and \ref{subsec:muonmenu}.
Dedicated triggers are added for specific analyses that are not covered by inclusive triggers.
The additional available HLT rate compared to \runii is dedicated to expanding the physics menu, in the physics and TLA streams, lowering trigger thresholds and including new triggers for previously unexplored phase space. The increase in the SFO capacity enables a larger recording bandwidth which is exploited with the delayed streams.
The breakdown of the approximate rates feeding these streams grouped by trigger signature,
as detailed in Section~\ref{sec:sigPerf}, is shown in Table~\ref{tab:signature_rates}.
 
In 2022 the baseline physics trigger menu was based on the L1Calo and L1Topo legacy systems.
This led to some limitations; in particular,
HLT chains seeded from L1Muon topological triggers could not be run until the new L1Topo and L1Muon were commissioned.
This mostly affected triggers for the ATLAS $B$-physics programme.
 
\begin{table}[htp]
\begin{center}
\caption{
Example breakdown of approximate total rates for physics triggers grouped by signature
at luminosity of $1.8\times\lumi{e34}$ and $\sqrt{s} = 13.6\,$\TeV.
Rates are quoted for the Main, Delayed and TLA streams subtracting off the contributions
from the less inclusive streams.
}
\label{tab:signature_rates}
\begin{tabular}{lccc}
\toprule
Signature & \multicolumn{3}{c}{Rate per stream [Hz]}\\
&	 Main & Delayed & TLA \\
\midrule
Electron	            &	270 &  &  \\
Photon		            &	120 &  &  \\
Muon		            &	290 &  &  \\
Tau			    &	160 &  &  \\
Missing transverse momentum &	140 &  &  \\
Unconventional Tracking	    &	40   &    &  \\
$B$-physics and light states&	    & 240 &  \\
Jet			    &   490 & 460 & 5000 \\
Jet with $b$-hadrons	    &	190 & 160 &  \\
Combined	            &	240 & 50 & 830 \\
\bottomrule
\end{tabular}
\end{center}
\end{table}
 
\begin{figure}[htbp]
\centering
\includegraphics[width=0.8\textwidth]{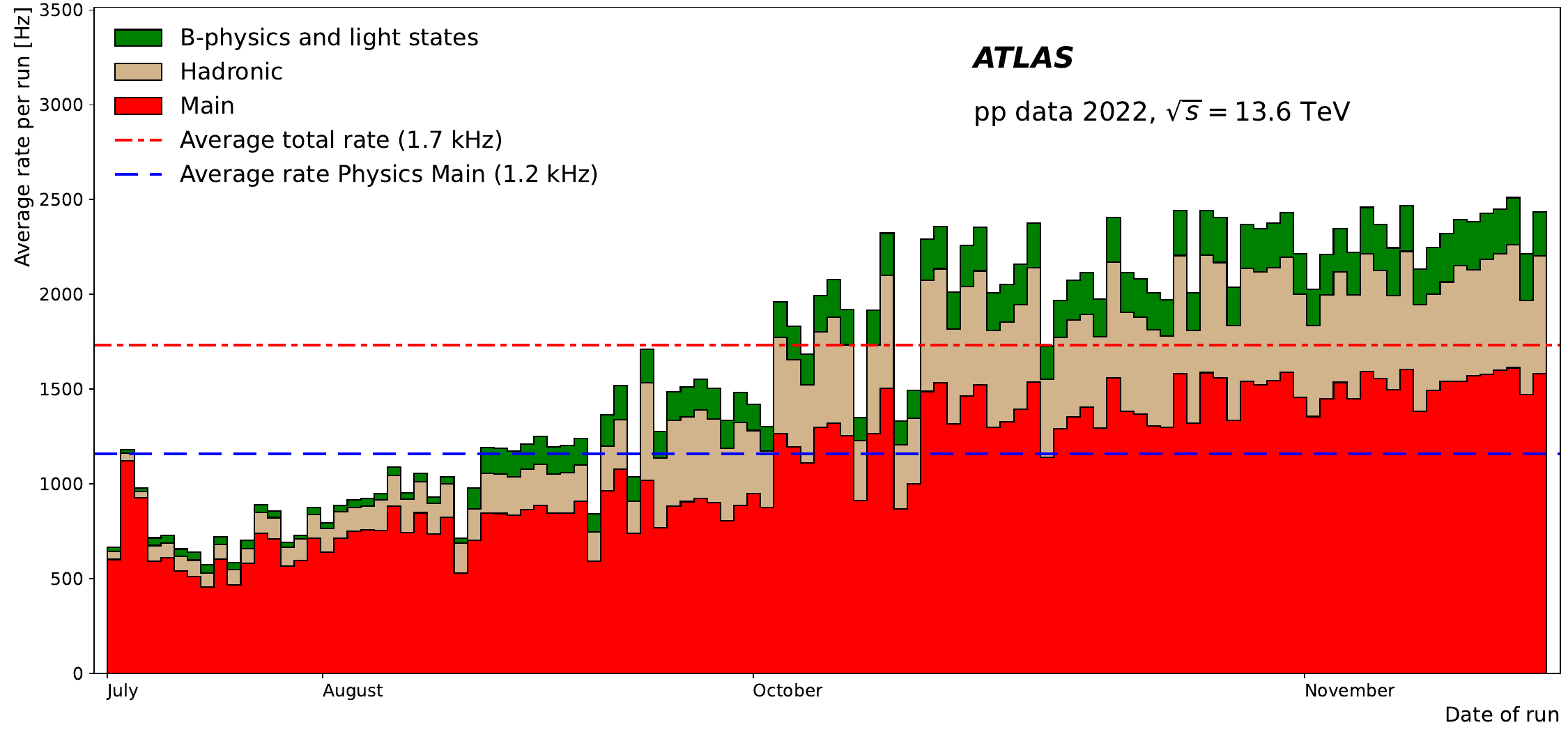}
\caption{
The average recording rate of physics data streams for the ATLAS \pp\ physics runs taken in 2022.
The average of all runs for these three streams is indicated as a dash-dotted line, and the average of the Main stream is indicated as a dashed line.
}
\label{fig:phys_stream_rate}
\end{figure}
 
The evolution of the average recording rates of the physics data streams in 2022 is shown in Figure~\ref{fig:phys_stream_rate}.
The average Run-3 rates of the Main, Hadronic and BLS streams are designed to be 1.6\,kHz,
1.1\,kHz and 0.5\,kHz, respectively. However, in 2022 they used only part of this allocation,
which in the case of BLS was dictated by the commissioning schedule.
 
The triggers with the largest rate contributing to the delayed streams are summarised in
Table~\ref{tab:delayed}. The $B$-physics triggers are strongly limited by the rate of L1 triggers.
In order to maximise the acceptance, L1 triggers with lower thresholds are enabled at the end of the fills,
once the luminosity is significantly below its peak value.
The rate of the BLS stream is kept approximately constant and thus the fraction of the BLS events in the recorded physics events
increases significantly at lower luminosities.
 
\begin{table}[htp]
\begin{center}
\caption{Summary of selected triggers in the delayed streams.
The VBF di-jet trigger selects events characteristic of Vector Boson Fusion, requiring the presence of a pair of jets satisfying kinematic correlations and an invariant mass greater than 1000\,GeV. All jets are within $|\eta|<2.5$ unless otherwise specified.
Rates are given at luminosity of $1.8\times\lumi{e34}$ and $\sqrt{s} = 13.6\,$\TeV.
}
\label{tab:delayed}
\begin{tabular}{lrr}
\toprule
Trigger                                    & \pt threshold [GeV]   & Rate [Hz] \\
\midrule
VBF di-jet                                 & 70 ($|\eta|<3.2$), 50 ($|\eta|<4.9$)                  & 270  \\
Two jets, two $b$-jets ($\epsilon=77$\%)   & 80, 55, 28, 20        & 160 \\
Six jets                                   & $6 \times 35$         & 140 \\
Five jets, one $b$-jet                     & $5 \times 35$, 25     & 50 \\
$B$-physics di-muon                         & 11, 6                 & 40 \\
$B \rightarrow K^*ee$                      & 5, 5                  & 170 \\
\bottomrule
\end{tabular}
\end{center}
\end{table}

The Run-3 TLA stream was expanded to include photon, muon and b-jet triggers, in addition to the jet triggers, that
were used in \runii~\cite{EXOT-2016-20}.
It is also now possible to record multiple trigger object collections at the same time, permitting the analysis of more complex final states.
The content of each individual event in the TLA stream is determined by specialised physics chains targeting these objects and their combinations,
as detailed in Sections~\ref{subsec:tla_photons},~\ref{subsec:tla_jets} and \ref{subsec:tla_bjets}.
 
The improvements to existing triggers and their Run-3 performance are detailed in Sections~\ref{sec:HLTreco}
and~\ref{sec:sigPerf}. In particular, the tracking improvements~\cite{ATL-PHYS-PUB-2021-012}
allowed for the introduction of triggers based on unconventional tracking which are
listed in Table~\ref{tab:llp} and
detailed in Section~\ref{sec:unconTrack}.
The shown rates are not unique and can have overlap with other physics triggers, as is the case for the displaced tau trigger, for example.
 
\begin{table}[htp]
\begin{center}
\caption{Summary of selected unconventional tracking trigger rates at luminosity of $1.8\times\lumi{e34}$ and $\sqrt{s} = 13.6\,$\TeV.
}
\label{tab:llp}
\begin{tabular}{lr}
\toprule
Trigger                                   & Rate [Hz] \\
\midrule
Isolated high \pt\ track & 1 \\
Large dE/dx & 5 \\
Disappearing track & 4 \\
Hit-based displaced vertex \met-seeded & 1 \\
Hit-based displaced vertex jet-seeded & 1 \\
Emerging jet (jet-seeded) & 10 \\
Emerging jet (photon-seeded) & 10 \\
Displaced single-jet & 15 \\
Displaced di-jet & 7 \\
Displaced electron & 13 \\
Displaced muon & 4 \\
\bottomrule
\end{tabular}
\end{center}
\end{table}


\subsection{Baseline physics trigger menu for heavy--ion collisions}
\label{sec:menuHI}
 
The characteristics of HI collisions of Pb+Pb are largely different from \pp collisions.
Apart from any hard scattering of interest, each HI collision is composed of multiple simultaneous
nucleon--nucleon interactions,
which generate a sizable underlying event (UE) contribution dominated by soft particle production.
As a result of this large UE contribution, the particle multiplicities and the total energy deposited into calorimeters in HI collisions are on average much larger than those in \pp collisions, and they also vary significantly from event to event.
The event-by-event variations correspond to the variations in the size and geometry of the overlap region of the two colliding Pb nuclei.
The variations are characterised by the centrality of Pb+Pb collisions, which is strongly correlated to total transverse energy measured in the forward calorimeter, $\Sigma \et^\text{FCal}$~\cite{HION-2016-04}.
Centrality classes are defined by dividing the $\Sigma \et^\text{FCal}$ distribution of minimum-bias inelastic Pb+Pb collisions into percentiles.
The 0--10\% interval includes events with the largest $\Sigma \et^\text{FCal}$, corresponding to the most central collisions (largest geometric overlap), while the 90--100\% interval includes the most peripheral collisions (smallest geometric overlap).
In the trigger, centrality can hence be mapped to the total energy measured in the full calorimeter system or in its forward part, both of which are accessible at L1.
 
Besides the variations of particle multiplicity from event to event, there is also an azimuthal anisotropy of particle production present in each HI event.
This is a result of the initial-state spatial anisotropy of the overlap region leading to sizable anisotropies of particle momenta in the final state.
 
The trigger menu for HI collisions must be designed to handle the large event-by-event variations and azimuthal anisotropy in each event.
On the other hand, HI collisions are operated at a lower centre-of-mass energy per nucleon pair and a lower luminosity compared to \pp collisions, so the hard scattering rate is lower in HI collisions.
The main goal of the Run-3 HI trigger menu design is to keep \pt thresholds for unprescaled triggers of different signatures as low as possible while minimizing the sensitivity of all triggers to UE contributions.
The unprescaled single electron and photon \pt thresholds are set at $15~\GeV$, while the unprescaled single muon trigger has a \pt threshold around $8~\GeV$.
This strategy ensures the collection of the majority of events with leptonic $W$ and $Z$ boson decays, which are the main source of events for the study of electroweak processes.
The lower \pt threshold for the single muon trigger allows for the selection of events with semi-leptonic decays of heavy-flavour hadrons.
The unprescaled single jet \pt threshold is set to $85~\GeV$ to collect events with jets for the study of QCD processes.
In addition to single physics signature triggers, dedicated multi-object triggers are added: an unprescaled di-muon trigger with \pt thresholds of $4~\GeV$ for both muons to collect events with di-muon decays of quarkonium states; an unprescaled muon and jet trigger with a muon \pt threshold of $4~\GeV$ and a jet \pt threshold of $60~\GeV$ to collect events with $b$-jets based on soft muons from $B$-hadron decays.
 
To reduce their sensitivity to UE contributions, calorimeter-based triggers -- such as electron, photon and jet triggers -- include a correction for the average energy contributed by the UE. The average UE energy is evaluated per calorimeter layer and cell following the iterative procedure used in offline reconstruction~\cite{HION-2020-09,HION-2017-10}. For each interval of width $\Delta\eta = 0.1$, the UE energy is averaged over $\phi$ taking into account the azimuthal anisotropy of deposited energy due to the collective motion of particles created in HI collisions. This estimated average UE energy is then subtracted from the cluster constituent cells. As a result, the cells in the \roi\ -- which are used as inputs to clustering and identification algorithms -- contain a reduced systematic bias due to the UE.
 
In cases where two nuclei do not interact hadronically due to a lack of geometric overlap, they can still interact electromagnetically, producing so-called ultraperipheral collisions (UPC). The large electric charge of Pb nuclei ($Z = 82$) results in a photon flux that is enhanced by a factor of $Z^2$ compared to \pp collisions. Therefore, cross-sections for electromagnetic (photon--photon and photon--nucleus) interactions are significantly increased. The events produced in these processes are typically accompanied by small multiplicities of produced particles, as well as forward rapidity gaps
defined by very low activity in the forward calorimeters.
 
The three physics streams for Pb+Pb data taking differ from those in \pp\ collisions and
aim to target different classes of events:
\begin{itemize}
\item \emph{physics\_HardProbes}: events produced by hard processes in inelastic Pb+Pb collisions and triggered with high-\pt\ electrons, photons, muons, jets and $b$-jets.
\item \emph{physics\_UPC}: events produced by various processes in ultraperipheral Pb+Pb collisions. Triggers in this stream select events with low-\pt\ electrons, photons, muons and jets as well as specific track multiplicities, in combination with vetos on the total calorimeter energy, and/or deposits in the ZDC.
\item \emph{physics\_PC/physics\_CC}: minimally trigger-biased (minimum bias) inelastic Pb+Pb collisions separated into peripheral (PC) and central (CC) events. Triggers for these streams are seeded off L1 total energy triggers.
\end{itemize}
The calibration and performance streams largely remain the same.
 
The HI triggers with muons rely on standard reconstruction described in Sections~\ref{sec:muons} and \ref{sec:bls}, and are not discussed further.
More details on other signature-specific HI configurations can be found in Sections~\ref{subsec:egamma_hi},
\ref{subsec:jets_hi}, \ref{subsec:bjet_menu} and \ref{sec:minbias}.


\section{High-level trigger reconstruction}
\label{sec:HLTreco}

Once an event is accepted at L1, it is processed by the HLT by making use of finer-granularity
calorimeter information, precision measurements from the MS 
and tracking information from the ID. 
The following sections describe the main HLT algorithms used for ID tracking, 
calorimeter clustering and muon reconstruction. 


\subsection{Inner detector tracking}
\label{sec:id}
 
The sequence of track reconstruction steps is briefly summarised below. More detailed information can be found in Ref.~\cite{TRIG-2019-03}.
The track reconstruction starts with the data preparation in the ID, which reconstructs clusters and space points using information from the pixel and SCT
data providers which fetch the raw detector data from the pixel and SCT read-out systems.
This step makes use of the RoI mechanism which allows the HLT to request only the data from those silicon modules inside the RoI.
The data preparation is then followed by the fast-tracking step, which runs a custom pattern recognition and a fast-track fit.
Following this, a precision tracking step takes the tracks reconstructed by the fast tracking and refits them
using the offline track reconstruction code. It does not use the slower offline pattern recognition code.
Additionally, these track candidates are extended into the TRT: using hits at larger radii leads to an improved track momentum resolution.
The TRT data preparation is performed only for the precision tracking following the extension of the tracks into the TRT.
 
The Run-3 tracking software is enhanced with respect to \runii with several new developments:
\begin{itemize}
\item Restriction of the \roi width along the beamline for the track reconstruction for the muon trigger isolation requirement (Section~\ref{sec:id_muon});
\item Implementation of the bremsstrahlung recovery for the electron triggers (Section~\ref{sec:id_electron});
\item Track reconstruction in the entire ID volume (full scan) for the jet- and \met-based triggers, followed by the vertex finding with these tracks (Section~\ref{sec:id_full});
\item Running of a preselection tracking stage in the $b$-jet trigger to allow the use of fast $b$-tagging algorithms prior to execution of the full scan tracking (Section~\ref{sec:id_bjet});
\item Implementation of \ac{LRT}~\cite{IDTR-2021-03} (Section~\ref{sec:lrt}).
\end{itemize}
 
The HLT minimum-bias triggers (Section~\ref{sec:minbias}) run the full offline pattern recognition and track reconstruction~\cite{PERF-2015-08}.
 
Except where stated, the standard selection\footnote{
With the addition that for the standard tight offline requirement on the number of holes (e.g. missing hits from active layers)~\cite{PERF-2015-08}, the tracks are required to
contain at least one pixel hit.} of offline tracks~\cite{PERF-2015-08} and objects~\cite{EGAM-2018-01,MUON-2018-03,JETM-2018-05,ATLAS-CONF-2018-023,FTAG-2019-07} is used for performance studies in this subsection.
Efficiencies are measured by taking the correlated ratio of the number of offline reference objects that
have a matched trigger track to the total number of offline objects passing the selection. The methodology for tracking performance studies
is described in detail in Ref.~\cite{TRIG-2019-03}.

\newcommand{\prap}{pseudorapidity\xspace}
\newcommand{\preselection}{preselection\xspace}
 
\subsubsection{Tau lepton tracking}
\label{sec:id_tau}
 
The tau lepton trigger tracking runs as a two-stage process, essentially unchanged since \runii and
described in more detail in Section~\ref{sec:tau} and in Ref.~\cite{TRIG-2019-03}.
First, the fast tracking runs in a core \roi of $0.2\times 0.2$ in $\eta - \phi$ space, fully extended along the beamline in the range $|z|<225\,$mm,
to identify the leading \pt\ tracks from the tau decay. This is followed by running both the fast tracking and the precision tracking
in a wider \roi ($0.8\times 0.8$ in $\eta - \phi$ space) to determine whether the tau candidates are isolated,
and to accommodate the wider opening angle for tau candidates with three tracks associated with them (3-prong).
This \roi is centred on the $z$ position of the leading \pT\
track identified by the first stage and limited to $|z|<10\,$mm relative to this leading track.
The efficiency of the tau lepton tracking is above 99\%.
 
\subsubsection{Muon tracking}
\label{sec:id_muon}
 
The muon triggers use the standard tracking sequence, which is essentially unchanged since \runii~\cite{TRIG-2019-03}
and has an efficiency better than 99\% for tracks originating from the beamline with an almost negligible dependence of the efficiency on the \pileup multiplicity.
Those triggers in which a muon track is required to be isolated
with respect to other tracks from the interaction
run the muon isolation ({\em muonIso}) tracking stage,
after the standard muon fast and precision tracking,
using a wider \roi ($0.7\times 0.7$) in $\eta - \phi$ space.
For Run 3, this wider RoI has a restricted full $z$-width of 20\,mm centred on the $z$ position at the beamline
of the muon candidate identified in the first round of precision tracking and after the muon reconstruction.
The restriction of the $z$-width reduces the execution time of the {\em muonIso} tracking, and is discussed in
more detail in Section~\ref{sec:execTimeTracking}.
 
The tracking efficiency in the widened \roi used for the muon isolation is shown in Figure~\ref{fig:muiso:eff}.
It is evaluated with respect to offline tracks within the wider \roi once the muon track candidate is removed.
The efficiency is greater than 99\% across the full \pt range shown.
There is a small dependence (less than 0.5\%) of the efficiency on pile-up.
 
\begin{figure}[tp]
\includegraphics[width=0.49\textwidth]{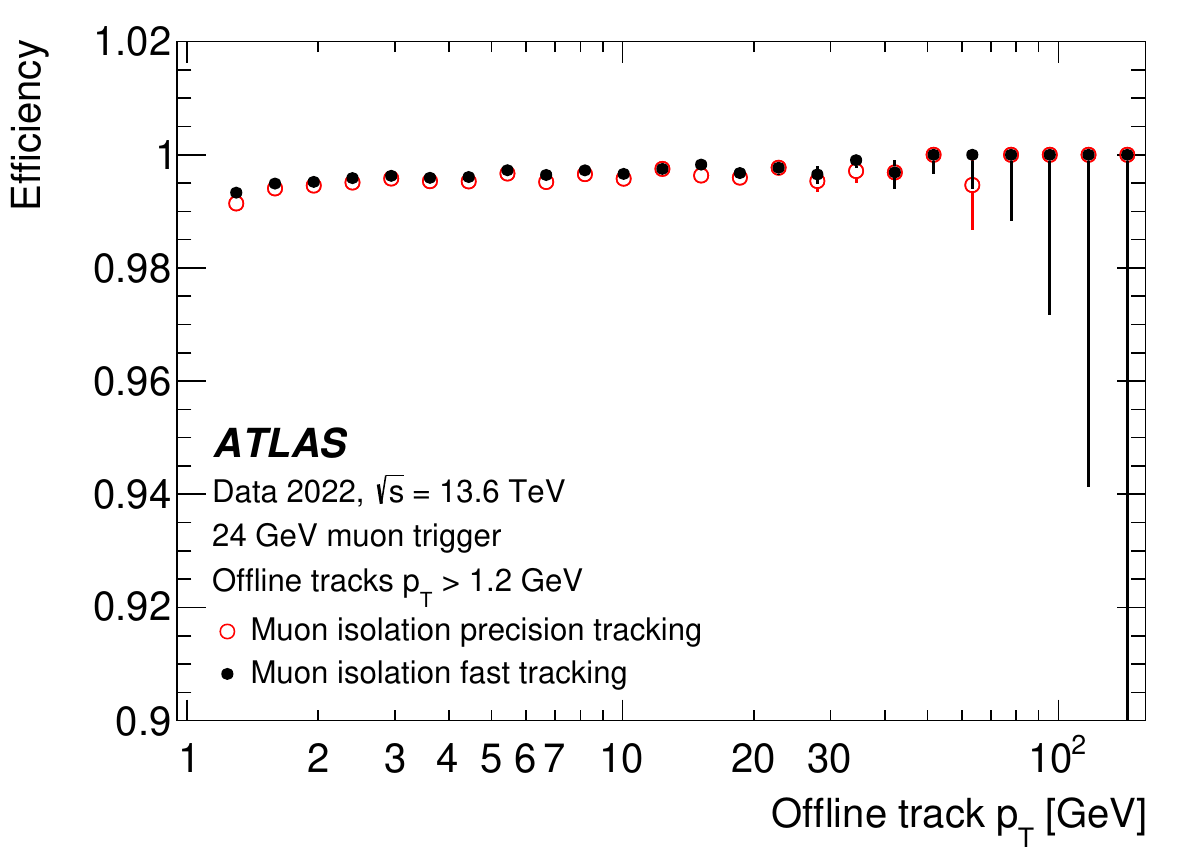}
\includegraphics[width=0.49\textwidth]{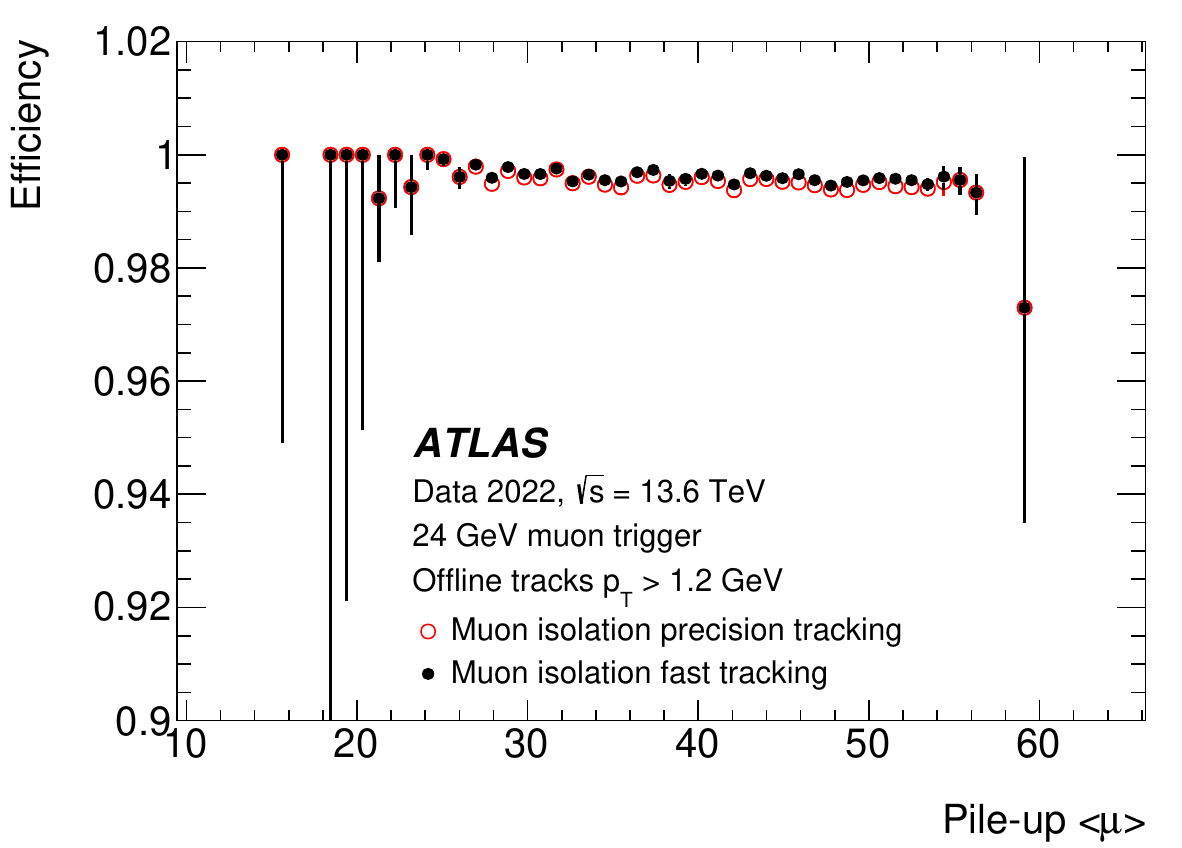}
\caption{The tracking efficiency for non-muon tracks reconstructed in the
{\em muonIso} \roi centred on the trigger muon selected by the 24\,\gev\ muon performance
trigger described in Section~\ref{sec:idperf}, but without selection on the trigger tracks from the {\em muonIso} reconstruction.
Efficiencies are shown for both the fast and precision tracking algorithms
as a function of (left) the offline muon transverse momentum and (right) the average pile-up.
Only statistical uncertainties are shown.
}
\label{fig:muiso:eff}
\end{figure}
 
\subsubsection{Electron tracking}
\label{sec:id_electron}
 
The electron triggers run both the fast, and precision tracking in the same \roi, produced following
the fast calorimeter reconstruction, as detailed in Section~\ref{sec:egamma}.
The dimensions of the \roi used for the tracking are $0.1\times 0.2$ in $\eta - \phi$ space,
with the $\eta$ width being reduced to half of that used in \runii, with no reduction of the physics performance.
 
The track reconstruction for electrons can be affected by bremsstrahlung which
causes tracks, in particular at lower \pt, to deviate from the expected helical path through the tracking detectors.
To improve the resolution of the electron track reconstruction in \runiii the offline Gaussian sum filter (GSF)
algorithm~\cite{ATLAS-CONF-2012-047} is added to the electron track reconstruction sequence in the HLT.
 
\begin{figure}[tp]
\includegraphics[width=0.49\textwidth]{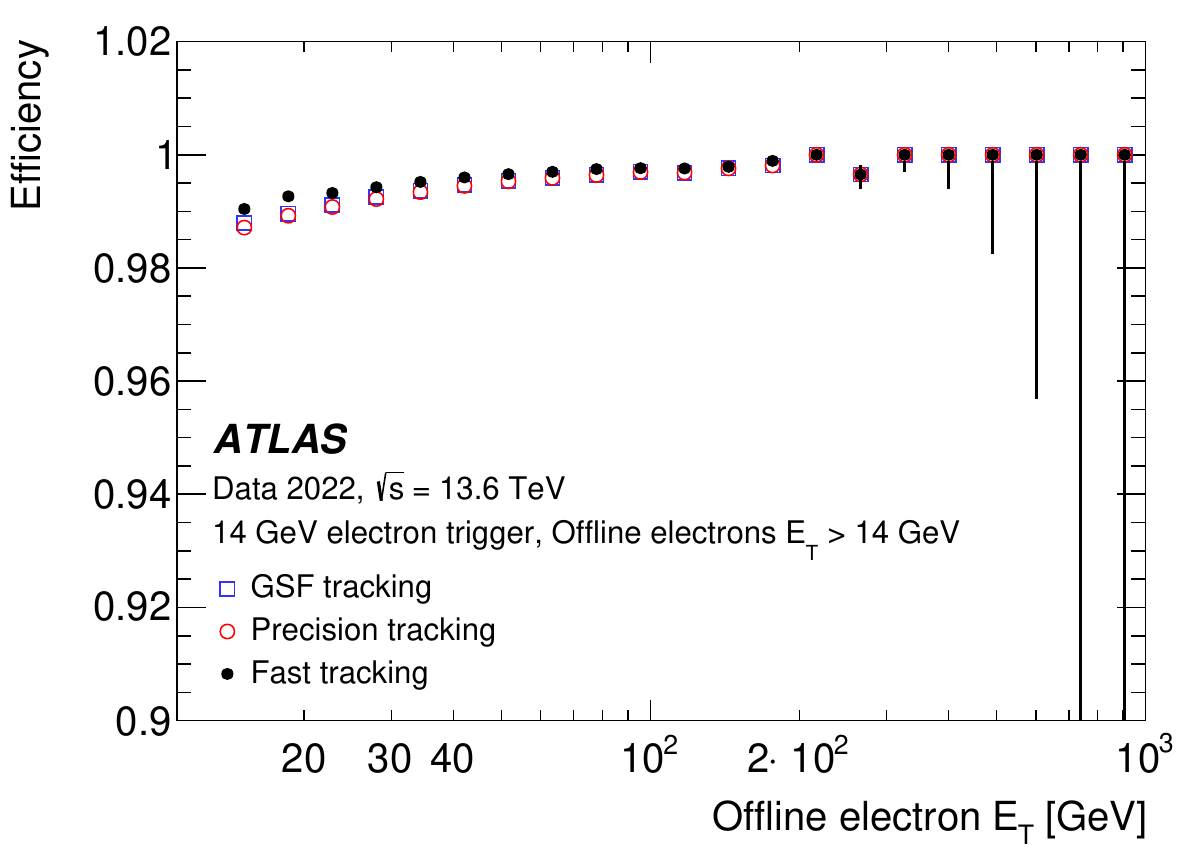}
\includegraphics[width=0.49\textwidth]{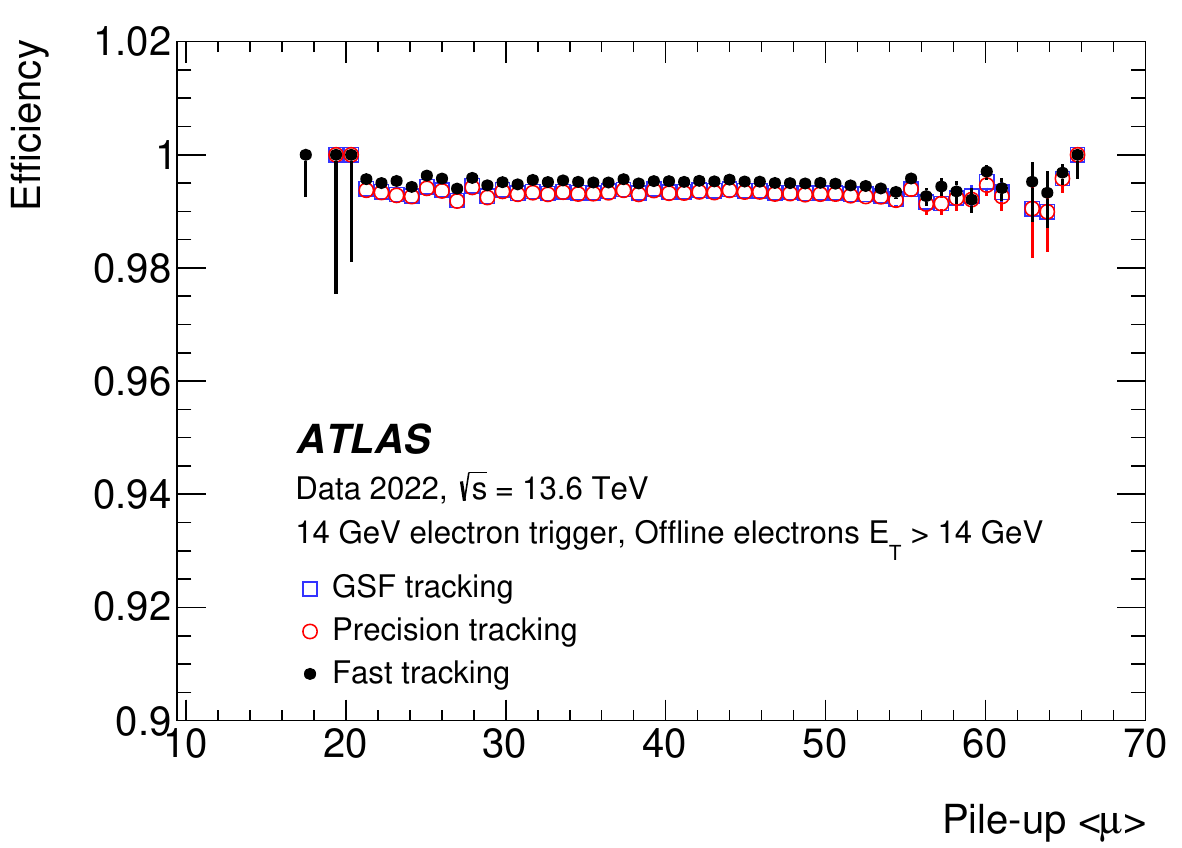}
\caption{
The ID tracking efficiency for electron candidates selected by the 14\,\gev\ electron performance trigger described in
Section~\ref{sec:idperf} which does not use ID tracking information.
The efficiency with respect to offline electrons is shown as a function of (left) the offline electron transverse energy and (right) the average pile-up.
Only statistical uncertainties are shown.
}
\label{figs:el:eff}
\end{figure}
 
Figure~\ref{figs:el:eff} shows the efficiency of the electron track reconstruction
as a function of the transverse energy and the pile-up for offline electron candidates with \et greater than 14~\gev.
To minimise the contribution from non-electron tracks, the tag-and-probe method is used for this study~\cite{TRIG-2019-03}.
The slow increase of the efficiency at low \et is due to the removal of some seed candidates that share hits with other seeds in the fast-tracking stage,
reducing the efficiency for bremsstrahlung candidates where the seeds may not lie so closely on a single helix.
By construction, the GSF efficiency can not be higher than the precision tracking efficiency since only tracks from
the precision tracking are used for the refit for the GSF tracks. The small apparent excess in the GSF
tracking efficiency in Figure~\ref{figs:el:eff}\,(left) arises since
the resolutions of the GSF tracks are better, and as such fewer tracks are excluded by failing the matching criteria.
In addition, different performance triggers are used for the GSF and precision tracking studies, such
that the event samples used for the calculation of efficiency are not exactly identical.
 
The effect of bremsstrahlung on the reconstruction of electron candidates is shown in Figure~\ref{figs:el:res}:
as the electron candidates are required to have $\et>14$~\gev, the track candidates with $\pt$ below this value would have undergone bremsstrahlung.
Because of this both the precision and fast tracking overestimate the $1/\pt$ with a very long tail to positive
values shown in Figure~\ref{figs:el:res}~(top left). This bias becomes progressively smaller at higher \pt\
as shown in Figure~\ref{figs:el:res}~(top right).
For the GSF tracking however, the distribution is more symmetric, and the bias is close to zero over the entire \pt range.
 
The resolution with respect to offline of the trigger electron $1/\pt$ is also seen in Figure~\ref{figs:el:res} (bottom)
and clearly shows the resolution improvement from the GSF tracking with respect to the precision tracking.
For reliable estimates of higher offline track transverse momentum, only candidates with
$\etovpt > 0.8$ are used for the determination of the resolutions. The GSF tracking improves the $1/\pt$ resolution
by nearly a factor of two. Similar improvement is observed for the resolutions of the azimuthal angle
and transverse impact parameter, but not in the track pseudorapidity and $z$ at the beamline
as the latter variables are less sensitive to the bending of the track in the magnetic field.
 
\begin{figure}[tp]
\includegraphics[width=0.49\textwidth]{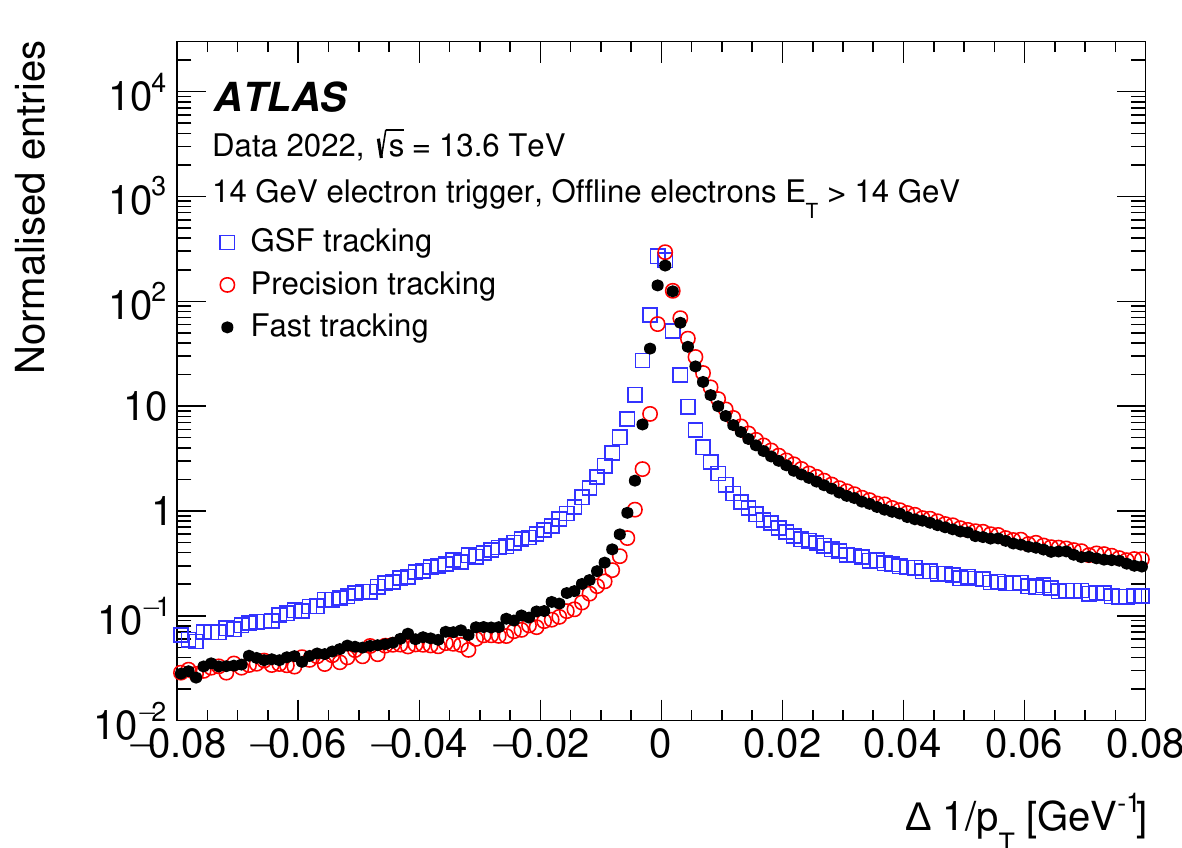}
\includegraphics[width=0.49\textwidth]{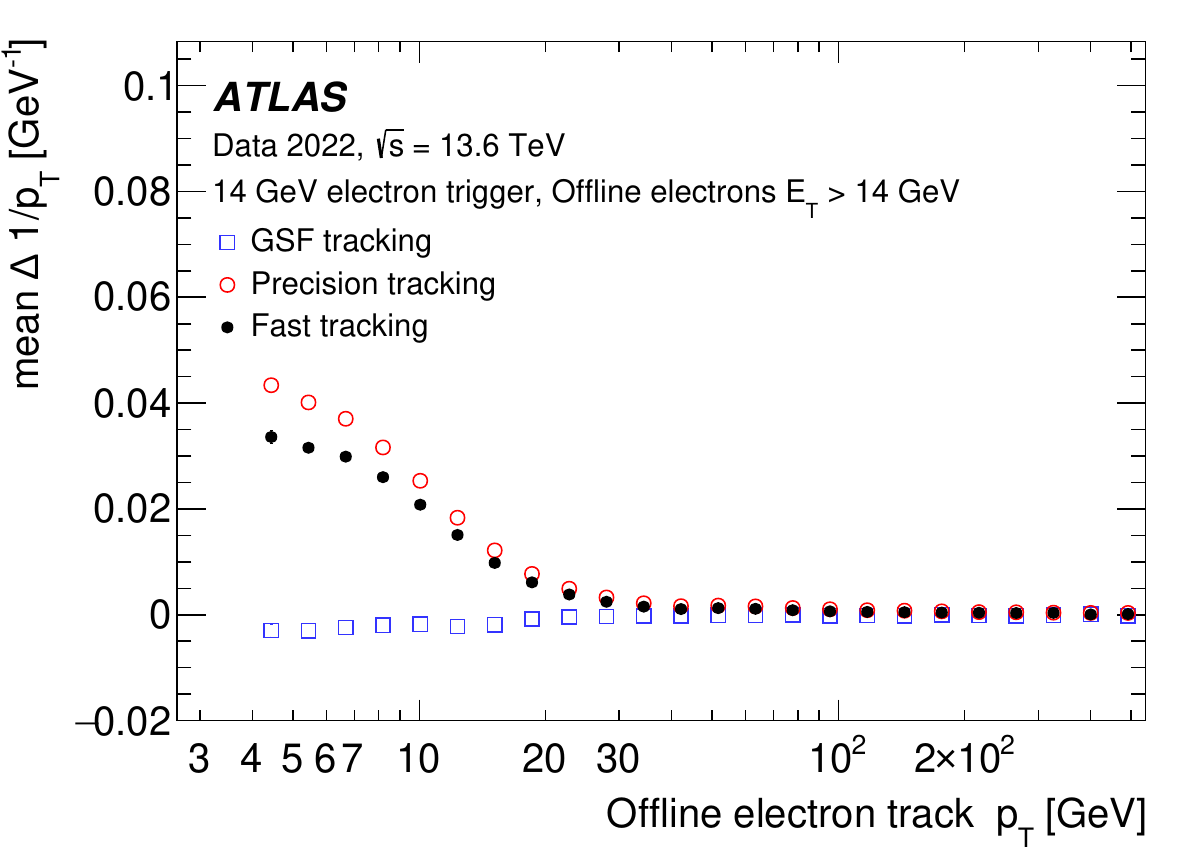}\\
\includegraphics[width=0.49\textwidth]{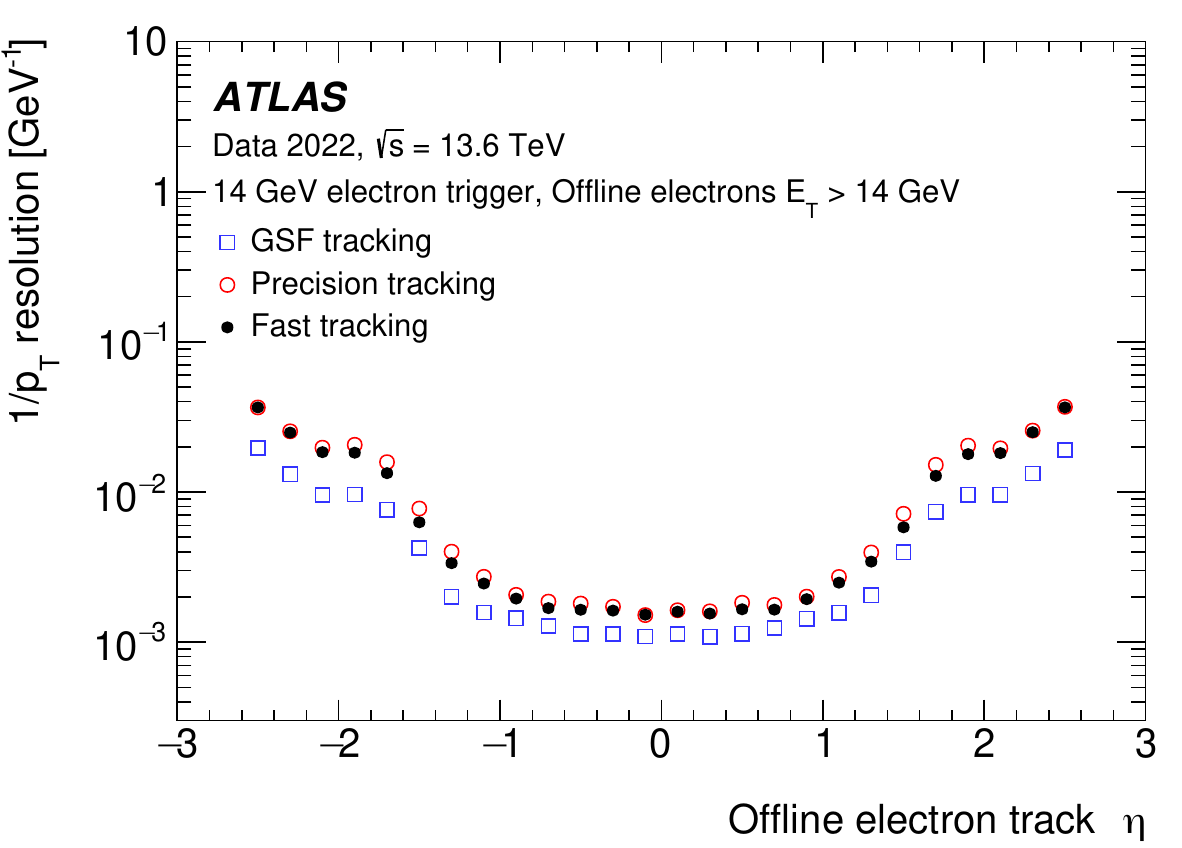}
\includegraphics[width=0.49\textwidth]{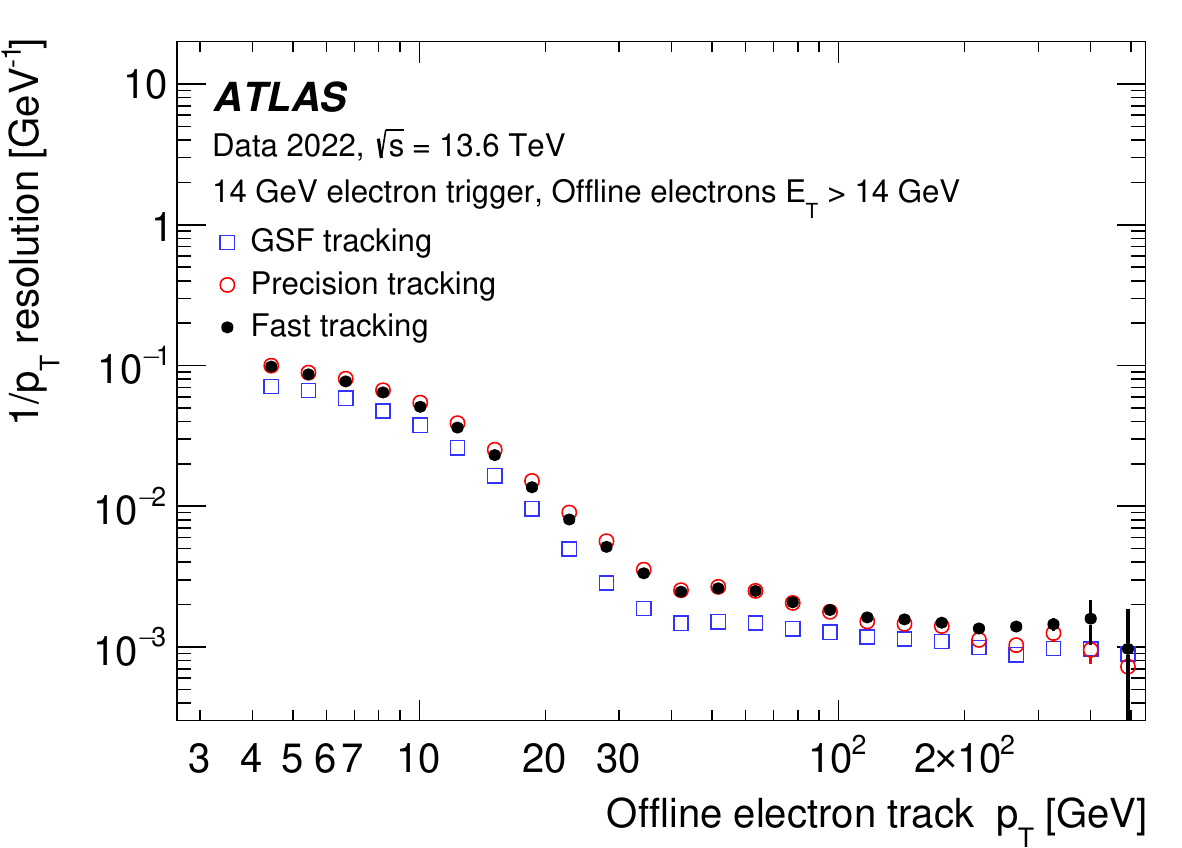}
\caption{(top left) The ID trigger residual in $1/\pt$ with respect to offline
and (top right) the mean of the residual as a function of the offline track \pt.
The resolution in the inverse transverse momentum as a function of offline track (bottom left) $\eta$
and (bottom right) \pt. Distributions are shown for the fast, precision and GSF tracking algorithms
for electron candidates selected with electron performance trigger with $\et>14$~\gev\
and passing the offline electron $\etovpt>0.8$ requirement.
Only statistical uncertainties are shown.
}
\label{figs:el:res}
\end{figure}

\subsubsection{Full scan tracking and vertex finding}
\label{sec:id_full}
 
To improve the trigger reconstruction of hadronic signatures (jets, $b$-jets, \met, etc.) with respect to the offline
reconstruction, tracks from the entire ID volume are combined with the calorimeter topological clusters, described in Section~\ref{sec:hltcalo}, to form
particle flow objects (PFO)~\cite{PERF-2015-09}. This approach improves the jet energy resolution at lower transverse momenta and better separates the hard interaction from pile-up.
 
The full scan tracking is executed only for the hadronic signatures. These algorithms are activated at a rate close
to 14\,kHz at $2\times\lumi{e34}$ after the preselections discussed in Sections~\ref{sec:jets}-\ref{sec:met}.
Even with the reduced input rate after the preselections and with all improvements described below,
the full scan fast tracking accounts for 26\% of the total event processing time of the HLT,
as described later in Section~\ref{sec:softwarePerf}.
An optimisation of the processing time at the cost of a reduced tracking efficiency was thus crucial for the implementation of the full scan tracking for \runiii.
 
One of the most time-consuming aspects of the full scan tracking is the seed making. In this case
the standard seed making~\cite{TRIG-2019-03} is reconfigured to use space-point triplets consisting of pixel-only or SCT-only hits.
Additionally, machine learning techniques~\cite{Lad,Long} for the seed selection are adopted to further reduce the processing time.
The assignment of SCT hits to the tracks is only performed during the track extrapolation into the SCT
with a reduced window to search for these hits in the subsequent layers.
The full scan tracking for the hadronic triggers is executed only once per event for the full ID volume, with the resulting tracks
and vertices used by each trigger that requires them.
 
\begin{figure}[tp]
\includegraphics[width=0.49\textwidth]{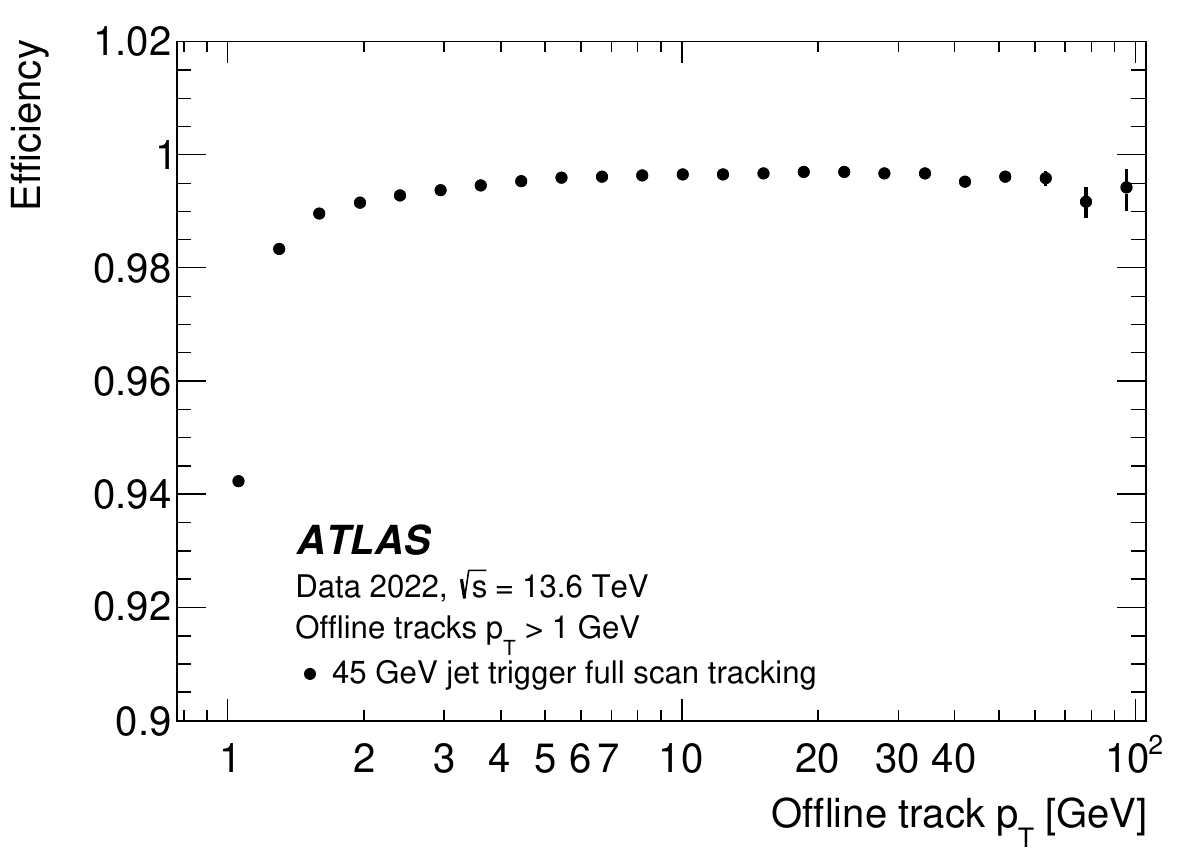}
\includegraphics[width=0.49\textwidth]{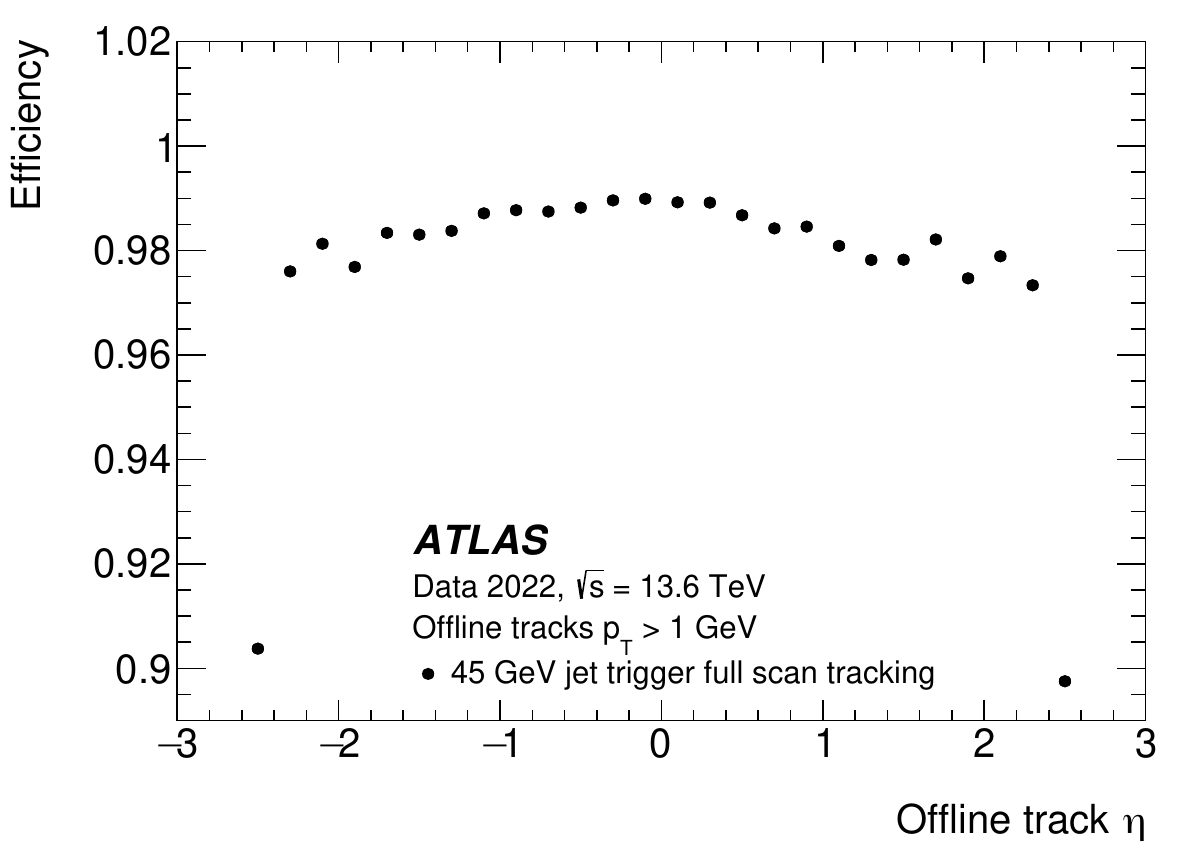}
\includegraphics[width=0.49\textwidth]{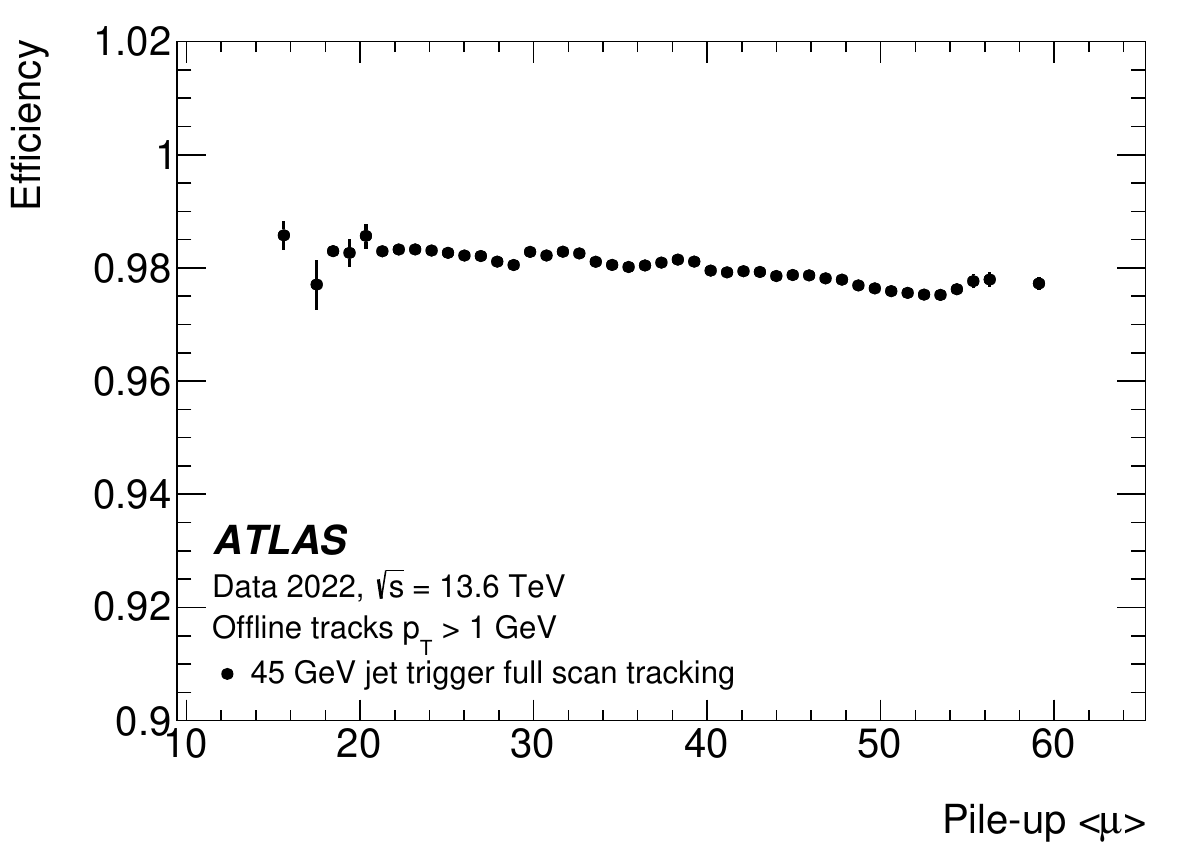}
\caption{
The full scan track finding efficiency with respect to offline tracks
versus the offline track (top left) \pt, (top right) $\eta$, and (bottom) the average pile-up.
The efficiency is evaluated for the 45~\GeV\ jet trigger.
Only statistical uncertainties are shown.
}
\label{fig:fsjets:eff}
\end{figure}
 
The full scan tracking efficiency for events recorded with a 45~\gev\ jet trigger is shown in Figure~\ref{fig:fsjets:eff}.
The efficiency is approximately 94\% at 1~\gev, the threshold used
in the pattern recognition, and reaches a plateau at approximately $\pt>5$~\gev.
A slight asymmetry of the efficiency between negative and positive $\eta$ is due to
the mean beam-spot position not being at $z = 0$. Since $\eta$ for the tracks
is always defined with respect to the $z_0$ position of the point of closest approach of the track to the beamline and not with respect to $z = 0$,
two tracks with identical $\eta$ but different $z_0$ pass through different parts of the detector.
The overall efficiency is approximately 98.5\%,
but with a large pseudorapidity dependence: it is only 90\% for $2.4<|\eta|<2.5$.
There is also a dependence on the pile-up, falling from 98.5\% at
$\langle\mu\rangle=20$ to approximately 97.5\% for $\langle\mu\rangle=54$.
 
\begin{figure}[tp]
\includegraphics[width=0.49\textwidth]{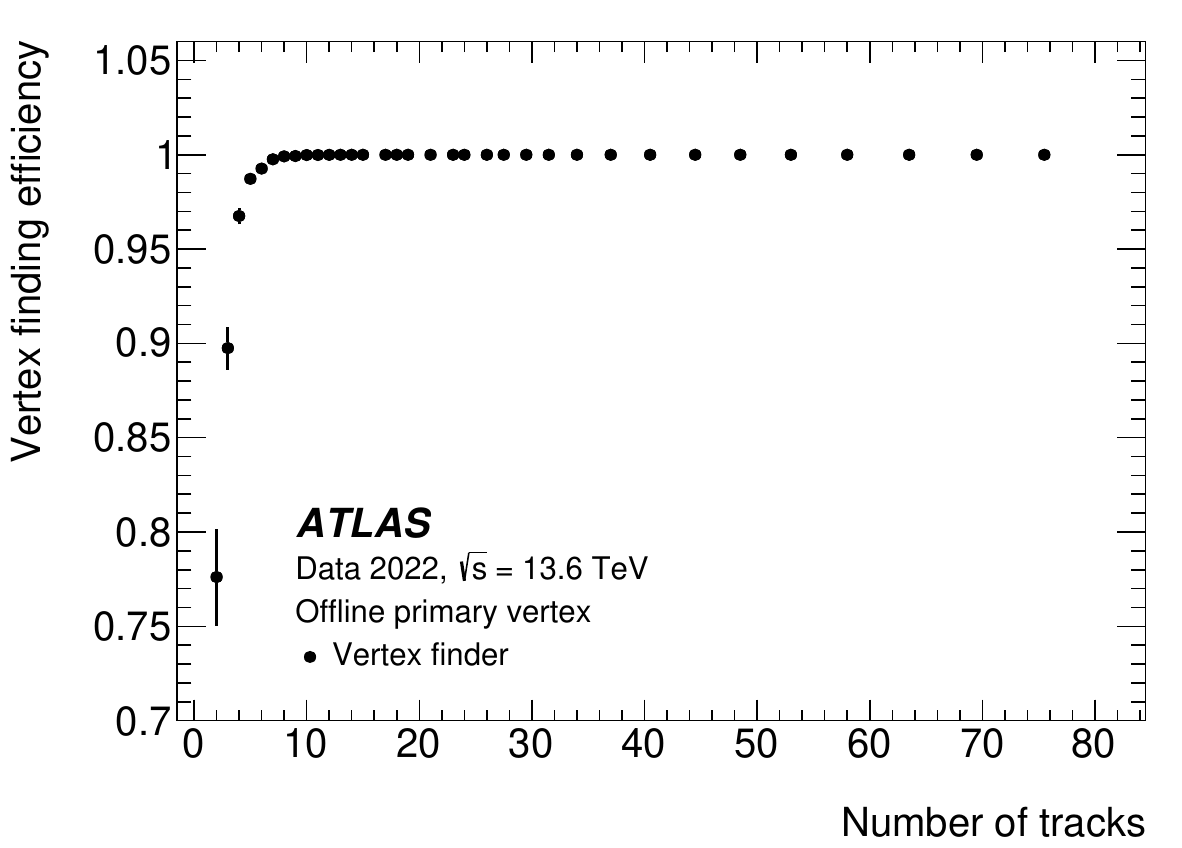}
\includegraphics[width=0.49\textwidth]{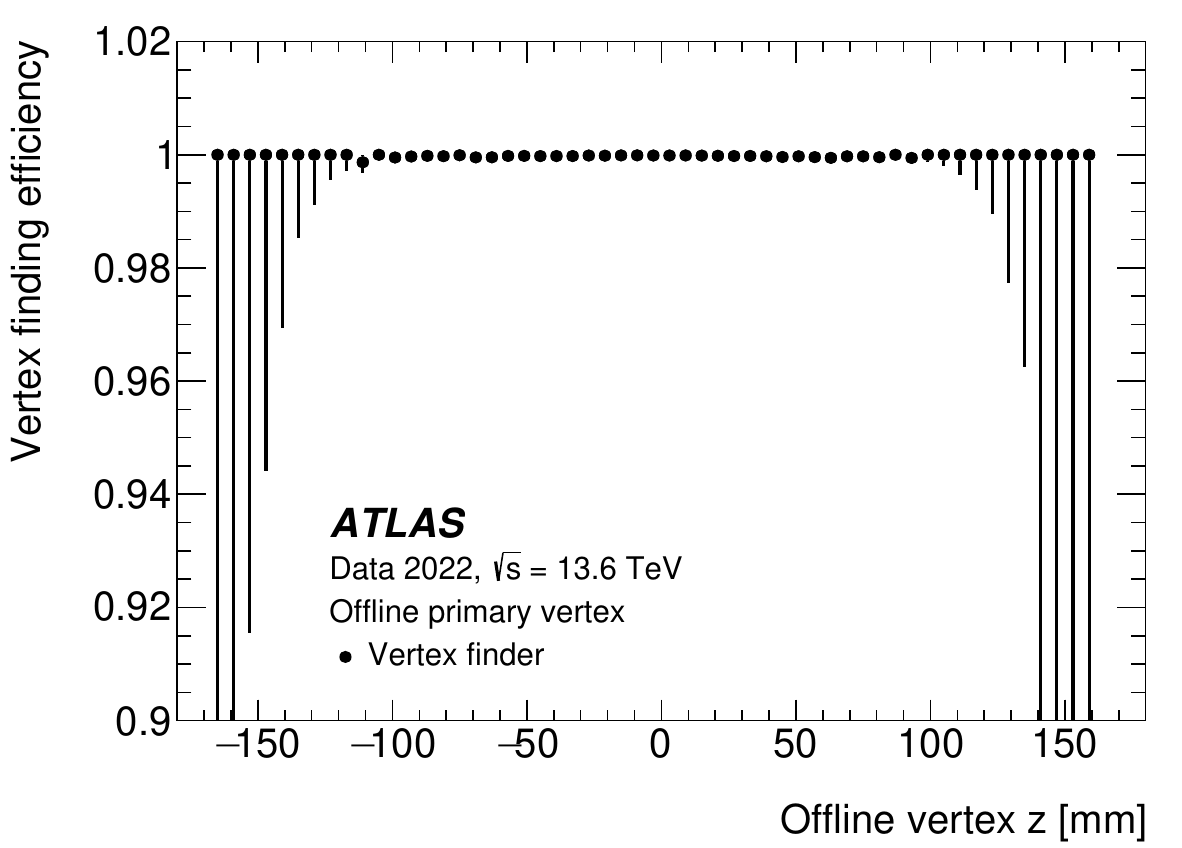} \\
\includegraphics[width=0.49\textwidth]{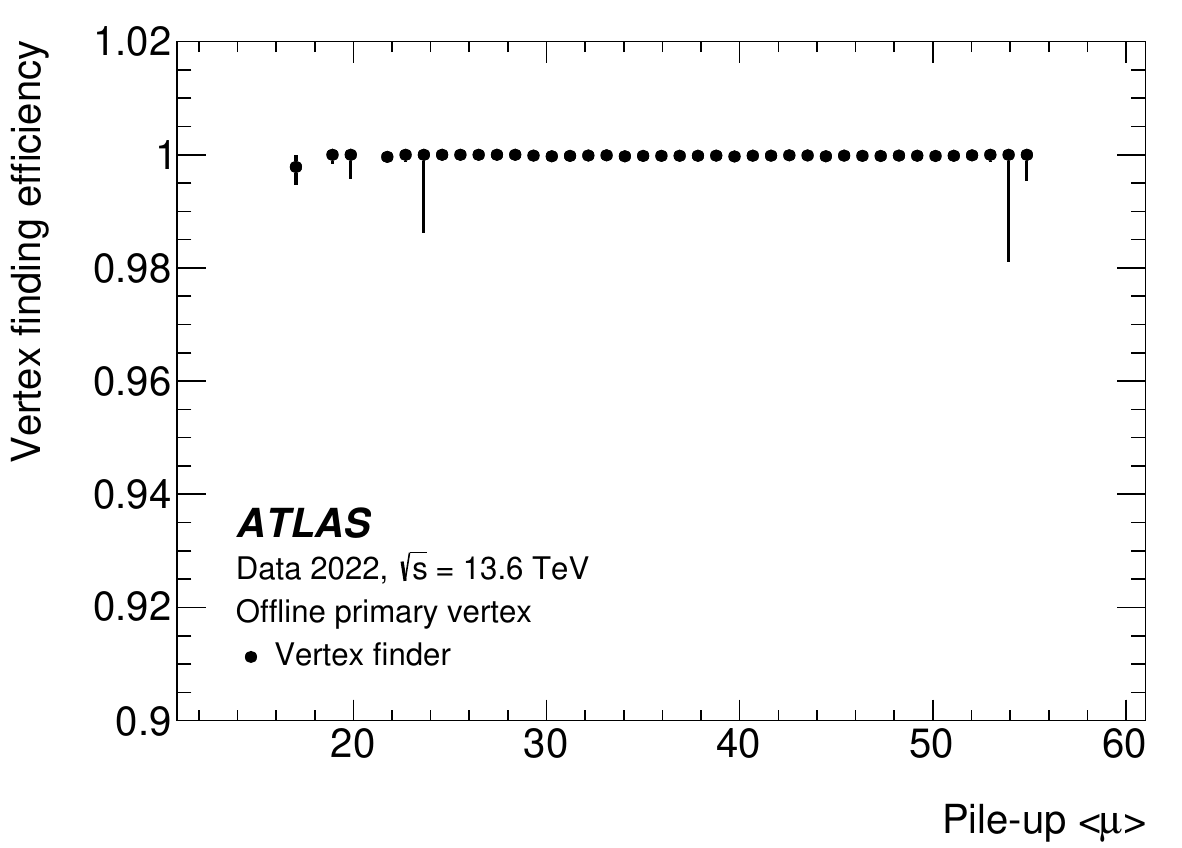}
 
\caption{
The trigger efficiency for finding the primary offline vertex, with respect to
(top left) track multiplicity,  (top right) $z$ position, and (bottom) the average pile-up.
The efficiency is evaluated for the 45~\GeV\ jet trigger.
Only statistical uncertainties are shown.
}
\label{fig:vtx:eff}
\end{figure}
 
The offline vertex algorithm is also used for the vertex reconstruction in the trigger.
It uses tracks from the full scan tracking
to identify the likely primary interaction as well as any additional interactions.
The vertex with the largest sum of the squared transverse
momenta for the tracks assigned to it is chosen as the primary interaction vertex, for both offline and in the trigger.
Only the primary offline vertex is considered as a reference for the trigger vertex efficiency study.
The efficiency for reconstructing the primary
interaction vertex in the trigger is shown in Figure~\ref{fig:vtx:eff}. For offline vertices with
more than six constituent tracks, the trigger vertex efficiency is better than 99.5\%.
Similarly, the vertex finding efficiency is close to 100\% for all $z$ values and all pile-up multiplicities.
 
\begin{figure}[tp]
\includegraphics[width=0.49\textwidth]{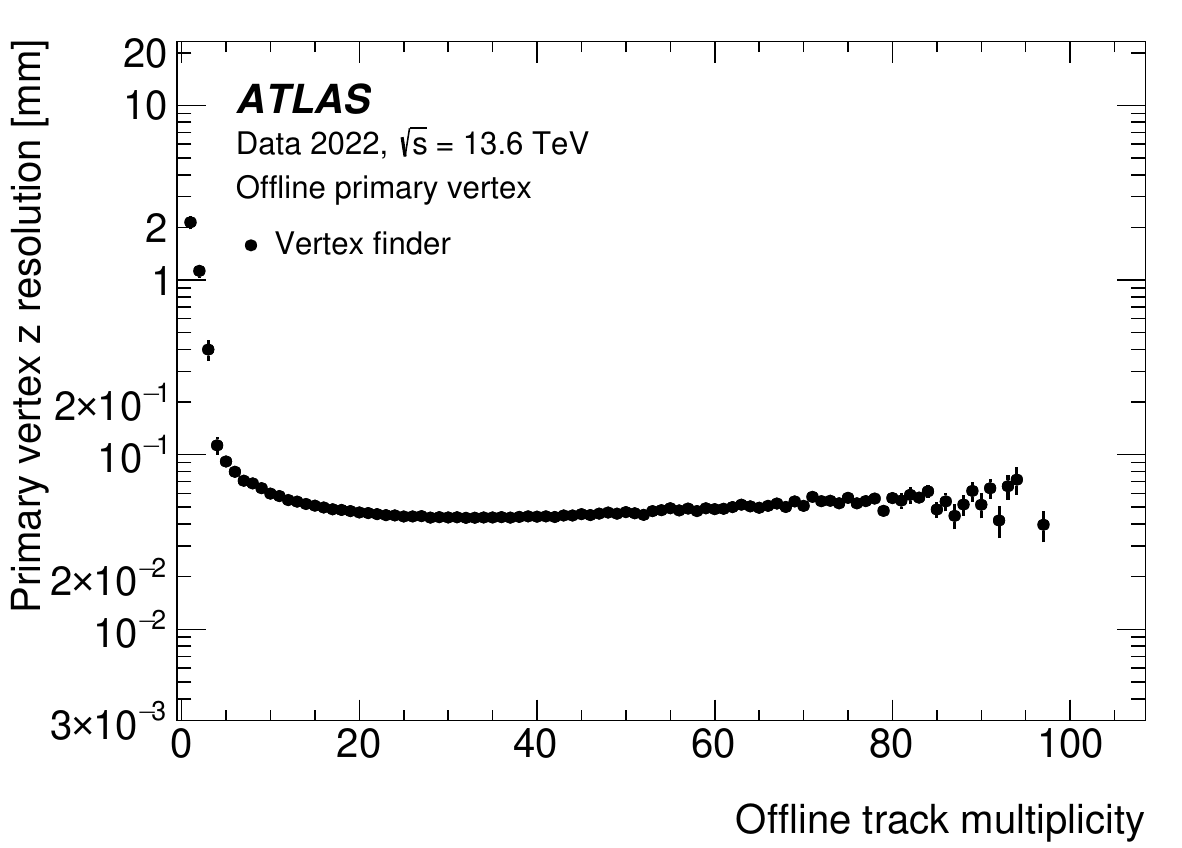}
\includegraphics[width=0.49\textwidth]{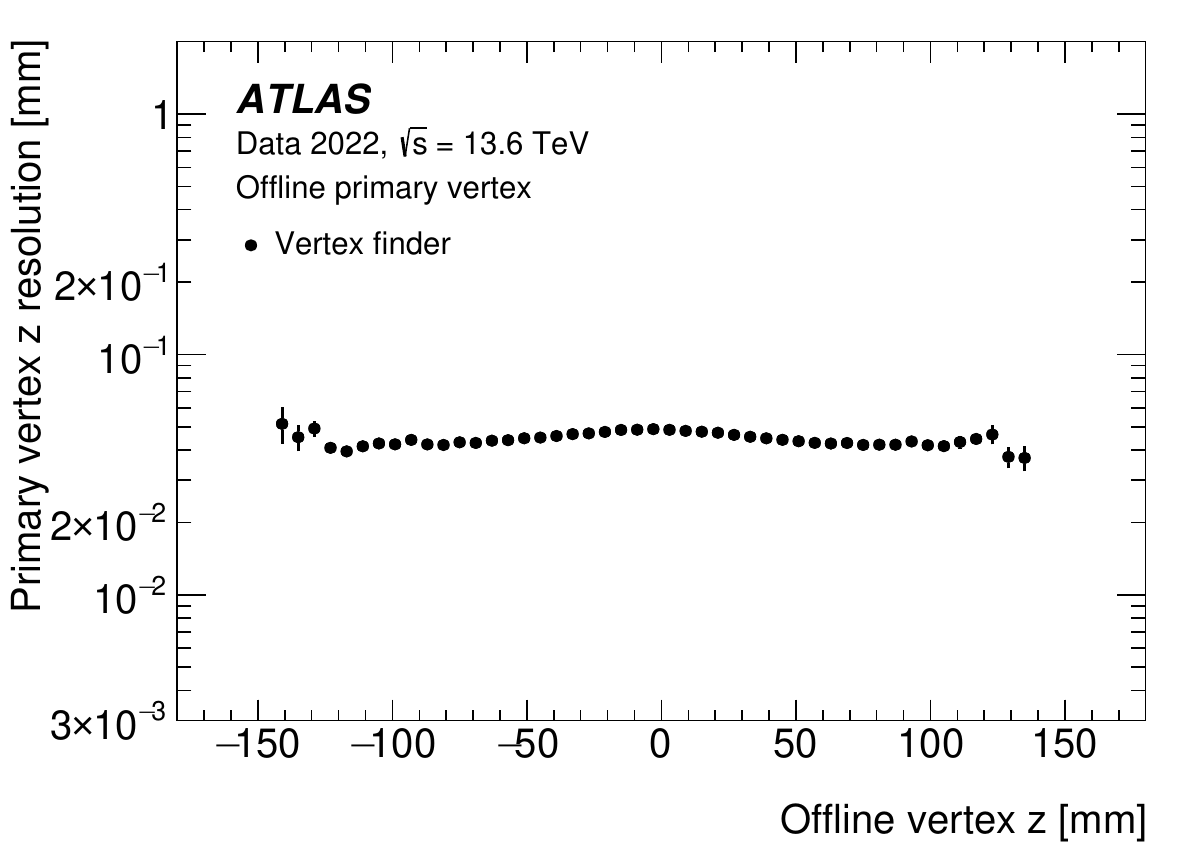}
\caption{The resolution of the primary trigger vertex with respect to the primary offline vertex (left) track multiplicity and (right) $z$ position.
The efficiency is evaluated for the 45~\GeV\ jet trigger.
Only statistical uncertainties are shown.
}
\label{fig:vtx:res}
\end{figure}
 
The resolution of the reconstructed $z$ position of the vertex is shown in Figure~\ref{fig:vtx:res},
illustrating the dependence on the number of tracks from which the vertex is constructed.
The resolution is approximately 1$\,$mm for very low track
multiplicities and can have values as low as 40$\,\mu$m at higher multiplicities.
The resolution is between 40 and 50$\,\mu$m over the full $z$ range.
 
\subsubsection{Tracking in the jets containing $b$-hadrons}
\label{sec:id_bjet}
 
\begin{figure}[tp]
\includegraphics[width=0.49\textwidth]{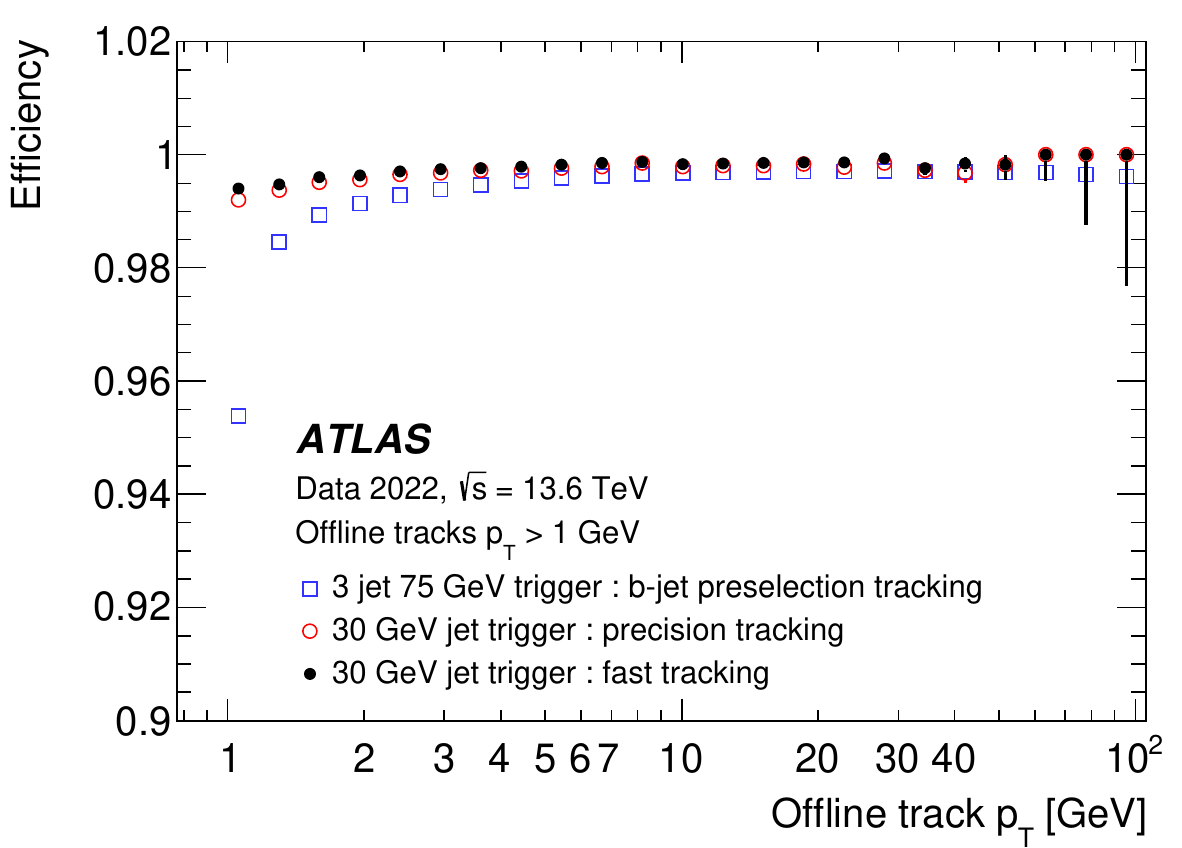}
\includegraphics[width=0.49\textwidth]{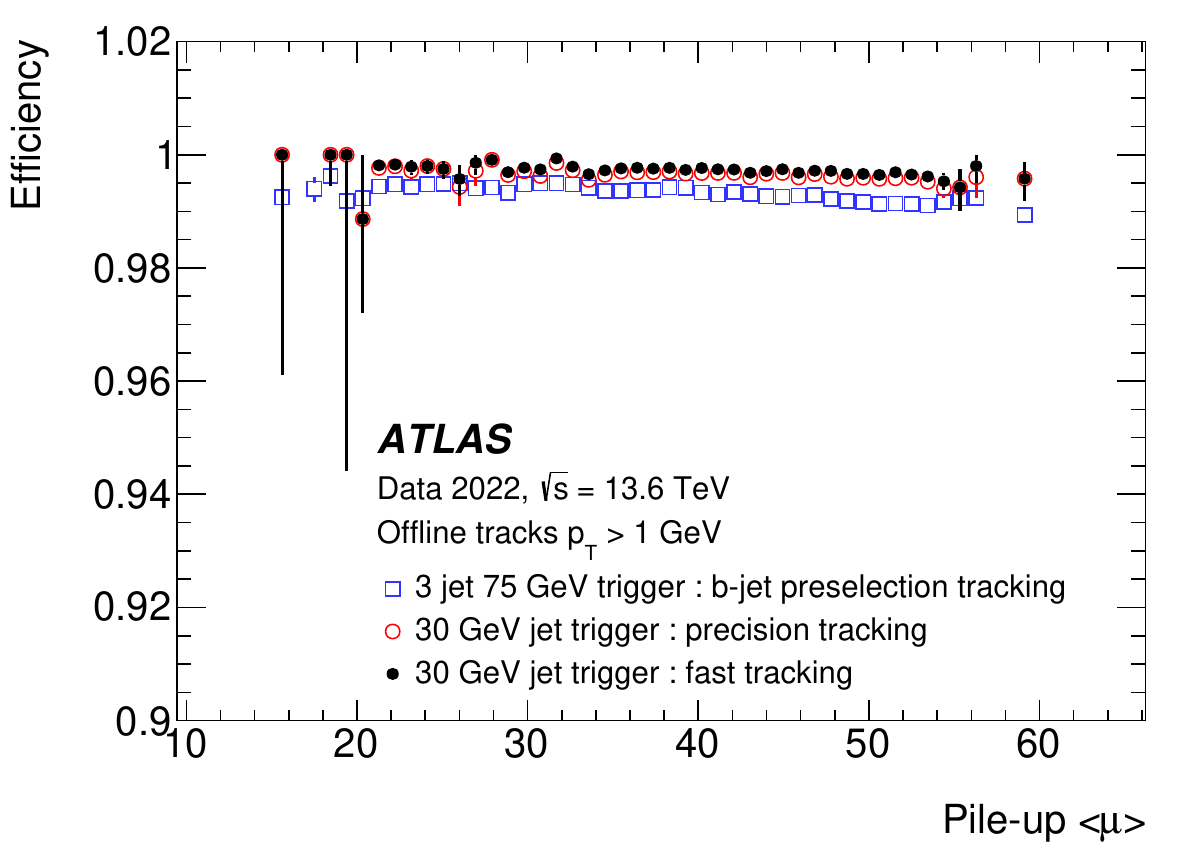}
\caption{The performance of the tracking for the $b$-jet signature versus (left) the offline track \pt and (right) the average pile-up.
Efficiencies with respect to offline tracks are shown for $b$-jet \preselection stage as well as for
the $b$-jet \roi fast and precision tracking stages. Only statistical uncertainties are shown.
}
\label{fig:bjets:eff}
\end{figure}
 
To reduce the rate of full scan tracking for $b$-jet triggers in Run 3, a calorimeter-only jet-finding preselection step is
implemented, followed by a $b$-tagging preselection stage running the fast tracking.
This instance of the fast tracking uses the same optimisations as the full scan tracking but executes in
a single composite RoI (super-RoI), which is constructed from all regions centred around
$|\Delta\eta|<0.4$, $|\Delta\phi| < 0.4 $ with respect to the jet axes for all jets with \et$>20\,$\gev\, identified in the jet preselection stage.
This \roi is extended along the beamline by $|z|<150$\,mm.
A fast $b$-tagging algorithm is then applied to the reconstructed tracks, as detailed in Section~\ref{sec:bjets} and Ref.~\cite{FastBjetPaperInProduction_ANA-TRIG-2022-03}.
Shown in Figure~\ref{fig:bjets:eff},
for the $b$-tagging \preselection stage the efficiency is only 95\% at the 1~\gev\ threshold
and does not reach the plateau until approximately 5~\gev. The \preselection tracking efficiency also shows
a slight reduction with increasing pile-up.
 
Following this preselection, after the rejection of events where no $b$-tag is found,
the full scan tracking is executed in the full ID volume to determine the primary vertex position
and reconstruct the tracks used for the particle flow (PFlow) jet reconstruction.
The need to reduce the processing time for the full scan tracking, described in the previous section, necessarily compromises the efficiency
to some degree. While this may be acceptable for the PFlow reconstruction,
for which it is predominantly intended, this would not be desirable for
the final $b$-tagging where the best possible efficiency for individual tracks is required.
Consequently, as in \runii, the standard RoI-based fast and precision track reconstruction is executed
in a separate \roi for each PFlow jet, centred on the jet direction
and with the \roi $z$-position determined from the primary vertex information.
These precision tracks are then used for the full $b$-tagging, described in Section~\ref{sec:bjets}.
For \runiii, the track \pt requirement used in the pattern recognition for the $b$-jet tracking stage is reduced to 0.8\,\GeV\ from the 1\,\GeV\ used in \runii.
As shown in Figure~\ref{fig:bjets:eff}, the \roi-based fast and precision tracking results in higher efficiencies for the tracks from $b$-jets.
The efficiency for the precision $b$-jet tracking is better than 99\% over the full phase space,
an improvement on the 84\% in the $1.0-1.2\,$\gev\ range seen in \runii~\cite{TRIG-2019-03}.


\subsubsection{Large radius tracking}
\label{sec:lrt}
 
\Ac{LRT} is a new feature introduced in the HLT for \runiii. It uses the same algorithms as the previously described standard tracking, but with modified configurations in order to reconstruct tracks at large radii or
impact parameter ($d_0$). These configurations are based on the offline \ac{LRT} reconstruction~\cite{IDTR-2021-03},
which was improved for Run 3 to reduce the number of fake tracks and processing time.
As with the standard tracking, \ac{LRT} is split into fast and precision track reconstruction steps, which can be performed either inside \rois or in the entire \ac{ID} volume (full scan \ac{LRT}).
In the case of \roi-based \ac{LRT}, e.g.\ for leptons, the \ac{LRT} is run by itself.
For the full scan \ac{LRT}, the tracking uses the remaining hits after a standard tracking pass, as is done for the offline \ac{LRT}.
 
The standard seeding step~\cite{TRIG-2019-03} is modified as follows: only hits from the SCT are used and the ordering of seeds by impact parameter is removed.
Tracking is expanded to cover $|z_0|<500$~mm and $|d_{0}|<300$~mm, but with generally stricter requirements on track quality to reduce the number of fake tracks, such as requiring at least eight hits on the track.
This limits tracks to originate before the first SCT layer in the barrel region, at a radius of approximately 300~mm.
The same momentum threshold of $\pt>1~\GeV$ as standard tracking is used.
For \ac{LRT} in electron and muon \rois, the size of the \roi is expanded in both $\eta$ and $\phi$, compared to the tracking for prompt leptons, in order to accommodate tracks that do not point to the beamline.
The size of the \roi in $\phi$ is a limiting factor in the efficiency to reconstruct tracks at large $d_{0}$, and a trade-off has to be made between tracking acceptance and computing cost.
In order to reduce the processing time for full scan \ac{LRT}, tracking is restricted to $|d_{0}|>2$~mm, tracks are not extended into the TRT, and track candidates are required to have $\pt>1~\GeV$ (on top of the $\pt$ requirement on the seeds).
 
\begin{figure}[htbp]
\centering
\includegraphics[width=0.49\textwidth]{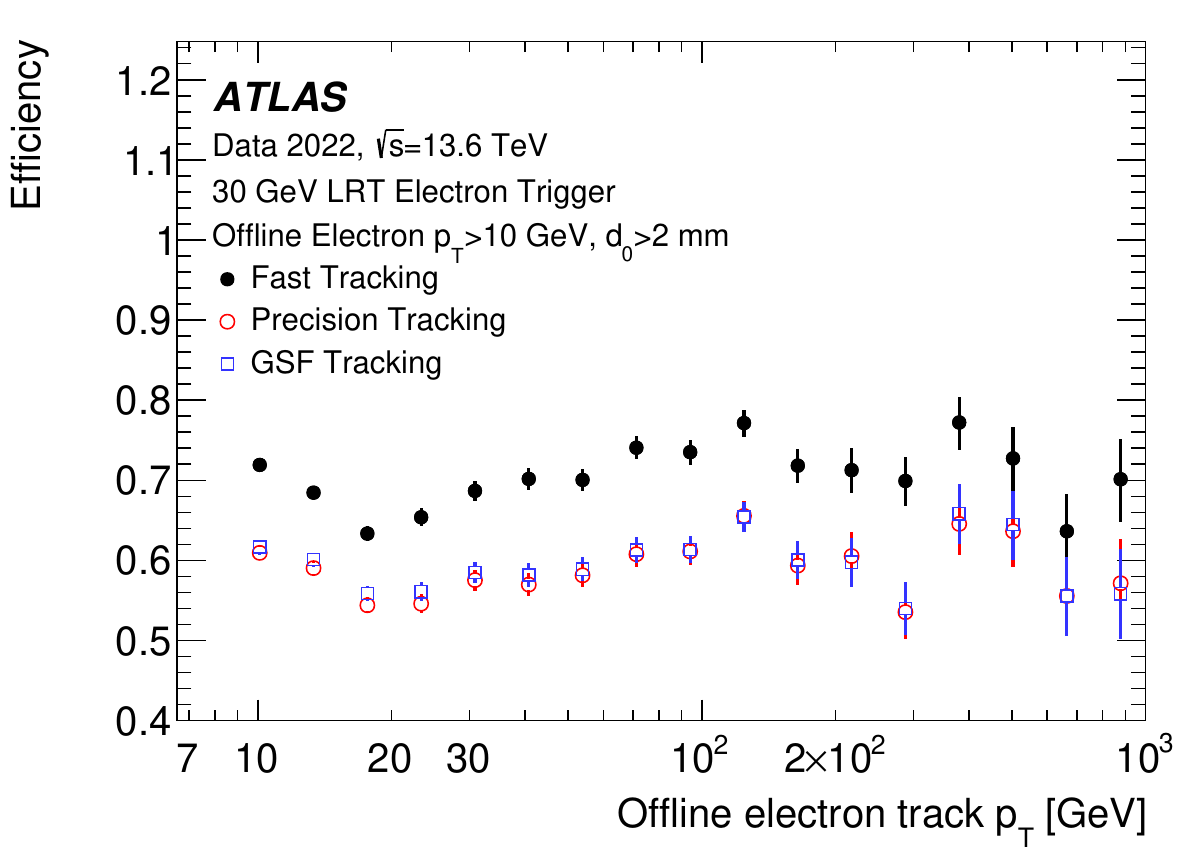}
\includegraphics[width=0.49\textwidth]{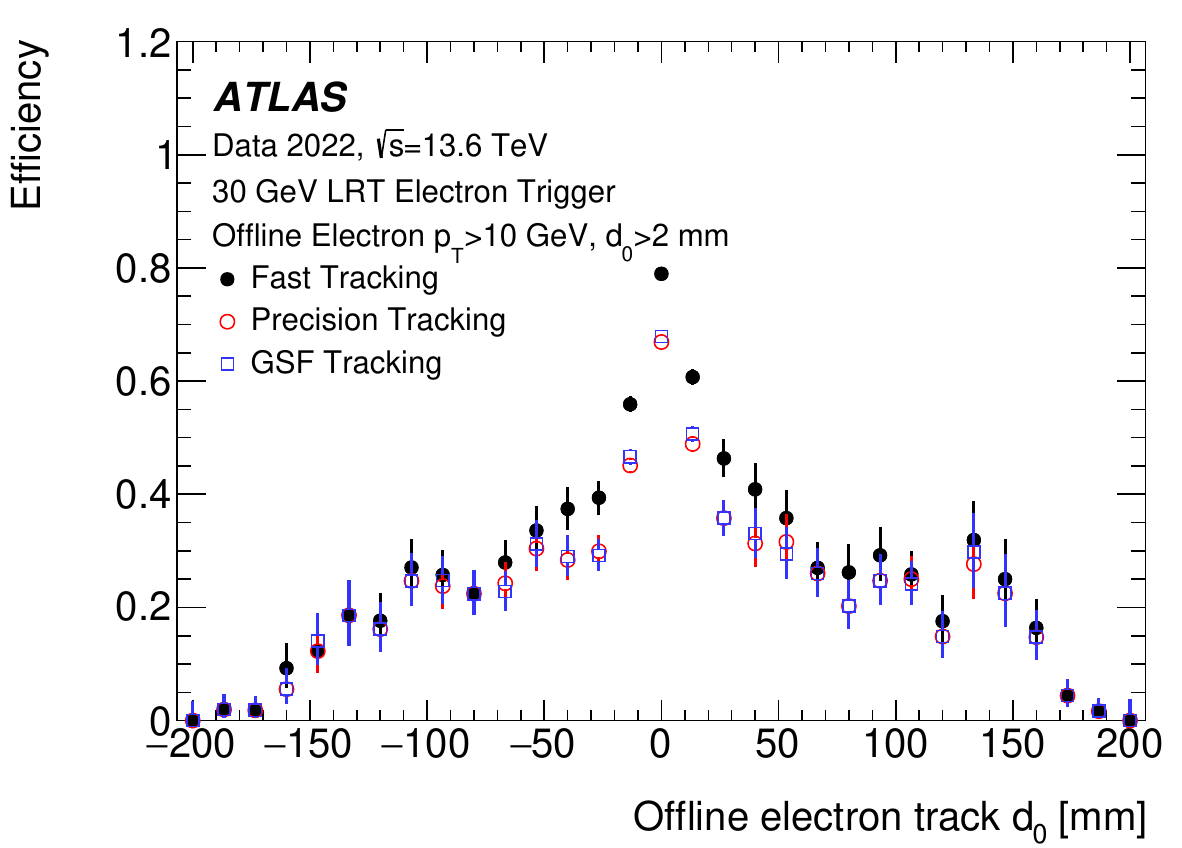}
\includegraphics[width=0.49\textwidth]{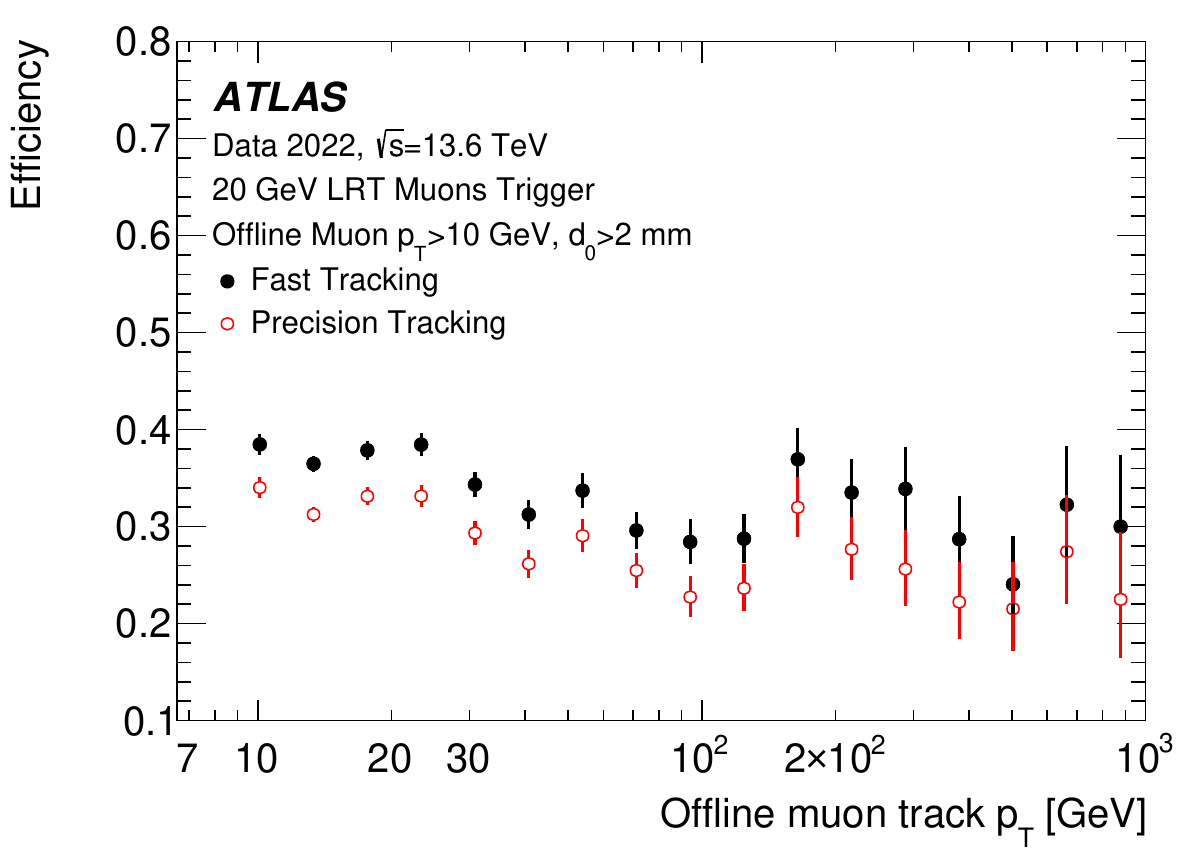}
\includegraphics[width=0.49\textwidth]{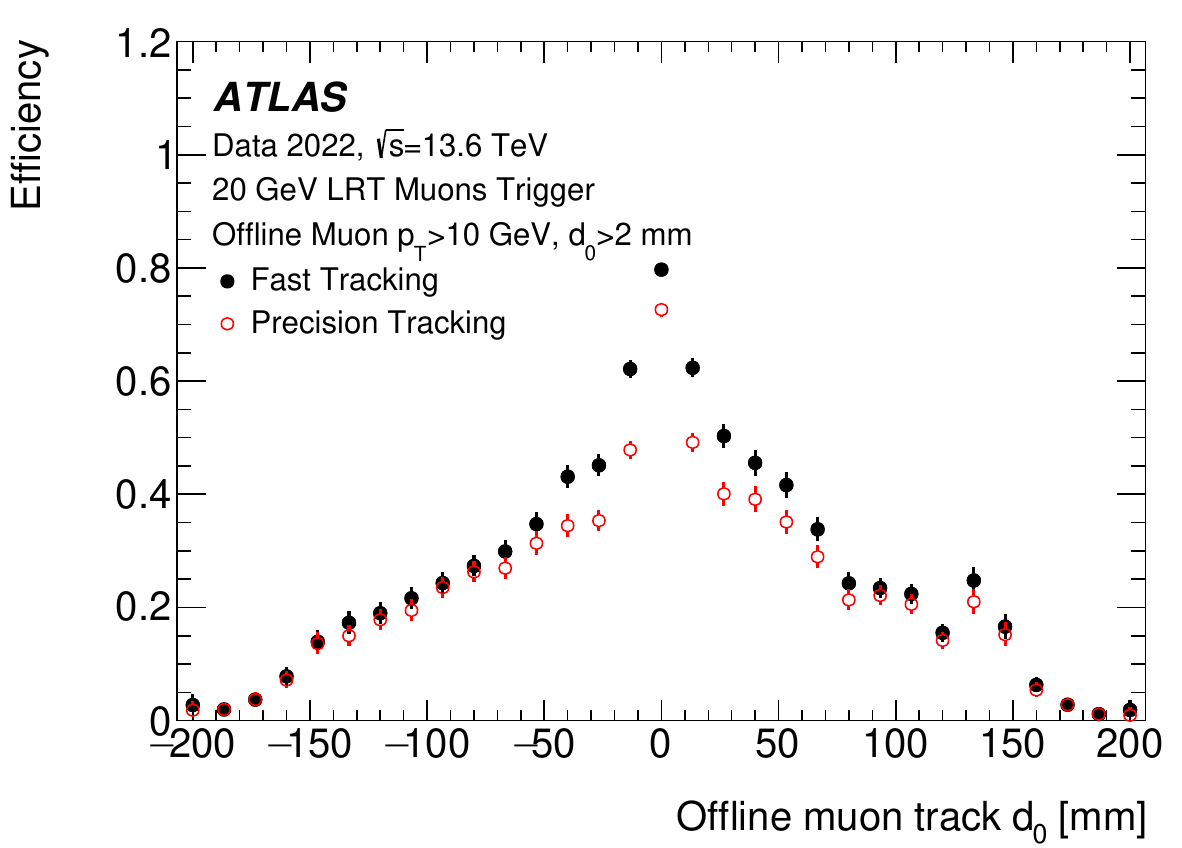}
\caption{Fast and precision LRT efficiencies for (top) electron and (bottom) muon triggers
as a function of offline electron and muon (left) track \pT\ and (right) $d_{0}$.
The corresponding merged collections of standard and large radius offline electron and muon tracks are used. The efficiency for the additional GSF step for electrons is also included. Only statistical uncertainties are shown.}
\label{fig:id:lrt_elmutau_perf}
\end{figure}
 
Figure~\ref{fig:id:lrt_elmutau_perf} shows the performance of \ac{LRT} in electron and muon trigger RoIs with respect to offline electrons and muons, respectively. 
No additional identification requirements are applied beyond the reconstruction.
Data were collected during 2022  using a trigger that does not apply any selection based on the tracking, described in Section~\ref{sec:idperf}. The offline lepton tracks are required to have at least eight silicon hits, $|d_0|>2$~mm, and $\pt>10~\GeV$.
Leptons using both the standard offline track reconstruction and large radius offline track reconstruction are used as the reference.
There is a small overlap where standard and \ac{LRT} tracks matched to the same calorimeter cluster or muon spectrometer segment that is not removed here.
Since there are very few \ac{SM} processes that produce displaced tracks, the tracks in these plots originate from a combination of photon conversions, heavy-flavour hadron decays, long-lived neutral kaon decays, and combinations of hits resulting in fake tracks, which are not necessarily representative of high momentum leptons.
 
The efficiencies in data are similar to that of tracks in \ttbar\ \ac{MC} simulation, while the efficiency for leptons in \ac{LLP} signal \ac{MC} simulation is much higher.
Figures~\ref{fig:id:lrt_el_mc_perf} and \ref{fig:id:lrt_mu_mc_perf} overlay the efficiency of electron GSF- and muon precision-tracking with respect to offline electrons or muons for these samples.
Tracking in the trigger for signal-like tracks is much more efficient out to large impact parameter values and less dependent on the amount of pile-up.
 
\begin{figure}[htbp]
\centering
\includegraphics[width=0.49\textwidth]{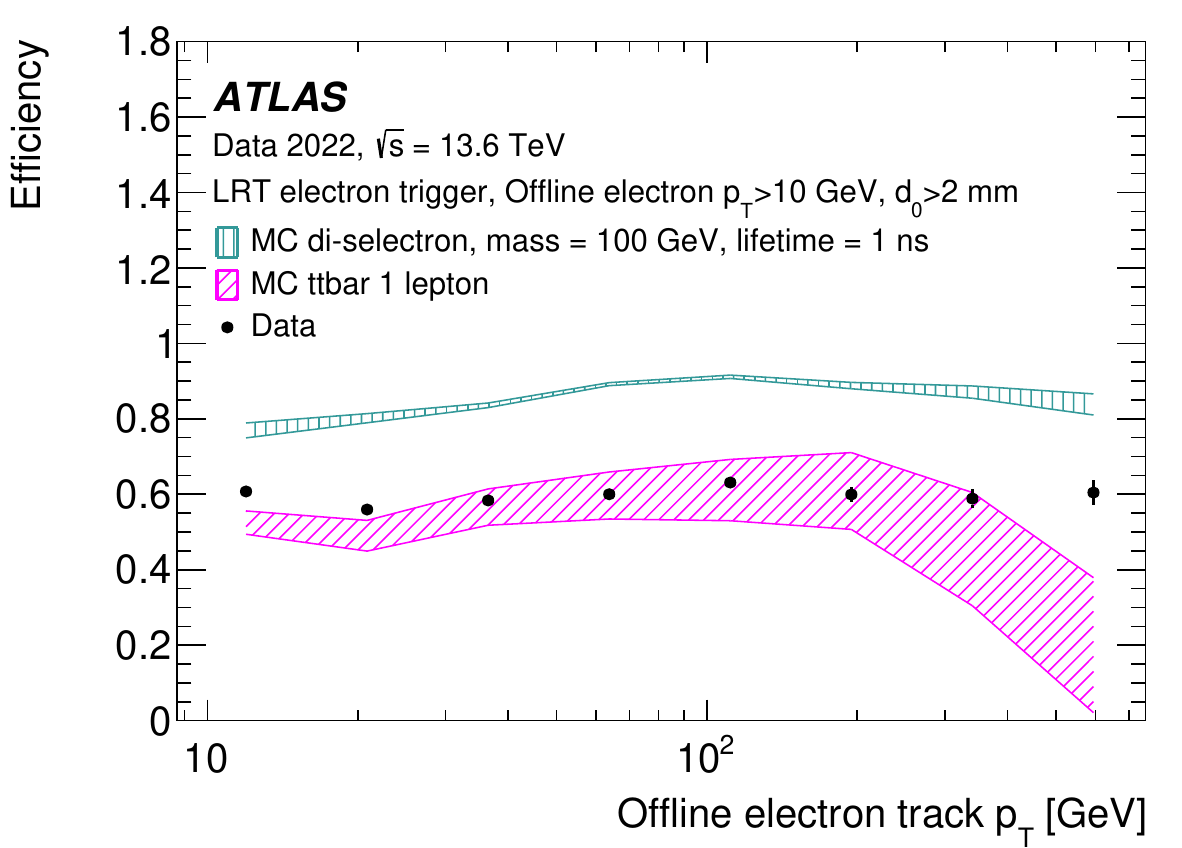}
\includegraphics[width=0.49\textwidth]{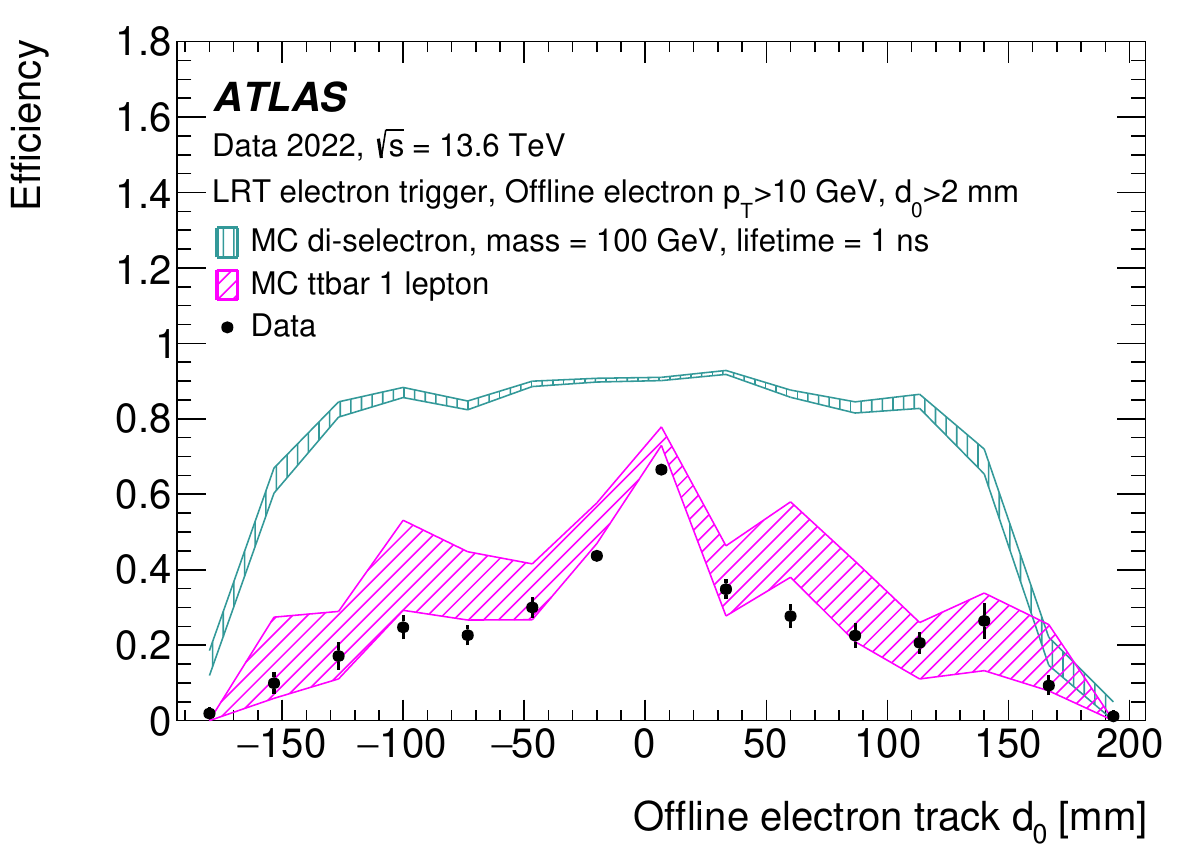}
\includegraphics[width=0.49\textwidth]{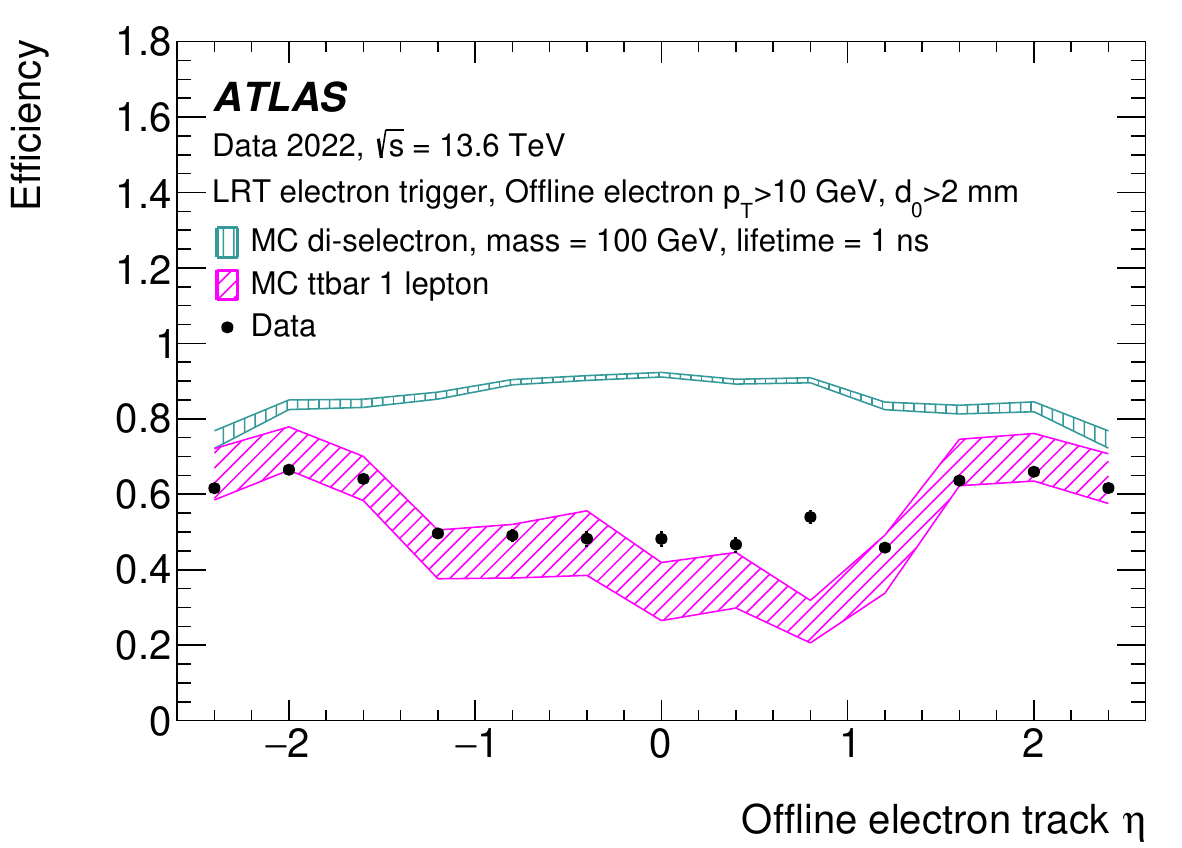}
\includegraphics[width=0.49\textwidth]{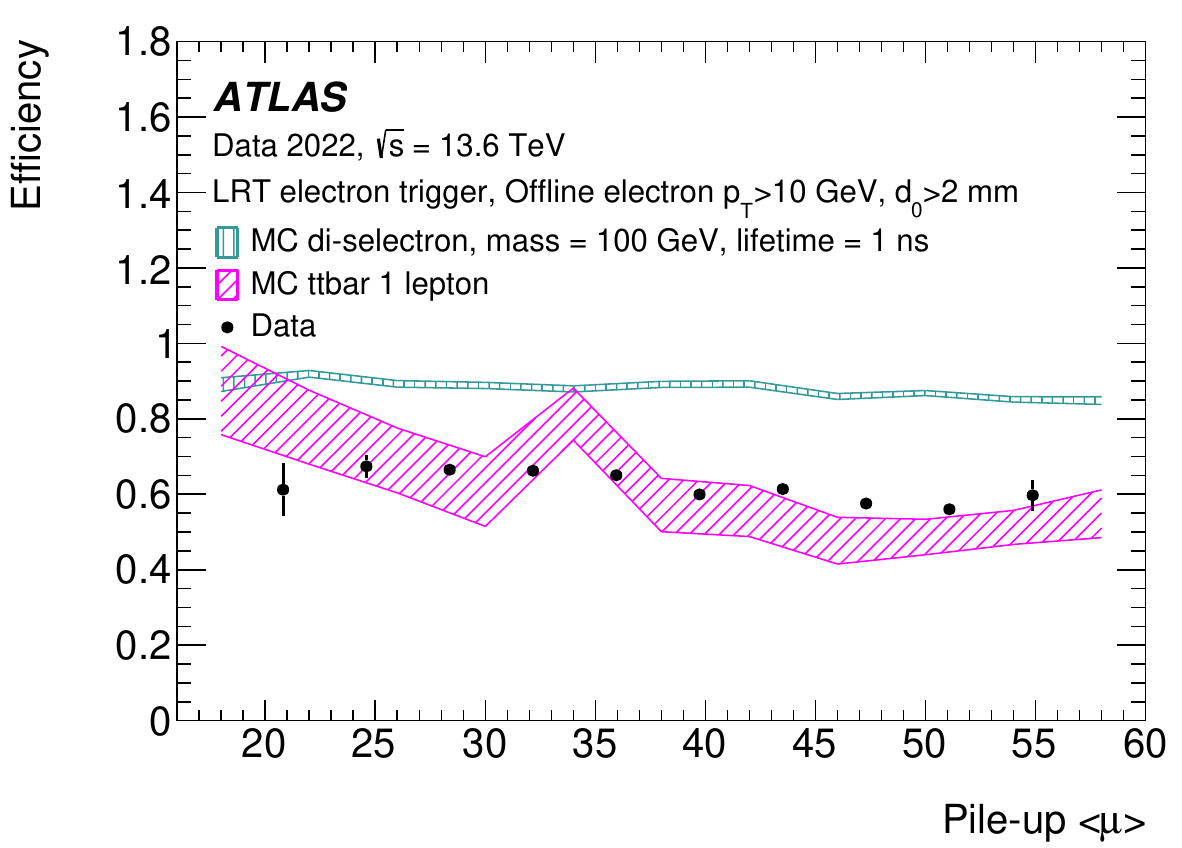}
\caption{The GSF tracking efficiency for the LRT electron trigger
with respect to the merged standard and large radius offline electron track collections
versus (top left) $\pt$, (top right) $d_{0}$, (bottom left) $\eta$, and (bottom right) the average pile-up
for data, semi-leptonic $\ttbar$ MC events, and simulated pair production of selectrons with
a 1\,ns lifetime and a mass of 100\,\gev. Only statistical uncertainties are shown.}
\label{fig:id:lrt_el_mc_perf}
\end{figure}
 
\begin{figure}[htbp]
\centering
\includegraphics[width=0.49\textwidth]{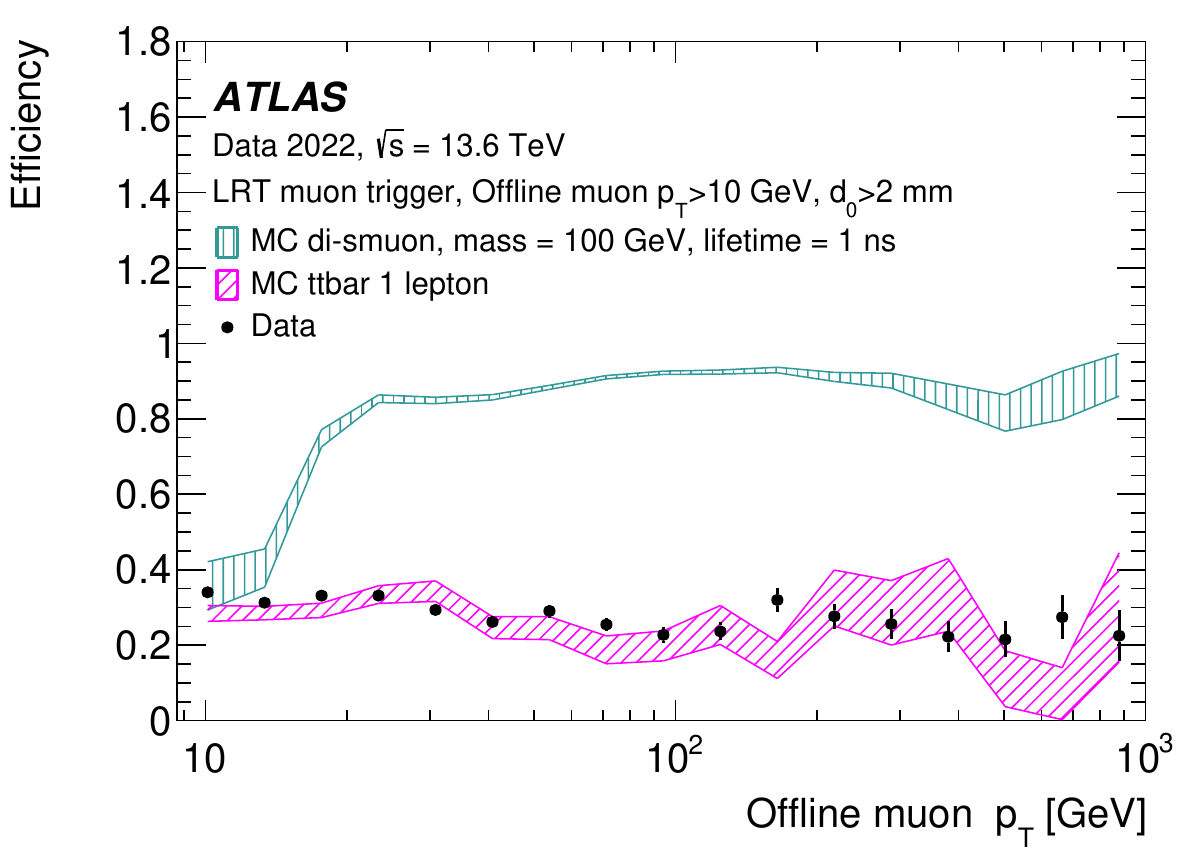}
\includegraphics[width=0.49\textwidth]{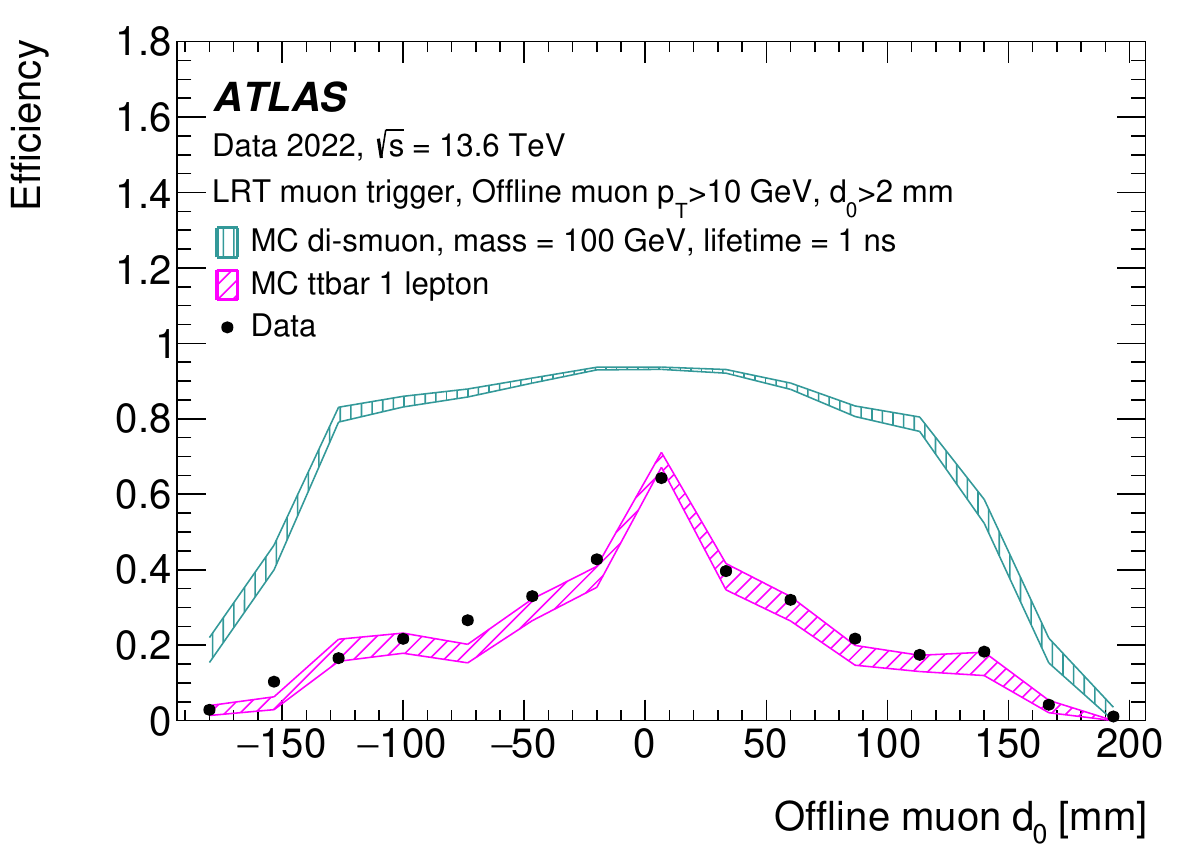}
\includegraphics[width=0.49\textwidth]{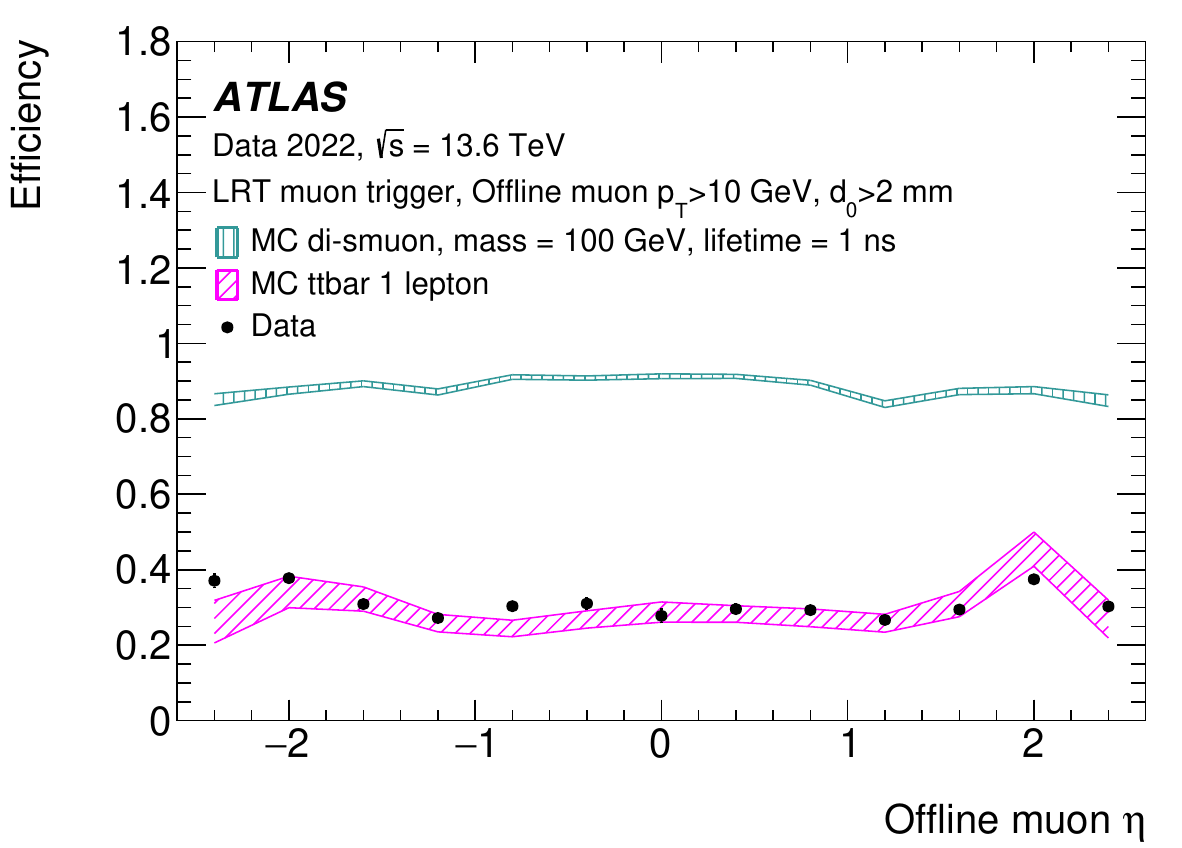}
\includegraphics[width=0.49\textwidth]{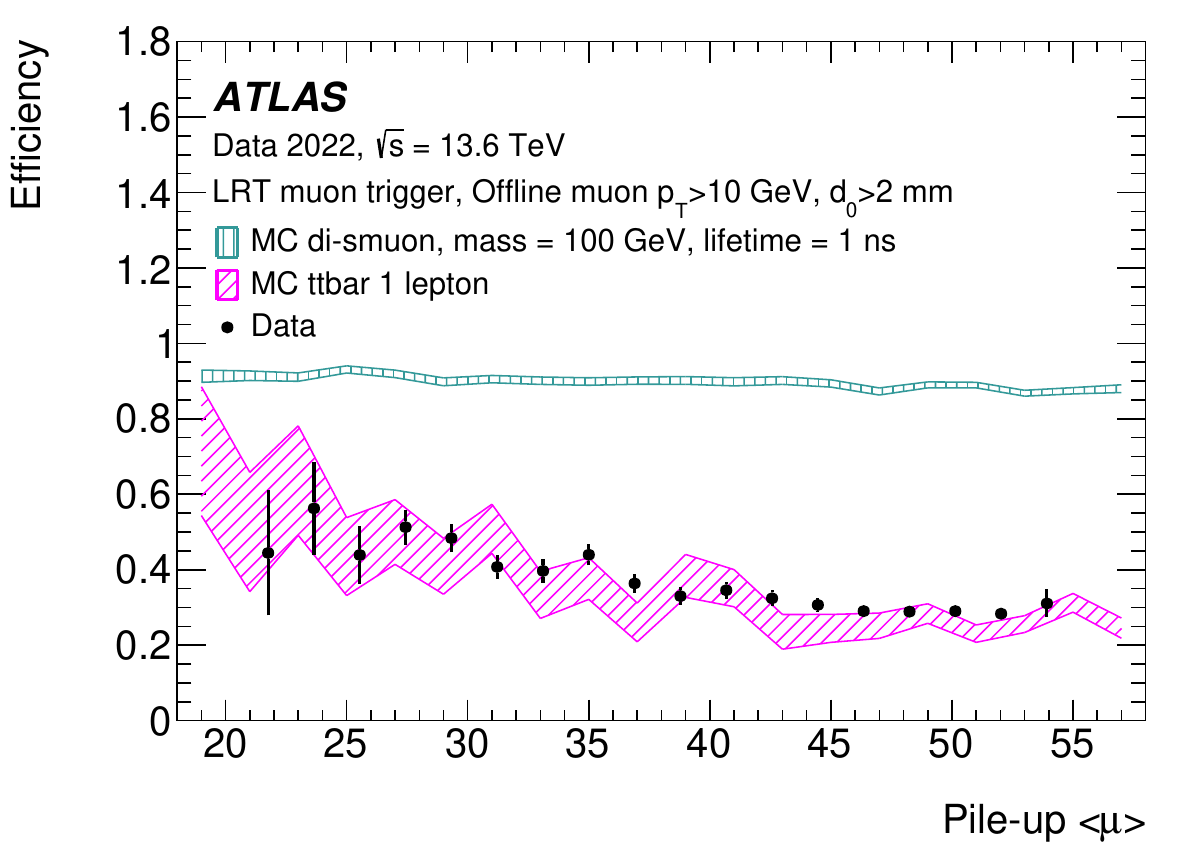}
\caption{The precision tracking efficiency for the LRT muon trigger with respect to the merged standard and large radius offline muon track collections
versus (top left) $\pt$, (top right) $d_{0}$, (bottom left) $\eta$, and (bottom right) the average pile-up
for data, semi-leptonic $\ttbar$ MC events, and simulated pair production of smuons with
a 1\,ns lifetime and a mass of 100\,\gev. Only statistical uncertainties are shown.}
\label{fig:id:lrt_mu_mc_perf}
\end{figure}
 
Data events collected in one run of 2022 were reprocessed to run the full scan trigger tracking configuration of \ac{LRT} for every event.
A collection of offline $K^0_S$ candidate vertices is produced by selecting both standard and \ac{LRT} offline tracks with $\pt>1~\GeV$
and opposite-charge that form a vertex with a mass within 25~\MeV\ of 497~\MeV~\cite{STDM-2010-09}.
The offline tracks associated to the $K^0_S$ candidates are then matched to online standard full scan tracks if they fall within $\Delta R < 0.05$ and $\Delta d_{0} < 2.5~$mm of each other.
Remaining offline tracks are used in the denominator to compute the online \ac{LRT} efficiency.
Figure~\ref{fig:id:lrt_fs_perf} shows the efficiency with respect to these remaining offline tracks versus the \pT\ and $d_0$ of the offline track.
The tracks from $K^0_S$ decays tend to have low \pT\ and the number of offline tracks falls off rapidly after a \pT\ of 5\,\GeV\ and a $|d_0|$ of 80~mm.
Due to the steeply falling \pT\ spectrum of the $K^0_S$ decay products, low momentum particles with poor \pt\
resolution make up a significant proportion of the high \pt\ offline reference sample. Many of these are not
reconstructed by the trigger leading to an apparent decrease in trigger efficiency at large \pt\ which is not representative of the actual efficiency for higher \pt\ particles.
The bottom of Figure~\ref{fig:id:lrt_fs_perf} shows the efficiency of matching both tracks of the offline vertices to online tracks, including standard and \ac{LRT}, versus the reconstructed radius of the $K^0_S$ vertex. The efficiency drops at radii corresponding to barrel layers of the Pixel detector, e.g.\ 88.5~mm and 122.5~mm.
 
\begin{figure}[htbp]
\centering
\includegraphics[width=0.49\textwidth]{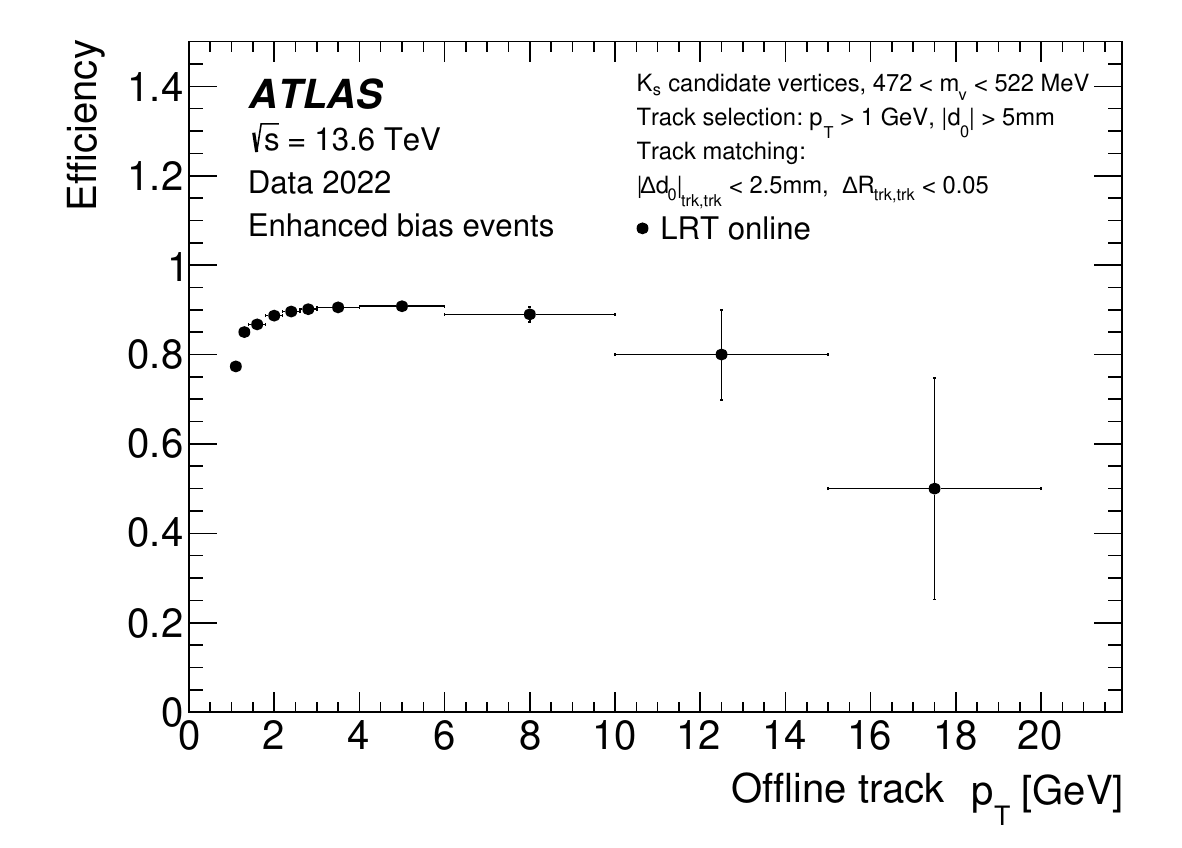}
\includegraphics[width=0.49\textwidth]{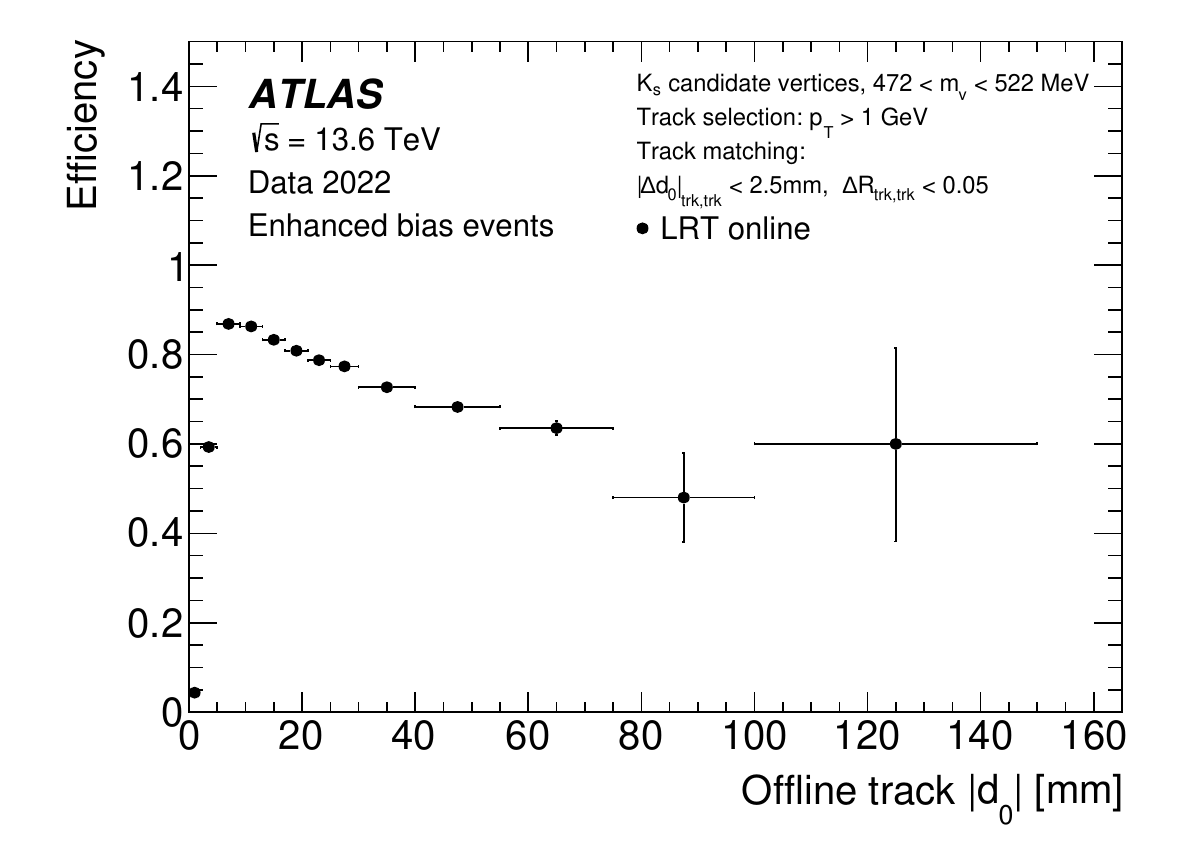}
\includegraphics[width=0.49\textwidth]{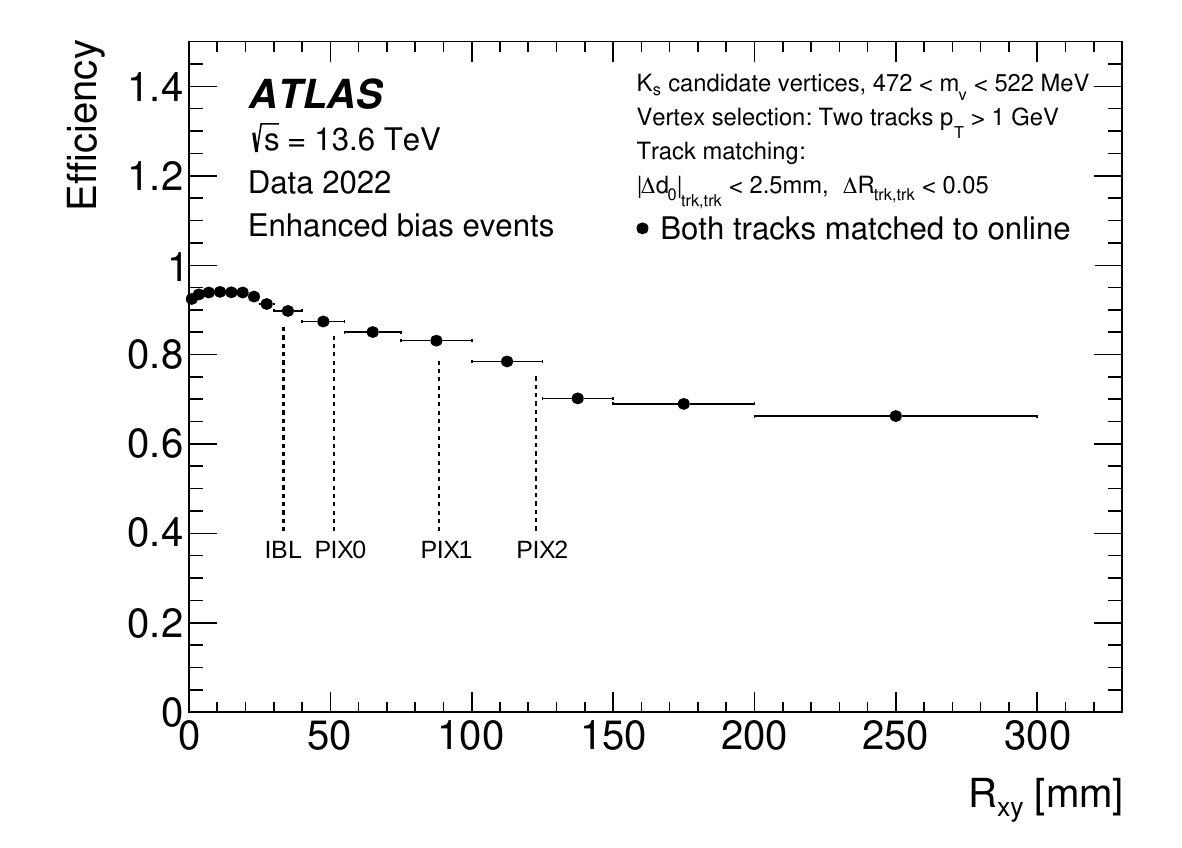}
\caption{Efficiency of online full scan \ac{LRT} with respect to offline tracks associated with $K^0_S$ candidates that have not been matched to standard online full scan tracks.
This efficiency is shown as a function of (left) the \pT\ and (right) $d_{0}$ of the offline tracks.
An additional requirement of $|d_0|>5$~mm is used when computing the efficiency versus \pt\ to remove tracks with low displacement, which are expected to be reconstructed by the standard tracking.
The bottom plot shows the efficiency, combining both standard and \ac{LRT}, to reconstruct both tracks of the offline vertex versus the radius of the vertex. Only statistical uncertainties are shown.}
\label{fig:id:lrt_fs_perf}
\end{figure}


\subsubsection{Execution time}
\label{sec:execTimeTracking}
\newcommand{\statonly}{Only statistical uncertainties are shown.}
 
The ID track reconstruction in the HLT  
comprises 59\% of the total event processing time in the trigger as detailed further
in Table~\ref{tab:hltprocessing_time_grouped} of Section~\ref{sec:menuPerf}.
The fast tracking is typically the most time-consuming algorithm due to the combinatorial nature of the pattern recognition stage which has a non-linear dependence on pile-up.
The precision tracking timing has a much smaller dependence on pile-up
since it depends only on the number of tracks passed to it from the fast tracking.
The precision tracking also includes hits from the TRT, which are obtained by the extension
of the tracks into the TRT before the final fit is performed.
 
\begin{figure}[thp]
\centering
\includegraphics[width=0.49\textwidth]{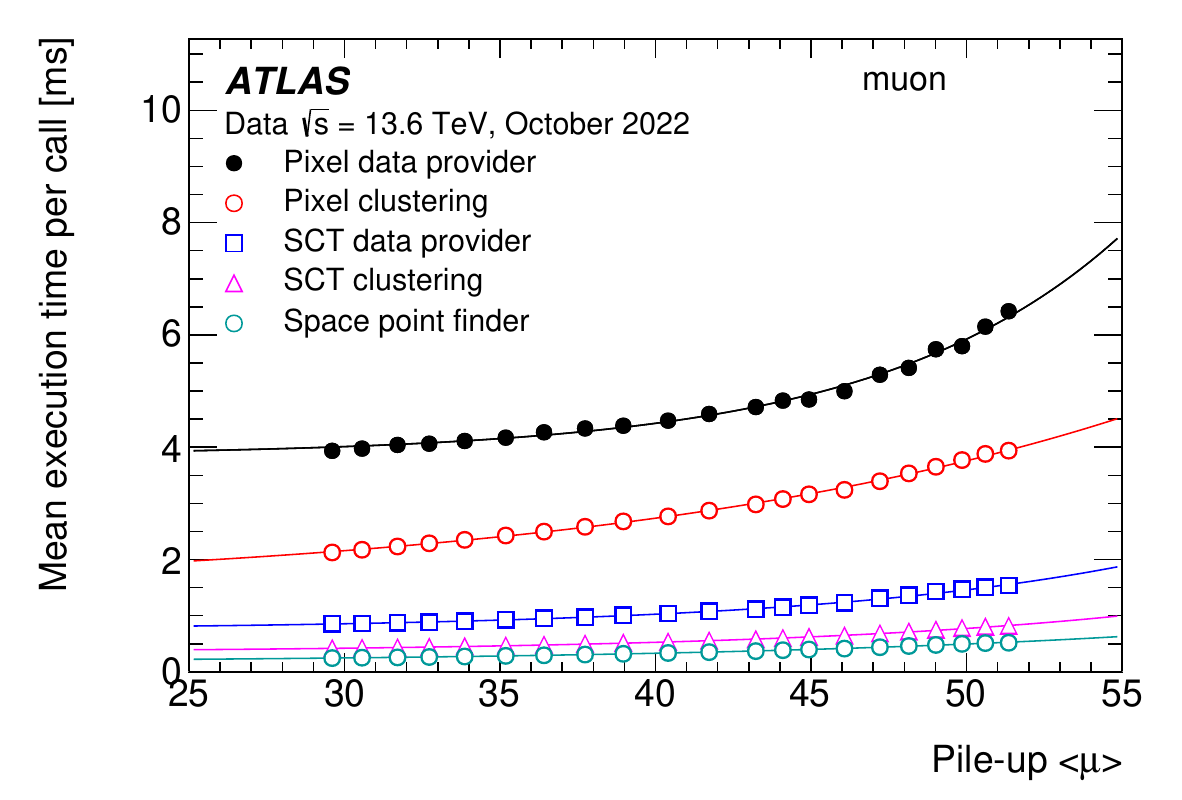}
\includegraphics[width=0.49\textwidth]{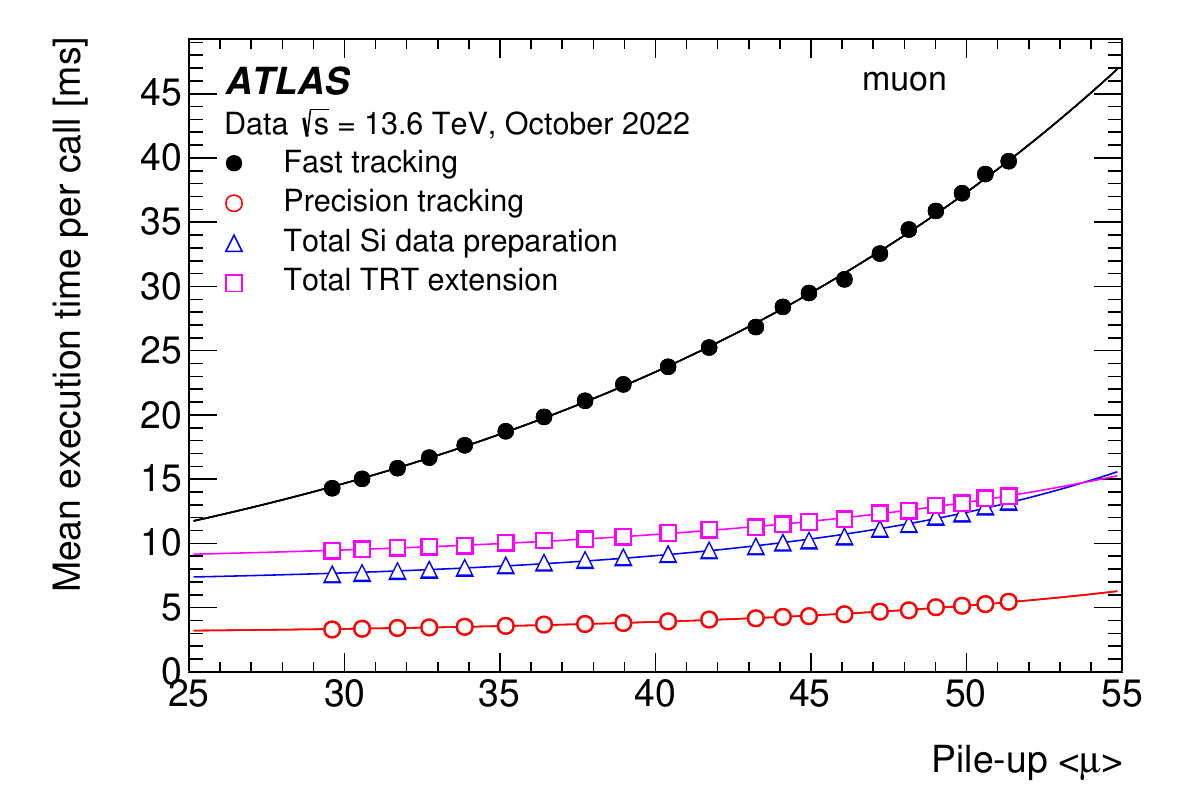}
\caption{Execution times for the muon track reconstruction in the trigger. Shown are the times for (left) the
pixel and SCT data preparation, and (right) the muon fast, and precision tracking, together
with the combined data preparation for pixel and SCT, and the total time spent in the TRT extension,
all as a function of the average pile-up. Solid lines show ad hoc fits to data to guide the eye. \statonly }
\label{app:fig:id:muon_cost}
\end{figure}
 
The timing measurements presented here were recorded from the HLT farm,
during a typical physics run from October 2022.
Figure~\ref{app:fig:id:muon_cost} shows the mean execution times for the ID trigger-related algorithms for the muon reconstruction
as a function of the \pileup interaction multiplicity.
Shown are the data preparation times for the silicon detectors prior to the fast tracking, and the execution times for the
fast, and precision tracking themselves together with the combined silicon data preparation time and the total time for the
full TRT extension used for the precision tracking. The combined pixel and SCT data preparation, at 12\,ms at high \pileup is slightly faster than
the full TRT extension, at approximately 13\,ms. The slowest component from the silicon data preparation is the pixel data
provider which fetches the pixel data from the pixel read-out. The slowest component overall, taking approximately 40\,ms
per \roi at high \pileup is the fast tracking, which exhibits a non-linear dependence on the \pileup multiplicity due to the
combinatorial nature of the pattern recognition. The precision tracking takes only approximately 6\,ms at high \pileup in the
muon \roi.
 
The timing of the {\em muonIso} tracking, shown in Figure~\ref{figs:muonIso_cost}, is significantly improved compared to Run 2~\cite{TRIG-2019-03}.
With the reduced $z$ width, the muon isolation tracking algorithms take less than 25\,ms
in total per \roi at a pile-up of 52. The total combined data preparation for the pixel and SCT takes  less than 5\,ms, with the
longest components being the pixel and SCT data providers, which fetch the data across the network, with the pixel and
SCT clustering being somewhat faster.
The TRT extension
has a combined execution time of approximately 20 ms, in this case, longer that the other algorithms. In contrast, the
standard muon fast and precision tracking with $|z|<225$~mm takes around 40\,ms per \roi with approximately 15\,ms for the silicon
data preparation.
The fast tracking alone for the muon isolation in \runii typically exceeded 115\,ms per \roi at similar pile-up~\cite{TRIG-2019-03}.
 
\begin{figure}[tp]
\includegraphics[width=0.49\textwidth]{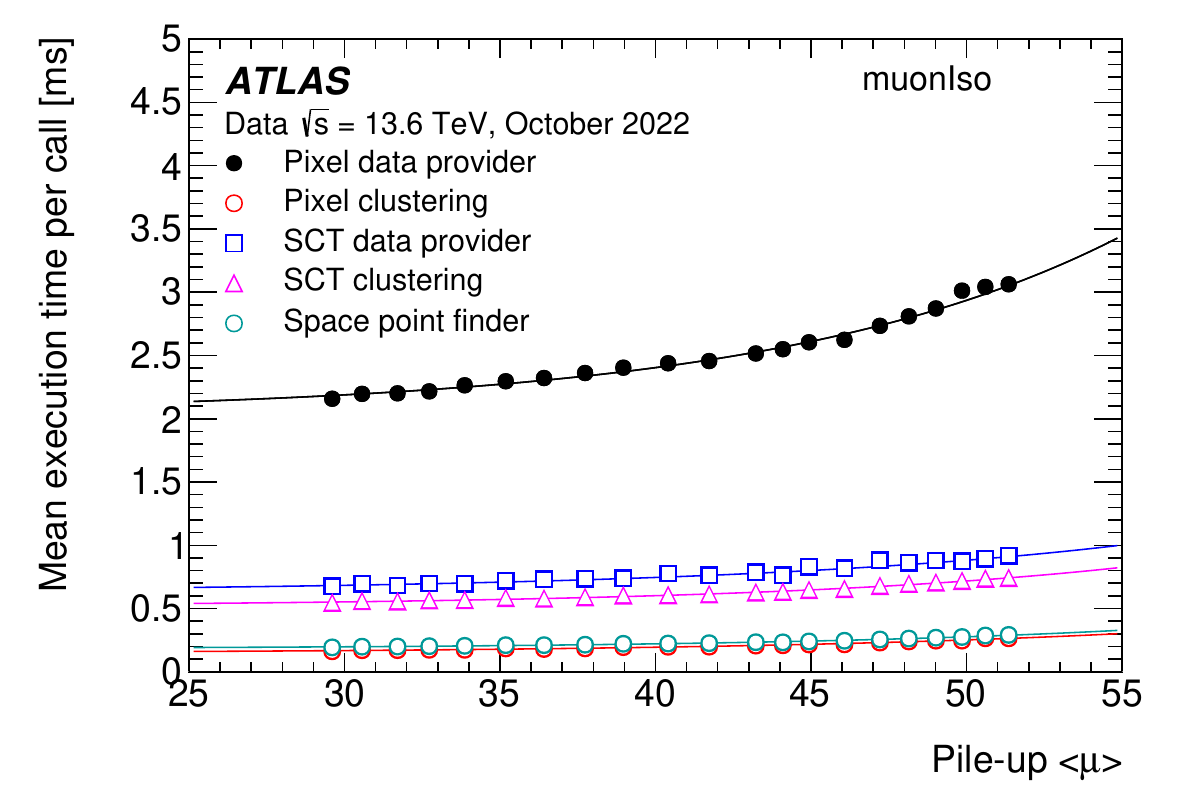}
\includegraphics[width=0.49\textwidth]{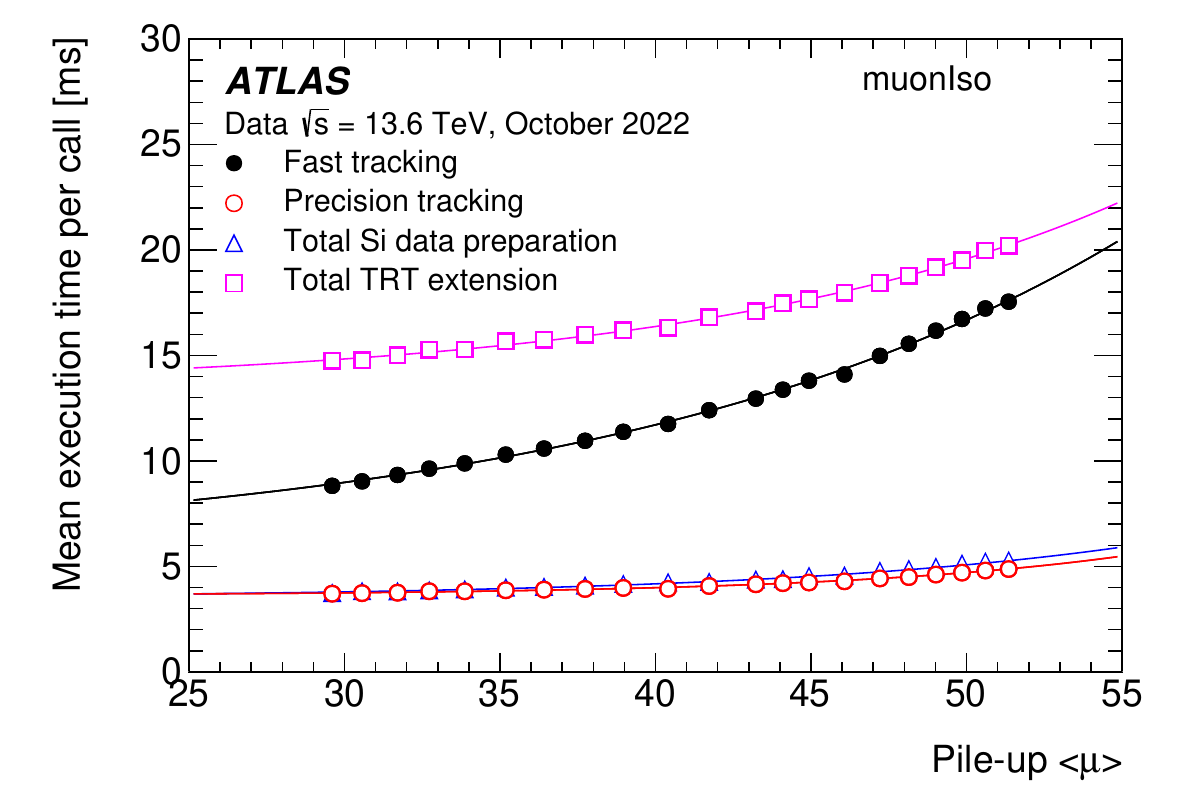}
\caption{Execution times for the muon isolation track reconstruction shown as a function of the average pile-up.
(left) Detailed breakdowns for the Pixel and SCT data preparation components.
(right) The muon isolation tracking, the combined silicon data preparation and TRT extension times.
Solid lines show ad hoc fits to data to guide the eye. \statonly }
\label{figs:muonIso_cost}
\end{figure}
 
\begin{figure}[tp]
\includegraphics[width=0.49\textwidth]{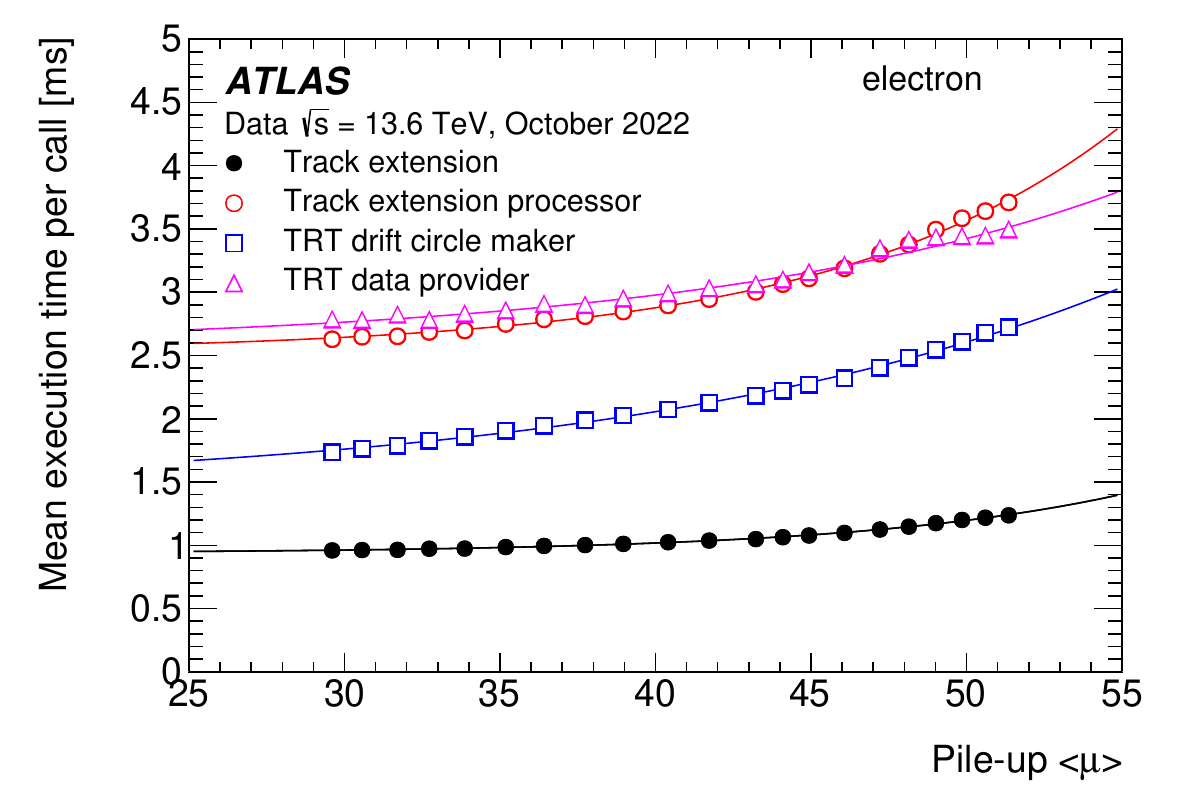}
\includegraphics[width=0.49\textwidth]{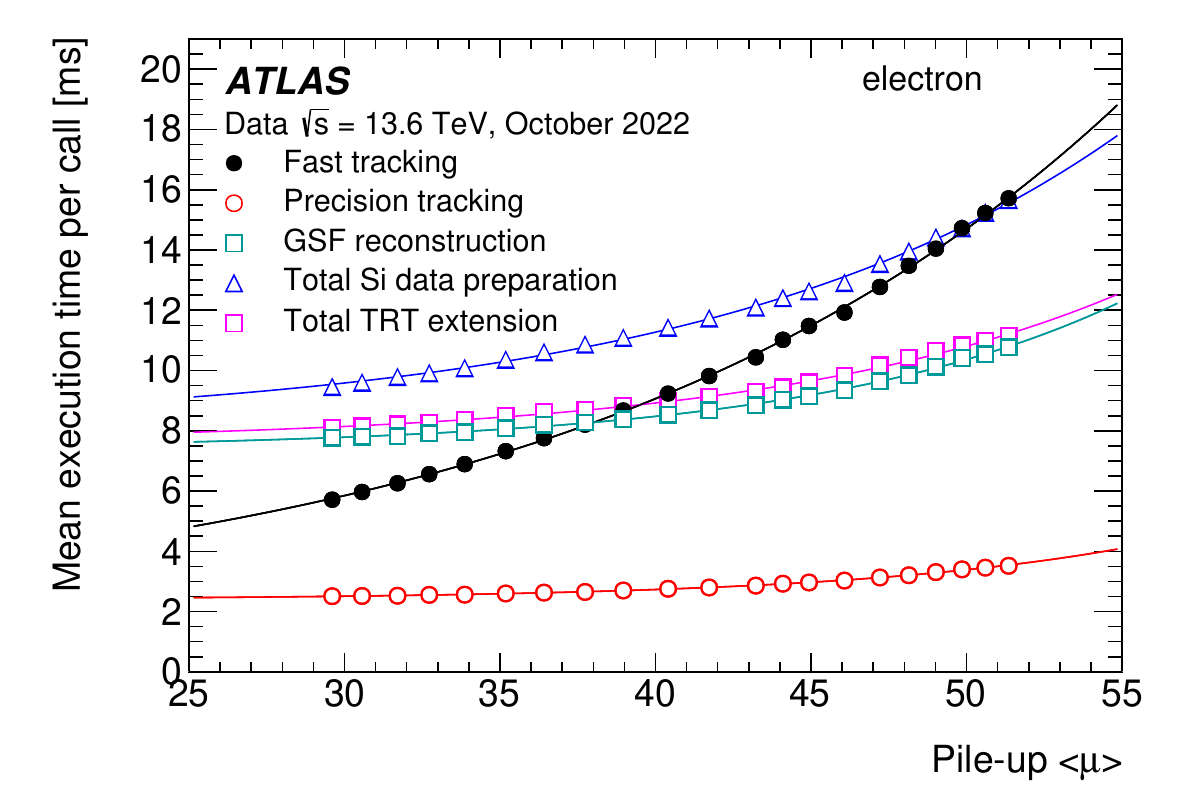}
\caption{Execution times for the (left) TRT extension and (right) tracking and data preparation for
the electron tracking, shown as a function of the average pile-up. Solid lines show ad hoc fits to data to guide the eye.
Only statistical uncertainties are shown.
}
\label{figs:elcost}
\end{figure}
 
The execution times for the electron-based algorithms are shown in Figure~\ref{figs:elcost}.
The small size of the electron \roi ($0.1\times 0.2$ in $\eta$--$\phi$ space) results in an execution time
for the electron pixel and SCT data preparation of less than 16\,ms at a pile-up of 52. 
The electron tracking takes an additional 20\,ms in total for the fast and precision tracking together.
The new GSF tracking takes approximately 11\,ms which is nearly three times slower than the precision
tracking due to the more complicated nature of the GSF reconstruction.
Since the GSF tracking is seeded by the tracks from the precision tracking, it has only a
weak dependence on the pile-up.
The extension of the electron tracks into the TRT and subsequent TRT data preparation and processing result in an additional 11\,ms per \roi,
as shown in Figure~\ref{figs:elcost} (left). The dominant contribution to the TRT processing is 4\,ms at high pile-up
from the TRT data provider, which includes the fetching of the TRT data from the ROS system.
 
\begin{figure}[tp]
\includegraphics[width=0.49\textwidth]{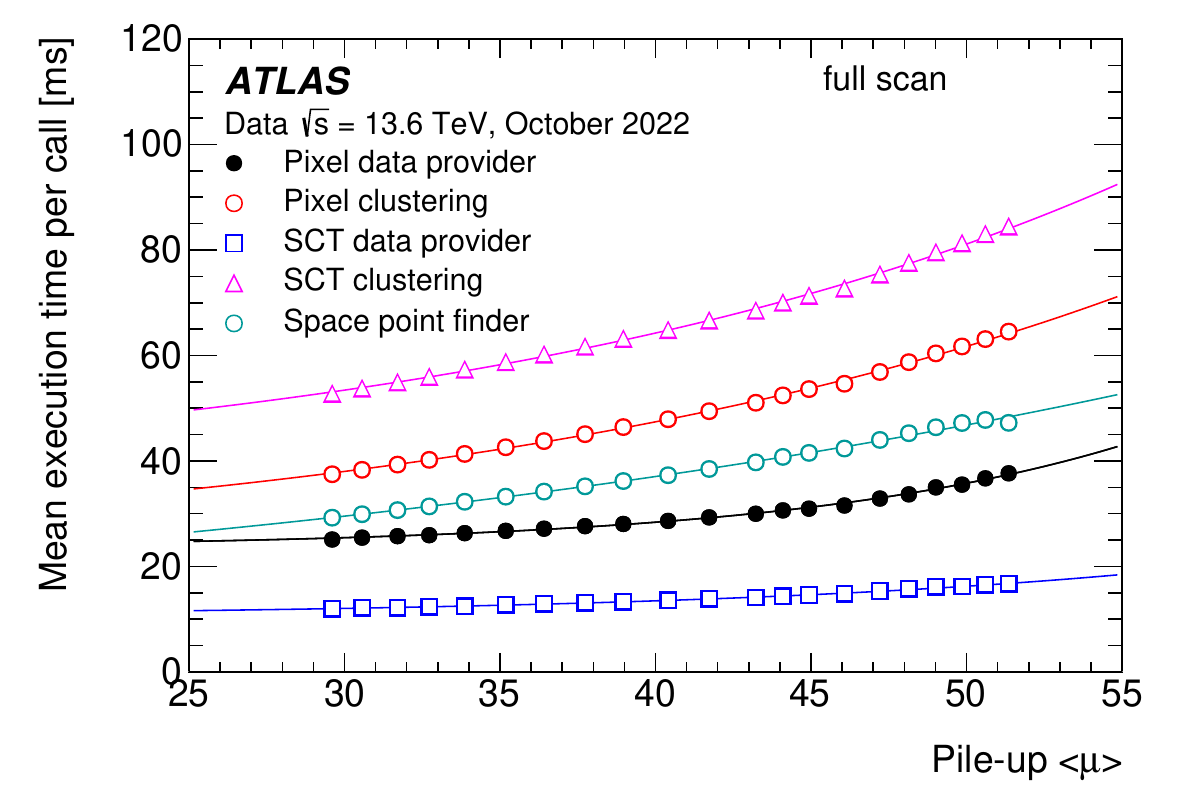}
\includegraphics[width=0.49\textwidth]{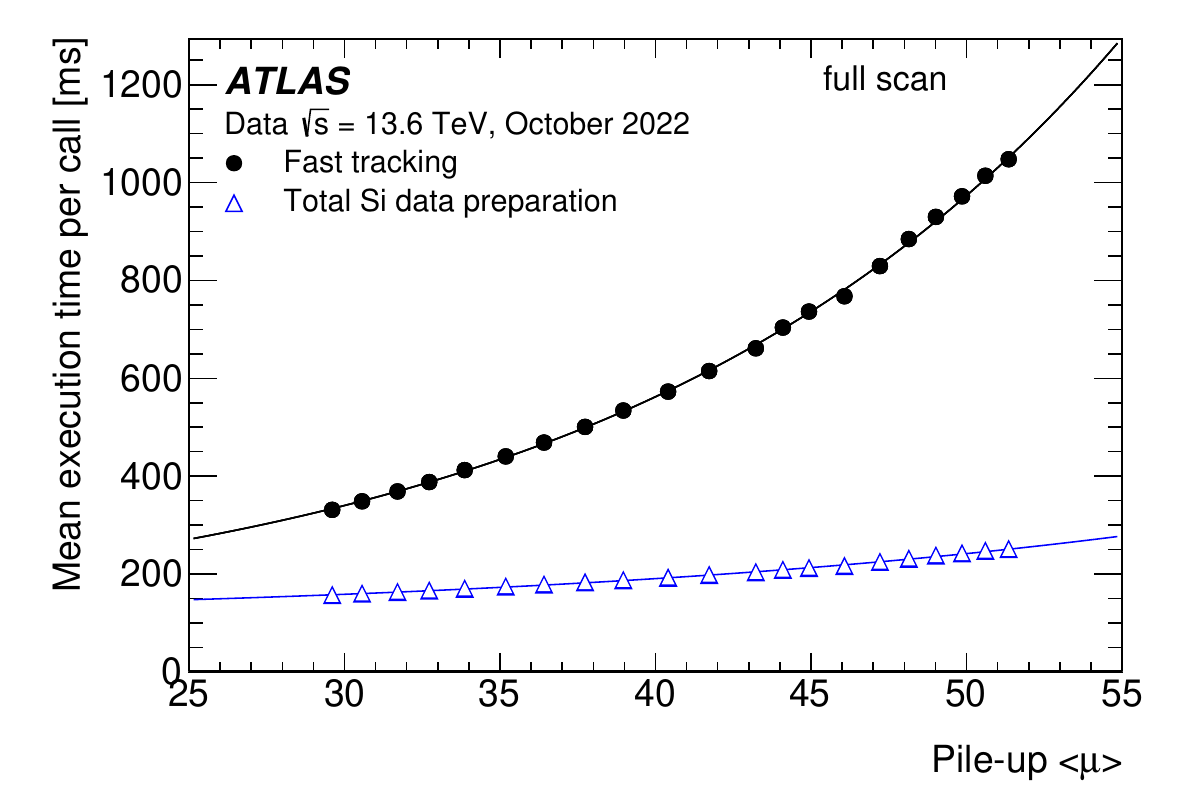}  \includegraphics[width=0.49\textwidth]{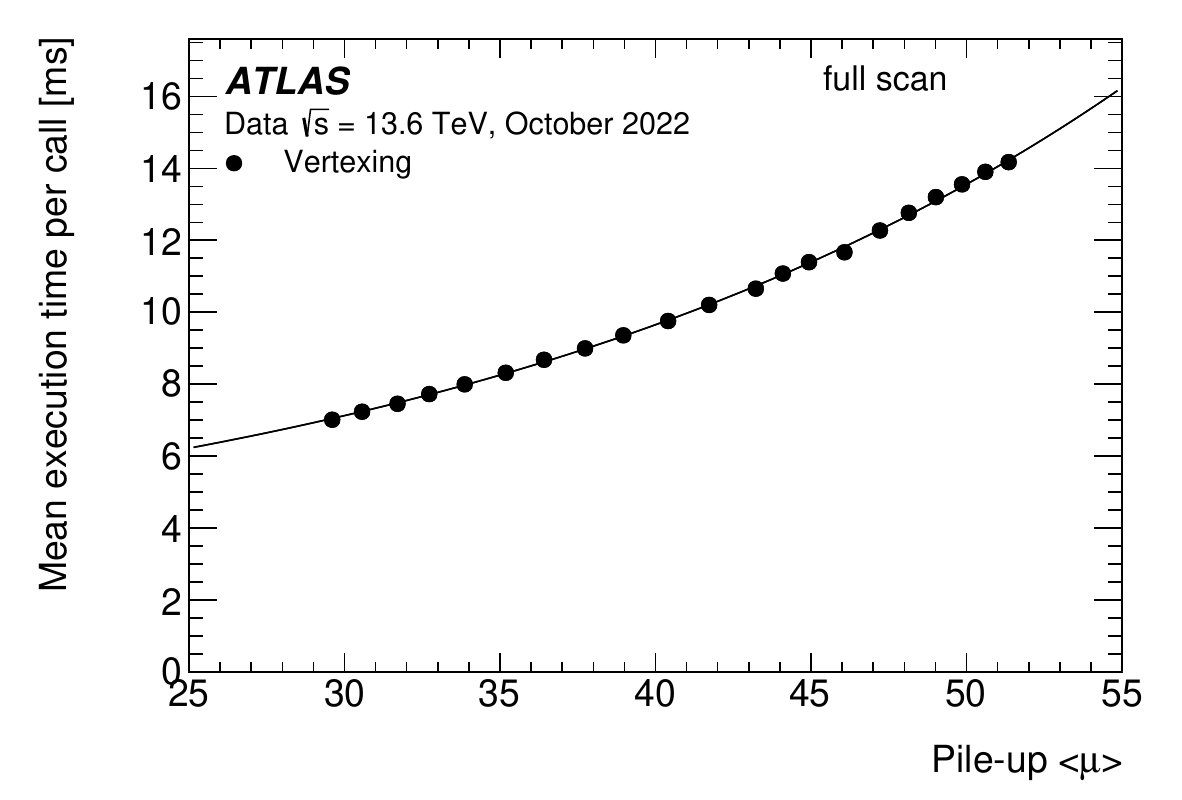}
\caption{ Execution times for (top left) the pixel and SCT data preparation, (top right) fast tracking and
total silicon data preparation time  for the full scan tracking in jet triggers, and (bottom)
for the vertex finding with the full scan tracks, all shown as a function of the average pile-up. Solid lines show ad hoc fits to data to guide the eye.
\statonly
}
\label{figs:jetcost}
\end{figure}
 
The execution times for the full scan trigger track and vertex reconstruction are shown in Figure~\ref{figs:jetcost}.
The combined data preparation for pixel and SCT is around 250\,ms. Even with the optimisation to reduce the processing time,
the full scan tracking still takes longer than 1.1\,s per event at pile-up of 52, while the vertex finding with the full scan
tracks takes only approximately 14~ms.
 
The execution times for the various tracking stages of $b$-jet triggers are shown in Figure~\ref{figs:bjetcost}.
The super-\roi tracking for the $b$-jet preselection stage takes 140\,ms at a pile-up of 52, with the data preparation
taking approximately 80\,ms in total. The precision tracking is not executed for the preselection.
The data preparation for the final \roi-based $b$-tagging stage is extremely fast (less than 4\,ms)
since only the data for those parts of the detector not already processed by the \preselection tracking need to be reconstructed.
The fast and precision tracking in the \roi for this final $b$-tagging stage takes less than 35\,ms per jet \roi at high pile-up.
As with the other signatures the fast tracking shows the largest dependence on the \pileup, although the longest contribution is
the TRT extension at low \pileup, which shows a similar trend to the precision tracking. It is slightly exceeded by the fast
tracking at higher \pileup.
 
\begin{figure}[tp]
\includegraphics[width=0.49\textwidth]{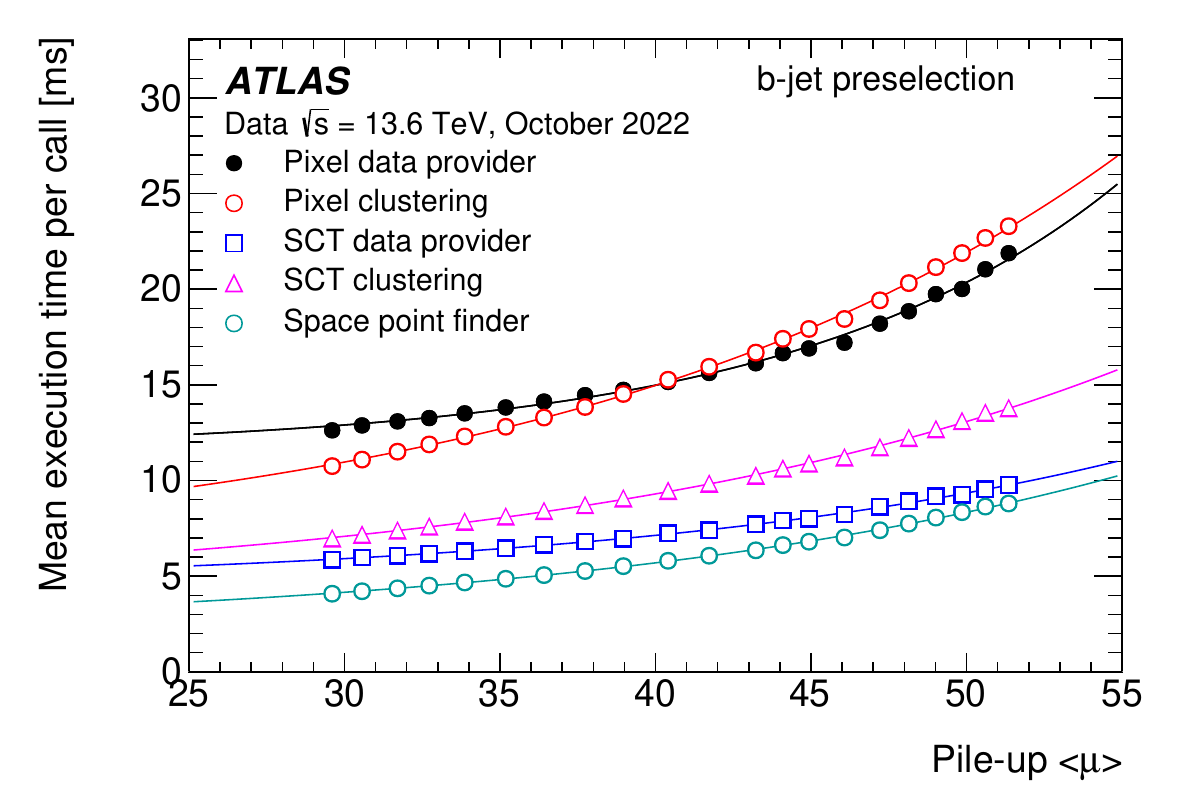}
\includegraphics[width=0.49\textwidth]{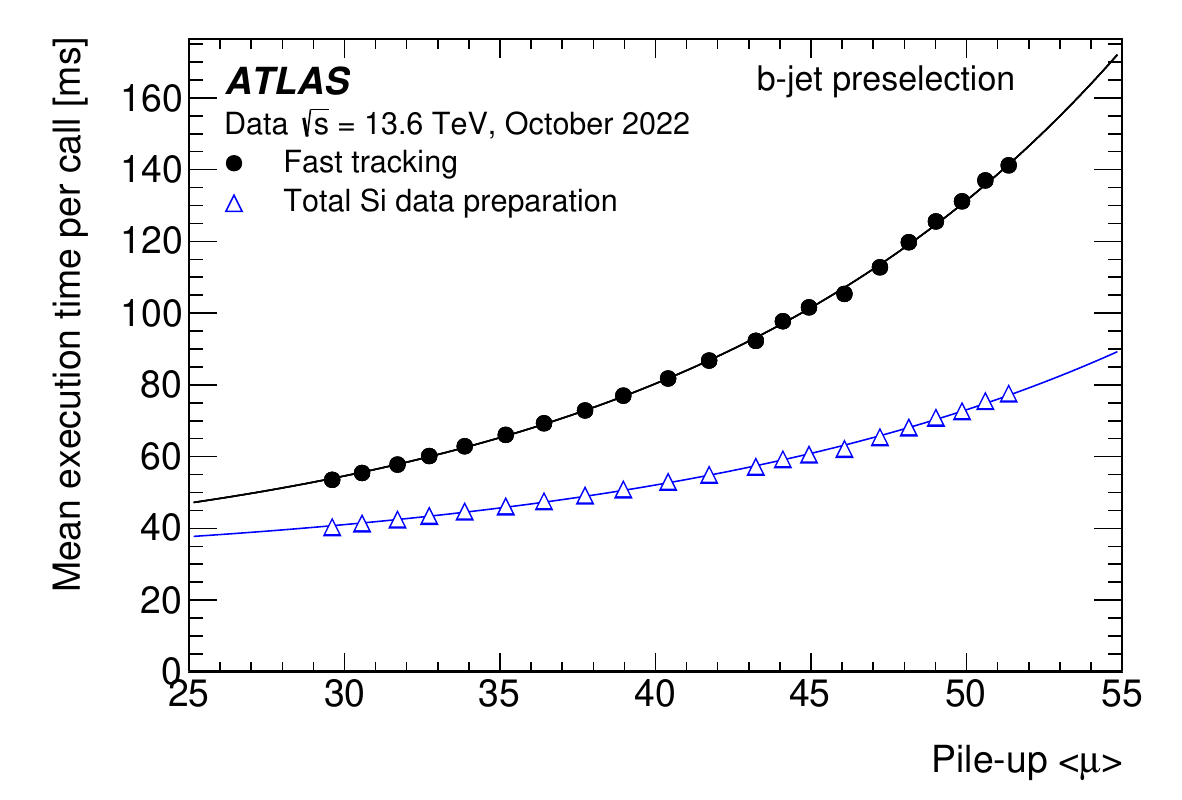} \\
\includegraphics[width=0.49\textwidth]{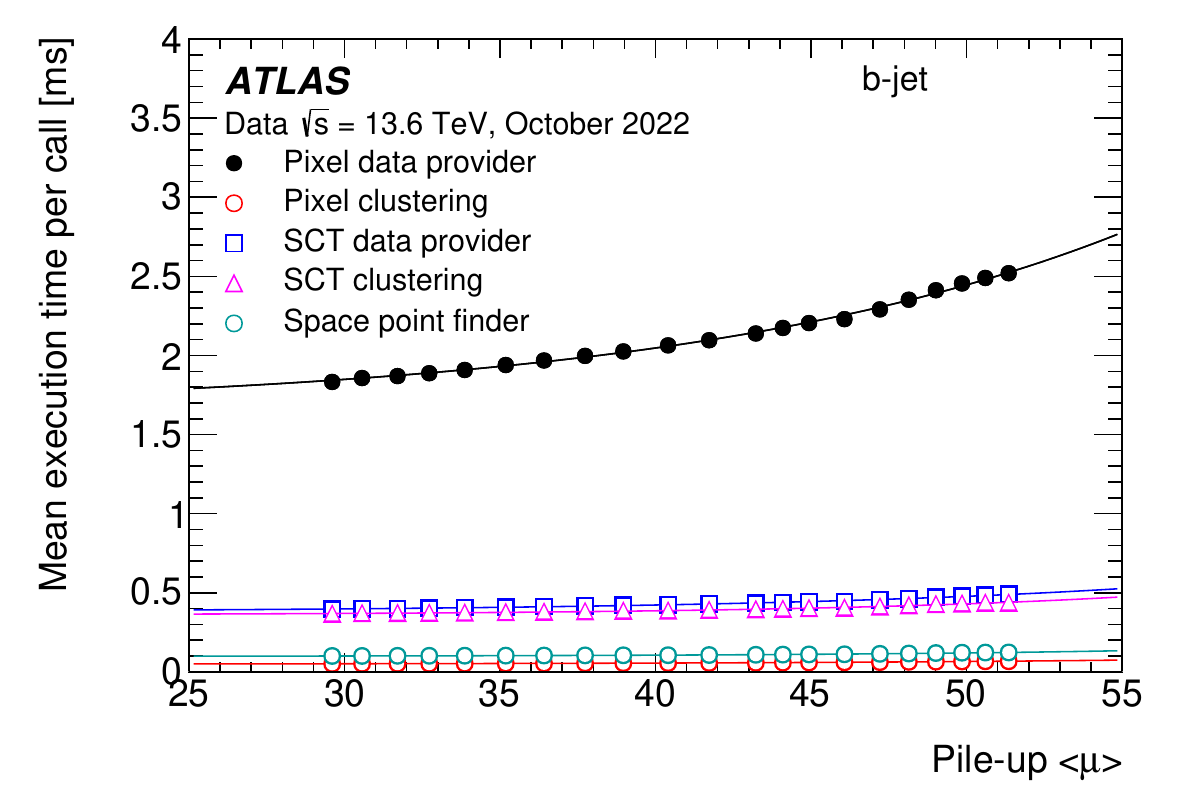}
\includegraphics[width=0.49\textwidth]{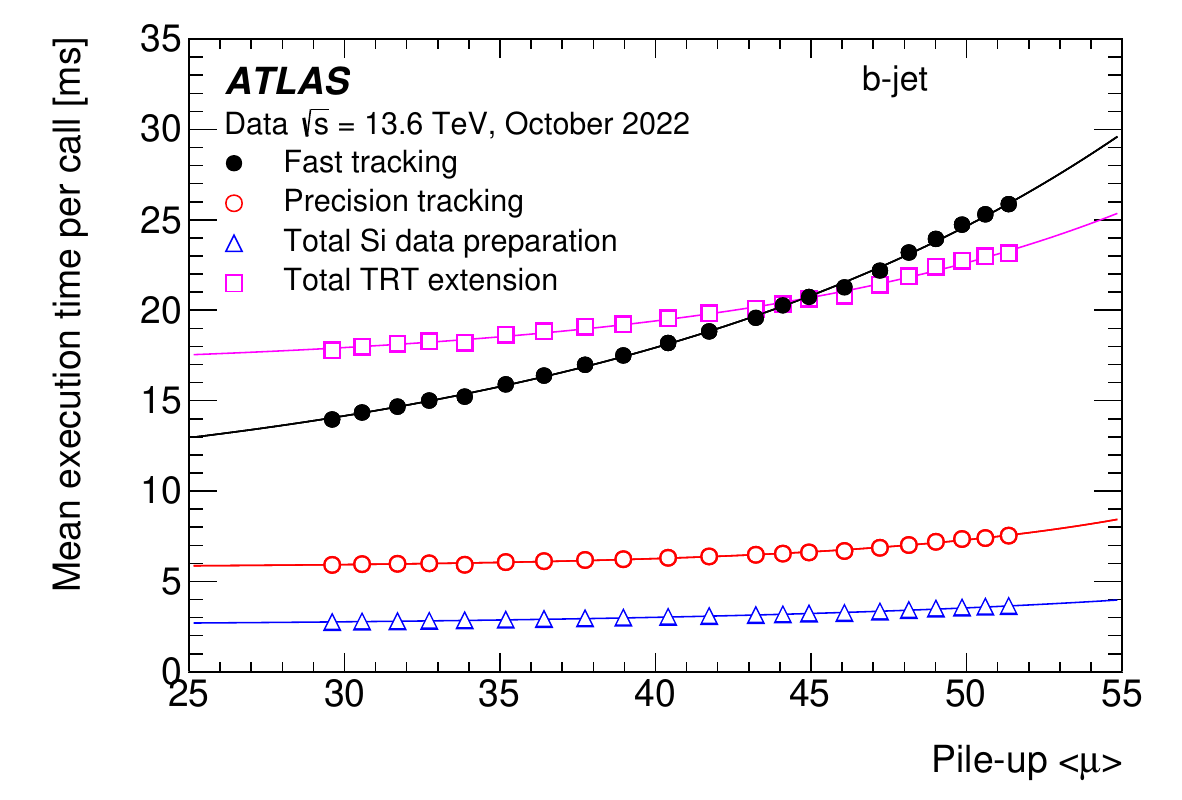}
\caption{Execution times for data preparation, tracking, and TRT extension for (top) the $b$-jet \preselection tracking
and (bottom) $b$-jet \roi as a function of the average pile-up. The detailed breakdowns of the silicon data preparation
components are shown on the left. Solid lines show ad hoc fits to data to guide the eye. \statonly}
\label{figs:bjetcost}
\end{figure}
 
Figures~\ref{fig:id:mu_lrt_cost} and~\ref{fig:id:el_lrt_cost} show the execution times in the \ac{HLT} for the muon and electron \ac{LRT}, respectively.
For muon LRT compared to standard muon tracking, the data preparation steps take approximately 1.5--3 times as long as standard tracking, except
for the pixel data provider which is similar.
The muon LRT fast tracking is 2--3 times faster than standard tracking, while the precision tracking is similar.
Differences in the processing times of the \ac{LRT} tracking with respect to the standard tracking
are due to the special configuration, such as using larger RoIs or seeding using only the SCT hits.
For electron LRT compared to standard electron tracking, the data preparation steps are a few milliseconds slower,
the fast tracking is approximately 1.5--2 times faster, while the precision and GSF tracking are 1.5--2 times slower.
The execution time of the track extension step is similar to standard tracking.  While the other steps to include the TRT are 1.5--2 times slower.

\begin{figure}[htbp]
\centering
\includegraphics[width=0.49\textwidth]{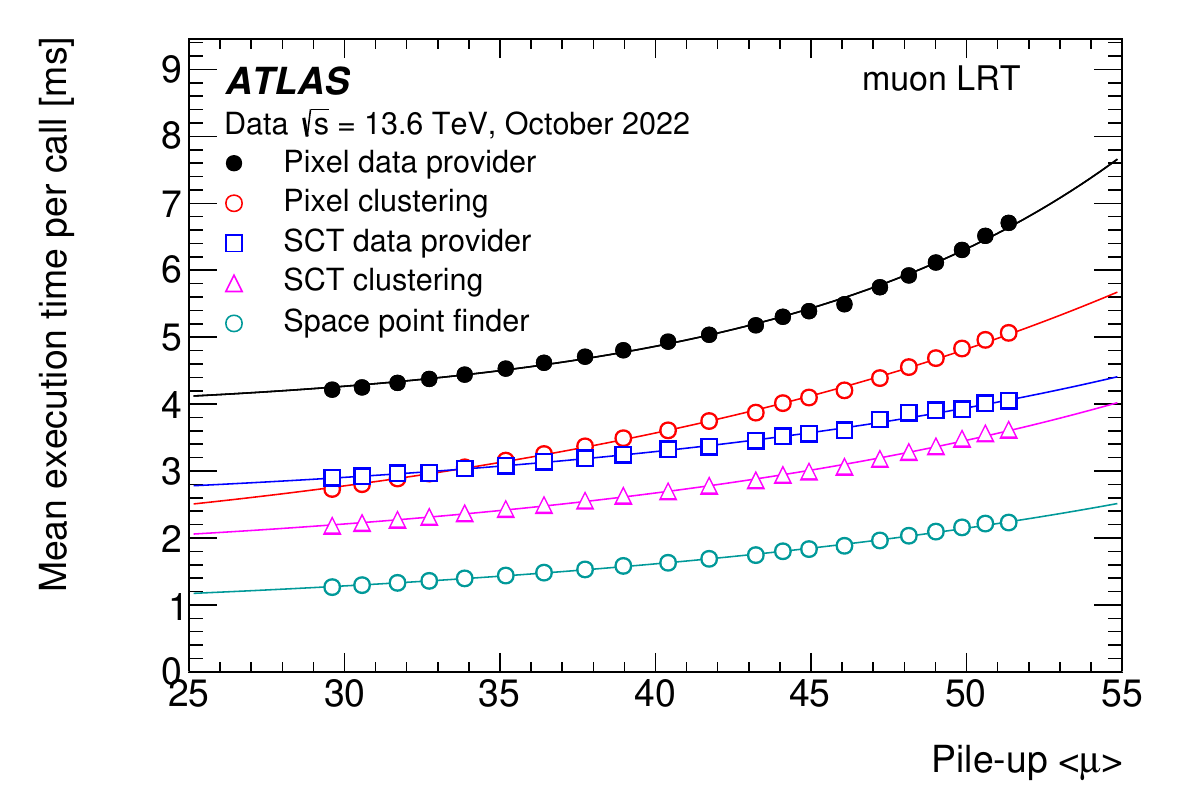}
\includegraphics[width=0.49\textwidth]{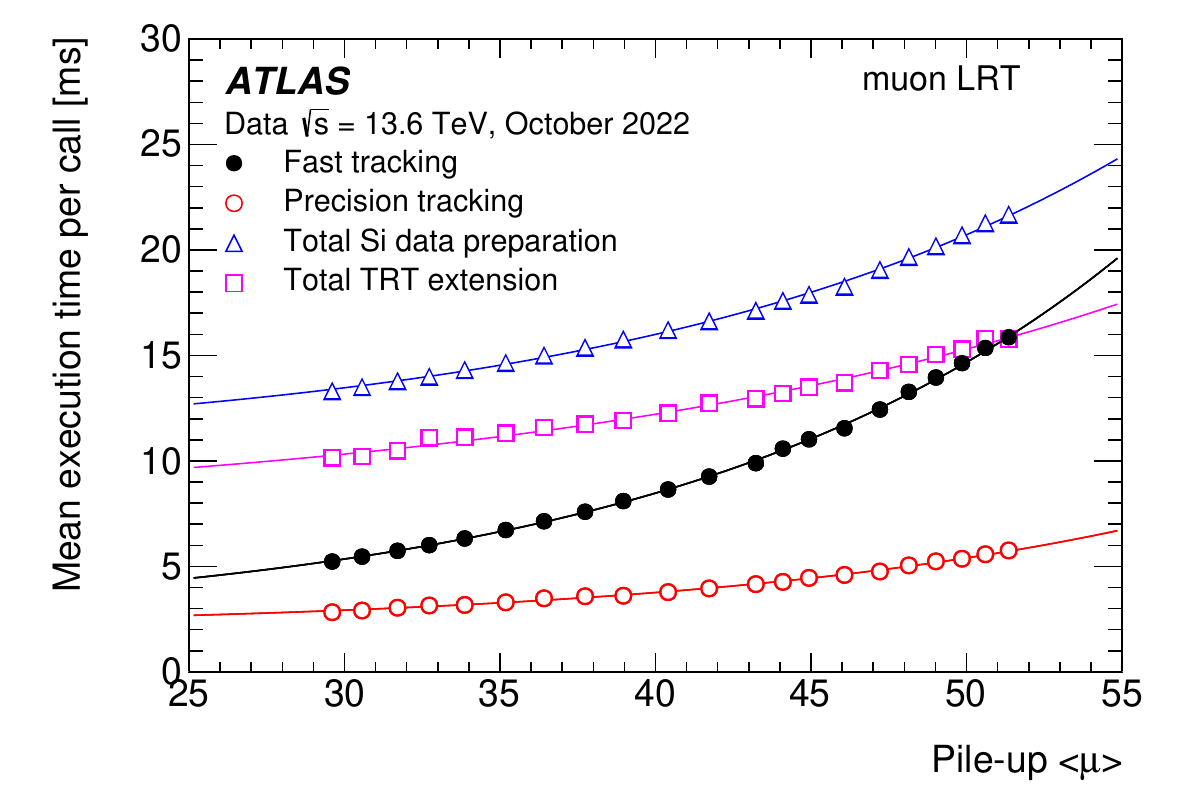}
\caption{The data preparation, TRT track extension, and tracking times for the muon large radius tracking algorithms as a function of the mean \pileup.
The detailed breakdown of the pixel and SCT data preparation stages are shown on the left, and the tracking times and combined
silicon data preparation time and total TRT extension on the right. Solid lines show ad hoc fits to data to guide the eye. \statonly }
\label{fig:id:mu_lrt_cost}
\end{figure}
 
\begin{figure}[htbp]
\centering
\includegraphics[width=0.49\textwidth]{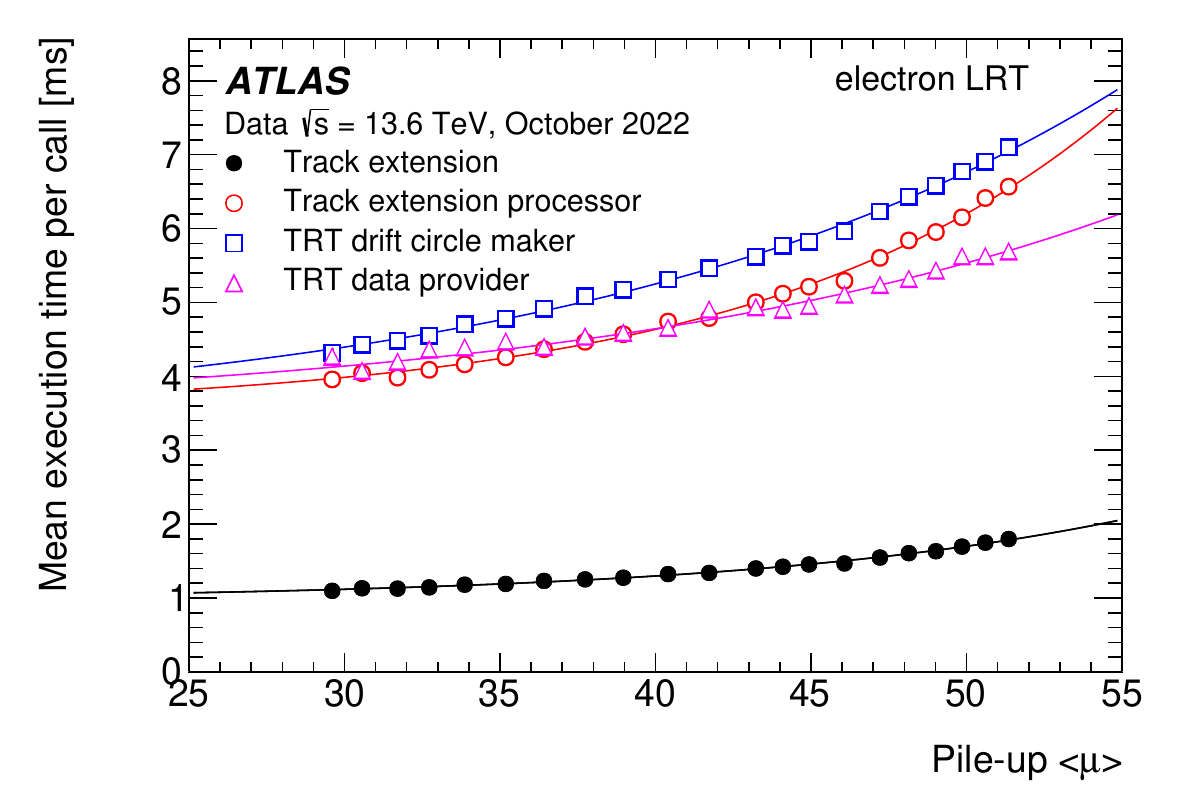}
\includegraphics[width=0.49\textwidth]{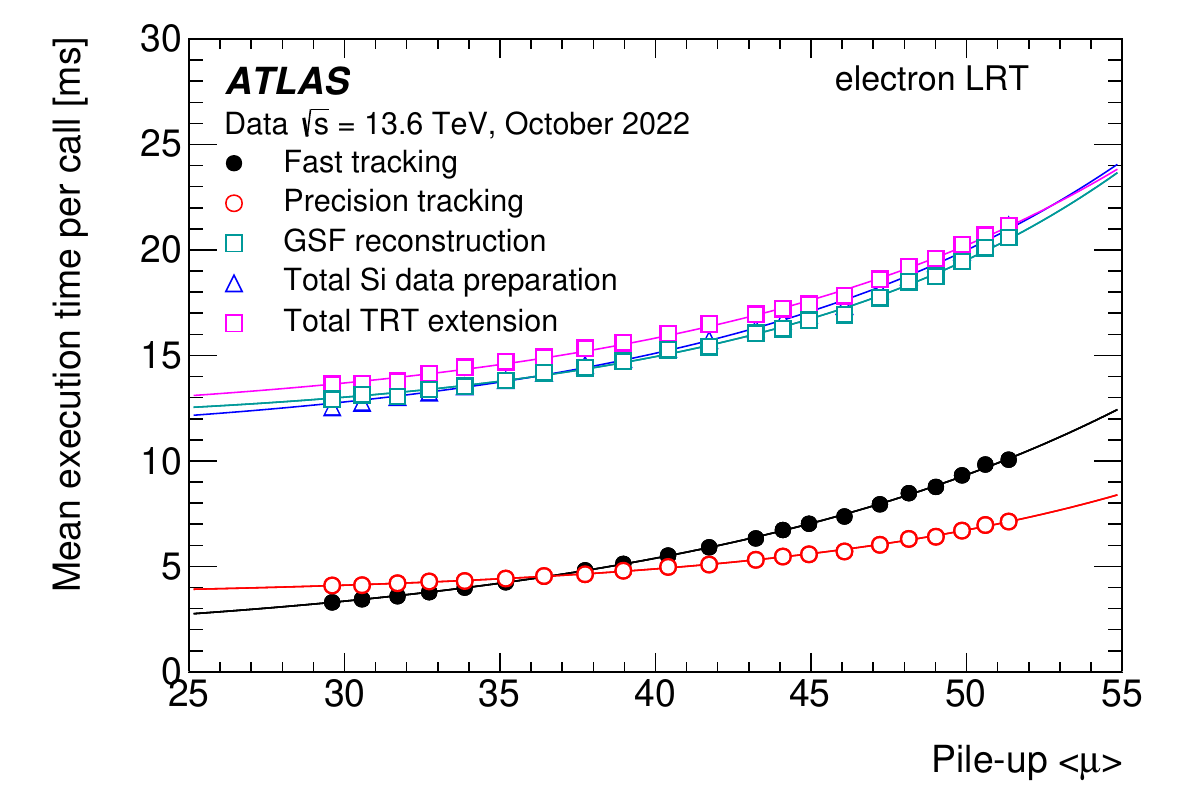}
\caption{The TRT track extension, data preparation and tracking times for the electron large radius tracking algorithms as a function of the mean \pileup.
The detailed breakdown of the TRT extension and data preparation stages are shown on the left, and the tracking times and combined
silicon data preparation time and total TRT extension on the right. Solid lines show ad hoc fits to data to guide the eye. \statonly }
\label{fig:id:el_lrt_cost}
\end{figure}


\subsection{Calorimeter reconstruction}
\label{sec:hltcalo}
 
The HLT Calorimeter (HLTCalo) software
performs the translation of the raw data read out from the LAr and Tile calorimeters
into the final software objects, the CaloCells.
There are 187,652 CaloCells for the entire detector.
CaloCells associate the geometrical information ($\eta$, $\phi$ and the longitudinal layer) of calorimeter cells
with their energy, pulse peaking time, hardware gain and quality factor as detailed in Ref.~\cite{denis}.
The calorimeter reconstruction algorithms use the CaloCells information
to reconstruct clusters of energy for candidate electrons, photons, taus and jet objects
as well as shower shape variables useful for particle identification.
The HLTCalo software handles both the high rate of regional data
requests, tens of kHz in multiple RoIs, and
a similar rate of data processing requests
which performs reconstruction in the entire volume of the
calorimeters for jets and missing transverse energy determination (full scan).
The HLTCalo software was adapted for the Run-3 AthenaMT framework,
though still not optimally as discussed in Section~\ref{sec:swperfscaling}.
 
Two different clustering algorithms are used to reconstruct the clusters of energy deposited in
the calorimeter: a sliding-window algorithm~\cite{TRIG-2016-01}
used for the fast electron and photon reconstruction step
and a topological-clustering ({\textit{topo-cluster}}) algorithm~\cite{PERF-2014-07}.
The topo-cluster algorithm begins with a seed-cell search and
iteratively adds neighbouring cells to the cluster if their energies\footnote{The LAr calorimeter electronics are designed such that signals from pile-up in earlier bunch crossings appear as negative energy.} ($E_\texttt{cell}$)
are above a given energy threshold that is a function of the
root-mean-square of the expected electronics and pile-up noise ($\sigma$). The seed
cells are first identified as those cells that have $|E_\texttt{cell}|/\sigma>4$.
All neighbouring cells with $|E_\texttt{cell}|/\sigma>2$
are then added to the cluster and, finally, all the
remaining neighbours to these cells with
$|E_\texttt{cell}|/\sigma>0$
are also added. These $\sigma$ thresholds (4,2,0) can, in principle, be adapted
but the numbers above are the same as used in the offline version of the
topo-cluster algorithm. The clusters are
grown by energy occupancy with no predefined shape. Given the number of
searches (via look-up table helpers) of the cell neighbours and calculations of energy
ratios, this algorithm is one of the most resource-consuming parts of the HLTCalo reconstruction.
 
One of the necessary conditions to calculate the energy in a given cell is related to
pile-up-dependent effects. There is a small and measurable bunch-by-bunch energy shift which
depends on the LHC luminosity profile (i.e. the structure of empty and filled bunches as well as their relative intensities, which is given by
the profile of $\langle\mu\rangle$ as a function of bunch-crossing identification).
The details of the offline correction procedure are provided in Ref.~\cite{PERF-2017-03}. The only difference
for the online application is that this profile
is updated only if
the pile-up value $\langle\mu\rangle$ changes by more than 5\%, which means that a few tens of
LBs can pass before the update.
This correction improves the resolution of many variables used for particle identification and energy estimation.
The calculation of this correction was adapted to the new AthenaMT environment of Run~3
by separating it into a run-long component and an event-by-event component.
 
\begin{figure}
\centering
\includegraphics[width=0.6\textwidth]{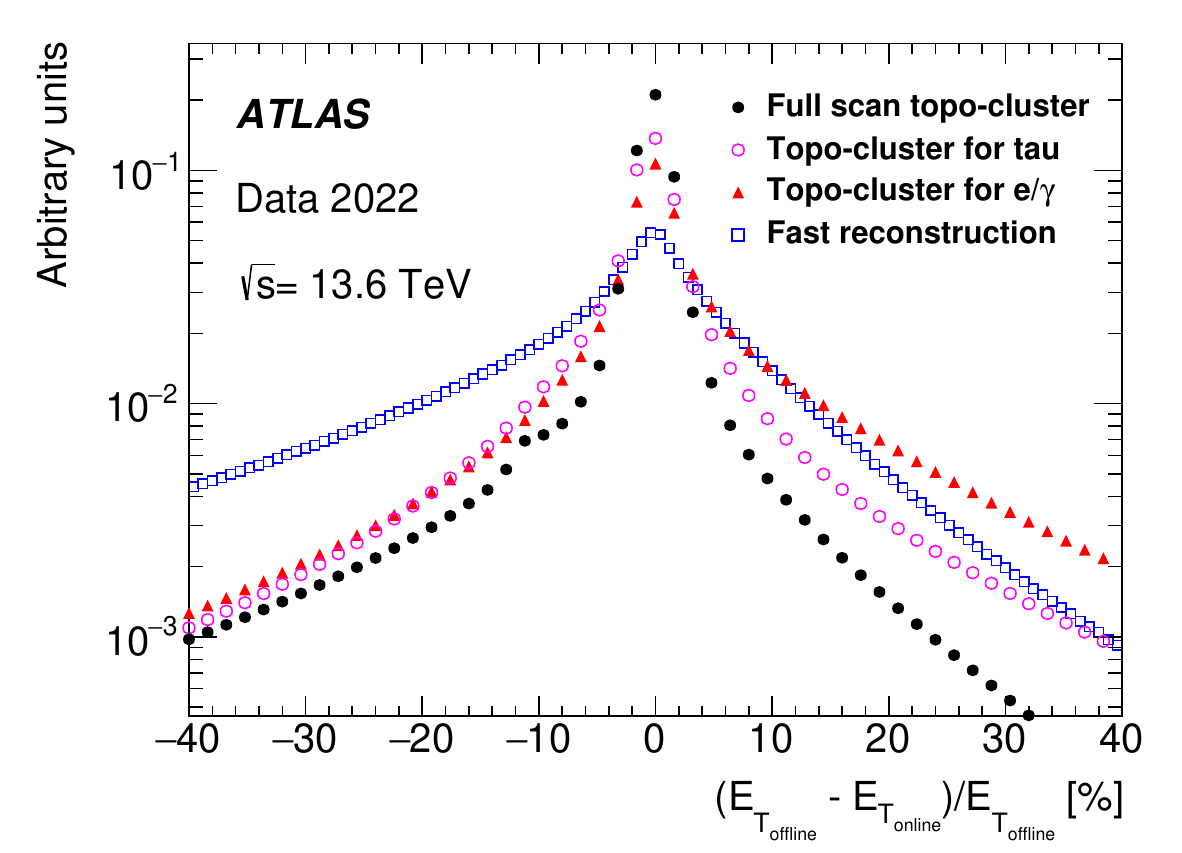}
\caption{Transverse energy resolution in percent for HLTCalo clusters obtained
with respect to calorimeter clusters reconstructed offline for 2022 data
for the following algorithms:
full scan topo-cluster for jets and \met, topo-cluster in RoI for taus,
topo-cluster in RoIs for electrons (e) and photons ($\gamma$),
fast sliding-window reconstruction for e/$\gamma$.
}
\label{fig:calo4}
\end{figure}
 
\begin{figure}
\centering
\includegraphics[width=0.49\textwidth]{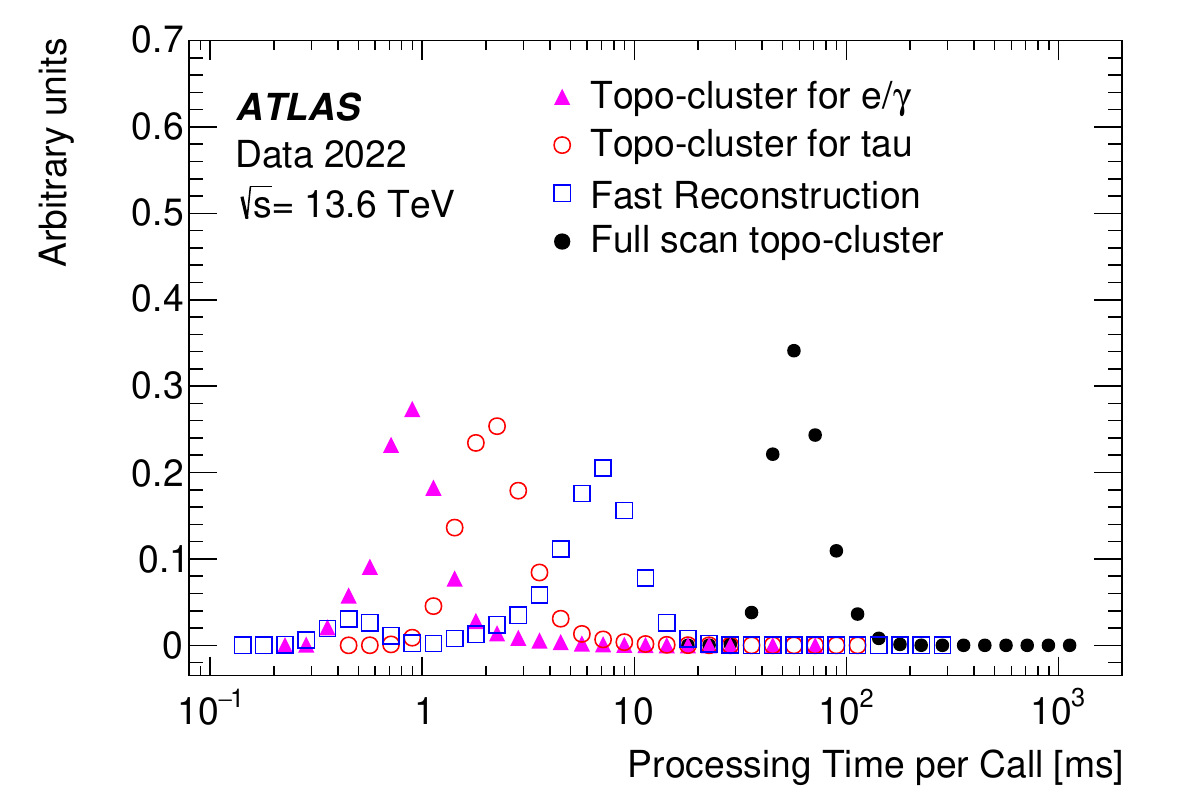}
\includegraphics[width=0.49\textwidth]{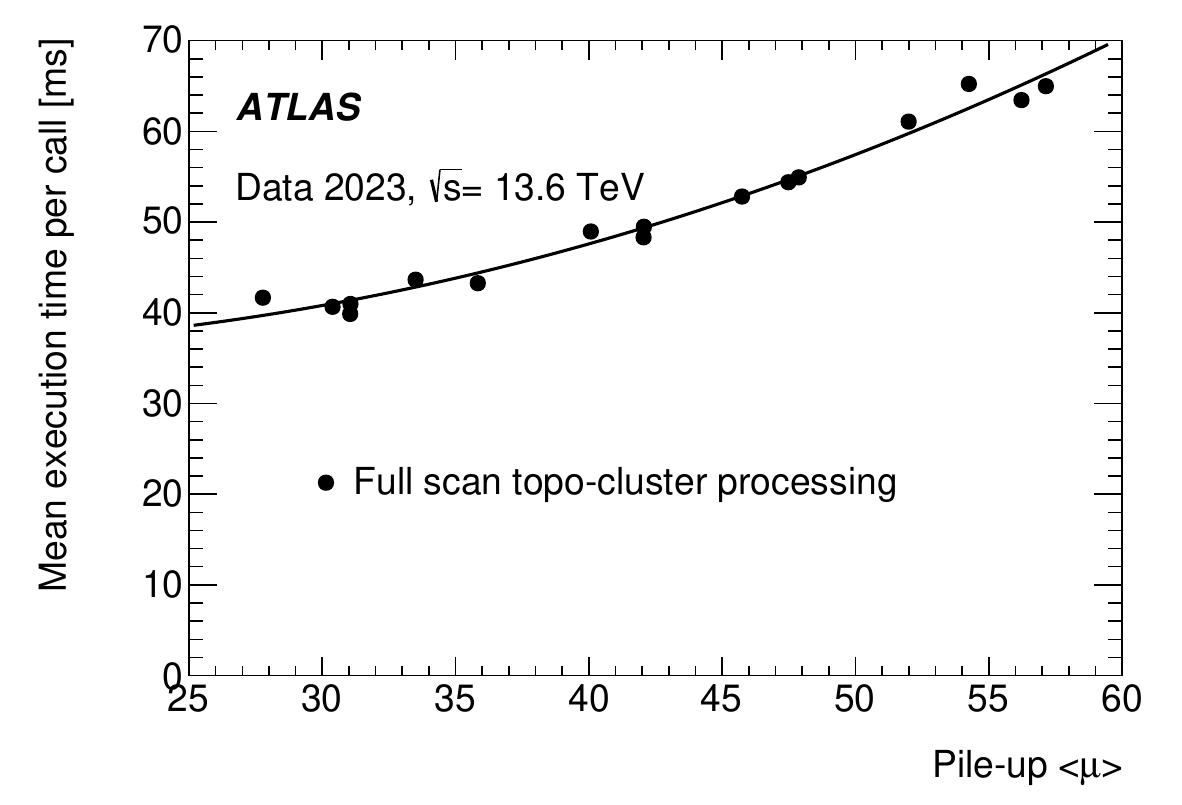}
\caption{(left) Processing time per call for the following algorithms:
topo-cluster in RoIs for e/$\gamma$,
topo-cluster in RoIs for taus,
fast sliding-window reconstruction for e/$\gamma$,
full scan topo-cluster for jets and \met.
(right) Processing time as a function of the average pile-up for the full scan topo-cluster algorithm.
A solid line shows an ad hoc fit to data to guide the eye.
}
\label{fig:calotime}
\end{figure}
 
The measurements presented here were done in the HLT farm during a typical physics run.
The \et\ resolution of HLTCalo clusters obtained with different reconstruction algorithms are shown in Figure~\ref{fig:calo4}.
The peaks of all four distributions are consistent with zero within 1\%.
The residual differences are due to different reconstruction and calibration algorithms.
The fast calorimeter reconstruction relies on a sliding window cluster algorithm, which allows for
a good electron energy estimate, despite the very short time available ($<5\,$ms).
It has an energy resolution of 4.3\% and a longer low-energy tail than the topo-cluster reconstruction
algorithms, which are shapeless and pick up much more low-energy activity.
The topo-clusters used for electrons, photons and taus are reconstructed in an \roi, as discussed in Sections~\ref{sec:egamma} and~\ref{sec:tau},
and those used for jets and \met are reconstructed in the full calorimeter volume (details in Sections~\ref{sec:jets} and~\ref{sec:met}).
Topo-clusters can be calibrated at the EM scale, for which the energy of
an isolated topo-cluster is the sum of its constituent cell energies.
Alternatively, local hadronic calibration~\cite{lc} can be applied as a cell-level correction to improve the mean response
to hadronic showers, and is primarily used for tau reconstruction, detailed further in Section~\ref{sec:tau}, as well as for large-radius jet triggers,
detailed in Section~\ref{sec:jets}.
The \et\ resolutions are 4.1\%, 3.7\% and 2.8\% for e/$\gamma$, tau and full scan topo-clusters, respectively.
The processing time of the main HLTCalo reconstruction algorithms per call is shown in Figure~\ref{fig:calotime} (left). It varies
from a few milli-seconds running at 22 (14)\,kHz for topo-cluster reconstruction in e/$\gamma$ (tau) RoIs to a few
tens of milliseconds at 23\,kHz for full scan topo-clustering. The latter also exhibits a known pile-up dependence as shown in Figure~\ref{fig:calotime} (right), unlike the other
algorithms. The fast calorimeter reconstruction runs at a total call rate of 132\,kHz\footnote{It runs at L1 EM rate of 46\,kHz for about 2.9 RoIs per event on average.} and its timing distribution has a double-peaked structure seen in Figure~\ref{fig:calotime} (left).
The lower peak in the processing time is due to a
fraction of calls which do not require time-consuming data requests as data was cached once by the first \roi
request and only the reconstruction is run.


\subsection{Tracking in the muon spectrometer}
\label{sec:muonrec}
The HLT muon reconstruction~\cite{TRIG-2018-01} consists of two steps: the first is fast and trigger specific, while the second is based on precision reconstruction.
The precision reconstruction makes use of the same software as the offline muon reconstruction with some adaptations for online running. Muon candidates are reconstructed from combined tracks in the MS and the ID subdetectors. Most of the muon triggers are based on combined muon candidates. MS-only candidates are used to trigger particular topologies, such as \acp{LLP} where there might be no corresponding ID track.
 
In the fast reconstruction stage, each L1 muon candidate is refined by including the precision data from the MDT chambers in the \roi defined by the L1 candidate. A track fit is performed using the MDT drift times and positions as well as the sTGC and MM chambers in the endcap regions, and a \pt measurement is assigned using lookup tables, creating MS-only muon candidates. The MS-only muon track is back-extrapolated to the interaction point and combined with tracks reconstructed in the ID to form a combined muon candidate with a refined track parameter resolution.
 
In order to recover the efficiency in low \pt di-muon topologies, e.g. $B$-meson decays where multiple muons arrive close together in the MS,
a new \emph{inside-out} algorithm was developed for \runiii.
The MS-only back-extrapolated tracks are used as an \roi to reconstruct the ID tracks,
which are then extrapolated to the MS and used as seeds for the fast MS-only muon reconstruction.
Collimated di-muon trigger candidates can hence be distinguished.
 
The precision stage follows the same strategy, combining tracks in the MS and the ID subdetectors to reconstruct the muon candidates. It starts from the refined \rois identified in the fast reconstruction step to form muon candidates using information from the MS detectors. They are then extrapolated to the interaction point and combined with ID tracks. If the MS track cannot be matched to an ID track, combined muon candidates are searched for by extrapolating ID tracks to the MS.
 
The full scan mode is used to find additional muons that are not found by the \roi-based method mainly due to
L1 inefficiencies (`\trig{noL1}' triggers). In the full scan mode, muon candidates are first sought in all muon detectors. Then, RoIs
are constructed around the found MS tracks and ID tracks are reconstructed within these RoIs. The same
combination procedure as for the RoI-based method is used to construct combined full scan muons.
Given the high CPU demand of the full scan reconstruction, it is only executed in multi-object triggers with at least
one of the trigger objects found by an RoI-based algorithm.
 
The performance of the muon tracks at the HLT level compared to the offline muon reconstruction is illustrated in Figure~\ref{fig:muon_resolution}.
The \pt resolution of the trigger muon candidates is measured with $Z\rightarrow \mu \mu$ candidate events using the tag-and-probe method~\cite{TRIG-2018-01}. Events are required to contain a pair of muons with opposite charge and an invariant mass within 10\,\GeV\ of the Z pole mass and the offline reconstructed muons are required to pass the \emph{medium} identification requirements. The distributions of the relative residuals between online and offline track parameters are constructed in bins of offline muon \pt and the width of the distribution in each bin is obtained by means of a Gaussian fit. The resolution for combined trigger muons is better than the resolution for MS-only muons as expected, thanks to the higher precision of the ID track measurements, especially at low transverse momentum. The relative \pt resolution is about 3--4\% (8\%) and 1--2\% (2--4\%) in the barrel (endcap) region for precision MS-only and combined muons,
respectively, and tends generally to degrade towards higher-\pt. The MS-only resolution in the endcap region is expected to improve after the NSW commissioning is completed.
 
\begin{figure}[htbp]
\centering
\includegraphics[width=0.49\textwidth]{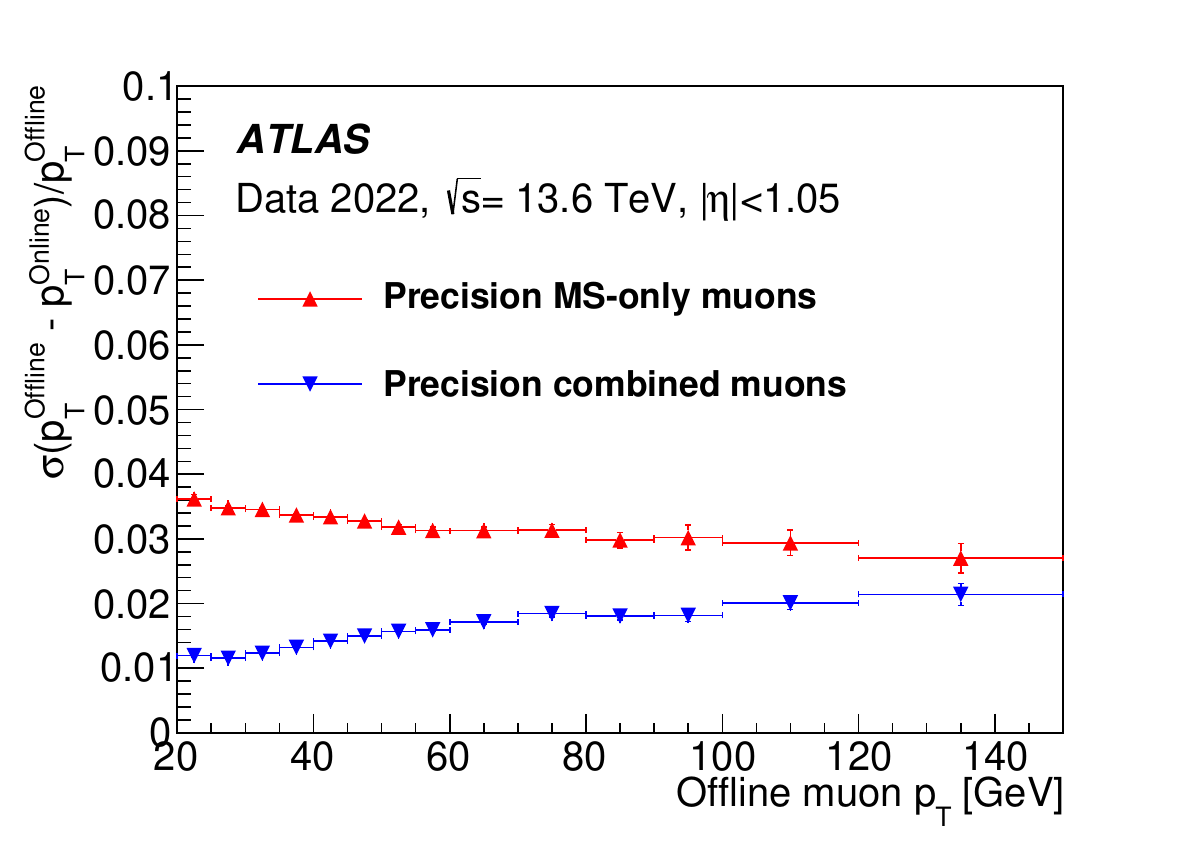}
\includegraphics[width=0.49\textwidth]{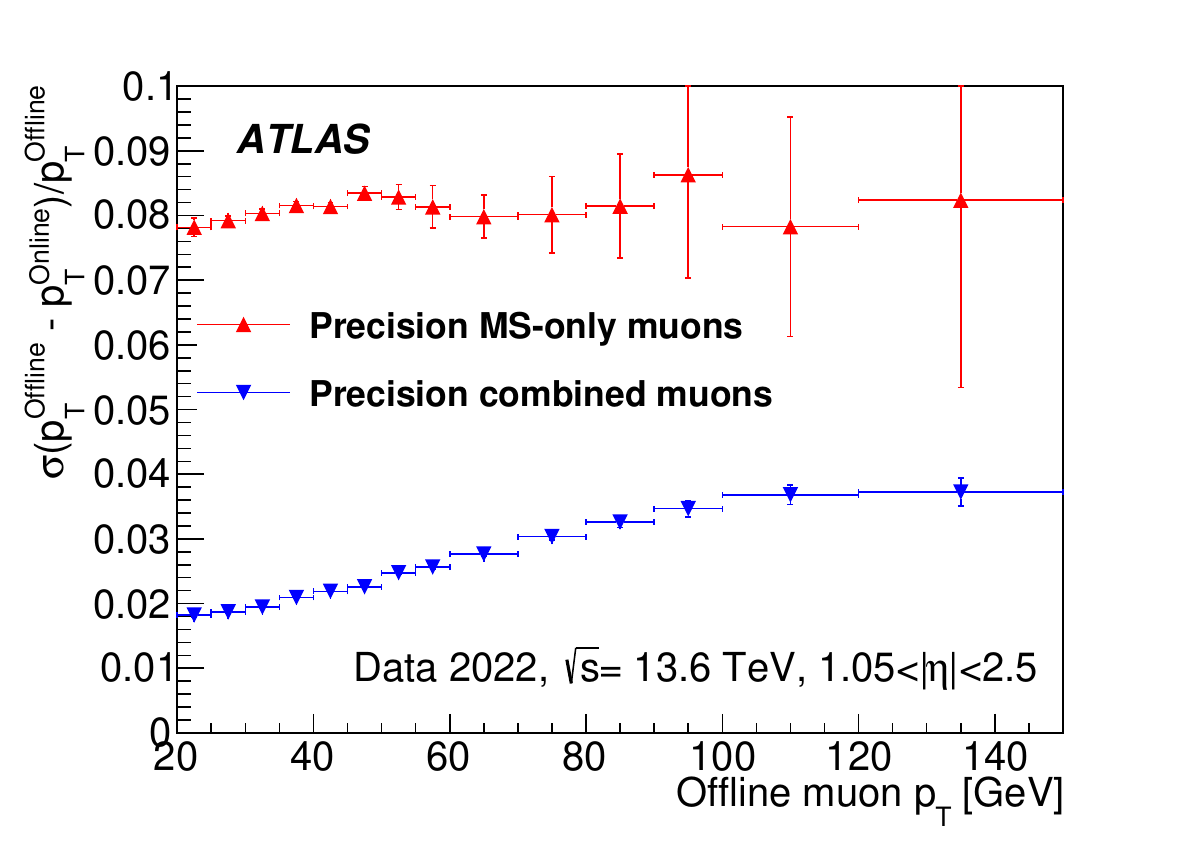}
\caption{Width of the residuals for \pt as a function of the offline muon \pt for the precision MS-only
and combined algorithms in (left) the barrel and (right) endcaps. Only statistical uncertainties are shown.}
\label{fig:muon_resolution}
\end{figure}
 
The timing measurements presented in were recorded from the HLT farm,
during a typical physics run from November 2022.
The processing times per \roi are shown in Figure~\ref{fig:muon_timing-fast} for the fast MS-only and fast combined algorithms.
The call frequencies are 37\,kHz and 23\,kHz, respectively. Figure~\ref{fig:muon_timing-precise} shows the most CPU consuming
steps of the precision reconstruction: segment-finding algorithm, MS-only track building, MS and combined muon candidate building and inside-out recovery.
The call frequency of the segment-finding, track-building and MS muon candidate building algorithms is 11\,kHz, while they are 2.7\,kHz and
0.34\,kHz for the combined muon candidate building and inside-out recovery, respectively.
The large tails of the distribution of the fast combined and precision combined algorithms are related to the algorithms not being fully
optimised for Run-3 conditions in 2022. The performance has since been optimised.
 
\begin{figure}[htbp]
\centering
\includegraphics[width=0.49\textwidth]{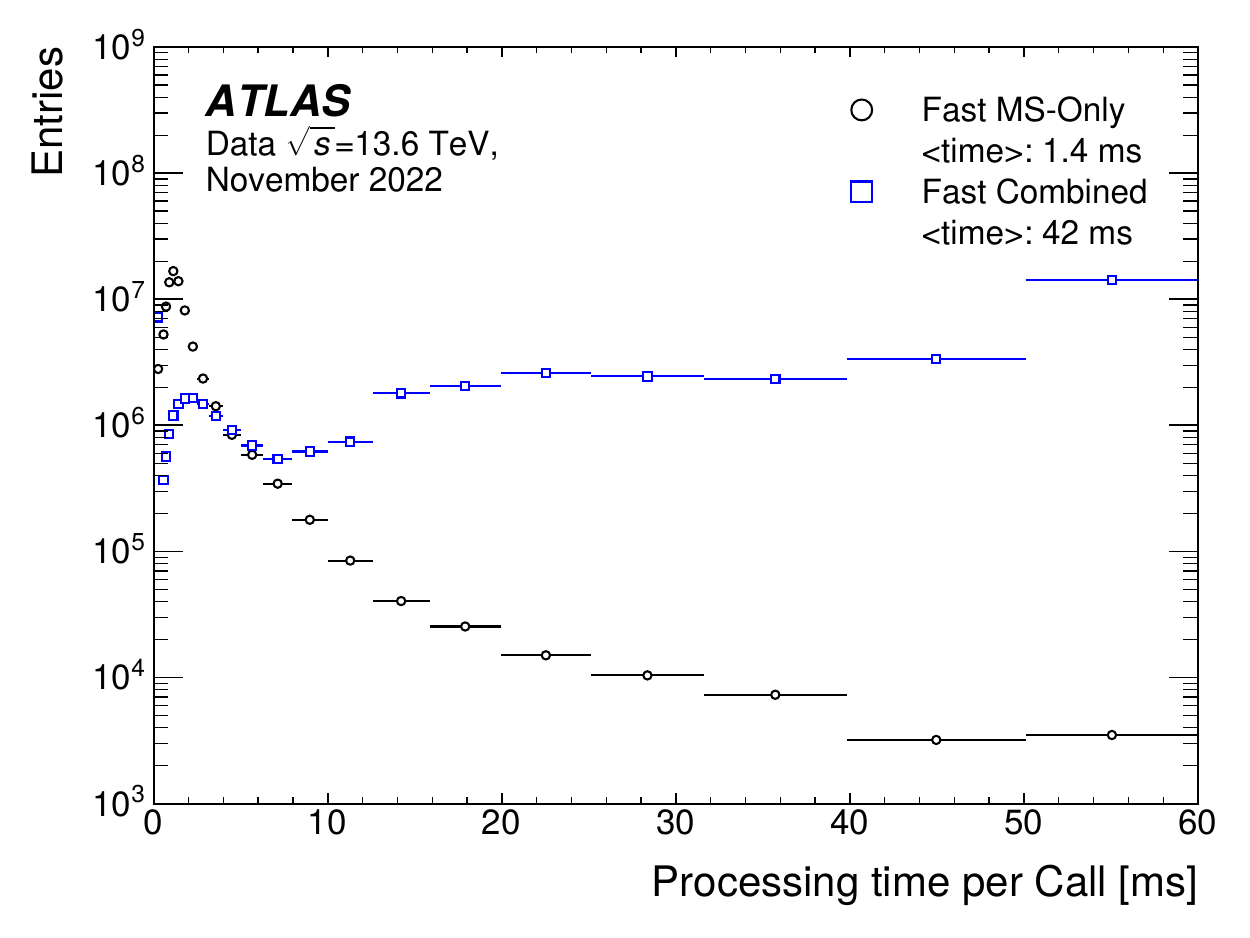}
\caption{Processing times per \roi for the fast MS-only and fast combined algorithms. The mean time of each algorithm is indicated in the legend.
The last bin of the distribution includes the overflow events.}
\label{fig:muon_timing-fast}
\end{figure}
 
\begin{figure}[htbp]
\centering
\includegraphics[width=0.49\textwidth]{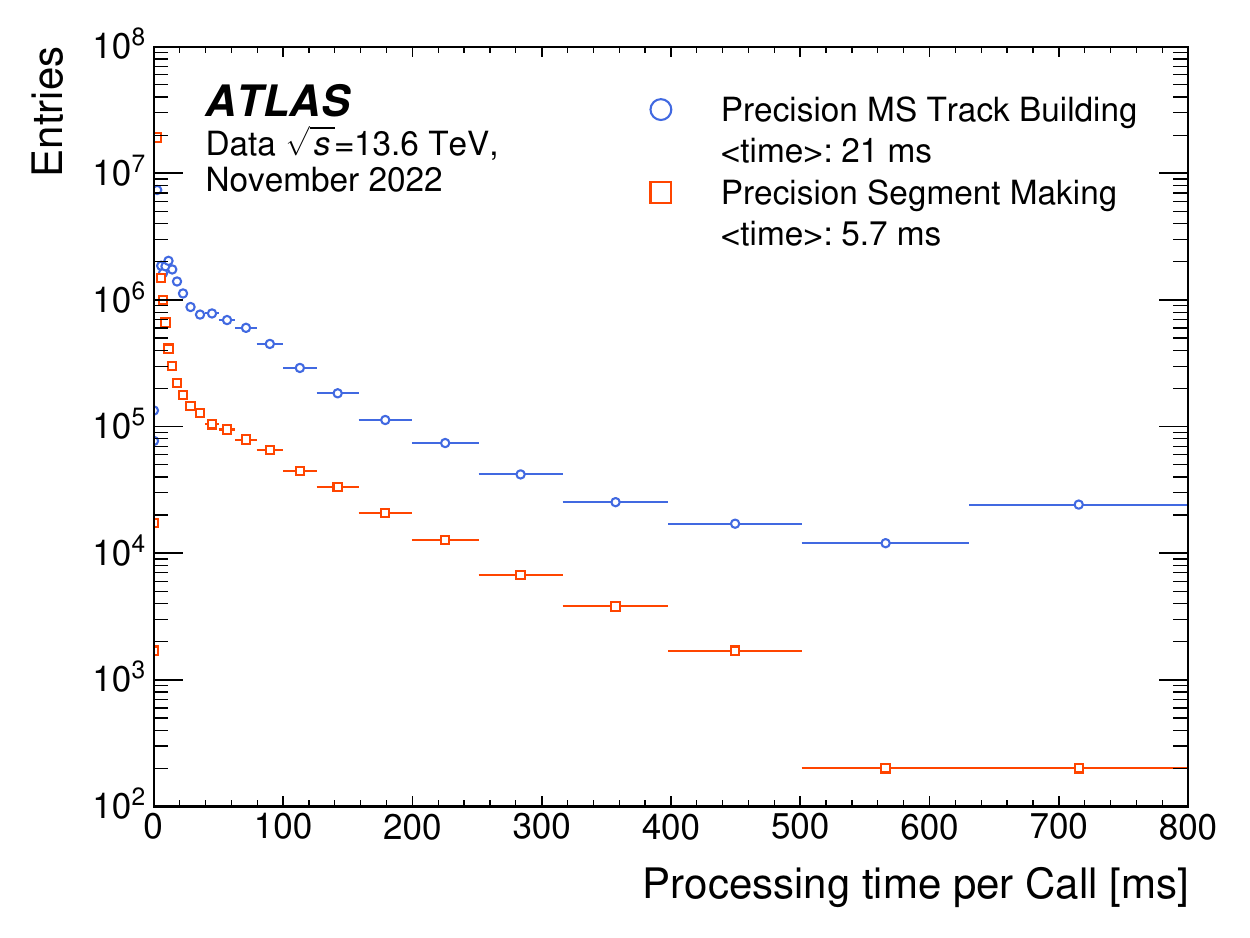}
\includegraphics[width=0.49\textwidth]{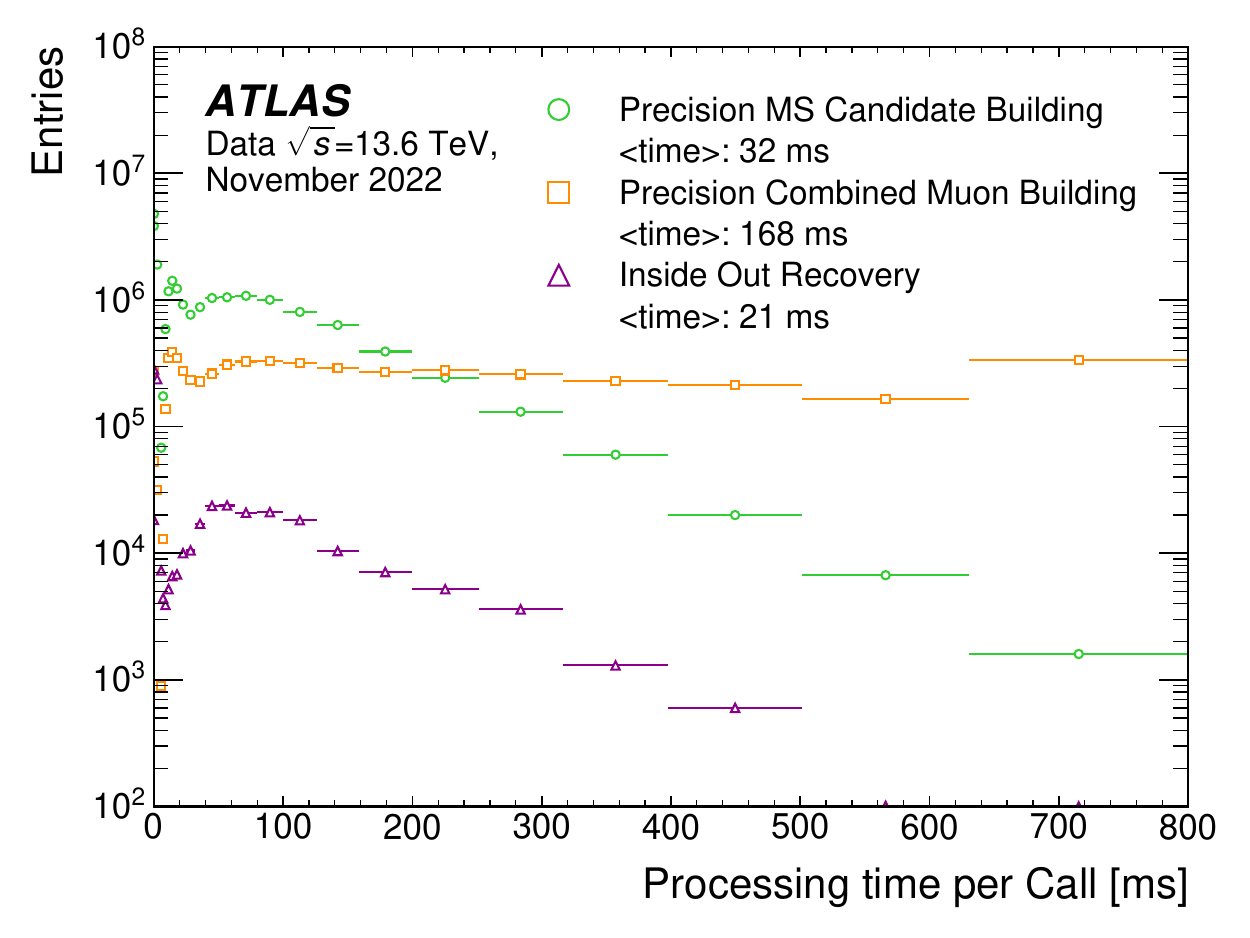}
\caption{Processing times per \roi for the most CPU consuming precision steps: segment-finding and track-building algorithm (left), muon candidate building and inside-out recovery (right). The mean time of each algorithm is indicated in the legend. The last bin of the distributions includes the overflow events.}
\label{fig:muon_timing-precise}
\end{figure}


\newpage
\section{Physics triggers}
\label{sec:sigPerf}

\FloatBarrier
The final event selection is based on trigger signatures, such as leptons, hadrons, and
global event quantities such as missing transverse momentum. They are formed by placing different selection
criteria on the various reconstructed objects.
The selection criteria and performance of the various trigger signatures is described
in this section, highlighting the differences with respect to the Run-2 selection and performance of primary triggers during 2022.


\pagestyle{plain}
\subsection{Electrons and photons}
\label{sec:egamma}
 
\subsubsection{Electron and photon trigger reconstruction and selection}
\label{subsec:ElePh_rec}
 
Electron and photon reconstruction at the HLT is performed on each EM \roi provided by L1,
which satisfies the \et requirement, and any other L1 selection requirements specified by the trigger menu (e.g. isolation) of at least one active HLT chain.
In the HLT, fast calorimeter algorithms are executed first, allowing precision algorithms to run at a reduced rate later in the trigger sequence.
The reconstruction of candidate electrons and photons uses the sliding-window algorithm with rectangular
clustering windows of size $\Delta\eta\times\Delta\phi = 0.075 \times 0.175$ in the barrel and $0.125 \times 0.125$ in the endcaps.
The fast calorimeter selection step has three implementations~\cite{TRIG-2018-05}.
The default fast calorimeter selection step for electrons uses a neural-network-based {\textit{Ringer}} algorithm~\cite{ringer},
which uses as input energy sums of all the cells in 100 concentric rings
centred around the most energetic cell in each calorimeter sampling layer.
In \runii it was used only for triggering electrons with \et$\ge 15$\,\gev, but it is applied from \et$\ge 5$\,\gev\ in \runiii.
The Ringer algorithm is optimised in two regions of \et, between 5--15\,\GeV\ with \jpsi$\rightarrow ee$ MC samples and for $\et>15$\,\GeV\ with $Z\rightarrow ee$ MC samples.
For electrons, it is optionally possible to use fast calorimeter selections
which take as an input either only the cluster \et (\et-based) or the cluster \ET with three shower shape parameters (cut-based)~\cite{TRIG-2018-05}.
The electron candidates are then required to have tracks within the \roi (obtained from the fast track reconstruction)
matching the corresponding clusters~\cite{TRIG-2018-05}.
In contrast, photon candidates are reconstructed using the calorimeter information only,
with cut-based selection criteria that are the same as in \runii~\cite{TRIG-2018-05}.
 
For \runiii, the HLT electron and photon precision reconstruction becomes closer to the offline reconstruction~\cite{EGAM-2018-01},
due to the implementation of a new super-clustering algorithm, described below, and the use of the GSF tracking algorithm for electrons, described in Section~\ref{sec:id_electron}.
 
Electron and photon super-clusters are reconstructed in two stages.
The first stage consists of finding the seed topo-cluster candidates in the same \roi as used
for the tracking: $0.1\times 0.2$ in $\eta - \phi$ space. These seed topo-cluster candidates form the basis of
super-clusters. The second stage is the identification of topo-clusters near the seed candidates which are identified as satellite
cluster candidates. They may emerge from bremsstrahlung radiation or topo-cluster splitting.
For a cluster to become an electron super-cluster seed, it is required to have a minimum
\et of 1~\GeV\ and match to a track with at least four hits in the silicon tracking detectors. For
photon reconstruction, a cluster must have \et greater than 1.5~\GeV\ to qualify as a super-cluster seed.
If a cluster meets the seed cluster requirements, the algorithm attempts to find satellite clusters,
which represent secondary EM showers originating from the same initial electron or photon.
A cluster is considered a satellite if it falls within a window of $\Delta\eta \times \Delta\phi = 0.075 \times 0.125$ around the seed cluster barycentre.
The energy of the clusters is calibrated using a multivariate technique such that the response of the calorimeter layers is corrected in data and simulation~\cite{EGAM-2018-01}.
 
At the precision step, the electron identification relies on a multivariate technique using a likelihood discriminant, while the photon identification is
cut-based. These identifications, as well as isolation requirements which are applied to some triggers, remain unchanged with respect to their Run-2 configurations detailed in Ref.~\cite{TRIG-2018-05}.
 
\subsubsection{Electron and photon trigger menu}
\label{subsec:egammamenu}
 
Data taken with electron and photon triggers are used in a wide range of ATLAS physics analyses, from SM
precision physics to searches for new physics. The various triggers cover the energy
range between a few \GeV\ and several \TeV.
The 2022 electron and photon trigger thresholds remain the same as in 2018~\cite{TRIG-2018-05}.
The minimum \et thresholds for the isolated single electron and non-isolated photon triggers are 26 and 140\,\GeV\ with
rates of 186\,Hz and 46\,Hz at luminosity of $1.8\times\lumi{e34}$, respectively.
To increase the single electron trigger efficiency at \et of 60\,\gev\ and 140\,\gev, triggers with no isolation requirements
and looser identification are present at rates of 20\,Hz and 2\,Hz, respectively.
Additional single electron triggers with \et thresholds at 140\,(300)\,\gev\, and the cut-based (\et-based) fast calorimeter selection, instead of the default Ringer algorithm,
run at 3\,(6)\,Hz improving sensitivity to merged electrons coming from decays of boosted dibosons~\cite{TRIG-2018-05}.
Triggering on low \et electrons and photons is very challenging because of the high rates at low trigger thresholds.
However, this can be mitigated by requiring the presence of
multiple electrons or photons in the event, which helps to reduce the trigger thresholds with respect to
single electron or photon triggers.
The primary di-photon trigger is mainly designed for the efficient selection of events with Higgs boson candidates
in the di-photon decay channel. It has a rate of 18\,Hz for medium photon identification and
trigger \et thresholds of 35 and 25\,\GeV\ for the leading and subleading photons.
A second symmetric di-photon trigger with loose identification and the \et requirement of 50\,\gev\ has a rate of 7\,Hz.
A di-electron trigger at lower \et requirement of 17\,\GeV\ each and loose identification has a rate of 12\,Hz.
 
\subsubsection{Electron and photon trigger performance}
 
The efficiency of the lowest unprescaled single-electron trigger with respect to offline electron candidates
is shown in Figure~\ref{fig:egammaPerf2} for each step of the online reconstruction.
This and the next measurements are performed using the tag-and-probe method in $Z\rightarrow ee$ events~\cite{TRIG-2018-05}.
The HLT inefficiency is due to
differences in the online and offline
electron identification and isolation~\cite{TRIG-2018-05} and a 1--2\% lower efficiency of
the trigger precision tracks, shown in Figure~\ref{figs:el:eff}.
The efficiency of the combination of all primary single-electron triggers is shown in Figure~\ref{fig:egamma2}.
Scale factors derived from the observed data/MC simulation differences are used to correct other MC simulation samples used in data analyses.
 
\begin{figure}
\centering
\includegraphics[width=0.49\textwidth]{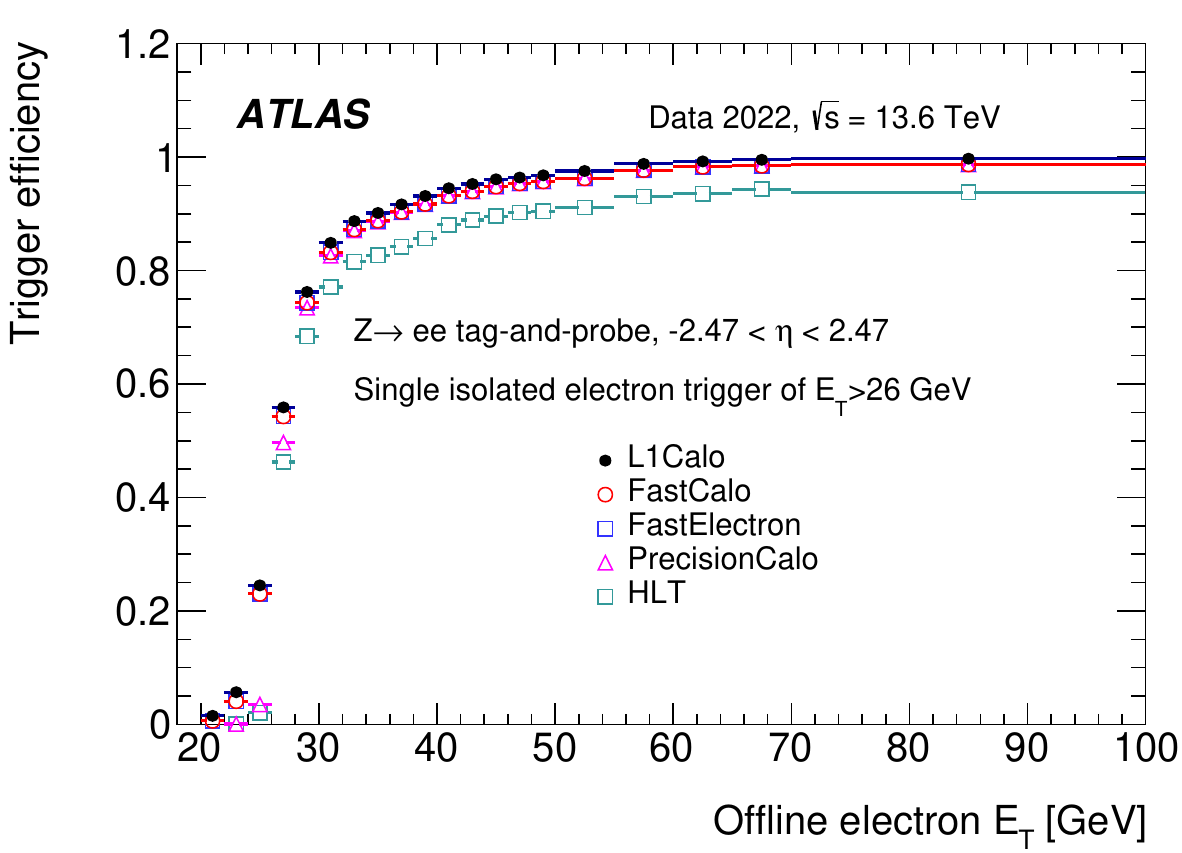}
\includegraphics[width=0.49\textwidth]{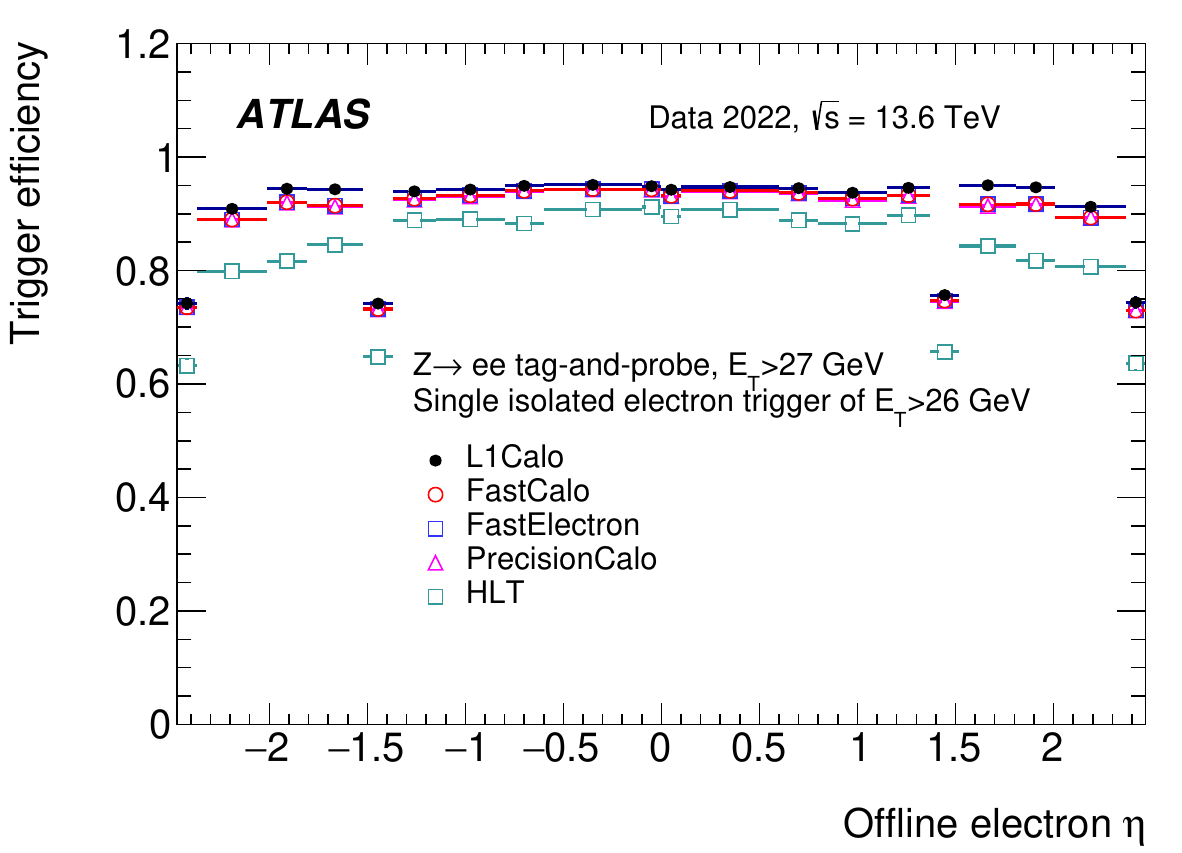}
\caption{Efficiency of the primary isolated electron trigger with \et$>$26~\GeV\ 
for all steps of the HLT electron reconstruction as a function of the offline electron
(left) \et and (right) $\eta$. Efficiency is given with respect to
the offline electrons which satisfy the tight identification and isolation criteria.
Only statistical uncertainties are shown.
}
\label{fig:egammaPerf2}
\end{figure}
 
\begin{figure}
\centering
\includegraphics[width=0.49\textwidth]{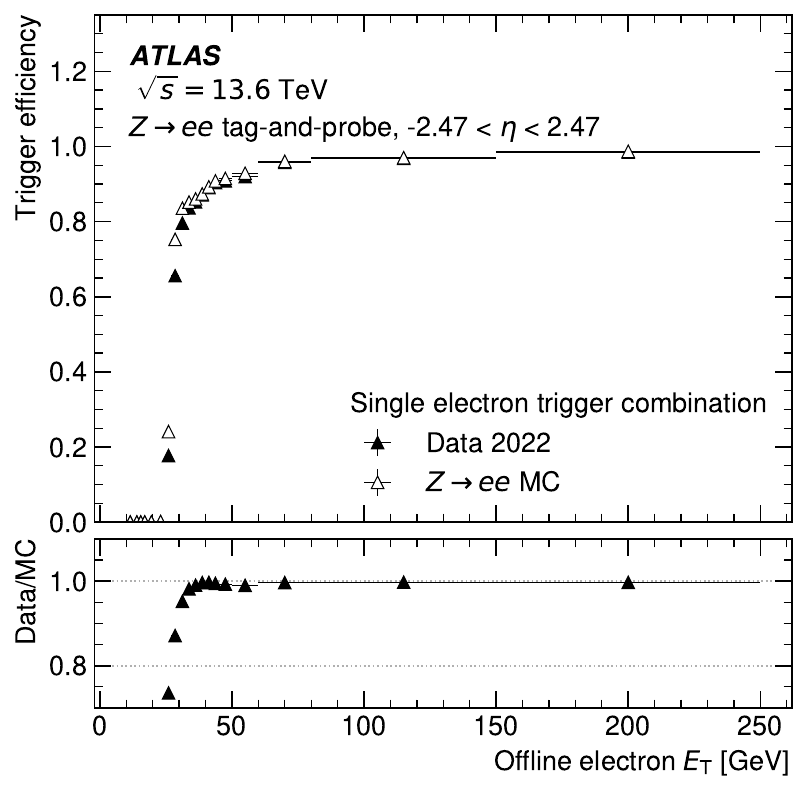}
\includegraphics[width=0.49\textwidth]{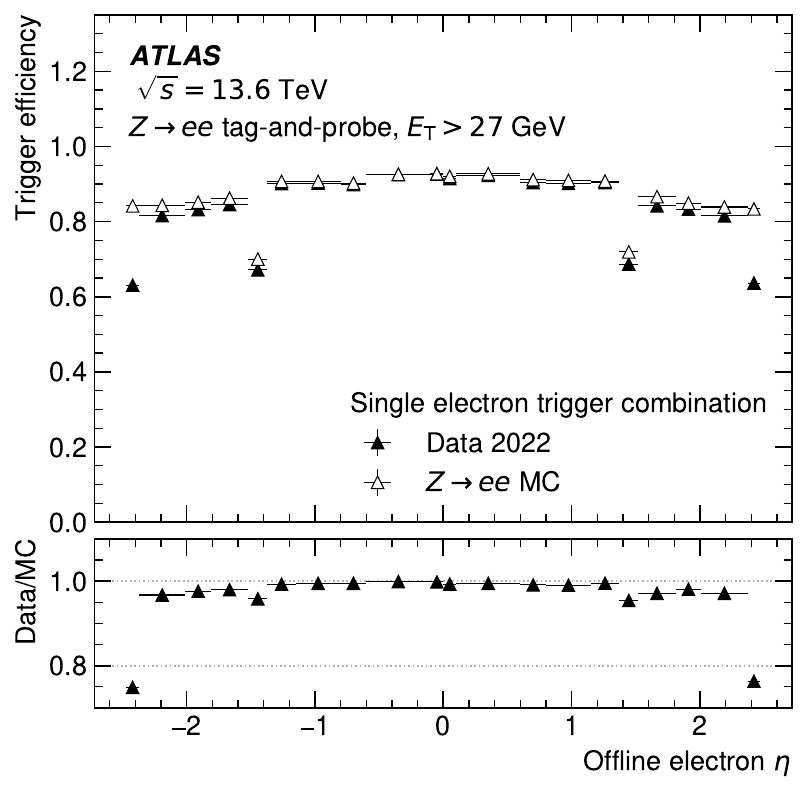}
\caption{Combined efficiency of three primary single electron triggers
as a function of the offline electron (left) \et and (right) $\eta$
for both MC simulation and 2022 data. Efficiency is given with respect to the offline electrons
which satisfy the tight identification and isolation criteria. Only statistical uncertainties are shown.
}
\label{fig:egamma2}
\end{figure}
 
Figure~\ref{fig:egamma4} shows the efficiencies of the primary di-photon triggers extracted with the bootstrap method~\cite{TRIG-2018-05}.
Photon trigger efficiencies are always very high, but while the 25\,\gev\ trigger is
fully efficient at 5\,\GeV\ above its threshold, higher \et\ triggers are not, as can be seen in Figure~\ref{fig:egamma4} (right).
As for the electron triggers, scale factors are used to correct for the observed data/MC simulation differences.
 
\begin{figure}
\centering
\includegraphics[width=0.49\textwidth]{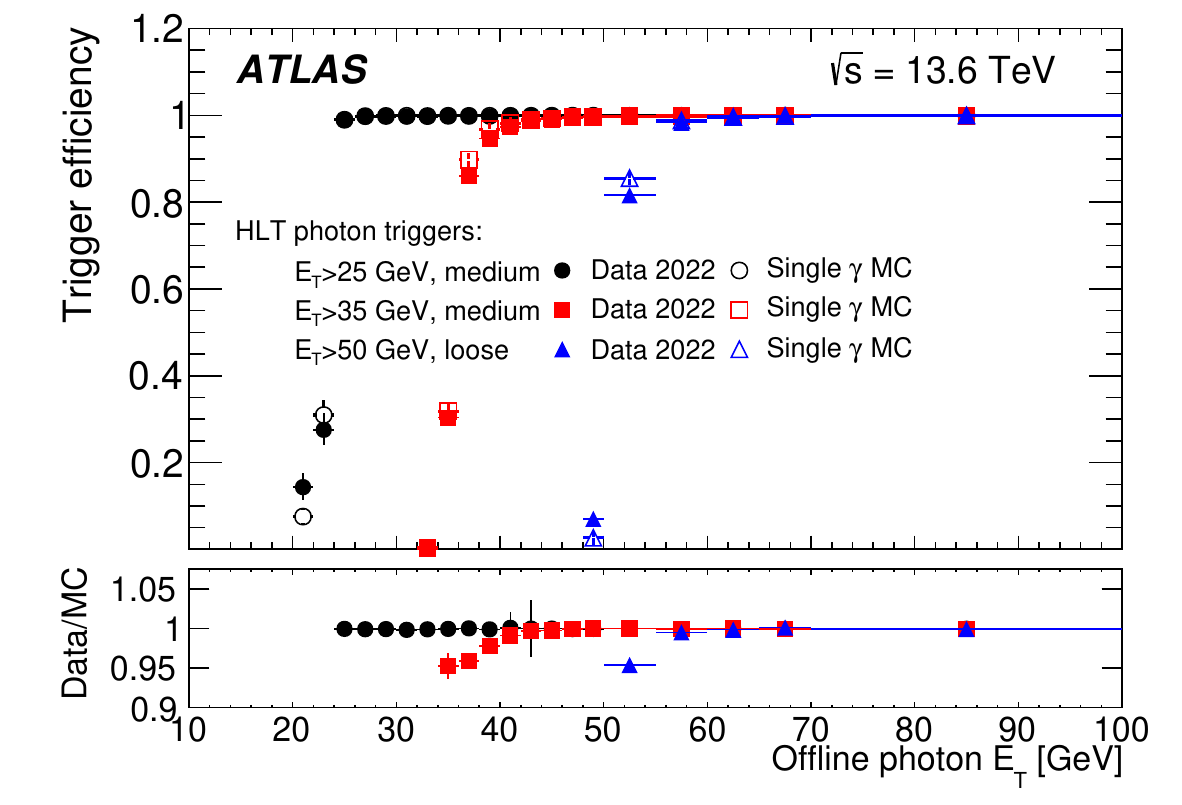}
\includegraphics[width=0.49\textwidth]{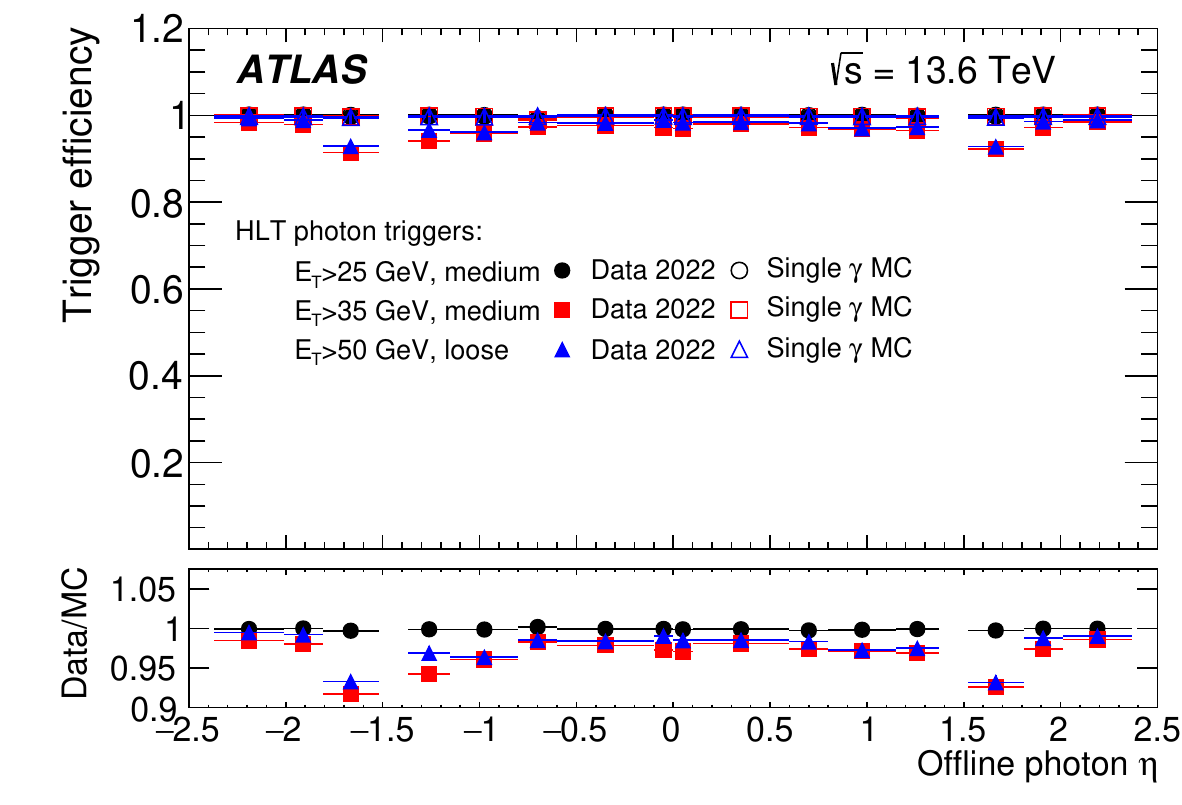}
\caption{Efficiency of the 25, 35 and 50~\gev\ legs of di-photon triggers as a function of the offline photon (left) \et and (right) $\eta$
for both MC simulation and 2022 data. The ratios of data to MC simulation efficiencies are also shown.
Efficiency is given with respect to the offline photons reconstructed with the tight offline identification and isolation requirements;
candidates in the calorimeter transition region $1.37<|\eta |<1.52$ are not considered.
Legacy non-isolated L1 EM trigger with $\et>20$\,\GeV\ defined in Ref.~\cite{TRIG-2018-05} is used as a seed.
For (right), only
offline candidates with \et\ values 5\,\gev\ above the corresponding trigger threshold are used.
Only statistical uncertainties are shown.
}
\label{fig:egamma4}
\end{figure}

\subsubsection{Trigger level analysis with photons}
\label{subsec:tla_photons}
 
Hadronically decaying low-mass resonances around the electroweak scale can be probed through composite trigger chains targeting
both the final-state di-jet pair and an additional feature, such as an initial-state radiation photon.
The Run-3 TLA workflow is upgraded to allow for the recording of HLT-reconstructed photons,
either standalone or in association with other TLA-compatible signatures to enable such searches at rates higher than
possible with the standard triggers. A new TLA trigger
records all jets in events with at least one tight photon~\cite{TRIG-2018-05}
with \et$>35$\,\gev\ and three additional jets with \pt$>25$\,\gev\ reconstructed with the PFlow algorithm, described in
Section~\ref{sec:jets}.
The photon will, in most events, be reconstructed as a jet as well, making this effectively a trigger for a signature with a photon and two jets.
A nominal HLT rate of such trigger is about 700\,Hz at
luminosity of $1.8\times\lumi{e34}$ at $\sqrt{s} = 13.6\,$\TeV.
This TLA approach allows
extended sensitivity of hadronic searches to resonance masses potentially as low as 100\,\GeV.
 
HLT photons saved to the TLA stream are decorated with additional variables computed within the HLT precision photon
reconstruction and calorimeter isolation trigger sequences. Such variables are intended
to enable the derivation of a custom residual photon energy calibration, as well as to
constrain the contribution from non-prompt, non-isolated photons in the TLA data set.
 
Figure~\ref{fig:tla_photon_performance} shows the energy response of HLT photons spatially matched to offline photons
in the $E_{T}$ range of interest for Run-3 TLA searches targeting photons. HLT photons calibrated with the
default 2022 calibration sequence are shown to be already within 2\% of offline-calibrated photons across the probed $E_{T}$ range.
 
\begin{figure}[htbp]
\centering
{\includegraphics[width=0.55\linewidth]{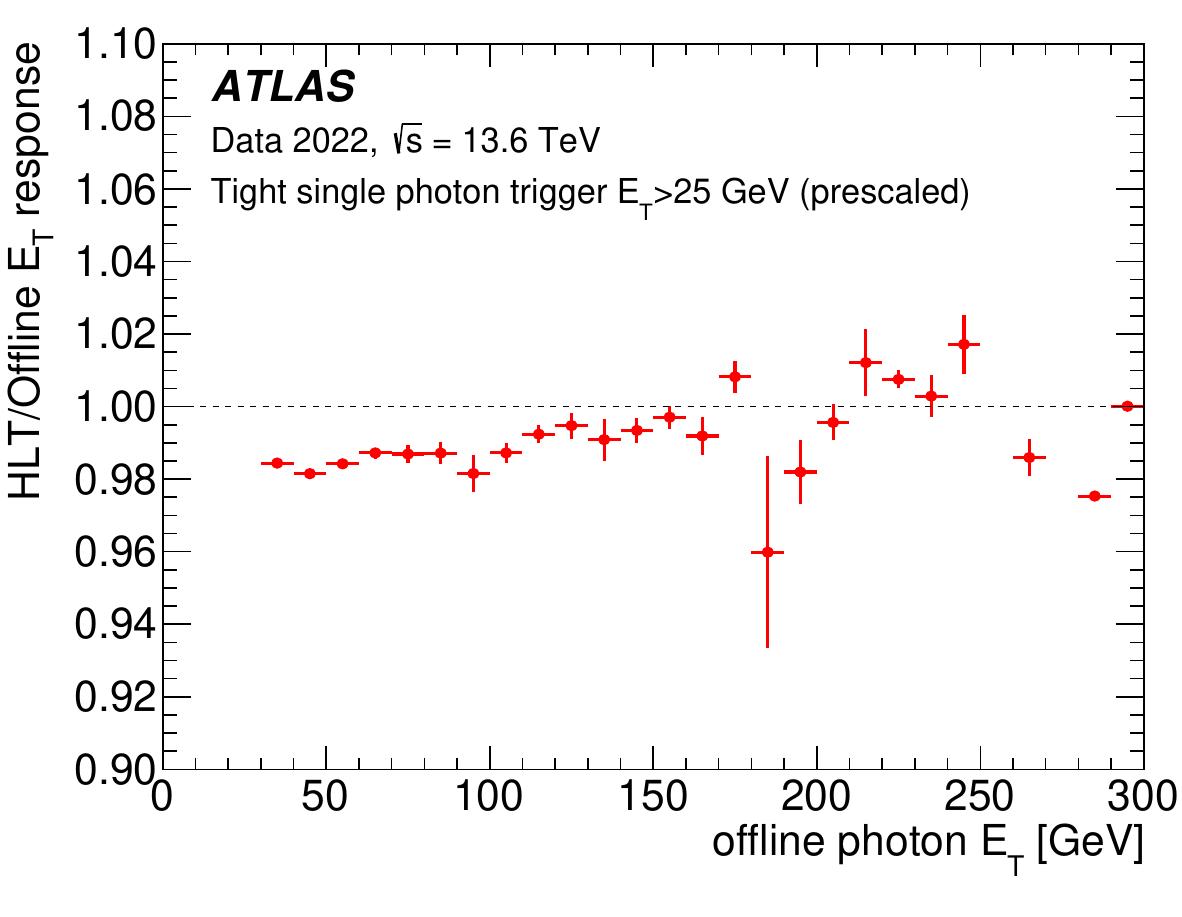}}
\caption{Energy response in TLA photons spatially matched to the leading offline photons.
HLT-level photons are calibrated with the default sequence employed within the HLT reconstruction for the 2022 data taking,
while reference offline photons are calibrated according to the Run-2 recommendations~\cite{EGAM-2018-01}.
Events are required to satisfy a 25~\gev\ tight single-photon prescaled support trigger.
Only statistical uncertainties are shown.
}
\label{fig:tla_photon_performance}
\end{figure}


\subsubsection{Electron and photon triggers for heavy-ion collisions}
\label{subsec:egamma_hi}
 
Due to the compactness of electron and photon showers in the detector, the standard L1 EM \rois can still be used in HI collisions to select electron and photon candidates with reasonable rate and purity.
Typical \et thresholds for L1 EM \rois are 12 and $16\,$\GeV.
However, the large UE contribution present in HI collisions leads to significant distortions of the shower shapes and subsequent inefficiency associated with the electron and photon identification requirements at HLT.
To improve the electron and photon trigger performance, the UE subtraction procedure is applied at the HLT.
 
Two electron HLT sequences are designed as primary physics triggers in the HI menu: a cut-based
trigger with an $\et > 20\,$\GeV\ requirement and a likelihood-based
trigger with an $\et > 15\,$\GeV\ requirement. For both triggers the UE subtraction is performed at the HLT as described in Section~\ref{sec:menuHI}.
For the likelihood-based electron identification~\cite{TRIG-2018-05}, the standard \pp probability density functions are evaluated using the UE-corrected variables.
An advantage of the likelihood-based approach is the significant reduction in the output rate in comparison to the cut-based trigger for a given \pt threshold.
The likelihood trigger has significantly better purity at the cost of a loss in trigger efficiency.
 
The primary photon HLT sequence in the HI trigger menu uses a cut-based
photon identification with an $\et >20\,$\GeV\ requirement
including the subtraction of the UE contribution.
After the UE subtraction procedure, the photon trigger efficiency remains high across the full range of centralities.


\subsection{Muons}
\label{sec:muons}
 
The reconstruction of muon candidates at the HLT is described in detail in Section~\ref{sec:muonrec}.
The criteria for selecting muon candidates are dependent on the algorithm used for their reconstruction.
The MS-only chains select on the \pt of the muon candidate measured solely by the muon spectrometer,
while the combined muon chains apply requirements on the matching between the ID and MS tracks and their
combined \pt.
 
\subsubsection{Muon trigger menu}
\label{subsec:muonmenu}
Muon triggers cover a wide momenta range between a few \gev, for $B$-meson-decay studies,
up to several \tev\, for new phenomena searches.
The primary triggers in the muon trigger menu
include single-muon triggers with and without isolation requirements, symmetric and asymmetric
di-muon and multi-muon triggers.
The Run-3 muon trigger menu is similar
to that used in \runii~\cite{TRIG-2018-01}, accounting for
the refinement of L1Muon thresholds,
discussed in Section~\ref{sec:L1Muon}.
 
The improved suppression of fake muons and the 2022 running conditions allowed for a lowered HLT threshold
of the lowest-unprescaled isolated single-muon trigger
by 2\,\gev\ to \pt$>24$\,\GeV\ maintaining its rate at about 200\,Hz at a luminosity of $1.8\times\lumi{e34}$.
As in \runii, combined muon candidates must fulfil the following track isolation requirement:
the scalar sum of the \pt values of tracks within a variable-size cone around the muon (excluding its own track)
must be less than 7\% of the muon \pt. The track isolation cone size for muons, $\Delta R$, is given by the smaller of
$\Delta R = 10$~GeV/\pt and $\Delta R = 0.3$.
A non-isolated trigger with $\pt>50$~GeV helps to increase the efficiency for high-\pt muons.
Additionally, a trigger that selects only muons in the barrel region ($|\eta|<1.05$)
reconstructed using MS-only information is available at \pt$>60$\,\gev.
 
There were no changes to the multi-muon thresholds in 2022 with respect to \runii.
Two combined muon candidates are required
with a \pt threshold of 14\,\gev\ each at the HLT with a rate of 24\,Hz at a luminosity of $1.8\times\lumi{e34}$.
To avoid an efficiency loss due to the limited acceptance of L1, di-muon and tri-muon triggers
seeded by the single L1 muon trigger 
are also present in the menu.
These sub-leading muons, with \pt greater than 8\,\gev\ and (4\,\gev, 4\,\gev) for di-muon and
tri-muon triggers, respectively,
are reconstructed by the muon full scan algorithm described in Section~\ref{sec:muonrec}.
Low-\pt di-muon triggers are further discussed in Section~\ref{sec:bls}.
 
\subsubsection{Muon trigger performance}
The muon efficiencies are determined using the tag-and-probe method with
$Z \rightarrow \mu \mu$ events.
The efficiencies of the combination of the two lowest-unprescaled single-muon triggers
are shown in Figures~\ref{fig:muon_eff:L1_HLT_barrel} and ~\ref{fig:muon_eff:L1_HLT_endcap}.
The HLT efficiency relative to L1 is close to 100\% both in the barrel and in the endcaps.
The L1 muon trigger efficiency is about 60\% in the barrel and 80\% in the endcap
regions for offline medium muons~\cite{MUON-2018-03} with \pt$>25$\,\gev.
 
The measured trigger efficiency in Figure~\ref{fig:muon_eff:L1_HLT_barrel} is lower than that of the expected efficiency in MC.
This is due to the L1 muon trigger in the barrel region being simulated with an optimistic lower-bound on chamber
efficiency to allow for a reasonable MC efficiency for chambers whose efficiencies are later recovered during data taking.
Scale factors derived from these observed differences are used to correct other MC simulation samples used in data analyses.
 
\begin{figure}[htbp]
\centering
\includegraphics[width=0.49\textwidth]{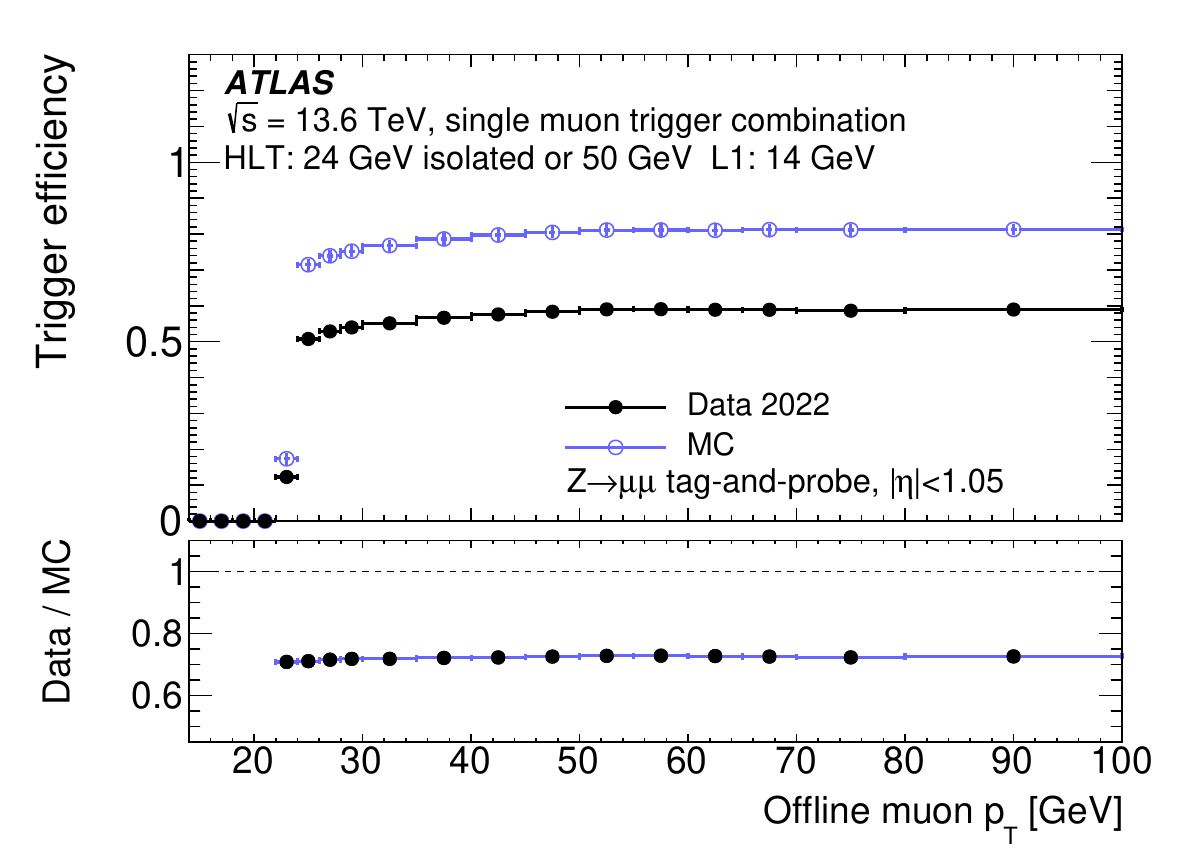}
\includegraphics[width=0.49\textwidth]{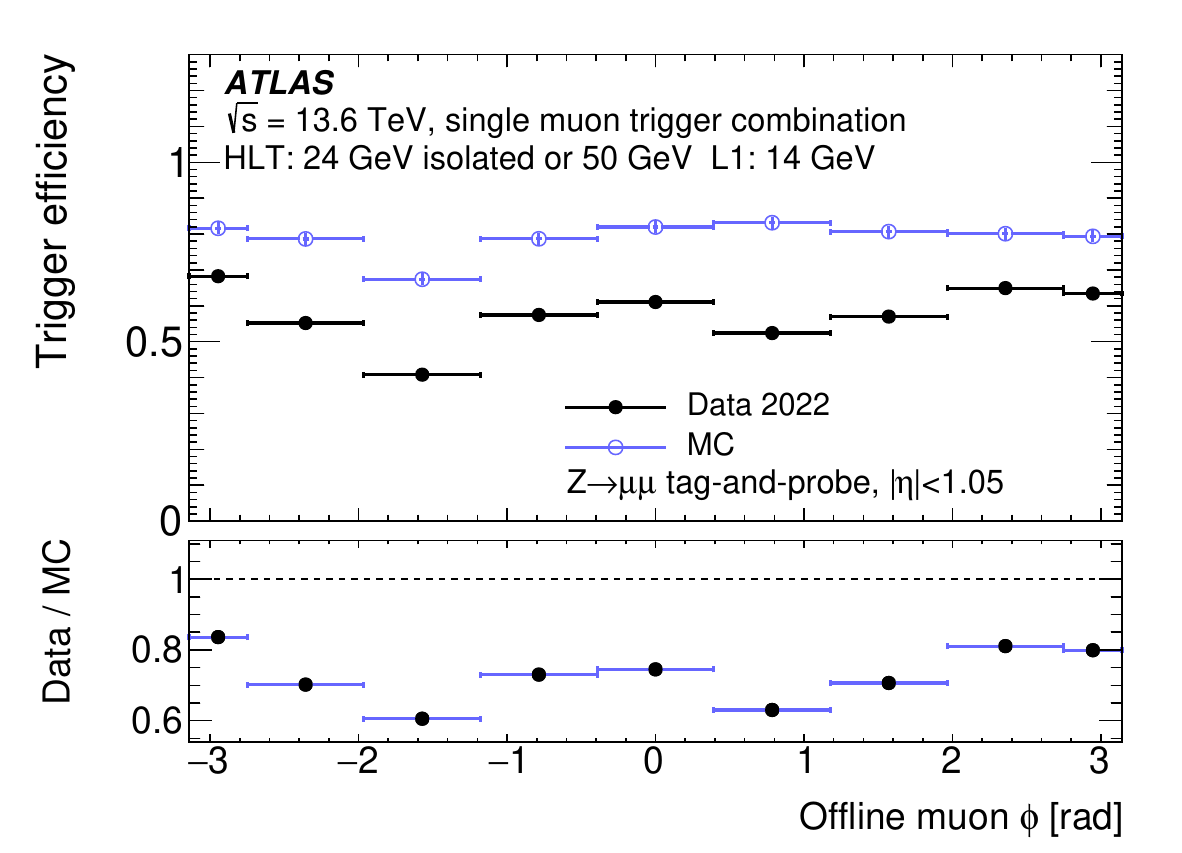}
\caption{
Muon trigger efficiencies as a function of (left) \pt and (right) $\phi$ for \pt$>25$\,\gev\
of the offline medium muon for the combination of two single-muon triggers in the barrel region.
The ratios of data to MC simulation efficiencies are also shown.
Only statistical uncertainties are shown.
}
\label{fig:muon_eff:L1_HLT_barrel}
\end{figure}
 
\begin{figure}[htbp]
\centering
\includegraphics[width=0.49\textwidth]{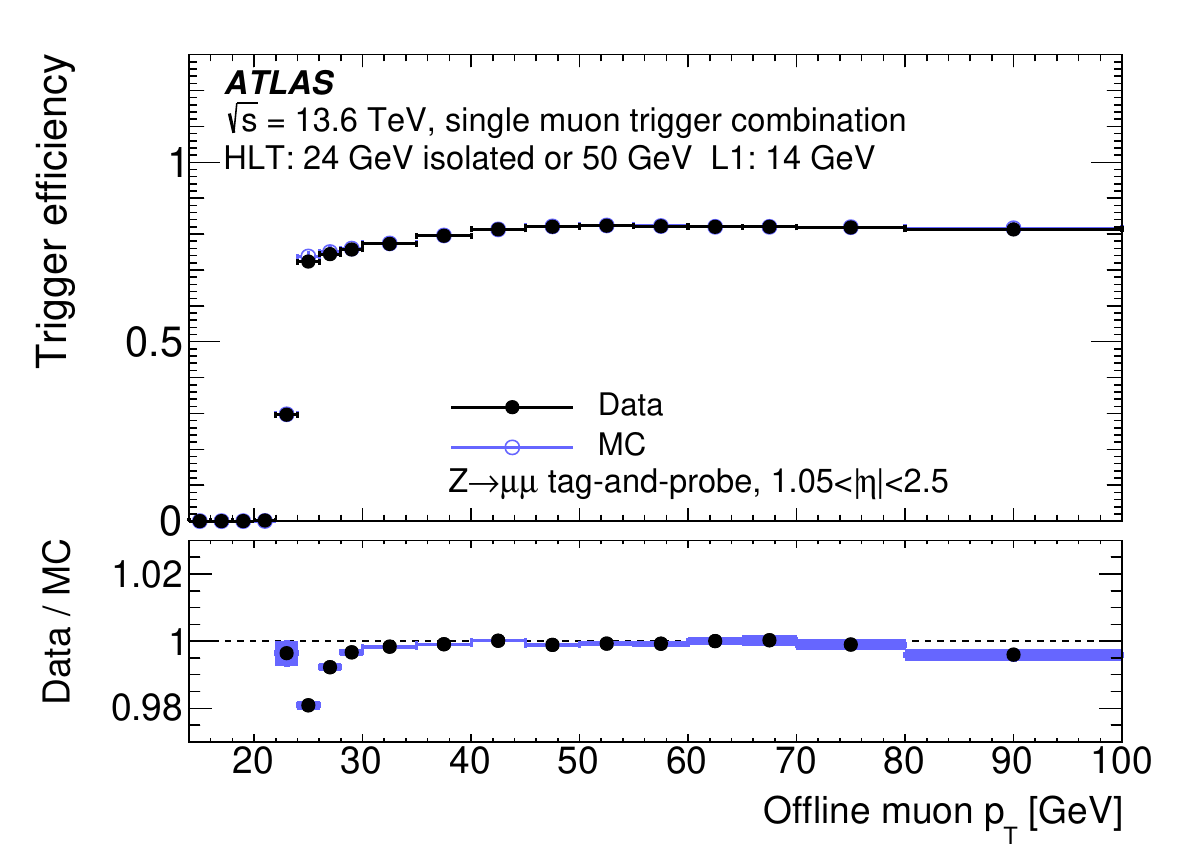}
\includegraphics[width=0.49\textwidth]{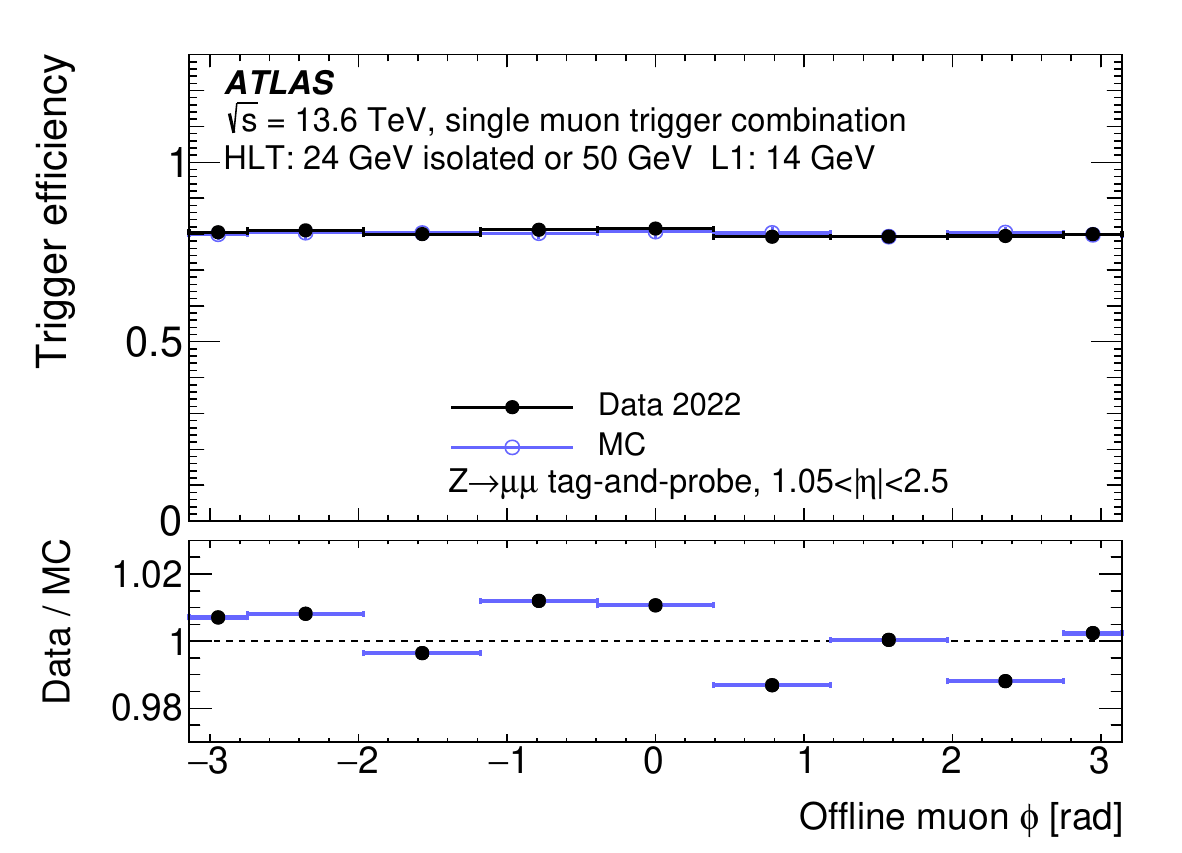}
\caption{
Muon trigger efficiencies as a function of (left) \pt and (right) $\phi$ for \pt$>25$\,\gev\
of the offline medium muon for the combination of two single-muon triggers in the endcap region.
The ratios of data to MC simulation efficiencies are also shown.
Only statistical uncertainties are shown.
}
\label{fig:muon_eff:L1_HLT_endcap}
\end{figure}


\subsection{Taus}
\label{sec:tau}
 
\subsubsection{Tau trigger reconstruction and selection}
 
The HLT tau trigger targets only hadronic decays of tau lepton candidates ($\tauhadvis$). Events with taus decaying leptonically are
recorded by electron and muon triggers described in Sections~\ref{sec:egamma} and \ref{sec:muons}, respectively.
The HLT tau trigger reconstruction is subdivided in two steps: a calorimeter-only preselection and
track reconstruction with an offline-like selection.
 
At the first step, the $\tauhadvis$ candidate is reconstructed purely from calorimeter information.
Calorimeter cells inside the \roi identified at L1 are retrieved and the topo-clustering algorithm,
described in Section~\ref{sec:hltcalo}, is executed.
Thanks to the full detector granularity and the bunch-by-bunch pile-up corrections, the energies of these reconstructed
topo-clusters are very close to the offline ones. These clusters are calibrated with the local hadron calibration
(LC)~\cite{lc} and their vectorial sum is used as a `jet seed' for the reconstruction of the $\tauhadvis$ candidate. The energy of the $\tauhadvis$
candidate is calculated from the LC clusters in a cone of $\Delta R < 0.2$ around the barycentre of the jet seed.
A dedicated $\tauhadvis$ energy calibration is applied to improve the precision of the energy measurement and follows
the offline procedure~\cite{ATL-PHYS-PUB-2022-044}. Then, a selection on the minimum \pt{} of the $\tauhadvis$ candidate is applied, and only the remaining candidates pass to the next step.
 
The second step of the HLT tau trigger first runs a fast-tracking algorithm,
followed by a precise measurement of the tracks associated with the $\tauhadvis$ candidate
and a final $\tauhadvis$ identification based on a Recurrent-Neural-Network (RNN) algorithm.
The fast tracking algorithm is a trigger-specific pattern recognition algorithm that runs in
two stages. In the first stage, the leading track is sought in a narrow
$\Delta \eta \times \Delta \phi$ around the $\tauhadvis$ candidate along the entire beamline.
In the second stage, additional tracks associated with the $\tauhadvis$ candidate are sought in a
larger $\Delta \eta \times \Delta \phi$ region but in a narrow range around the leading track
along the beamline. This strategy is CPU-efficient as it minimises the volume in which the
pattern recognition algorithm is executed, as discussed in Section~\ref{sec:id_tau}.
 
A precision-tracking algorithm similar to the offline one is run using the tracks identified by the second step of the fast tracking as seeds to measure their properties more precisely. Using these tracks as well as the calorimeter information, the input variables for the $\tauhadvis$ identification are computed. Three sets of variables for zero-prong, one-prong and multi-prong $\tauhadvis$ candidates are used depending on whether the number of precision tracks associated to the $\tauhadvis$ in $\Delta R < 0.2$ is zero, one, or more than one, respectively. The architecture implementation of the RNN follows closely its offline counterpart, as described in Ref.~\cite{ATL-PHYS-PUB-2022-044}. Finally, tau candidates are required to have up to three tracks within $\Delta R < 0.2$ and up to one track within $0.2 < \Delta R < 0.4$ around the $\tauhadvis$. In addition, they are required to be identified by the RNN. Given the small increase in event rate and potential efficiency gain, the identification working point is loosened for the $\tauhadvis$ $\pt > 280\,$\GeV, and both criteria (identification and number of tracks) are completely dropped when the $\tauhadvis$ $\pt > 440\,$\GeV.
A different RNN is used to trigger on \ac{LLP} and trained for zero-, one- and multi-prong taus using a MC sample of \ac{LLP}.
 
\subsubsection{Tau trigger menu}
 
The tau trigger menu selects a wide spectrum of final states that involve hadronically decaying tau leptons.
There are four main categories of triggers: single-tau, di-tau, tau+X (X=light leptons, \met) and
events for tag-and-probe performance studies. For each of these categories, the definition of
the triggers varies based on (i) the identification requirement, (ii) the \pt{} threshold applied and,
for the multi-object final states, (iii) the eventual presence of a topological cut on the angular
distance between the two objects selected in the final state. Dedicated triggers for identifying
tau leptons originating from the decay of \ac{LLP} are included. The lowest threshold of the
single-tau unprescaled triggers is 160\,\GeV\ and 180\,\GeV\ for the standard and LLP taus.
The individual rates are about 40\,Hz and 50\,Hz, respectively, at a luminosity of $1.8\times\lumi{e34}$. The overlap fraction of the higher threshold LLP trigger with the 160\,\GeV\ standard trigger is 90\%.
Lower threshold single-tau triggers with or without looser requirements are prescaled at L1 and/or the HLT
and are used to support the primary triggers. There are two primary di-tau triggers, which run at rates of 25\,Hz and 10\,Hz, respectively.
The first, main, trigger requires two tau candidates of \pt{}$>35$ and 25\,\GeV\ and the second, LLP, trigger requires \pt{}$>80$ and 60\,\gev.
LLP tau triggers are discussed further in Section~\ref{sec:unconTrack}.
 
\subsubsection{Tau trigger performance}
The tau trigger efficiency is determined using a tag-and-probe
analysis, which selects events in two complementary signal regions~\cite{TRIG-2016-01},
one enriched in $Z\rightarrow \tau\tau$ events and another one
enriched in dileptonic $t\bar{t}\rightarrow W(e\nu)bW(\tau\nu)b$ decay events. The events for the
efficiency measurement are selected using single light lepton triggers
as well as dedicated tag-and-probe trigger chains that use as ``tag'' a
single light lepton trigger other than the tau. The combination of the
two signal regions allow for the measurement of the tau trigger
performance with sufficient statistics both at the low and high \pt{}
regimes up to a tau \pt{} of about 200\,\GeV. The background events in
these signal regions mostly come from events where a jet has
been misidentified as a hadronic $\tau$, and these are estimated using
either data-driven methods or simulated events. In the
$Z\rightarrow \tau\tau$ enriched region the background from
misidentified jets come from QCD or $W$+jets processes, while in
the $t\bar{t}$ enriched region the background comes from QCD or
semi-leptonic $t\bar{t}$ decay events. Other minor background sources
are due to diboson or $Z\rightarrow ee$ events where an electron is
mis-reconstructed as a hadronic $\tau$.  Figure~\ref{fig:tau_pt} shows
the comparison between data and \ac{MC} simulation of the tau \pt{} in the
one-prong case for the $Z\rightarrow \tau\tau$ and $t\bar{t}$
selections obtained with the electron tag and tau probe trigger.
 
\begin{figure}[htb!]
\centering
\includegraphics[width=0.49\textwidth]{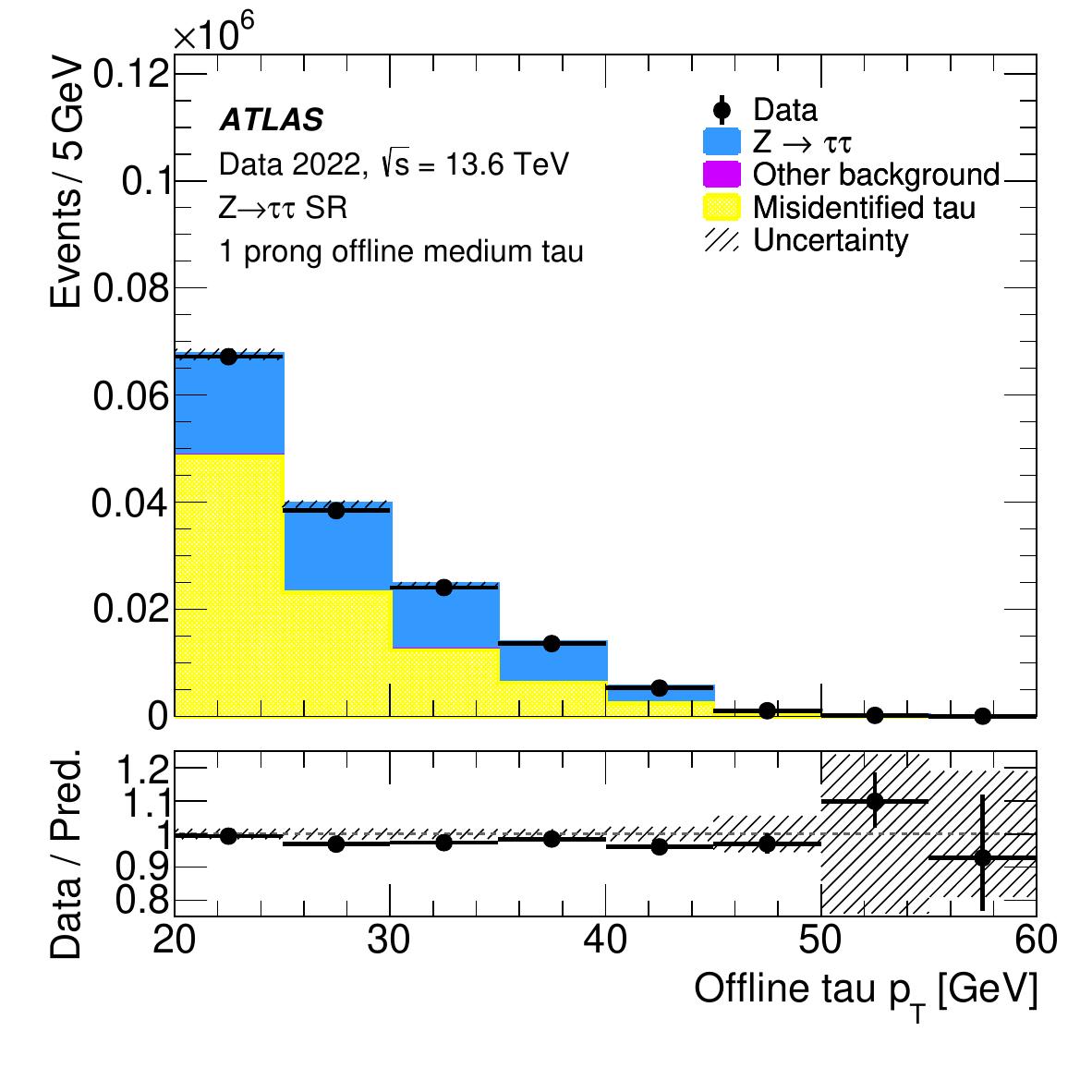}
\includegraphics[width=0.49\textwidth]{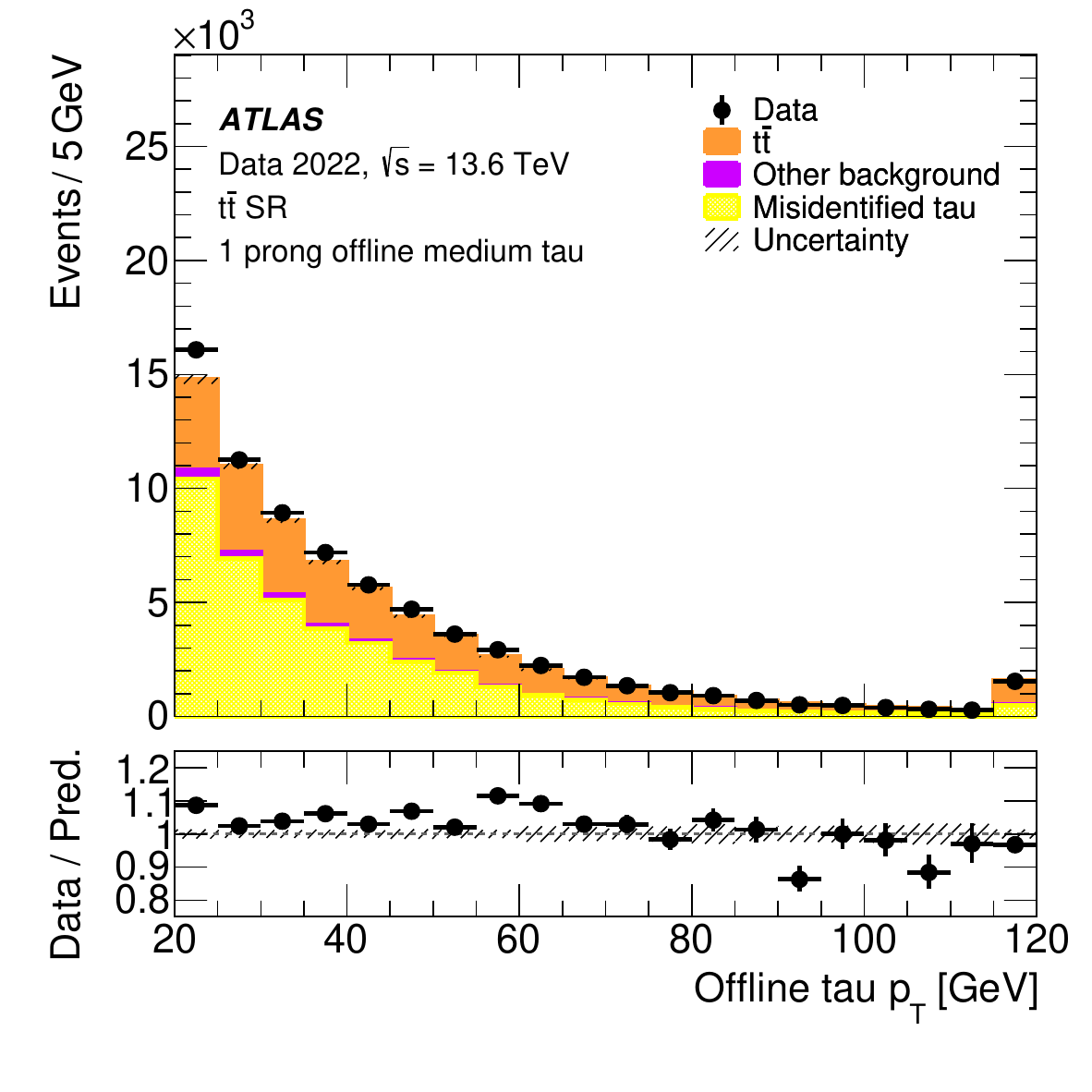}
\caption{Distributions for data and MC simulation of the \pt{} of
one-prong taus in (left) the $Z\rightarrow \tau\tau$ and (right)
the $t\bar{t}$ signal regions (SRs). The data set 
was recorded with the $e-\tau$ tag-and-probe trigger.
The bottom panels show the ratio
between data and the signal-plus-background prediction. Only
statistical uncertainties are shown.}
\label{fig:tau_pt}
\end{figure}
 
The tau trigger efficiency is calculated with respect to the reconstructed offline tau candidates and is separated between one-prong and three-prong tau cases. Figure~\ref{fig:taueff} shows the resulting efficiency as a function of the offline tau \pt{} for one-prong and three-prong tau candidates using 2022 data recorded 
with a combined electron tag and tau probe trigger.
 
\begin{figure}[htb!]
\centering
\includegraphics[width=0.49\textwidth]{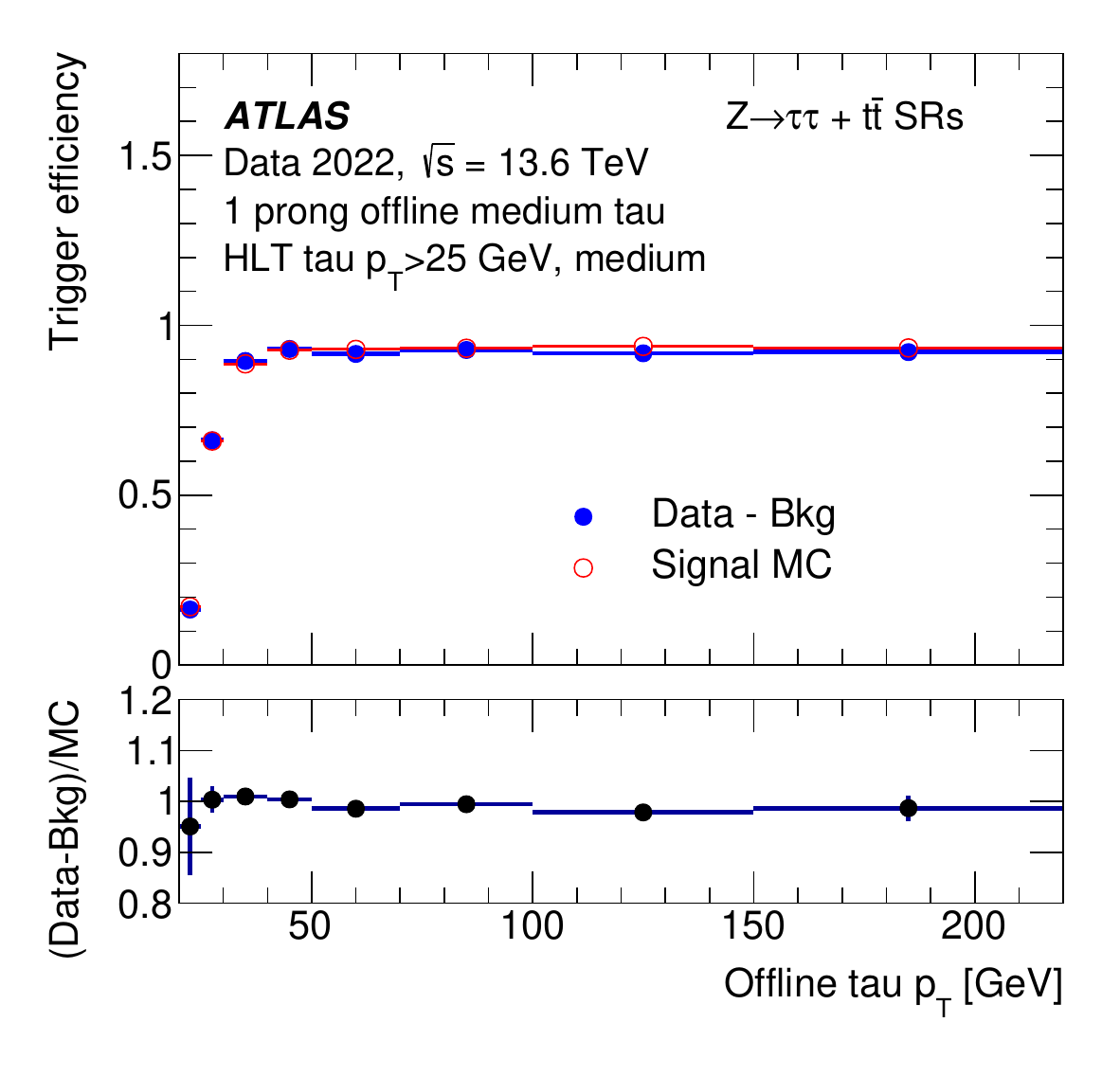}
\includegraphics[width=0.49\textwidth]{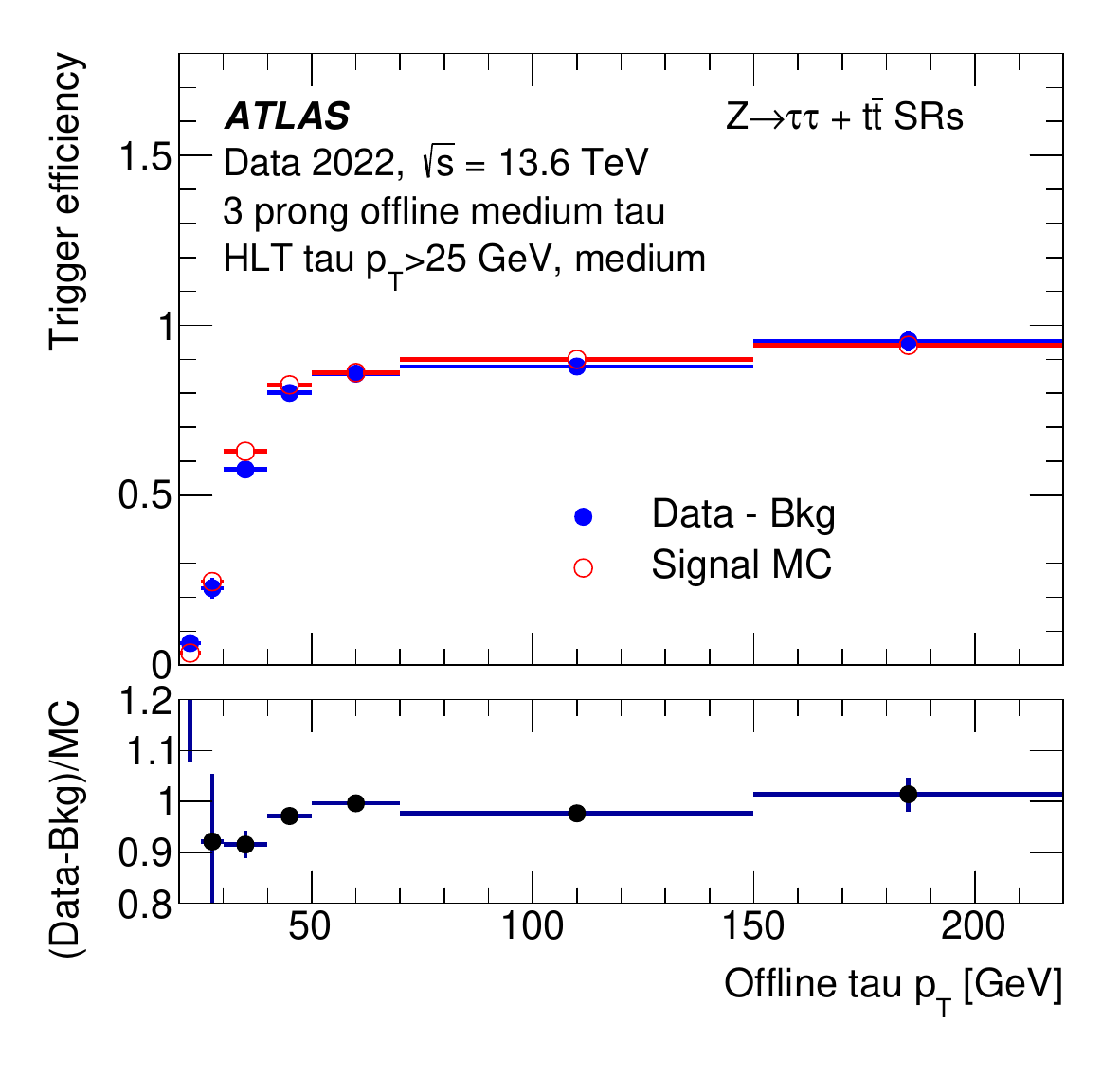}
\caption{Tau trigger efficiencies as a function of offline tau
\pt{} for (left) the one-prong and (right) three-prong tau
candidates. The legacy isolated L1 tau trigger with $\et>12$\,\GeV\ defined in Ref.~\cite{TRIG-2016-01} is used as the seed.
The measurement is performed using a combination of
$Z\rightarrow \tau\tau$ and $t\bar{t}$ selections~\cite{TRIG-2016-01} in events 
recorded with the $e-\tau$ tag-and-probe trigger. The efficiency from the sum
of the $Z\rightarrow \tau\tau$ and $t\bar{t}$ signal processes
estimated with MC (data after subtracting the background) is shown with open (filled)
markers. The bottom panels show the ratio of the efficiencies.
Only statistical uncertainties are shown.}
\label{fig:taueff}
\end{figure}
 
In the one-prong selection, the resulting efficiency of the tau
trigger reaches a plateau slightly below 95\% for tau candidates with
$\pT>40\,$\GeV, while it is about 90\% for the three-prong
case. The small differences between the data and MC signal prediction observed in Figure~\ref{fig:taueff}
are used to derive scale factors to correct other MC simulation samples used in data analyses.


\subsection{Jets}
\label{sec:jets}
 
\subsubsection{Jet trigger reconstruction}
 
A detailed description of jet triggers in \runii can be found in Ref.~\cite{TRIG-2016-01}. At the HLT, small- and
large-radius jets are reconstructed using the \antikt algorithm~\cite{Cacciari:2008gp} with a radius parameter
of $R=0.4$ and $R=1.0$, respectively. While jet triggers in \runii employed only calorimeter topo-clusters 
as input, Run-3 triggers are able to rely on PFO 
thanks to the extended tracking capabilities offered by the HLT in \runiii.
 
Despite the large reductions in tracking time at the HLT, the CPU time usage of jet chains is still
high due to full scan tracking, as discussed in Section~\ref{sec:id_full}.
For high-\pt jet chains such as single- and multi-jet chains, the large gap between L1 and HLT thresholds allows for an early reduction in event rate via the application of a calorimeter-only jet selection (\emph{calorimeter preselection}) before any kind of tracking is executed. It is based on topo-cluster jets,
described in Section~\ref{sec:hltcalo}. Minimal HLT jet \pt requirements are adjusted to obtain the maximal CPU-time reduction which does not impact trigger efficiencies.
The preselection stage is found to reduce the output rates of HLT jet trigger chains by less than 2\%.
Table~\ref{tab:jet_presel} shows the impact of different preselections on the event rates of various jet triggers.
 
\begin{table}[h!]
\centering
\caption{Impact of calorimeter preselections on the event rates of jet triggers in which tracking reconstruction is performed. 
The percentage of events passing the preselection step is estimated with respect to the original L1 trigger rate from a reprocessing of Run-2 Enhanced-bias data \cite{ATL-DAQ-PUB-2016-002}.
The HLT selection column shows the lowest jet trigger threshold which this preselection is applied for.}
\label{tab:jet_presel}
\begin{tabular}{cccc}
\toprule
L1 selection & Preselection & HLT selection     & Events passing preselection \\
\midrule
1 jet, $\pt>100\,$\gev\ & 1 jet, $\pt>225\,$\gev\ & 1 jet, $\pt>275\,$\gev\  & 23\% \\  
3 jets, $\pt>50\,$\gev\ & 4 jets, $\pt>85\,$\gev\  & 4 jets, $\pt>115\,$\gev\  & 15\% \\  
4 jets, $\pt>15\,$\gev\  & 5 jets, $\pt>50\,$\gev\  & 5 jets, $\pt>70\,$\gev\  & 7\% \\  
4 jets, $\pt>15\,$\gev\ &  6 jets, $\pt>40\,$\gev\ & 6 jets, $\pt>55\,$\gev\  & 6\% \\  
\bottomrule
\end{tabular}
\end{table}
 
Even though this strategy works well for baseline jet triggers, it is not optimal for triggers
whose HLT jet thresholds are very close to the L1 thresholds. One example of such triggers is the $b$-jet triggers,
where a more complex preselection step 
detailed in Sections~\ref{sec:id_bjet} and \ref{sec:bjet_sel} is included in the HLT reconstruction.
 
After the preselection step, fast full scan tracking is executed as detailed in Section~\ref{sec:id_full}
and the resulting tracks are extrapolated to the calorimeter. The PFO formation starts by matching
full scan tracks and topo-clusters taking into account the extrapolated track position and topo-cluster location.
The topo-clusters which are not matched to any track are referred to as neutral PFOs.
The matched topo-clusters have their energy removed according to the expected calorimeter energy
deposited by the matched track. The tracks are considered as charged PFOs. Any topo-cluster surviving
the energy subtraction procedure becomes a neutral PFO. Charged PFOs not matched to the primary
vertex are discarded,
which is the dominant means of pile-up suppression.
Outside the geometrical acceptance of the tracker, only the calorimeter information is available. Hence, in the forward region
the topo-clusters alone are used as inputs to the PFlow jet reconstruction.
 
After reconstruction, small-$R$ jets are calibrated through a procedure similar to that used
offline~\cite{JETM-2018-05}. A correction accounting for pile-up contamination is applied on
an event-by-event basis to the jets. This is followed by another correction, compensating for
the energy response of the calorimeter. Finally, a sequence of calibrations is applied to
PFlow triggers to correct for the residual discrepancies between reconstructed and simulated
jets and accounts, for example, for energy differences resulting from the different showering of quarks and gluons.
 
For \runiii, large-$R$ jet triggers are also extended 
to use PFOs as inputs.
In addition, the Constituent Subtraction~\cite{Berta_2019} and SoftKiller~\cite{Cacciari_2015}
algorithms are applied to neutral PFOs to subtract energy contamination originating from pile-up interactions.
After this step,
Soft Drop grooming~\cite{Larkoski_2014} with parameters $\beta=1.0$ and $z_\text{cut} =0.1$ is
applied to mitigate the contamination from initial state radiation and underlying event.
Finally, a calibration similar to small-$R$ jets is used to correct for the energy response of
the ATLAS calorimeter and adjust the reconstructed mass of the large-$R$ jet.
 
\subsubsection{Jet trigger menu}
 
Jet triggers are used for a wide set of measurements, ranging from precision physics measurements to detector performance studies.
The inclusive jet, di-jet and multi-jet production measurements rely on the events selected
by small-$R$ single- and multi-jet triggers. Events selected by these triggers are also employed for the calibrations of the jet
energy scale and resolution, as well as for new physics searches such as supersymmetry. Another important class of jet
triggers is represented by large-$R$ triggers, selecting final states with boosted weak vector bosons (W/Z) or Higgs bosons. These triggers
are generally employed by analyses searching for heavy resonances predicted, for example, by theories of extra-dimensions and the two-Higgs-doublet model.
 
The jet trigger chains are initiated by L1 algorithms selecting single jets, multi-jets, or, for very low-\pt{} HLT thresholds,
a random trigger at L1.
The primary, unprescaled jet chain applies a threshold of 420\,\gev{} and has a rate of 42 Hz (at $1.8\times\lumi{e34}$).
Small-$R$ single jet chains with \pt{} thresholds below 420\,\GeV{} are prescaled to provide a complete jet \pt{} spectrum, with a constant rate per chain ranging from 1 to 4\,Hz; large-$R$ single jets with thresholds between 110~\GeV{} and 460~\GeV{} are also collected using prescaled triggers at a rate range between 2 and 3 Hz per chain.
Trigger chains, which select events based on the scalar sum of the transverse momentum of all jets (\HT), contribute with a rate of 34 Hz.
Topo-cluster-based jet triggers are also available in \runiii as a backup for PFlow jet triggers.
 
\begin{figure}
\centering
\includegraphics[width=0.49\textwidth]{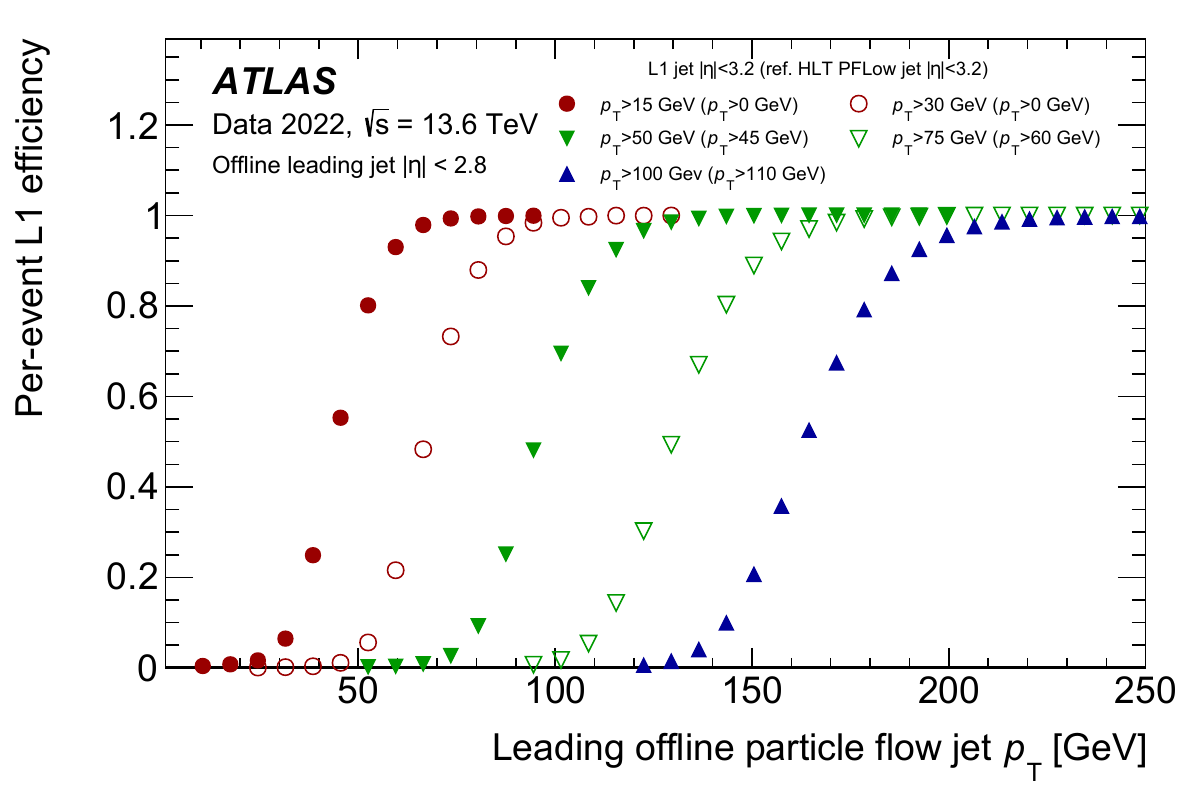} 
\includegraphics[width=0.49\textwidth]{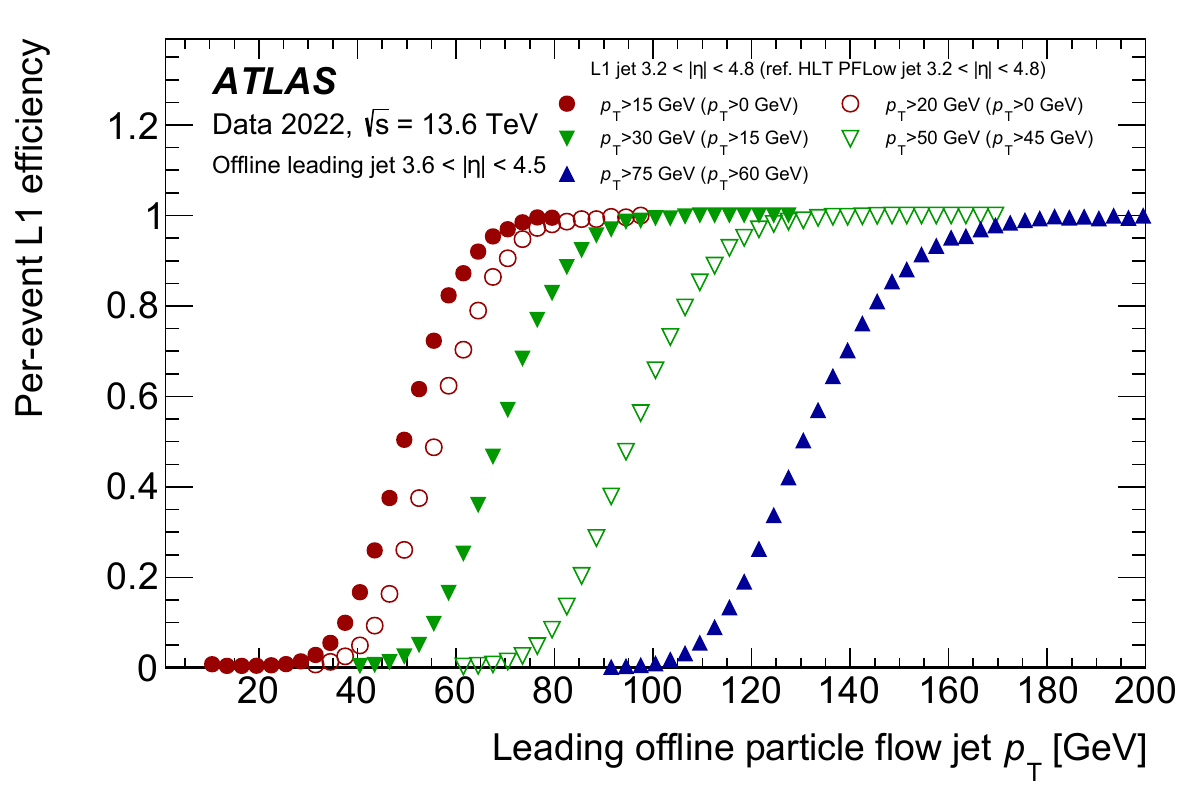} 
\caption{Efficiency of the legacy L1 single jet triggers in the (left) central
and (right) forward
regions. The efficiency is computed using the bootstrap method with respect to events taken by an independent trigger, shown in brackets, that is 100\% efficient in the relevant region.
Only statistical uncertainties are shown.}
\label{fig:jet:l_1}
\end{figure}
 
\begin{figure}
\centering
\includegraphics[width=0.49\textwidth]{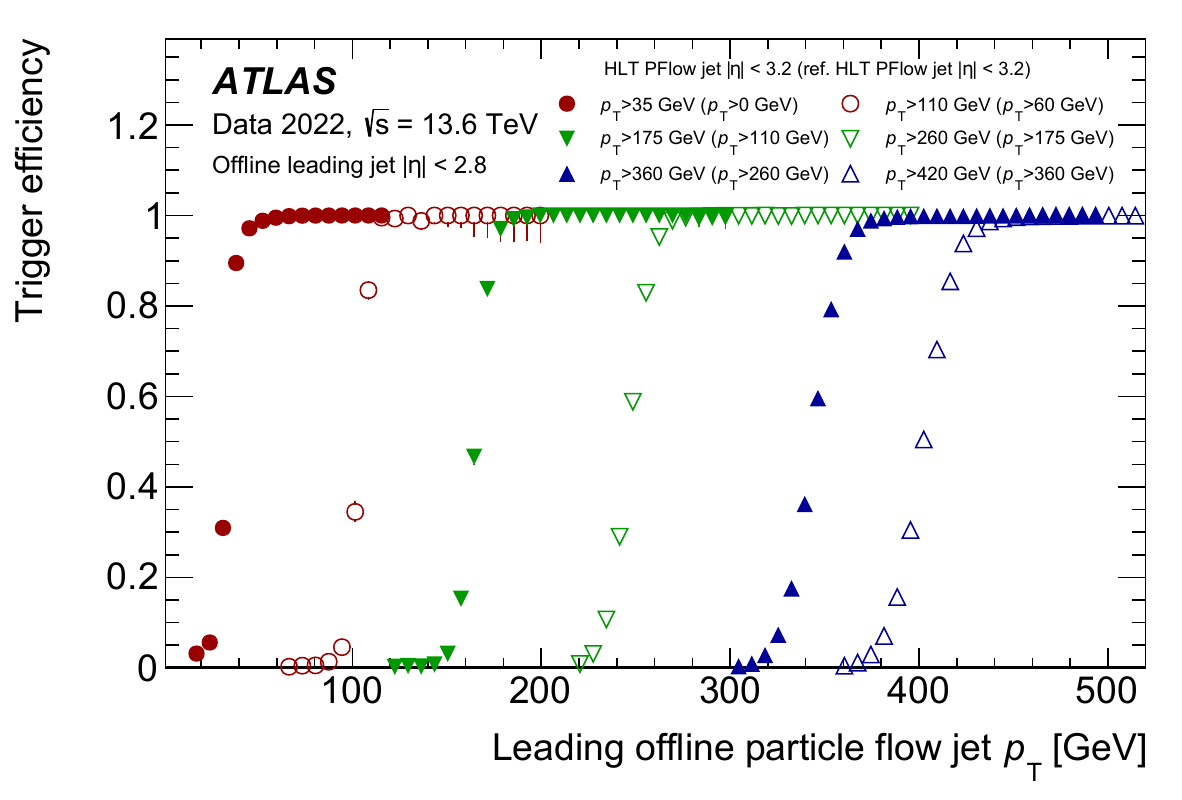}  
\label{fig:jet:hlt_c}
\includegraphics[width=0.49\textwidth]{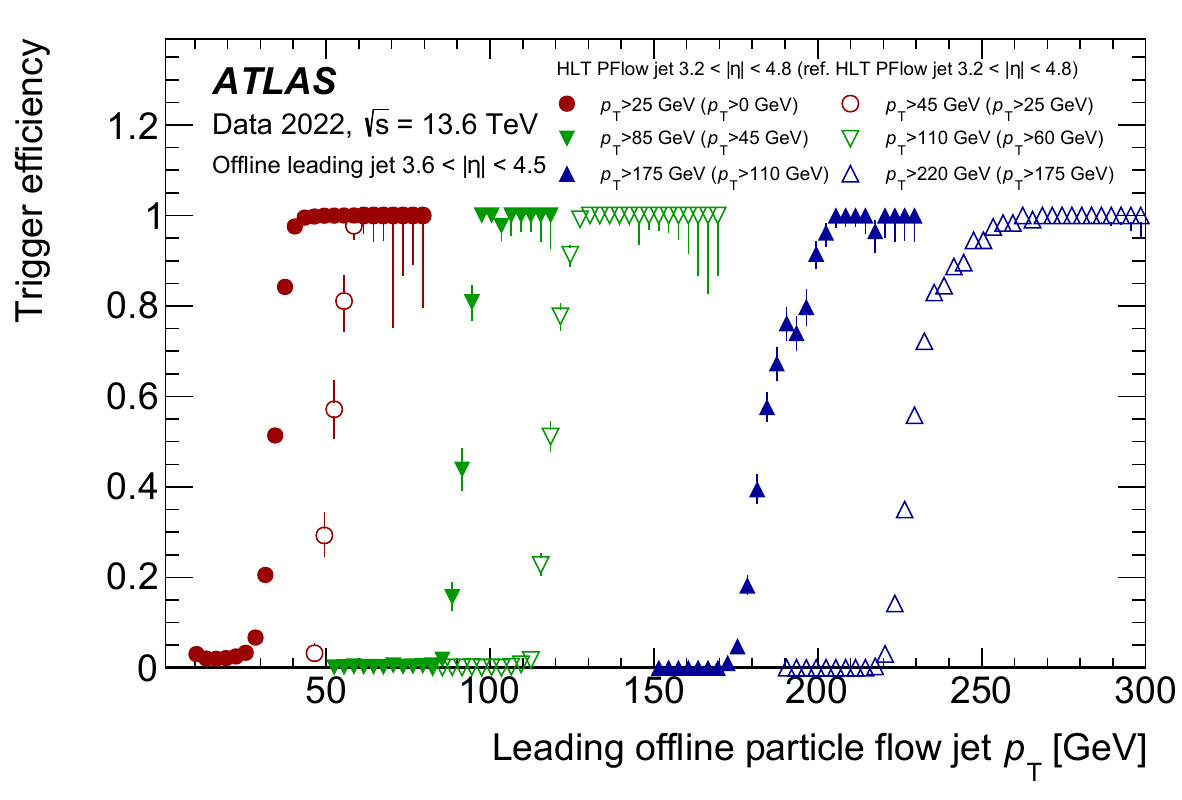} 
\label{fig:jet:hlt_f}
\caption{Efficiency of single jet trigger selection by the HLT in the (left) central 
and (right) forward
regions. Central PFlow jet triggers exploit the particle flow reconstruction with calorimeter clusters and tracks,
while forward PFlow jet triggers rely on the topo-cluster-based reconstruction
to trigger events outside the ID acceptance ($|\eta| < 2.5$). The efficiency is computed using the bootstrap method with respect to events taken by an independent trigger, shown in brackets, that is 100\% efficient in the relevant region.
Only statistical uncertainties are shown.}
\label{fig:jet:hlt}
\end{figure}

\begin{figure}
\centering
\includegraphics[width=0.49\textwidth]{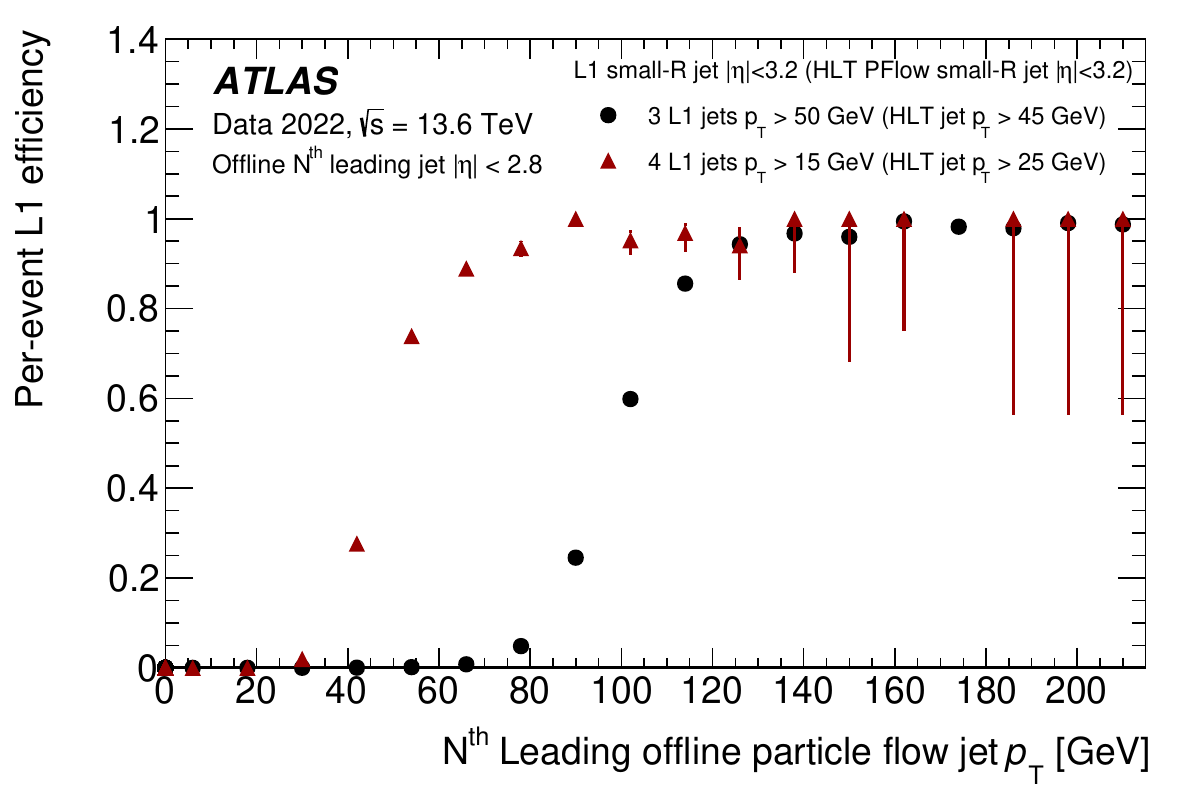} 
\includegraphics[width=0.49\textwidth]{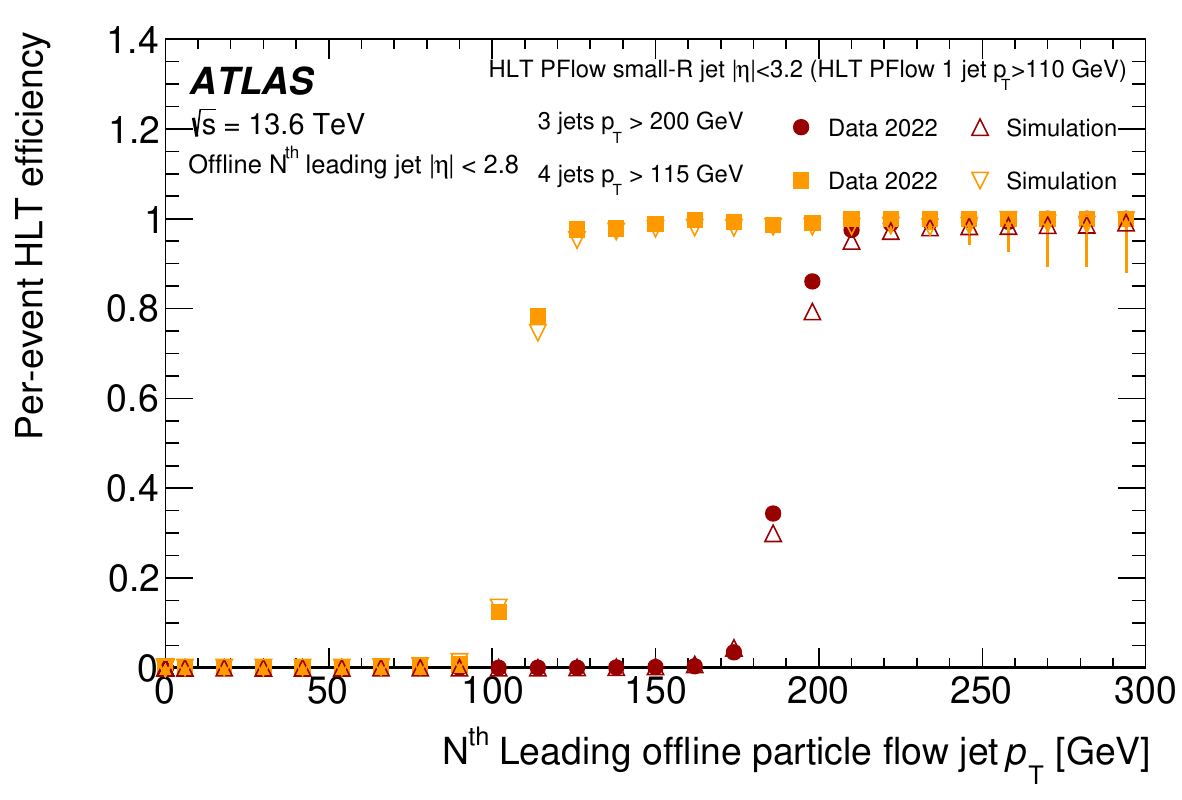} 
\caption{Multi-jet trigger efficiency as a function of the offline reconstructed $N^\text{th}$ jet \pt (left) at L1 
and (right) in the HLT. The efficiency is computed using the bootstrap method with respect to events taken by an independent single jet trigger,
shown in brackets, that is 100\% efficient in the relevant region. Only statistical uncertainties are shown.}
\label{fig:jet:mj}
\end{figure}
 
\begin{figure}
\centering
\includegraphics[width=0.49\textwidth]{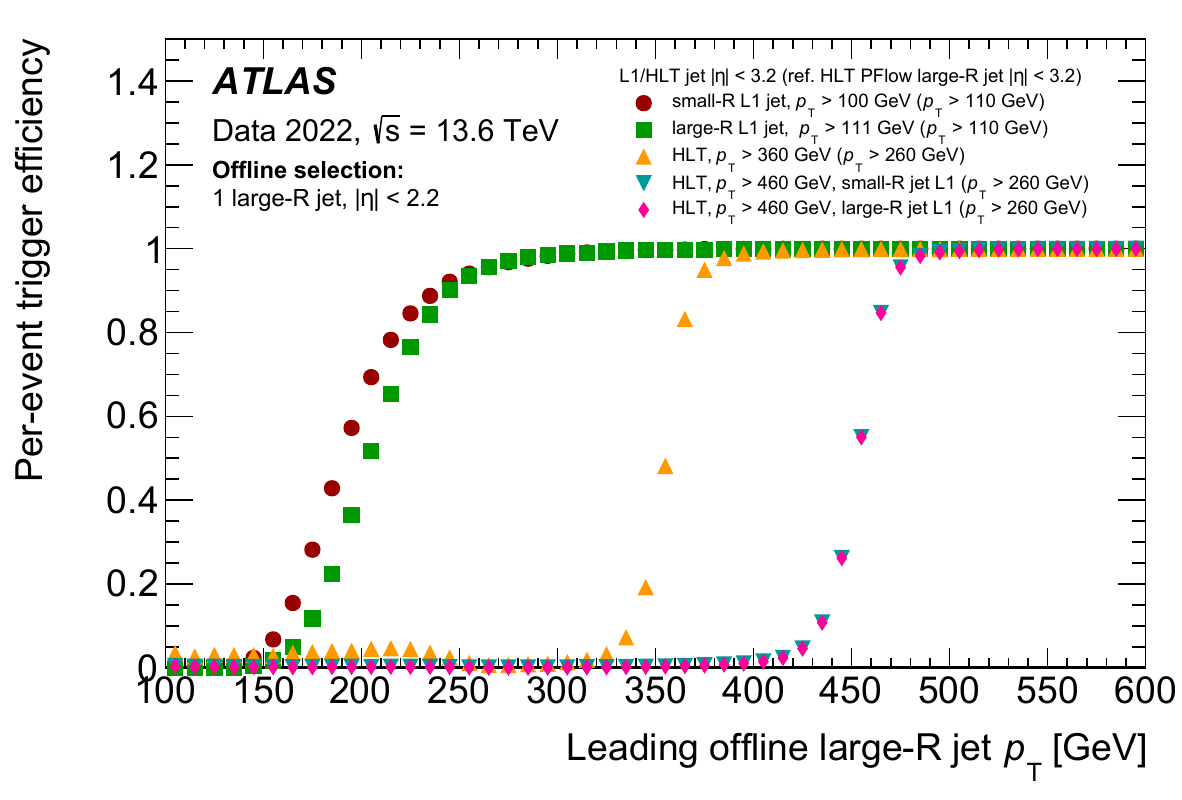} 
\includegraphics[width=0.49\textwidth]{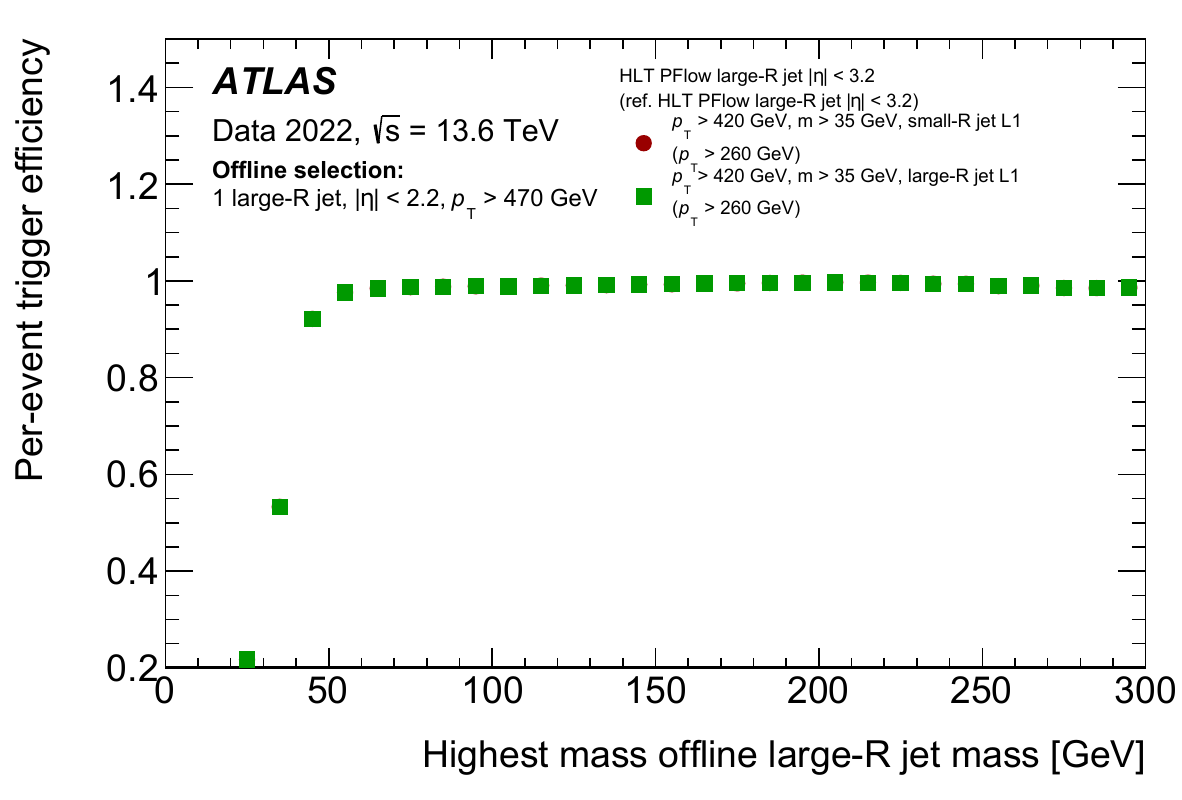} 
\includegraphics[width=0.49\textwidth]{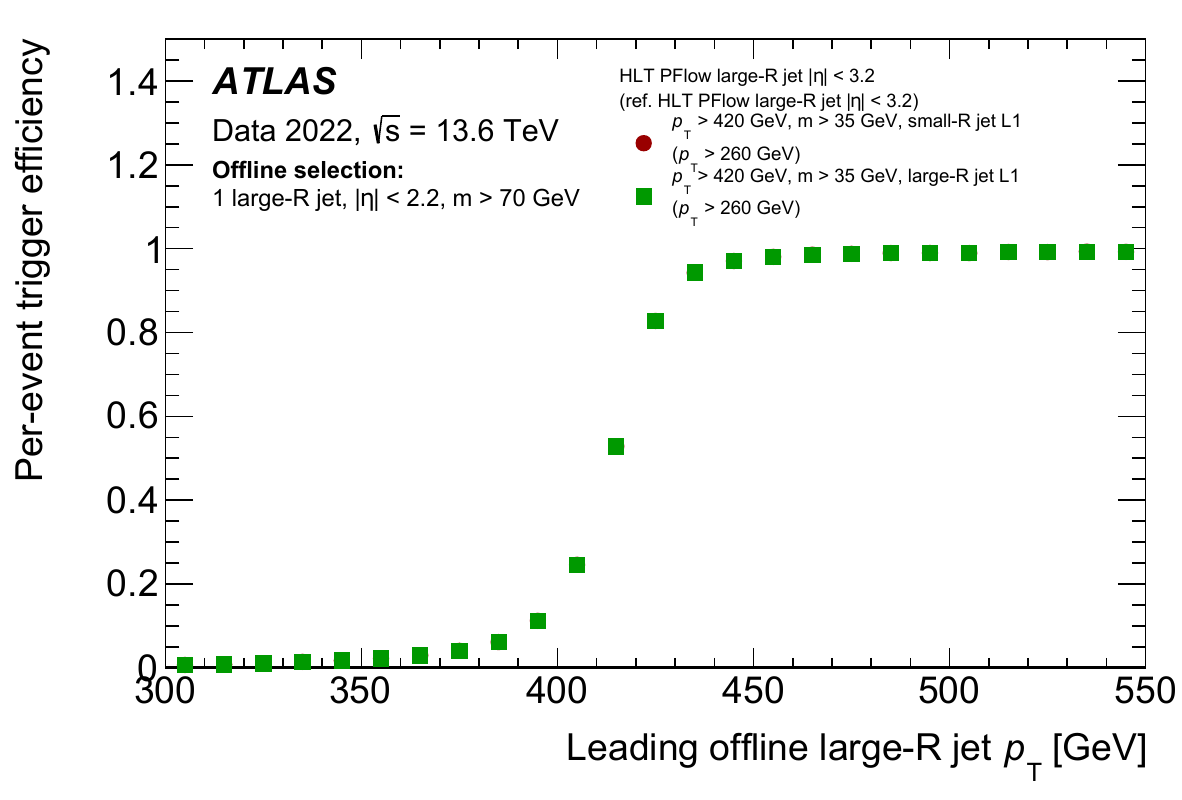}
\includegraphics[width=0.49\textwidth]{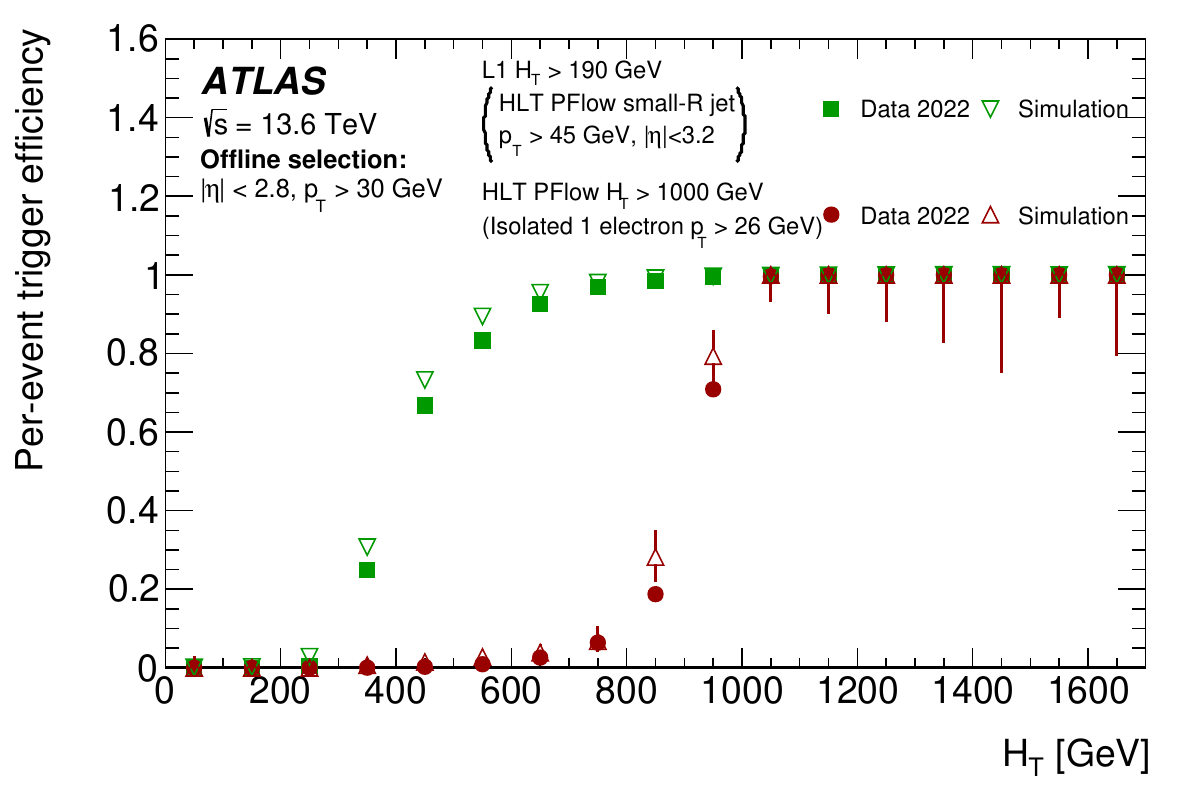} 
\caption{L1 and HLT efficiencies are shown for
single large-$R$ jet triggers as a function of (top left) the leading offline large-$R$ jet \pt\
and (top right) highest mass offline large-$R$ jet mass (m), (bottom left) large-$R$ single jet triggers with a mass cut as a function of leading offline large-$R$ jet \pt\ and (bottom right) \HT{} triggers as a function of offline \HT{}.
The efficiency of triggers is computed using the bootstrap method with respect to events taken by an independent trigger, shown in brackets,
that is 100\% efficient in the relevant region.
Only statistical uncertainties are shown.
}
\label{fig:jet:lvl1}
\end{figure}
 
\subsubsection{Jet trigger performance}
 
The efficiency of jet triggers is primarily a function of the \pt{} of the leading jet in the event, and is measured using the bootstrap method.
In addition to the selection criteria specified in the following figures, the offline jets are required to satisfy a timing cut of less than
12.5\,ns to minimise the contamination from out-of-time pile-up jets.
 
The efficiencies of L1 and HLT single jet triggers as a function of offline reconstructed jet \pt{} are shown in Figures~\ref{fig:jet:l_1} and~\ref{fig:jet:hlt}, respectively.
The reference trigger selections reported on the legend have a looser L1 and HLT selection than the probed triggers to ensure an unbiased reference dataset.
It must be noted that the jet-energy-scale correction is applied at the HLT but not at L1,
meaning that the L1-jet thresholds are effectively 50-100\% higher when considered at the jet-energy scale of the HLT.
A steep rise in efficiency near the nominal thresholds reflects the compatibility of energy scales for jets reconstructed in the trigger and offline.
 
The efficiency of a multi-jet trigger requiring \(N\) jets depends mostly on the \pt{} of the \(N^\text{th}\) \pt{}-ranked jet. The relevant efficiency measurements are shown in Figure~\ref{fig:jet:mj} as a function of the offline reconstructed \(N^\text{th}\) jet \pt{} for the primary L1 and HLT multi-jet triggers.
Lower efficiency at high \pt\ in the L1 four-jet trigger is due to the merging of the close-by jets at L1, which should be improved in the upgraded L1Calo system.
 
Efficiencies of the large-$R$ jets as a function of the leading offline jet \pt{} are shown in Figure~\ref{fig:jet:lvl1} (top left), for both HLT and L1 trigger chains.
Figures~\ref{fig:jet:lvl1} (top right) and (bottom left) present the efficiencies as a function of the offline jet mass and \pt, respectively, for two main HLT chains with mass cuts.
The L1 seed is shown not to have an impact on the efficiency of these triggers due to the large gap between L1 and HLT thresholds.
The triggers shown are fully efficient at high jet \pt and mass values, and the steep rise in efficiency near the nominal thresholds reflects a good energy resolution.
 
The \(H_T\) trigger efficiency as a function of offline \HT{} is presented in Figure~\ref{fig:jet:lvl1} (bottom right) for both L1 and HLT triggers,
which also rises sharply. The observed data/MC differences will be corrected later-on in physics analyses by dedicated scale factors applied to MC simulation samples.
 
\subsubsection{Trigger level analysis with jets}
\label{subsec:tla_jets}
 
HLT topo-cluster-based and PFlow jets with transverse momentum larger than 20\,\GeV\ are saved to the TLA stream described in Section~\ref{sec:menu}.
To remain within the CPU constraints of the HLT farm, full scan tracking and PFO reconstruction are executed only
if the event passes the calorimeter preselection step.
 
In 2022, jet chains populating the TLA stream were seeded by two L1Calo legacy triggers, L1 jet $\pt>100\,$\gev\ (which reaches 50\% efficiency at around 160\,\GeV) and
$\HT>190\,$\gev, 
with a total rate of approximately 4\,kHz.
This allows for the recording of orders of magnitude more events containing lower-\pT\
jets with respect to standard ATLAS jet triggers, extending the sensitivity of hadronic searches to resonance masses as low as 400\,\GeV~\cite{EXOT-2016-20}.
 
As shown in Figure~\ref{fig:tla_jets}, the response of trigger jets with the default calibration is already within 2\% of offline jets across the momentum range of interest to TLA studies.
For \runiii, a custom residual energy scale calibration is to be derived and applied at the analysis level in order to further improve the response and resolution of TLA jets.
In order to derive such a calibration, an extensive set of jet variables, beyond the jet four-momentum, is saved in the TLA stream.
This includes the energy fraction deposited in
the EM and hadronic endcap calorimeters, the momentum density of soft radiation in the event, the number of primary vertices reconstructed at the HLT, the jet active area~\cite{JETM-2018-05},
and the number of constituents encompassing 90\% of the jet transverse energy.
 
\begin{figure}[htbp]
\centering
\includegraphics[width=0.7\linewidth]{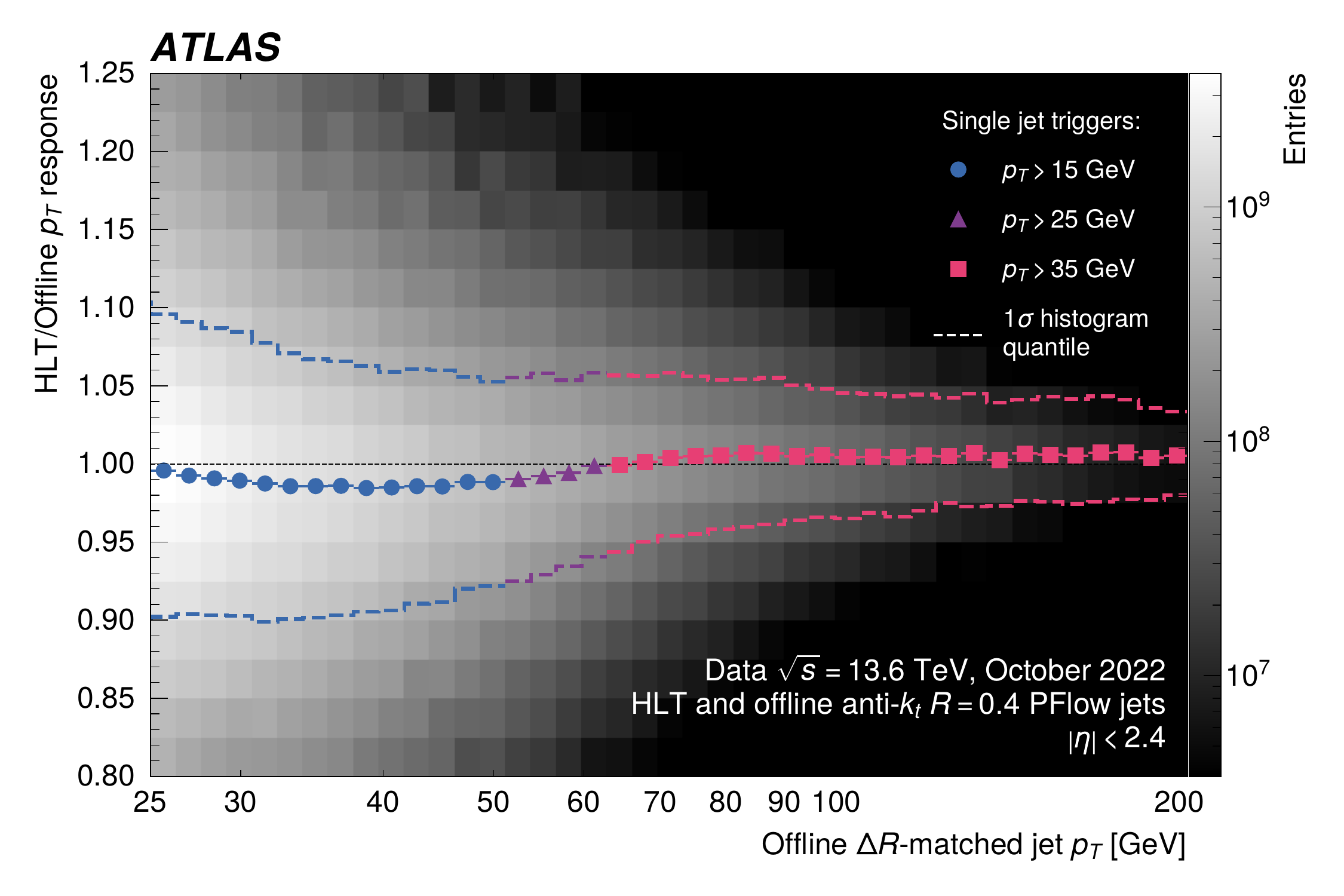}
\caption{
The response curve for HLT-level PFlow jets in the momentum range of interest for Run-3 TLA.
The online-to-offline response is evaluated at the final energy scale in the calibration sequence
of the jets. HLT-level jets are calibrated with the default sequence used at the HLT for the 2022 data taking.
Offline jets are corrected with the calibration available during 2022 data taking. Events are required to pass the single
jet $p_{\text{T}} > 15\,$\GeV\ (circle markers), $p_{\text{T}} > 25\,$\GeV\ (triangle markers) or $p_{\text{T}} > 35\,$\GeV\ (square markers) random-seeded triggers depending on the
$p_{\text{T}}$ of the leading reference jet. For each event, only HLT jets matched ($\Delta R < 0.3$) to the leading and
sub-leading offline jets reconstructed within $\left|\eta\right| < 2.4$ are considered. The overlaid heat map refers
to the number of events in the analyzed sample for each configuration of the reference jet momentum and response.
}
\label{fig:tla_jets}
\end{figure}
 
\subsubsection{Jet triggers for heavy-ion collisions}
\label{subsec:jets_hi}
 
Jets are direct probes of the quark--gluon plasma produced in HI collisions, and studying how they are
modified in such collisions compared to \pp collisions provides insight into their interactions with this QCD medium.
The main challenge for triggering on jets in HI collisions is the presence of a large amount of UE.
The HLT jet algorithm uses the HI UE subtraction procedure described in Section~\ref{sec:menuHI}
and selects events containing jets with transverse energies exceeding a threshold ranging from 60 to $85$\,\GeV.
Jets are reconstructed at the HLT across the entire calorimeter using the anti-\kt algorithm with distance parameter
$R = 0.4$, from projective towers of size $\Delta\eta \times \Delta\phi = 0.1 \times 0.1$ formed from the summation of calorimeter cell energies with UE subtraction applied.


\subsection{Jets containing $b$-hadrons}
\label{sec:bjets}
 
\subsubsection{$b$-jet trigger reconstruction and selection}
\label{sec:bjet_sel}
 
Triggers based on jets containing $b$-hadrons, so-called $b$-jet triggers, are designed to allow for highly efficient recording of fully hadronic events
with predominantly heavy flavour content jets. The detailed description of the $b$-jet triggers in \runii can be found in Ref.~\cite{TRIG-2018-08}, the changes for \runiii are discussed here.
The basic inputs to $b$-tagging are reconstructed jets, reconstructed tracks and the position of the
primary vertex.
The jet reconstruction is described in Section~\ref{sec:jets}.
The primary vertex finding and $b$-jet track reconstruction are discussed in Sections~\ref{sec:id_full} and~\ref{sec:id_bjet}, respectively.
 
The main $b$-jet identification steps are as follows:
\begin{itemize}
\item A fast $b$-tagging algorithm is run on super-\rois and uses the tracks inside them to provide the first $b$-jet preselection. To reduce the rate of the full scan tracking for $b$-jet triggers, the $b$-jet preselection stage is run with lower track \pT\ thresholds and 
with a super-\roi composed of wider regions around jets than were used for the vertex tracking in \runii~\cite{TRIG-2019-03,TRIG-2018-08}.
The details of the fast $b$-tagging algorithm used at this stage are provided in Ref.~\cite{FastBjetPaperInProduction_ANA-TRIG-2022-03}.
\item The final $b$-tagging algorithm, detailed below, uses precision tracking and primary vertex information.
\end{itemize}
 
The $b$-jet identification relies on the properties of $b$-hadrons: long lifetime (about 1.5\,ps), hard fragmentation, a relatively large mass of about 5\,\GeV,
and a displaced (secondary) vertex (SV) formed a few millimeters away from the primary vertex.
Tracks associated with $b$-jets are characterised by larger transverse ($d_0$) and longitudinal ($z_0$) impact parameters.
In addition, $b$-hadrons decay semileptonically, either promptly, or via a subsequent $c$-hadron decay, to electrons or muons.
The branching ratio of these semileptonic decays is about 20\% each and results in the presence of a low-\pt lepton close to the $b$-jet.
 
The flavour tagging identification is done in two steps~\cite{FTAG-2019-07}.
Low-level algorithms reconstruct characteristic features of the heavy flavour jets
based on track properties, such as impact parameters, or combine those tracks to reconstruct the SV. The outputs of these low-level algorithms are then
combined into high-level algorithms, usually using some multivariate technique.
 
Low-level taggers used at the HLT are the IP2D algorithm, which utilises the signed transverse impact parameter significance of tracks,
and the IP3D algorithm, which uses, in addition, the longitudinal impact parameter significance~\cite{FTAG-2019-07}.
 
The secondary vertex algorithm, SV1, uses tracks associated with jets (after rejecting those compatible with $K_s$ or
$\Lambda$, photon conversions or interactions with detector material) to assign decay products from
$b$- or $c$-hadrons to a single common SV. Several discriminating variables associated with the SV are then used as inputs to the high-level tagger.
Finally, the JetFitter algorithm exploits the topology of weak $b$- and subsequent $c$-hadron decays inside a jet to reconstruct the full decay chain,
recreating the approximate $b$-hadron path from the primary vertex via bottom and charm vertices.
 
Another low-level algorithm is the Deep Impact Parameter Sets (DIPS)~\cite{ATL-PHYS-PUB-2020-014}, based on the Deep Sets architecture. It uses
impact parameter information, accounting for correlations between the track features, among other variables.
DIPS considers tracks in the jet as an unordered and variable-sized set, which is physically better motivated than the algorithm based
on recurrent neural network~\cite{TRIG-2018-08} used in \runii, given that the $b$-hadron decay products do not exhibit any intrinsic sequential ordering.
The performance of the DIPS tagger is investigated using \ttbar\ MC events and 2022 data
collected with a calibration trigger with one electron, one muon and two jets.
The final DIPS discriminant distribution is shown in Figure~\ref{fig:bjet_dips_disc} (left).
In the case of SM $HH\to\bbbar\bbbar $, a key signature relying on $b$-jet triggers,
the $b$-jet preselection step lowers the input rate to the remaining HLT
by a factor of five at the cost of reducing the overall signal efficiency by roughly 2\%.
More details about DIPS, its training and usage in $b$-jet triggers are given in Ref.~\cite{FastBjetPaperInProduction_ANA-TRIG-2022-03}.
 
The newly developed algorithm based on deep feed-forward Neural Networks, the so-called DL1 series~\cite{FTAG-2019-07}, replaced
boosted-decision-tree based taggers utilised in \runii~\cite{TRIG-2018-08}.
The particular instance of the algorithm used in 2022 is called DL1d and takes as inputs the kinematic variables (\pt\ and $\eta$) of the jet
as well as final discriminants from lower-level taggers (IPxD, SV1, JetFitter and DIPS).
 
From 2023 the $b$-jet trigger relies on a novel algorithm GN1, which is based on Graph Neural Networks (GNNs)~\cite{ATL-PHYS-PUB-2022-027}.
Unlike the DIPS and DL1d, the GN1 utilises a single neural network
taking the tracks and jet information directly as inputs and is thus independent of
other flavour tagging algorithms. The GN1 combines a GNN
with the two auxiliary training objectives: the grouping of tracks originating from a common vertex
and the prediction of the underlying physics process from which each track originated.
This approach leads to a better understanding of the jet's internal structure and, thus, a better algorithm performance.
 
The expected trigger rate as a function of the $b$-tagging efficiency using the DL1d and GN1 algorithms is shown in Figure~\ref{fig:bjet_dips_disc}~(right).
The expected light jet rejection as a function of the $b$-tagging efficiency for various $b$-taggers as well as their operating points
are shown in Figure~\ref{fig:bjet_ROC}. The GN1 tagger performance exceeds that of DL1d, which was the main tagger in 2022.
 
\begin{figure}[hbtp]
\centering
\raisebox{0.1\height}{\includegraphics[width=0.49\textwidth]{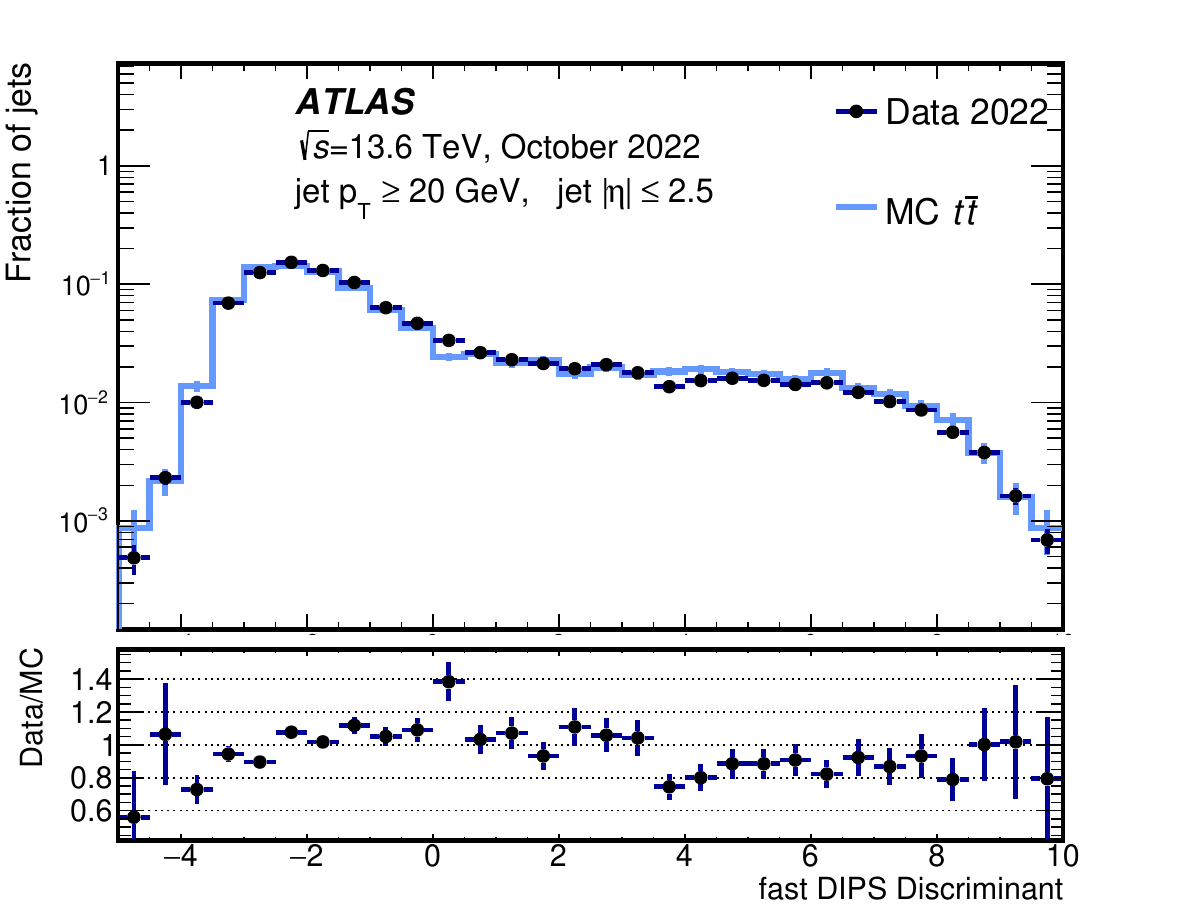}} 
\includegraphics[width=0.49\textwidth]{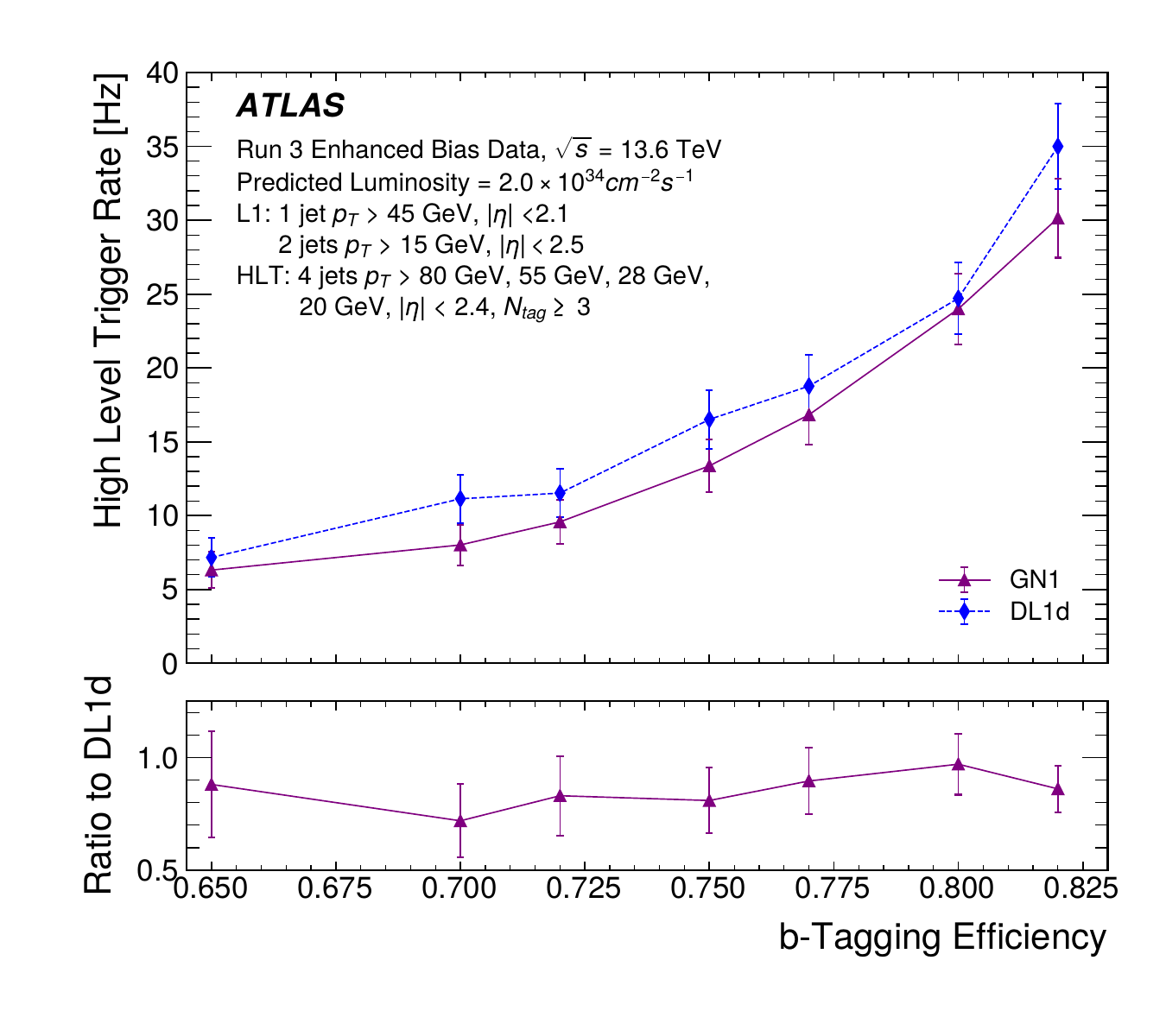} 
\caption{(left) Distribution of the $b$-jet discriminant score for the DIPS algorithm shown in data from selected runs and \ttbar\ MC events
collected with a calibration trigger with one electron, one muon and two jets.
Only statistical uncertainties are shown.
(right) Expected trigger rates as a function of $b$-tagging efficiency using the
DL1d and GN1 algorithms, while requiring at least four HLT PFlow jets, three of which are
required to be above the $b$-tagging threshold. Rates are estimated with
Run-3 Enhanced-Bias data, and the $b$-jet efficiencies are estimated for $t\bar{t}$
samples using PFlow jet. The relative errors on trigger rates are between 8(7)\% and 18(20)\% for DL1d (GN1).
}
\label{fig:bjet_dips_disc}
\end{figure}
 
\begin{figure}[hbtp]
\centering
\includegraphics[width=0.65\textwidth]{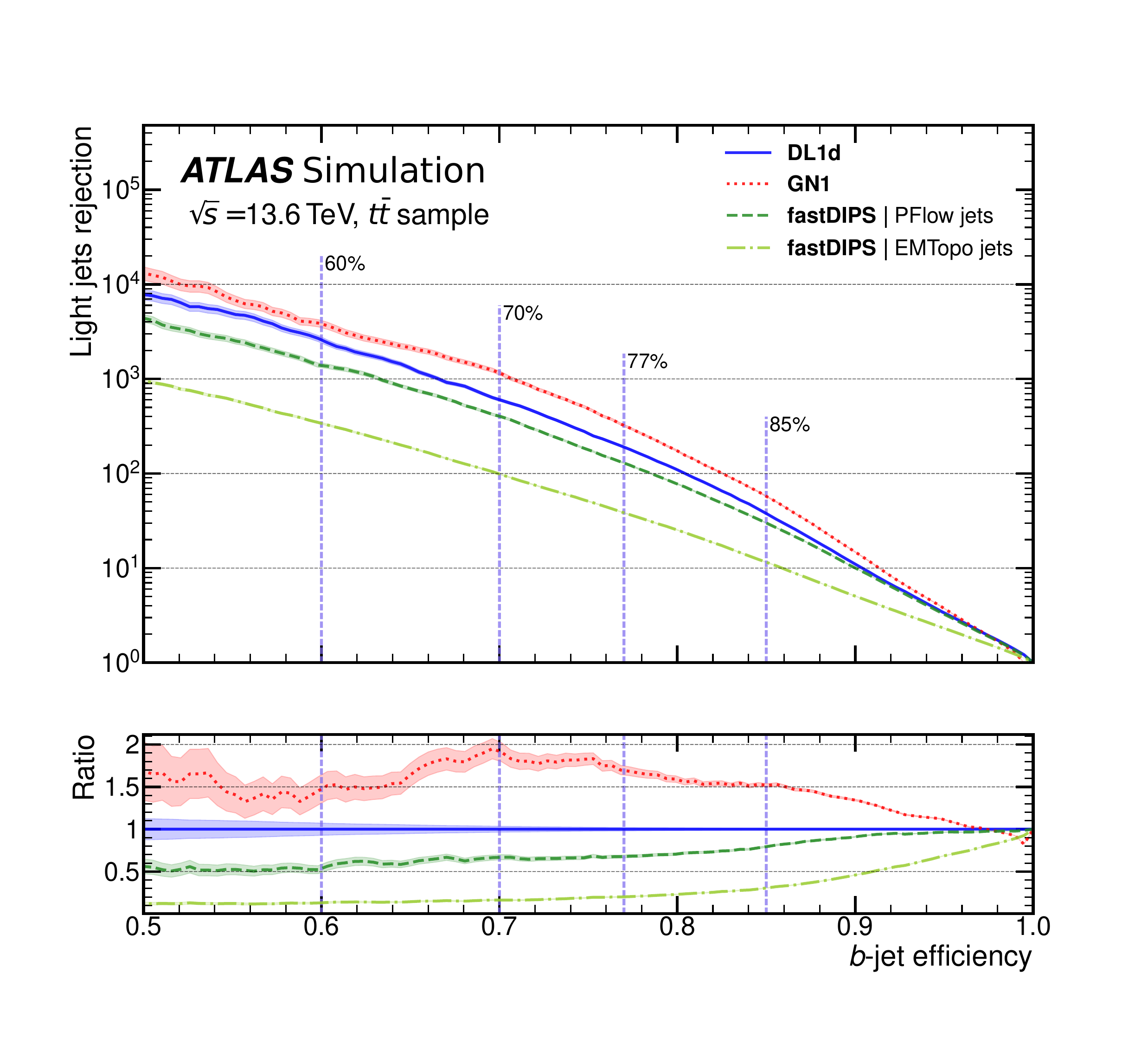}
\caption{The light jet rejection as a function of the $b$-tagging efficiency (ROC curve) for the DIPS algorithm used on topo-cluster jets for preselection stage (dash-dotted line), the DIPS algorithm used on PFlow jets for TLA triggers (dashed line), the DL1d algorithm used as the final tagger in 2022 (solid line) and the GN1 algorithm used as the final tagger from 2023 (dotted line) obtained on a \ttbar\ MC sample. Statistical uncertainties for each ROC curve are represented with shaded regions around the curves. The vertical dashed lines represent the operating points for $b$-tagging used at the HLT. The bottom panel displays the ratio of all the ROC curves with respect to the DL1d performance.
}
\label{fig:bjet_ROC}
\end{figure}

\subsubsection{$b$-jet trigger menu}
\label{subsec:bjet_menu}
 
$b$-jet triggers are crucial for several precision measurements and searches for new particles such as resonant and
non-resonant $HH\to\bbbar\bbbar $~\cite{HDBS-2018-41, HDBS-2019-29}, VBF $H\to\bbbar $~\cite{HIGG-2019-04, HIGG-2020-14},
$\ttbar H\to \ttbar \bbbar$ with both top quarks decaying hadronically~\cite{HIGG-2015-05}, $b\phi\to b\bbbar$~\cite{HIGG-2016-32},
third generation squarks~\cite{EXOT-2019-04}, low mass di-$b$-jet resonances~\cite{EXOT-2016-33}, etc.
 
The $b$-jet trigger menu consists of the following physics triggers: single-$b$-jet triggers,
multi-$b$-jet triggers and specialised $b$-jet triggers designed for specific physics processes which often occur in combination with other signatures.
Single-$b$-jet triggers and multi-$b$-jet triggers are seeded from dedicated single-jet or multi-jet L1 objects.
At the HLT various combinations of the $p_T$ thresholds and $b$-tagging operating points are employed.
In multi-$b$-jet triggers, between one and four jets are tagged, while other jets are not tagged.
An asymmetric chain requiring four jets with \pT\ greater than
80\,\gev, 55\,\gev, 28\,\gev, and 20\,\gev\ of which two are $b$-tagged (with efficiency of 77\%) has a rate around 150 Hz,
the lowest unprescaled single $b$-jet trigger (\pt$>225\,$\gev) around 50 Hz and
one jet of \pt$>150\,$\gev\ plus two $b$-jets with \pt$>55\,$\gev\ and $b$-tagging
efficiency of 70\% around 10 Hz at luminosity of $1.8\times\lumi{e34}$.
 
Flavour-tagging calibration triggers~\cite{TRIG-2018-08} include a single lepton (electron or muon)
with one or two jets with \pt$>20$\,\GeV, electron-plus-muon with one or
two jets with \pt$>20$\,\GeV, and muon-jet matched triggers,
that are also used for physics studies in HI collisions.
These latter triggers were seeded from either a single-muon or a muon-plus-jet trigger item at L1 in 2022.
New chains seeded from L1 objects which use topological information,
such as $\Delta R$, are planned to be introduced later in \runiii.
At the HLT, a muon is required to be matched to a jet by fulfilling a requirement of
$\Delta R(\mu,\mathrm{jet})<0.5$, with an additional requirement for \pp collisions
$ \Delta z (\mu,\mathrm{jet})<2$\,mm, where the $ z$-position of the jet is taken to be the primary vertex $z$-position.
 
\subsubsection{Trigger level analysis with $b$-jets}
\label{subsec:tla_bjets}
 
The Run-3 TLA HLT infrastructure allows for the full outcome of flavour tagging
algorithms to be written out in events triggered by TLA chains targeting $b$-jet signatures.
In every TLA event, each HLT-level jet (for which flavour tagging is performed) is linked to
the corresponding $b$-tagging object in which the values of low and high-level taggers are stored.
Furthermore, the value of the DIPS tagger is added to every PFlow jet saved to the TLA stream,
regardless of whether the triggering chain is configured to perform dedicated flavour tagging or not.
This setup allows for the recording of orders of magnitude more events containing lower-\pT
$b$-jets with respect to standard $b$-jet triggers, extending the sensitivity of low mass di-$b$-jet resonance searches to resonance masses
down to 100~\,\GeV.


\subsection{Missing transverse momentum}
\label{sec:met}
 
\subsubsection{\met trigger reconstruction}
\label{sec:met1}
 
The missing transverse momentum (\met) trigger is used to identify events with
particles that do not interact with the ID or calorimeters because of the absence
of strong or electromagnetic interactions and that have lifetimes large enough to leave the detector without decaying into detectable particles.
Examples of such processes include SM Z boson decays to neutrinos~\cite{HIGG-2018-04}, as well as searches
for beyond SM decays involving dark matter~\cite{EXOT-2020-11} or supersymmetric~\cite{SUSY-2019-08} particles.
 
The \met can be computed from the magnitude of the vector sum of constituents, c:
$$\met = \left| \sum_c \Vec{p}_{T,c} \right|.$$
In addition to the \met, the scalar sum of the constituents is also computed. The main variations in how \met is computed in the trigger involve details of which constituents to utilise in the sum in the equation above.

The following HLT \met algorithms are available in Run 3:
\begin{itemize}
\item \texttt{cell}: The \met is formed from a sum over all calorimeter cells passing the selection $E_i>-5\sigma_i$, $|E_i|>2\sigma_i$, where $\sigma_i$ is the estimated noise in that cell.
This is to protect against spurious large negative energy signals, discussed in Section~\ref{sec:hltcalo}.
The cell energy magnitude must be greater than $2\sigma_i$ to reduce the effect of noise from electronics and pile-up.
\item \texttt{tcpufit}:
The \met is formed using topo-clusters 
as inputs,
which are combined in $\eta-\phi$ patches with dimensions of approximately $0.7\times 0.7$ (roughly the size of an $R=0.5$ jet).
The energy contribution from pile-up to these patches is estimated by a fit over them, requiring that pile-up events have
no true \met and are approximately evenly distributed over the calorimeter~\cite{TRIG-2019-01}.
The estimated pile-up contribution to each patch is subtracted, and the remaining patch
transverse momenta are summed to obtain the \met. The full algorithm is described in Appendix A of Ref.~\cite{TRIG-2019-01}.
\item \texttt{trkmht}: The \met is formed over all jets passing a Jet Vertex Tagger (JVT) selection~\cite{PERF-2014-03} where applicable.
The vector sum of the \pt\ of tracks from the primary vertex that are not associated with a passing jet
defines a track soft term which is added into the \met calculation.
\item \texttt{pfopufit}: The \texttt{pfopufit} algorithm uses the same techniques as \texttt{tcpufit},
but the input topo-clusters are modified to use PFO tracks~\cite{PERF-2015-09}.
As well as the improved momentum resolution of the PFOs,
the vertex information provided by the charged PFOs is used to improve the categorisation
of deposits into hard-scatter and pile-up.
\item \texttt{mhtpufit}: The \met is formed from a sum over jets passing a JVT selection~\cite{PERF-2014-03}.
A similar technique to \texttt{tcpufit} is used to correct these jets for the impact of pile-up.
Two variants are used: \texttt{mhtpufit\_pf} uses PFOs and jets formed from them
to estimate the pile-up contributions, \texttt{mhtpufit\_em} uses jets formed from
EM scale topo-clusters and the hadronic scale topo-clusters to estimate the pile-up contributions.
\item \texttt{pfsum}: The \met is formed from a sum over the PFOs in the event.
Two variants exist: in \texttt{pfsum\_vssk} the PFOs have their energies reduced according to their
Voronoi areas~\cite{ATLAS-CONF-2017-065}, whereas in \texttt{pfsum\_cssk} the constituent
subtraction method~\cite{Berta_2019} is used instead.
In both cases, the SoftKiller algorithm~\cite{Cacciari_2015} is used to remove PFOs from
low-energy areas of the calorimeter.
\end{itemize}


In 2022, the primary L1 item for \met was the L1Calo legacy trigger with \met$>50\,$\gev~\cite{TDAQ-2019-01, TRIG-2019-01}. 
Before any kind of tracking is executed, an early reduction of rate at HLT is achieved by a calorimeter preselection requirement
that the \trig{cell}-based \MET\ is greater than 65\,\GeV.
The choice of default Run-3 algorithm is made based on the performance in terms of background rejection versus signal efficiency and is discussed in the following paragraphs.
 
\subsubsection{\met trigger performance and menu}
 
Since muons are treated as invisible by the \met\ trigger algorithms described above,
events with a boosted \Zmm\ can be used for the \met performance studies with
\(p_T(\mu\mu)\) serving as a proxy for \met.
A background acceptance vs. trigger efficiency curve is shown in Figure~\ref{fig:MET-ROC} (left) comparing the performance of a selection of \met algorithms considered for \runiii.
The efficiencies are calculated as the fraction of events passing a given \met\ requirement for data events with
an actual number of interactions per bunch crossing of at least 38, and a \Zmm\ event selection passing
a single muon trigger, with \(76<M(\mu\mu)<106\)\,\GeV\ and \(p_T(\mu\mu)>175\)\,\GeV.
The background rate is obtained from an offline trigger reprocessing of the 2022 data collected with zero-bias triggers described in section~\ref{sec:zb}.
The new Run-3 default \met\ trigger is based on \trig{pfopufit} algorithm, which shows the best ability to retain signal, whilst rejecting background,
of all the algorithms listed in Section~\ref{sec:met1}.
The Run-2 default \trig{tcpufit}
trigger is maintained as backup and for analyses where the primary vertex may not match the online primary vertex.
These triggers had unprescaled HLT thresholds of 90~\gev\ and 115~\gev\ in 2022 and rates of about 70\,Hz and 30\,Hz
at luminosity of $1.8\times\lumi{e34}$, respectively.
Requiring the presence of additional objects in the event, allows for decreased thresholds for \met triggers, as discussed in Section~\ref{sec:unconTrack}.
 
A signal efficiency curve is shown in Figure~\ref{fig:MET-ROC} (right) comparing the performance of the primary L1 \met trigger alone to the default HLT \met\ trigger chains used in \runiii and \runii.
The efficiencies are calculated as the fraction of events passing a given \(p_T(\mu\mu)\) requirement for data events with a \Zmm\ event selection passing a single muon trigger, with \(76<M(\mu\mu)<106\)\,\GeV. The 2018 L1 efficiency matches that of 2022 within a few percent, so it is not shown.
The improved efficiency seen for the Run-3 \trig{pfopufit}-based chain compared to the Run-2 chain reflects the \trig{pfopufit} algorithm's enhanced momentum resolution and hard-scatter/pile-up categorisation.
 
\begin{figure}
\centering
\includegraphics[width=0.49\textwidth]{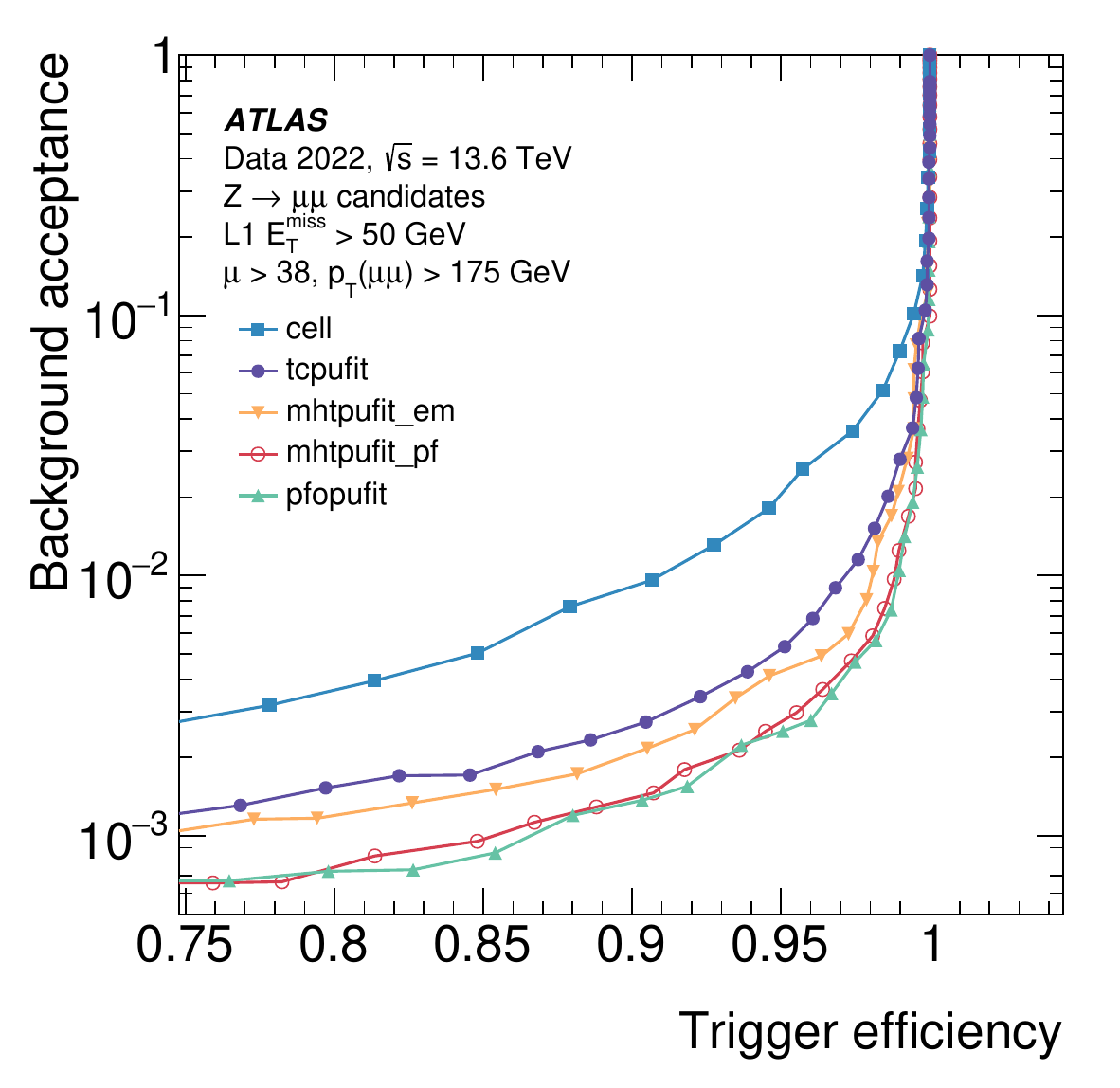}
\includegraphics[width=0.49\textwidth]{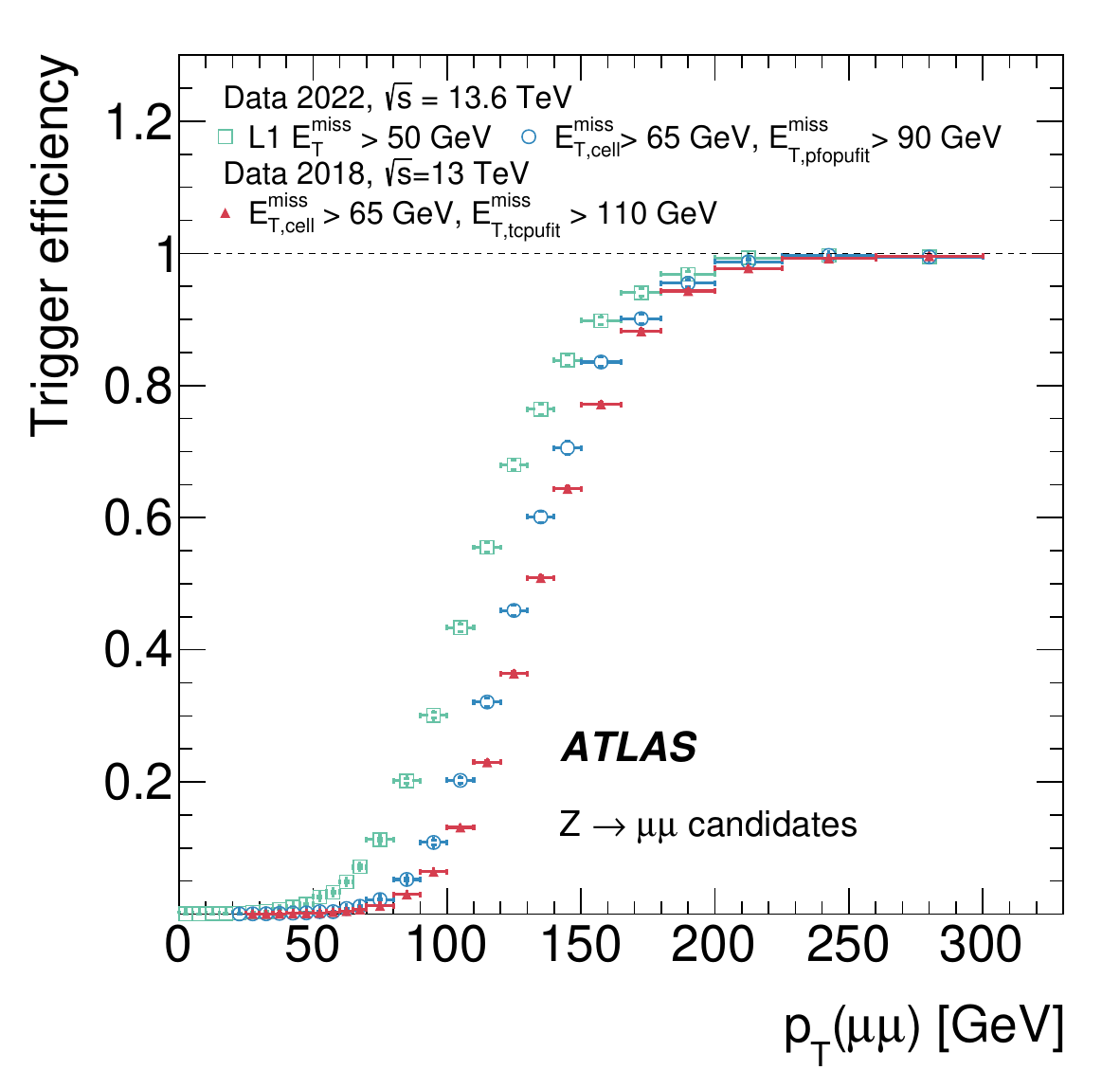}
\caption{
(left) Background acceptance vs. trigger efficiency curves for a selection of Run-3 \met\ trigger algorithms. 
(right) Trigger efficiency as a function of \(p_T(\mu\mu)\) for the primary L1 \met trigger and the full trigger chain, with specific thresholds listed.
The trigger rates for the HLT chains shown are approximately the same rate (to within 10\%).
The data are presented in comparison to 2018 efficiency.
Only statistical uncertainties are shown.
}
\label{fig:MET-ROC}
\end{figure}


\subsection{$B$-physics and light states}
\label{sec:bls}
 
The trigger selection of events for $B$-physics analyses is primarily based on the identification of $b$-hadrons through decays with a muon pair in the final state. However, one or more muons or electrons could also be present in specific selections. Examples are decays with charmonium, $B\rightarrow (\Jpsi$ or $\psi')X \rightarrow \mu\mu X$, rare decays $B^0_{(s)}\rightarrow \mu\mu$, and $B \rightarrow \mu\mu X$ decays. Decays of prompt charmonium and bottomonium are also identified through their di-muon decays, and are therefore similar to $B$-meson decays, apart from the lack of measurable displacement from the $pp$ interaction point. As the BLS topologies are significantly different from the majority of the ATLAS physics triggers, they are recorded to a separate BLS stream, as discussed in Section~\ref{sec:menuBase}. In 2022 this stream collected data with an average rate of approximately 240\,Hz.
 
\begin{figure}[htbp]
\centering
\includegraphics[width=0.49\textwidth]{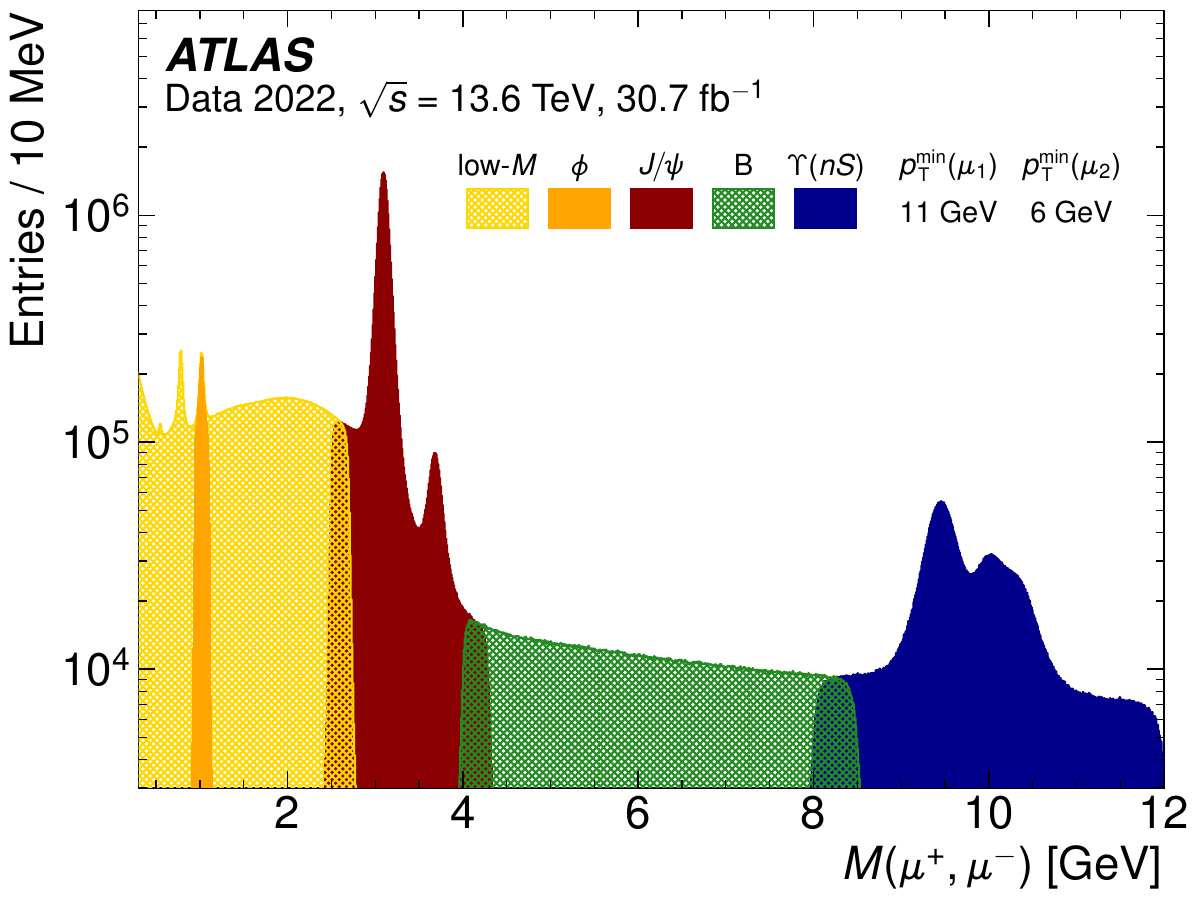}
\caption{Invariant mass distribution of offline-selected di-muon candidates passing the lowest thresholds of di-muon $B$-physics triggers. Triggers targeting different invariant mass ranges are illustrated with different colours, and the differing thresholds are shown with different shadings. No accounting for overlaps between triggers is made, and the distributions are shown overlaid, and not stacked.}
\label{fig:bls:massSpect}
\end{figure}
 
As $B$-mesons are light, the resulting muon momenta are rather soft.
To control the input rate to the HLT, most of the $B$-physics triggers require two muons at L1. Their rate is substantially reduced compared to single-muon L1 triggers. Depending on the mass of the resonance this can result in the muons being within a single \roi or in separate \rois. At the HLT, muons are reconstructed using the same algorithms as described in Section~\ref{sec:muonrec} with the additional requirement that the combined muons have opposite charges and form a good vertex (where a fit is performed using the ID track parameters) within a certain invariant mass window. For example, the mass range of the \trig{bJpsimumu} is between 2.5 and 4.3\,\GeV\ while \trig{bUpsi} is between 8 and 12\,\gev. Examples of these mass spectra can be seen in Figure~\ref{fig:bls:massSpect}.
 
Di-muon trigger rate restrictions at L1 define the lowest muon transverse momentum thresholds for primary $B$-physics triggers. HLT triggers based on two L1 muons passing a 3\,\GeV\ \pt threshold need to be prescaled for most data-taking luminosities.
The prescales are adjusted to maximise the recorded number of low-\pt di-muon events while remaining within operational constraints of the ATLAS TDAQ system as discussed in Section~\ref{sec:menuBase}.
Higher threshold HLT triggers seeded from two muons passing 5\,\gev\ or 8\,\gev\ thresholds at L1
are unprescaled when the L1 and HLT bandwidths allow it.
 
Additional primary and supporting triggers are also implemented. Triggers requiring three muons at L1 help to maintain the lowest muon \pt thresholds for certain event signatures with a likely presence of a third muon. For semileptonic decays, such as $B^0 \rightarrow \mu\mu K^{*0} \left(\rightarrow K^+\pi^- \right)$, searches for additional ID tracks and a combined vertex fit are performed, assuming a few exclusive decay hypotheses. For the start of Run 3 a so-called \trig{bBmuX} selection is implemented to perform the partial reconstruction of the B-hadron decay final state $D^{*+} \mu^- X$ with the $D^{*+}$ reconstructed through the cascade hadronic decay chain $D^{*+} \to \pi^+ D^0(\to K^-\pi^+)$.
 
\begin{figure}[htbp]
\centering
\includegraphics[width=0.49\textwidth]{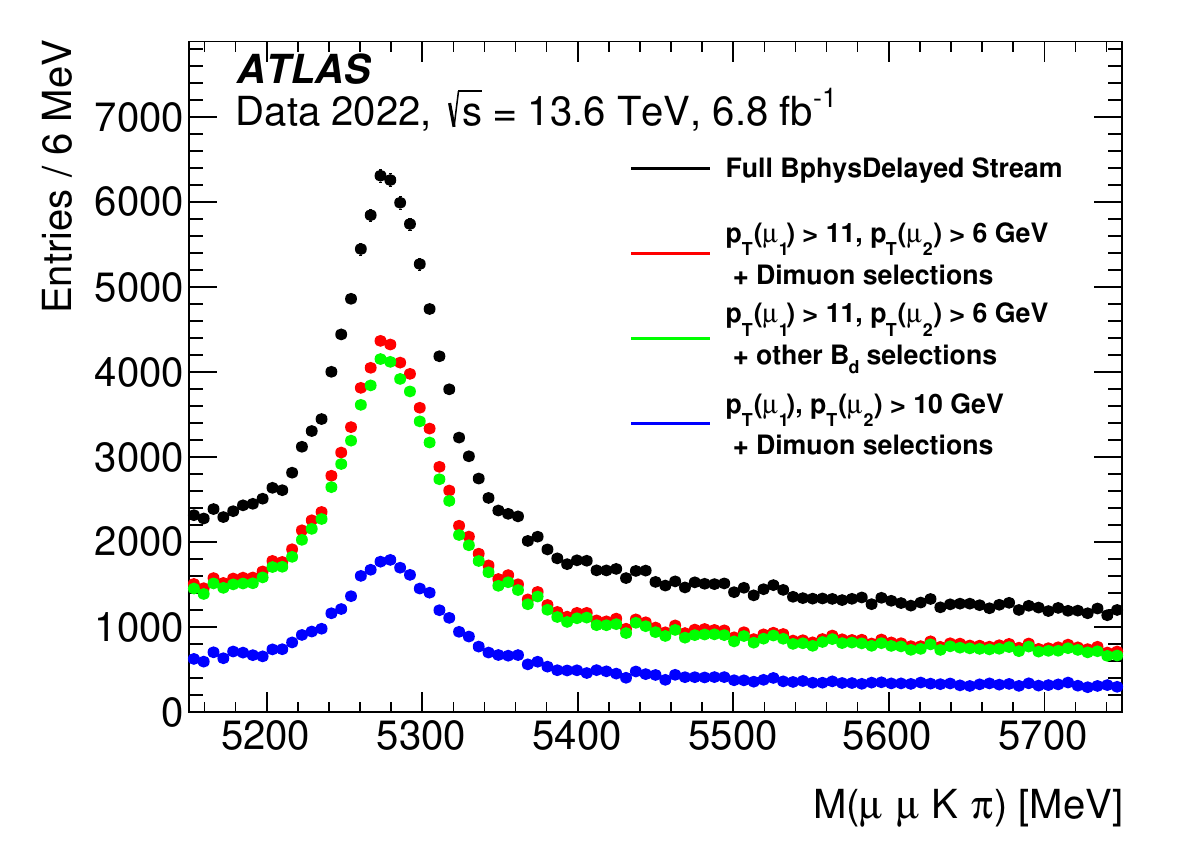}
\includegraphics[width=0.49\textwidth]{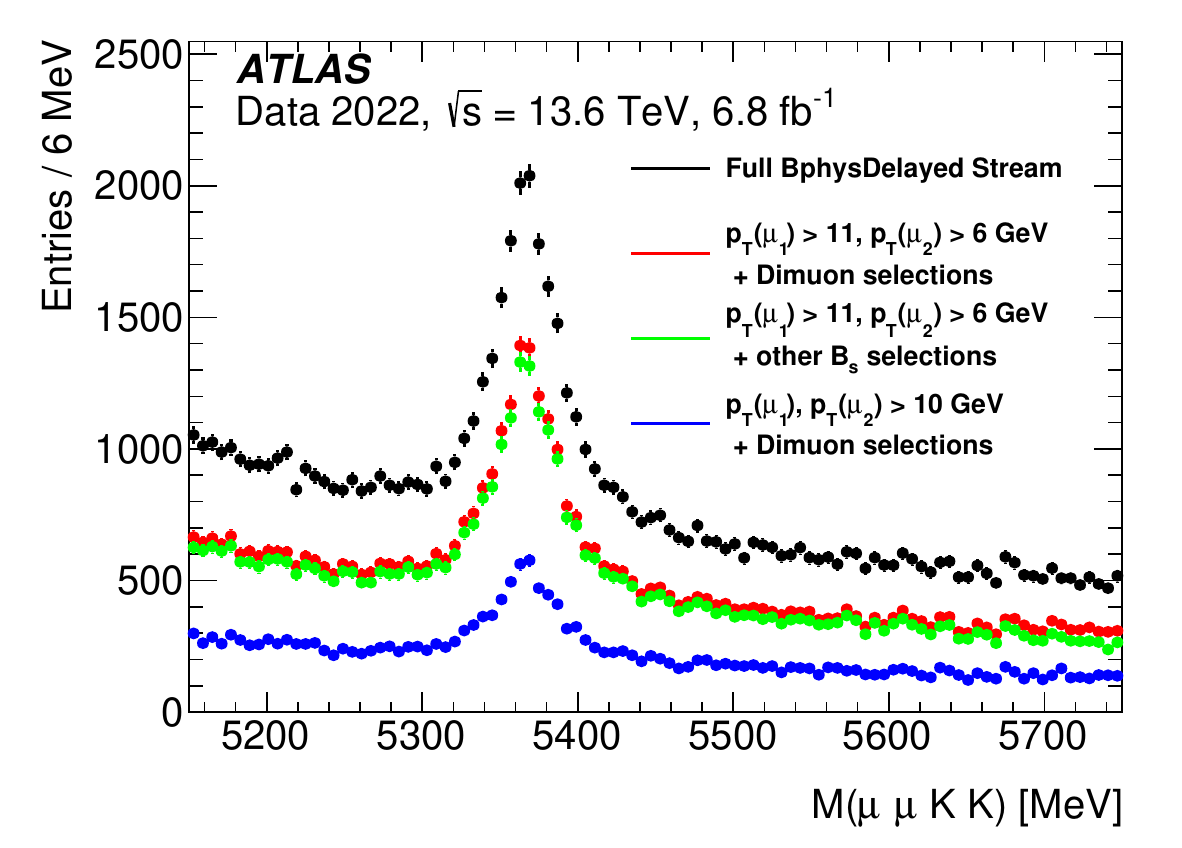}
\caption{To validate the selection of the \trig{bBmumux} the mass spectra of the selected (left) $B_d \rightarrow J/\psi K^*$ and (right) $B_s \rightarrow J/\psi \phi$ events from the BLS Delayed stream are plotted along with the spectra for a variety of \trig{bJpsimumu} and \trig{bBmumux} trigger decisions.
Only statistical uncertainties are shown.
}
\label{fig:bls:massplot}
\end{figure}
 
To evaluate the $B$-meson selection at the HLT, supporting triggers without vertex or charge requirements are used.
The mass spectra of various \trig{bBmumux} trigger decisions are plotted and fitted using events from the BphysDelayed stream. The consistency of the mass peaks in Figure~\ref{fig:bls:massplot} with PDG averages and previous measurements~\cite{PDG} demonstrate the validity of the HLT reconstruction and selection.


\subsection{Minimum-bias and forward signatures}
\label{sec:minbias}
To study diffraction, soft-QCD and similar topics, highly efficient triggers are necessary to select \pp inelastic, diffractive and scattering interactions with
the least possible trigger bias. These data also serve as references for heavy-ion measurements. The minimum-bias triggers discussed below are enabled only during dedicated low-$\mu$ runs.
 
\subsubsection{Minimum-bias triggers}
Depending on the number of collisions per bunch crossing, $\mu$, the minimum-bias triggers play various roles.
At very low values of $\mu$ ($\mu \ll 1$) the triggers require one or two signals on any side of the Minimum Bias Trigger Scintilator (MBTS) detector to select actual collision events and to ensure that
the data sample is not dominated by empty events with no \pp interactions. The MBTS detector was replaced for \runiii, as discussed in Ref.~\cite{atlas-det-run3}.
It is a two-armed large-area plastic scintillator with a very high light yield covering $2.0<|\eta|<4.0$.
On each side, the scintillator is divided into two concentric rings, each consisting of 8 octants covering the full azimuth.
Altogether, the MBTS deliver 32 signals to the CTP, which can be used independently or combined.
Further event selection is possible at the HLT by requiring matching timing between MBTS signals or the reconstruction of tracks in the ID.
As a side benefit, reconstructing the MBTS signals at the HLT allows for the monitoring of the timing and energies of individual channels.
 
At $\mu$ values close to unity the need for MBTS as a source for the hardware trigger is less relevant and the minimum-bias sample can be collected by a random trigger at L1
followed by the requirement of a track reconstructed at the HLT.
 
At $\mu\sim1-3$, every bunch crossing contains an inelastic collision and the L1 random trigger is sufficient for minimum-bias triggering.
In these conditions, the focus shifts to collecting events with high multiplicity tracks (HMT) or events with a high momentum track
for analyses like Bose-Einstein correlations or azimuthal correlations similar to what is done in HI physics~\cite{HION-2017-02}.
The possibility of using a trigger with the total (transverse) energy deposit in the calorimeter is also planned.
 
At the start of 2022 data taking no track selection was applied at the HLT.
This allows for the comparison in performance of the minimum-bias tracking selection in offline versus the trigger.
In \runiii the track reconstruction for the minimum-bias trigger is based on offline algorithms.
The minimum-bias online tracks are required to satisfy the minimum-bias track selection: $\pt>0.1\,$\gev\ and $|\eta|<2.5$.
In addition to the online selection above, the offline tracks are required to satisfy the standard minimum-bias
analysis selection~\cite{STDM-2015-02}: $|d_0|<1$\,mm, $|z_0\sin\theta|<1.5$\,mm, a measurement in the IBL if it is expected,
at least one pixel measurement and the number of measurements in the SCT above 2, 4, and 6 for tracks of \pt above 0.1, 0.3, and 0.4\,\GeV, respectively.
Furthermore, the online and offline conditions are not exactly identical, e.g. the map of dead pixel and SCT modules is known only after data have been taken.
A comparison of transverse momentum, \pt, and $\eta$ for online tracks and offline tracks undergoing
corresponding minimum-bias selections is shown in Figure~\ref{fig:minbias:tracks_perf}.
Despite the looser online track selection, the purity of minimum-bias triggers is quite high.
 
\begin{figure}
\centering
\includegraphics[width=.49\textwidth]{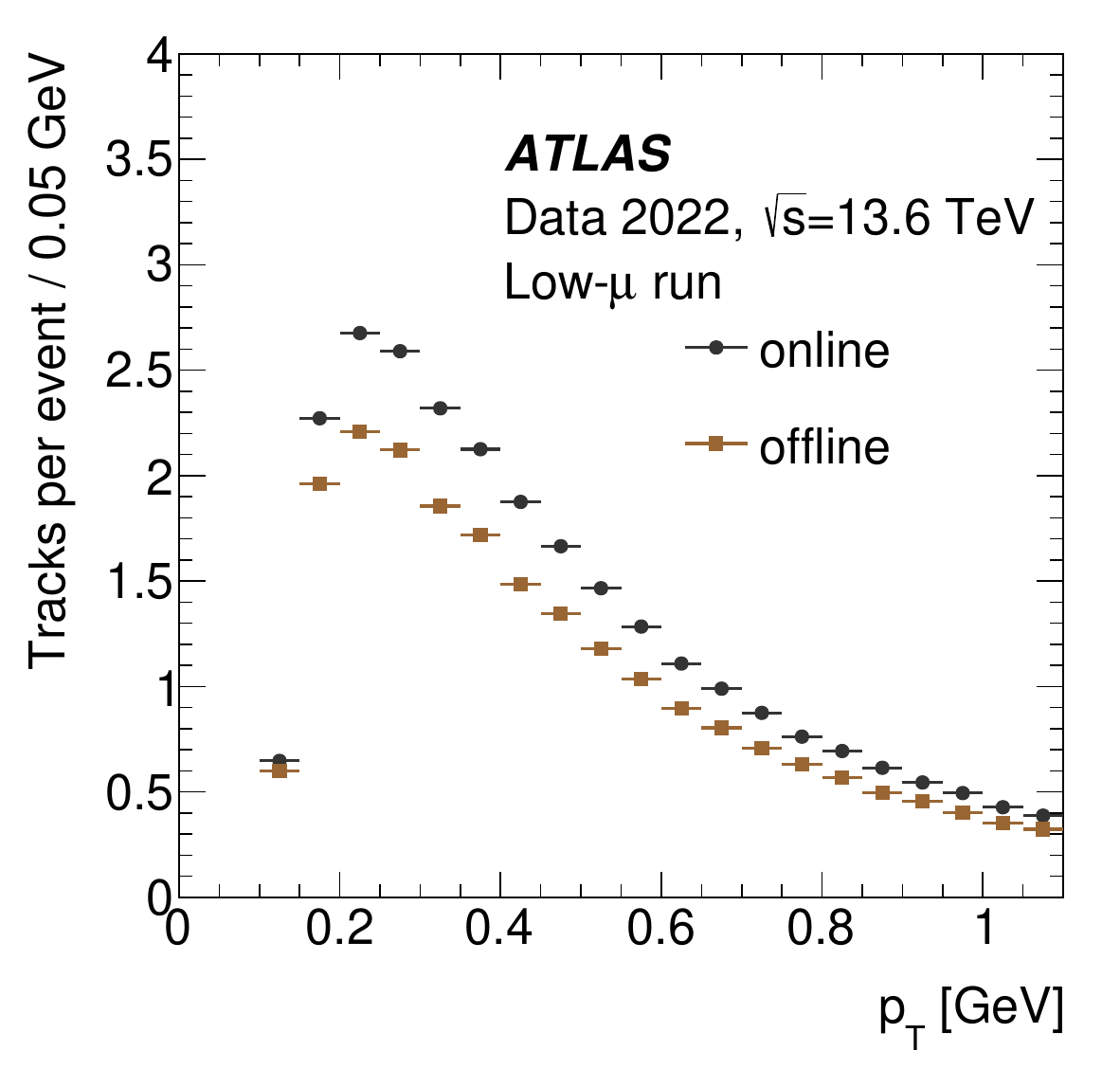}
\includegraphics[width=.49\textwidth]{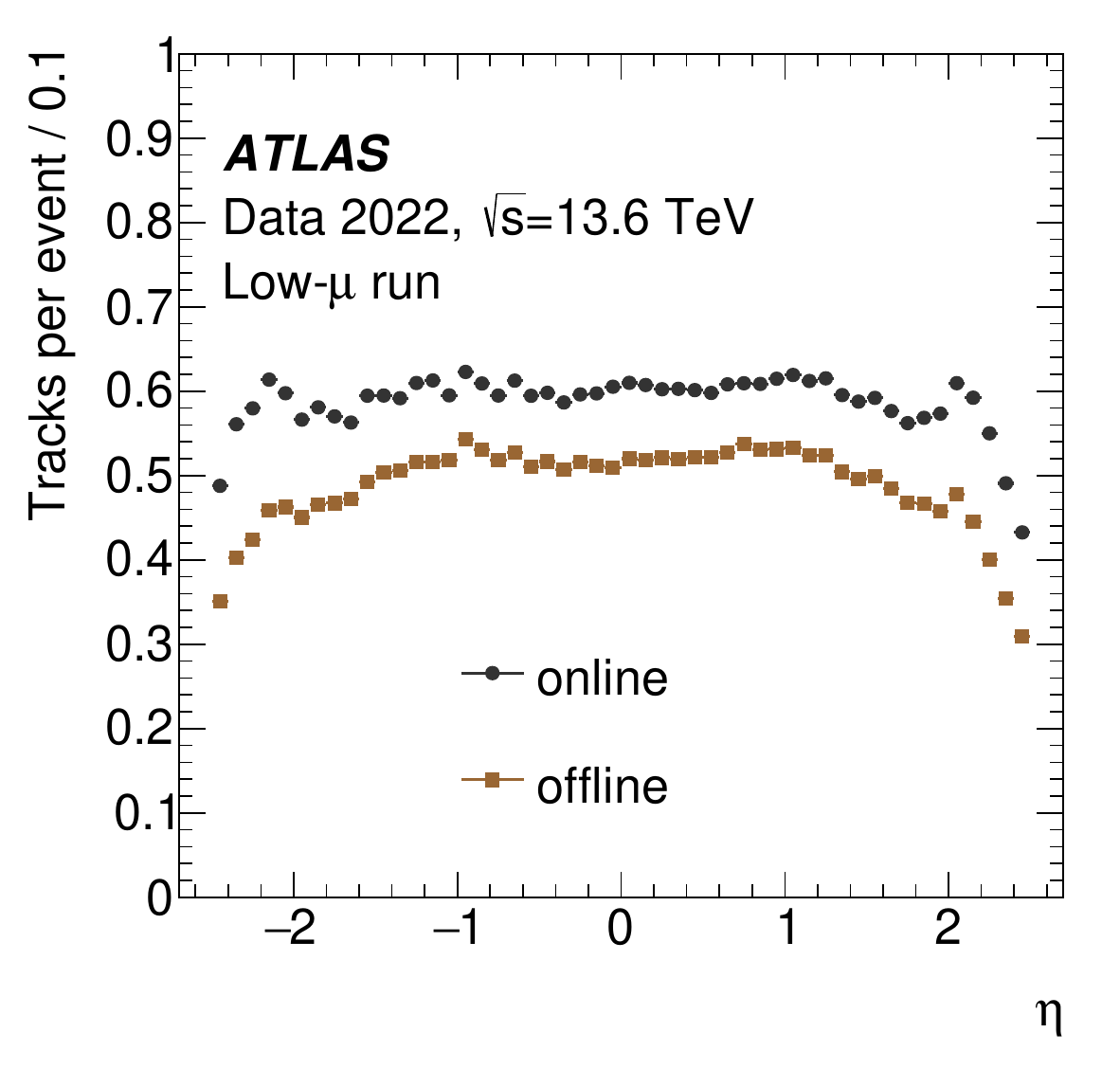}
\caption{Comparison of (left) \pt\ spectra and (right) $\eta$ of tracks reconstructed offline and passing
minimum-bias quality selection and tracks reconstructed by the HLT online during low pile-up run in 2022.}
\label{fig:minbias:tracks_perf}
\end{figure}
 
\begin{figure}
\centering
\includegraphics[width=.49\textwidth]{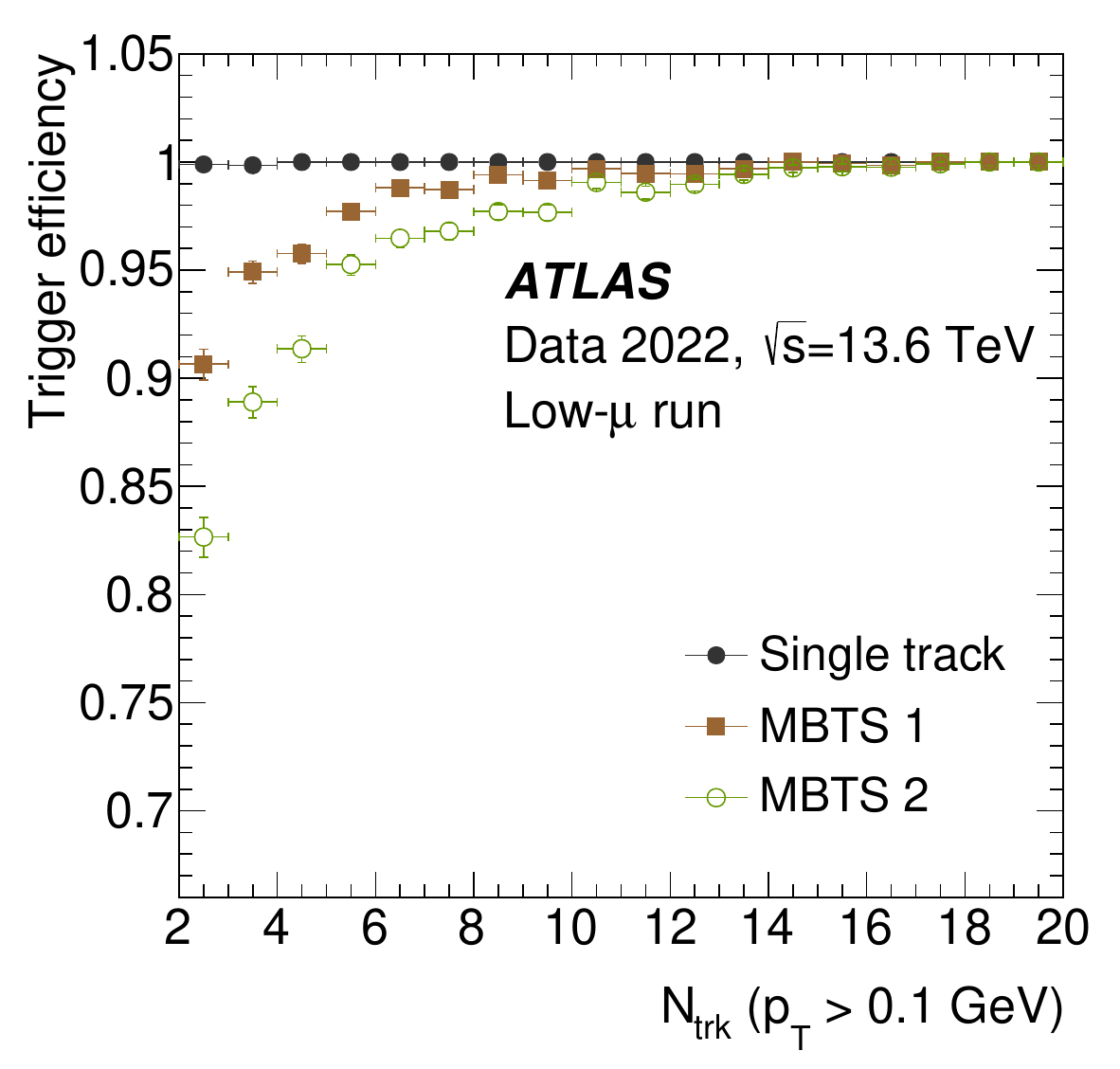}
\includegraphics[width=.49\textwidth]{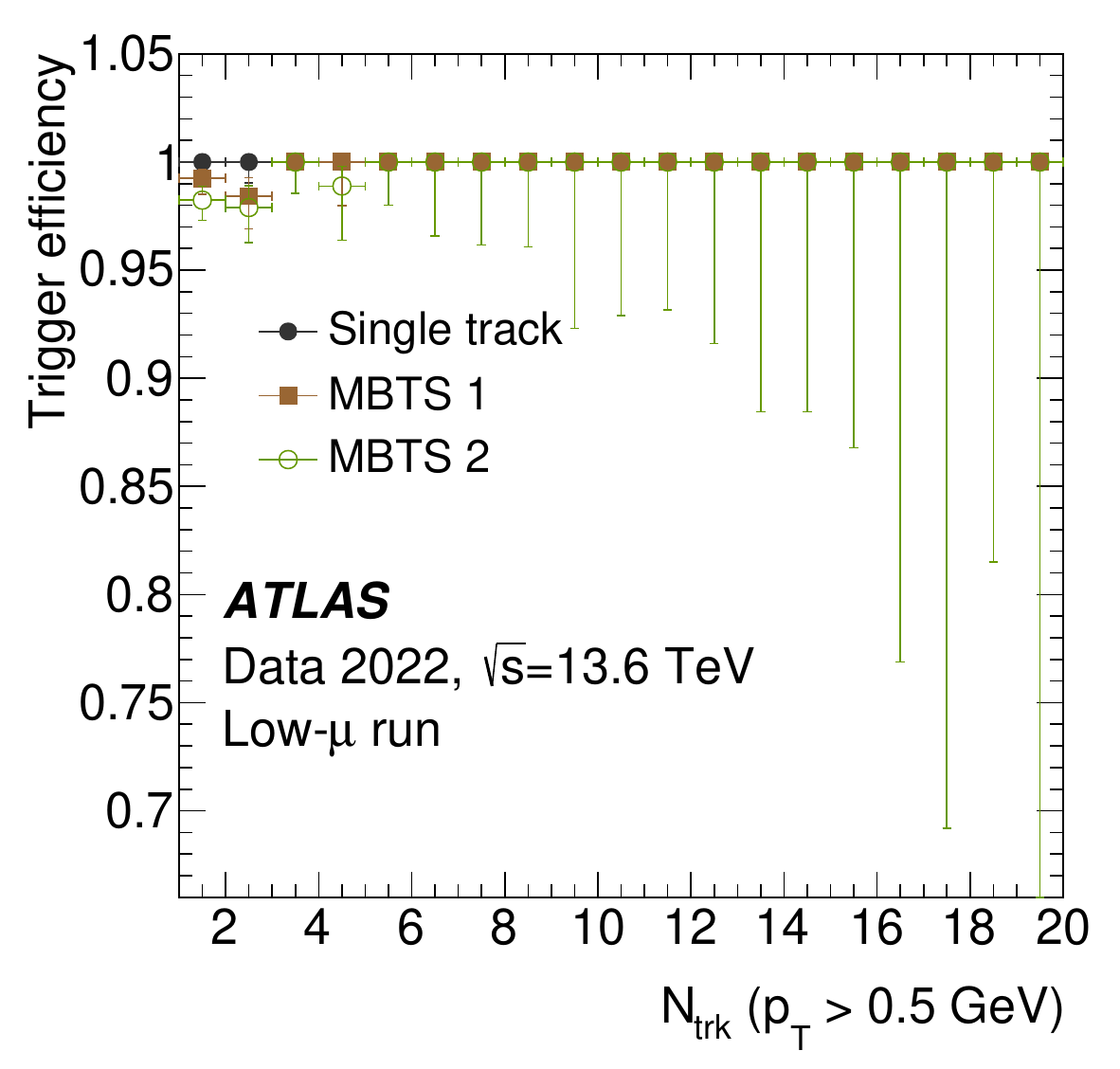}
\caption{(left) Efficiency of triggering for events with at least two tracks of $\pt>0.1\,$\GeV\ and
passing minimum-bias quality track criteria (see text) by single track trigger and MBTS triggers requiring single (MBTS 1) or two hits (MBTS 2) in the scintillator.
(right) Efficiency of triggering for events with at least one track of $\pt > 0.5\,$\GeV\ passing minimum-bias
track quality criteria by the same triggers as shown on the left panel. The efficiency is measured as a function of the number of offline tracks, $N_\mathrm{trk}$,
with respect to the fully efficient trigger requiring only two clusters in the pixels detector and three space-points in SCT seeded from random bunch crossing at L1.
Efficiencies of the MBTS single and two hit triggers are measured with respect to the single track trigger. Only statistical uncertainties are shown.
These triggers were active during low pile-up runs in 2022.}
\label{fig:minbias:eff}
\end{figure}
 
Triggers with track reconstruction include a preselection step based on the count of pixel clusters and SCT space-points (coinciding pairs of hits on both sides of an SCT module).
In particular, the single-track trigger preselection requires two-pixel clusters and three SCT space points.
This preselection step reduces the input rate prior to execution of the tracking algorithms without any efficiency loss.
Random-seeded triggers with only the preselection step and no further HLT requirements are used to collect unbiased samples for performance studies.
Figure~\ref{fig:minbias:eff} shows the efficiency of HLT tracking selection and the L1 MBTS trigger as a function of the number of tracks reconstructed offline passing minimum-bias selection,
as obtained with a data set collected with the random-seeded trigger at L1.
The HLT tracking selection is nearly fully efficient if at least two tracks in the event are present (a required minimum multiplicity for the $\pt > 0.1\,$\GeV\ working point).
The inefficiency of the MBTS triggers is attributed to disabled modules (one on each side) and only a partial geometric overlap between the MBTS and the inner tracking volume.
For track \pt above 0.5\,\GeV, the track trigger is again fully efficient and the MBTS trigger performance mostly recovered.
This is because the presence of at least one track of this momentum is usually correlated with additional activity resulting in signals in a few MBTS counters and thus corresponding triggers.
 
For HMT triggers the requirement on pixel clusters is removed and only the SCT space-points are counted and subject to a threshold requirement, optimised to have full efficiency at a given number of tracks.
For HMT triggers at moderate values of $\mu$, an additional pile-up mitigation strategy is required. This is because
the counting of all tracks in the event instead of only tracks belonging to the highest multiplicity vertex impacts the purity of this type of trigger.
In addition, performing tracking in the full volume of the ID becomes more time-consuming with rising $\mu$, and so any preselection is beneficial.
An algorithm to approximate the vertex position along $z$ and the count of tracks originating from it was developed and optimised for expected data-taking conditions.
It uses the triplets of clusters from the Pixel detector that are used to construct a linear extrapolation to the luminous region.
Extrapolated positions are histogrammed along $z$ with the binning optimised so that the coincidental combinations form a negligible background and real vertices form a well-pronounced peak.
The threshold cut is applied on the count in the peak in the step preceding the tracking step.
Because of the availability of the vertex $z$ position, the tracks in HMT triggers with pile-up suppression are counted only if they are within 10\,mm along $z$ from the approximate vertex position.
The impact of the additional pile-up suppression procedure on the efficiency of HMT triggers is shown in Figure~\ref{fig:minbias:eff_hmt}: here again the offline tracks are required to pass a minimum-bias selection and the triggers are seeded from the random L1 trigger.
At moderate $\mu$ values, these chains are planned to be seeded by a trigger that sums up all energy in the calorimeters, which was not yet commissioned at the time of the 2022 low-$\mu$ runs.
 
\begin{figure}
\centering
\includegraphics[width=.49\textwidth]{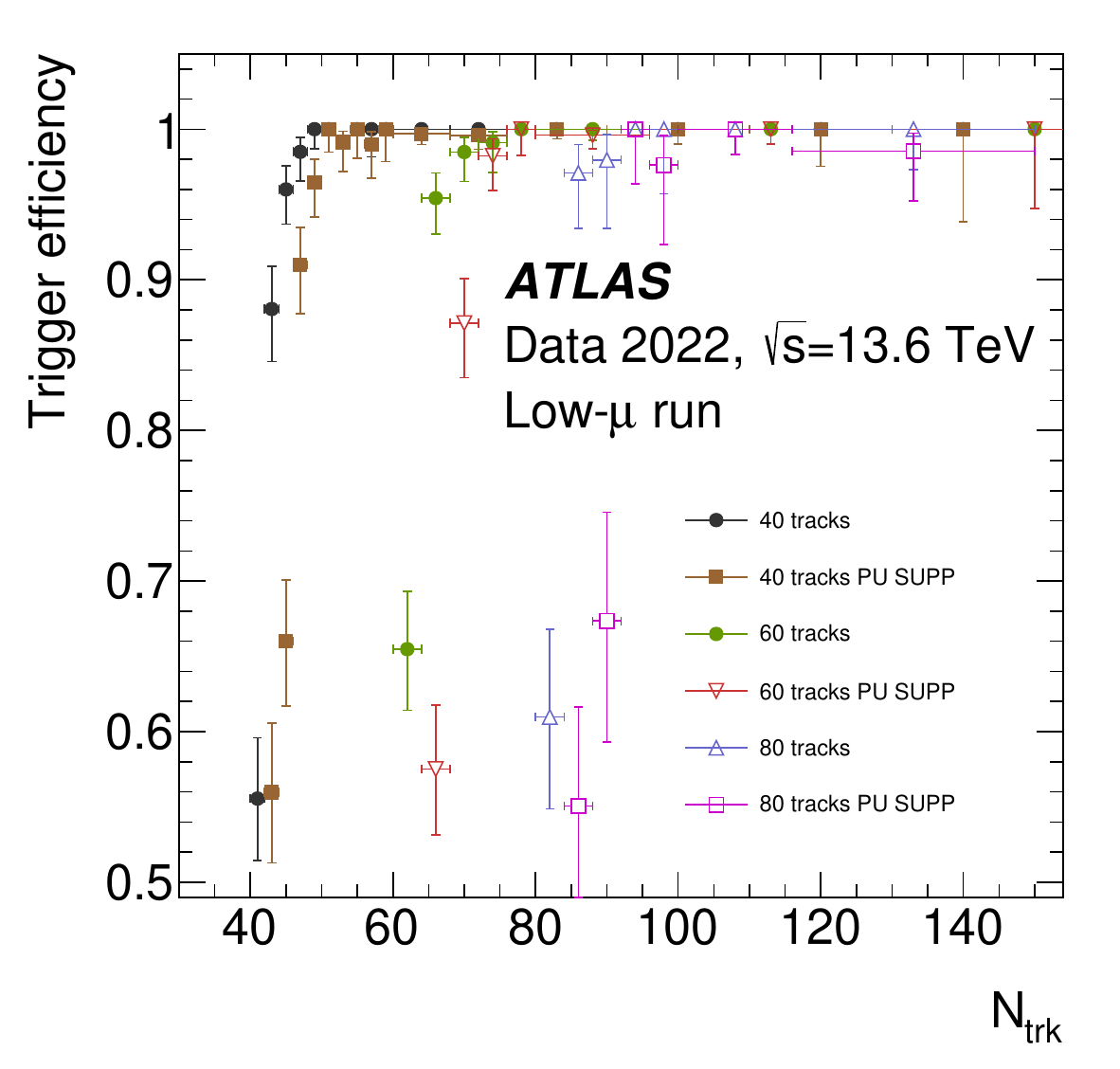}
\caption{Efficiencies of high multiplicity triggers targeting selection of events with 40, 60, and 80 tracks as
a function of number of tracks passing minimum-bias selection quality, $N_\mathrm{trk}$.
For each requirement the performance of the pile-up suppression (PU SUPP) variant of the trigger, designed to be robust in higher values of pile-up, is shown.
Only statistical uncertainties are shown.
These triggers were active during low pile-up runs in 2022. }
\label{fig:minbias:eff_hmt}
\end{figure}
 
\subsubsection{Forward triggers}
In addition to the minimum-bias triggers above, a diverse set of triggers aims to record events with elastic, diffractive or central-exclusive interactions. Their usage depends on the data-taking conditions such as pile-up or LHC beam optics~\cite{beam_optics}. The latter is usually defined by the value of the betatron function at the collision point, $\beta^*$~\cite{beam_optics}.
During the high-$\beta^*$ runs the focus is on elastic scattering events triggered solely by the ALFA detector, although there are also some triggers
combining ALFA detector information with that of other ATLAS subdetectors which target the soft diffractive events. A detailed description of ALFA triggers can be found in Ref.~\cite{ALFA}.
During low-$\beta^*$ (`standard') runs triggers are based on signals from various stations of the AFP detectors with or without coincidence with standard signatures from the central detector.
A detailed description of AFP can be found in Refs.~\cite{atlas-det-run3, AFP1}.
The AFP can deliver trigger signals from two of its detector systems: the Silicon Trackers (SiT) and the Time of Flight (ToF) detectors. SiT trigger signals are expected to be more efficient (98-99\% per station) than the signals from the ToF system (about 80\% per side),
but are also known to suffer from a 400\,ns dead time after each hit~\cite{SiT_testbeam}, causing the efficiency to significantly decrease
for the later bunches in a bunch train when the pile-up exceeds one.
As the SiT trigger dead time depends on the beam intensity and train structure, the following logic is used:
\begin{itemize}
\item During high-$\mu$ runs, when the presence of protons in AFP is expected to be in every second bunch crossing\footnote{In the first approximation this depends on pile-up. \textit{E.g.} for $\mu = 50$ and a probability of registering pile-up proton originating from a single \pp interaction of 2\%, the chance of observing a proton in AFP is $1 - (1-0.02)^{50} \approx 64\%$.}, the trigger items are based on ToF,
\item during low-$\mu$ runs, when the probability of observing a proton in consecutive bunch crossings is small, the trigger items are based on SiT.
\end{itemize}
 
During high-$\mu$ data taking, the physics programme using proton tagging is focused on the measurements of exclusive and two-photon exchange processes, typically in conjunction with high-$p_T$ object(s) produced in the ATLAS detector to keep the rate low. An example of such a process is exclusive jet production when un-prescaled jets with $p_T$ of about 150~\GeV\ are required.
In order to trigger such events, the presence of a jet with a minimum threshold of 50~\GeV\ and a proton(s) in AFP (ToF trigger) is required at L1. At the HLT, matching is required between the kinematics of the centrally reconstructed di-jet system and the scattered protons as reconstructed using the SiT. Depending on the settings of
the algorithm selection criteria, the efficiency is expected to be $60-85\%$. In addition, the match between the di-jet vertex and the vertex $z$ location reconstructed using the ToF data can be applied.
Such trigger chains may obtain a rate reduction of a factor of 100 when compared to a nominal jet trigger with a similar $p_T$ threshold.
 
The composition of the trigger menu for low-$\mu$ runs depends on the exact data-taking conditions. The number of colliding bunches, pile-up and beam optics~\cite{beam_optics} play a key role in the expected event rate.
Studies of soft diffractive processes are usually realised using the least biased triggers.
For AFP, this translates into the requirement of a proton to be detected in both stations on the same side of ATLAS (single tagged events) or all four
AFP stations (both ATLAS sides; double-tagged events). AFP triggers can be combined with other ATLAS trigger objects.
For example, for diffractive charm meson production, in addition to the presence of protons in AFP, a track with a certain transverse momentum is additionally required.


\subsection{Unconventional tracking signatures}
\label{sec:unconTrack}
 
The search for LLPs is an important part of the Run-3 physics programme, as they appear in many motivated scenarios of phenomena beyond the SM.
\acp{LLP} which decay within the \ac{ID} volume or pass entirely through it result in a variety of unconventional tracking signatures in the detector.
Several new triggers presented below were developed to target such unique signatures: long-lived charged particles, displaced jets, jets with displaced tracks and displaced leptons.
They make use of both standard tracking and \ac{LRT}, described in Section~\ref{sec:id}.
The use of tracking for these signatures leads to lower background rates, which allows for lower particle momentum or \met requirements in the trigger, resulting in
large gains in sensitivity for Run-3 searches. Rates for the triggers discussed below are given in Table~\ref{tab:llp}.
 
\subsubsection{Long-lived charged particles that partially or fully traverse the inner detector}
 
Three new triggers are developed targeting long-lived charged particles that partially or fully traverse the ID.
They make use of standard prompt full scan tracking executed after a calorimeter \met\ or jet preselection is applied to reduce
the full scan tracking rate, as described in Section~\ref{sec:id_full}.
In \runii, these searches generally relied on the \met\ trigger \cite{SUSY-2016-32,SUSY-2018-42,SUSY-2016-06}
to select events, which resulted in a low acceptance for models not producing large \met.
 
\textbf{Isolated high \pt track}\newline
Heavy charged particles (such as charginos, sleptons, and $R$-hadrons) with sufficiently long lifetimes can leave an isolated, high-\pt\ track in the \ac{ID}.
A new isolated high-\pt\ track trigger 
is introduced for \runiii
to increase sensitivity to \ac{LLP} signatures with low \met in the event.
It requires a \trig{tcpufit} \met\ threshold of 80\,\gev, 30\,\gev\ below the lowest unprescaled \met\ trigger, to reduce the amount of CPU-time spent on full scan tracking.
The trigger uses tracks from the fast tracking step to 
select events with at least one isolated track with $\pt>120$\,\GeV\
that also passes additional track quality requirements. For the track to be isolated,
the scalar sum of the track \pt\ within $\Delta R<0.3$ must be less than 10\,\GeV.
Figure~\ref{fig:utt:lrt_isohighpt_sig} shows
the expected performance of the isolated high-\pt\ track trigger 
with respect to the HLT \met calculated with \texttt{cell} and \texttt{tcpufit} algorithms,
compared to the ``or'' of the lowest unprescaled \met\ triggers based on \texttt{tcpufit} and \texttt{pfopufit} algorithms.
All of these algorithms are described in Section~\ref{sec:met}.
SUSY di-stau \ac{LLP} \ac{MC} simulation for a stau mass of 600~\GeV\ and a lifetime of 10~ns is used to evaluate the efficiency.
The new trigger increases the acceptance to signal-like events at lower values of \met.
 
\begin{figure}[htbp]
\centering
\includegraphics[width=0.49\textwidth]{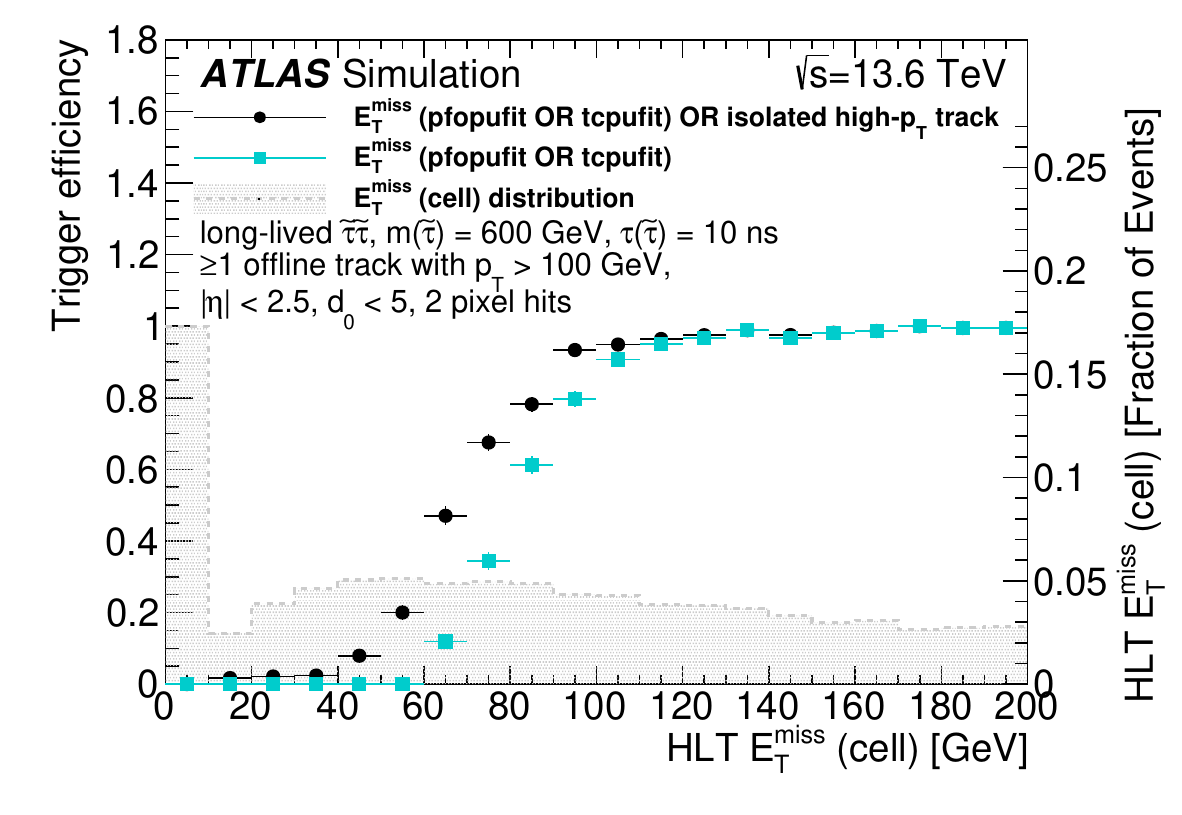}
\includegraphics[width=0.49\textwidth]{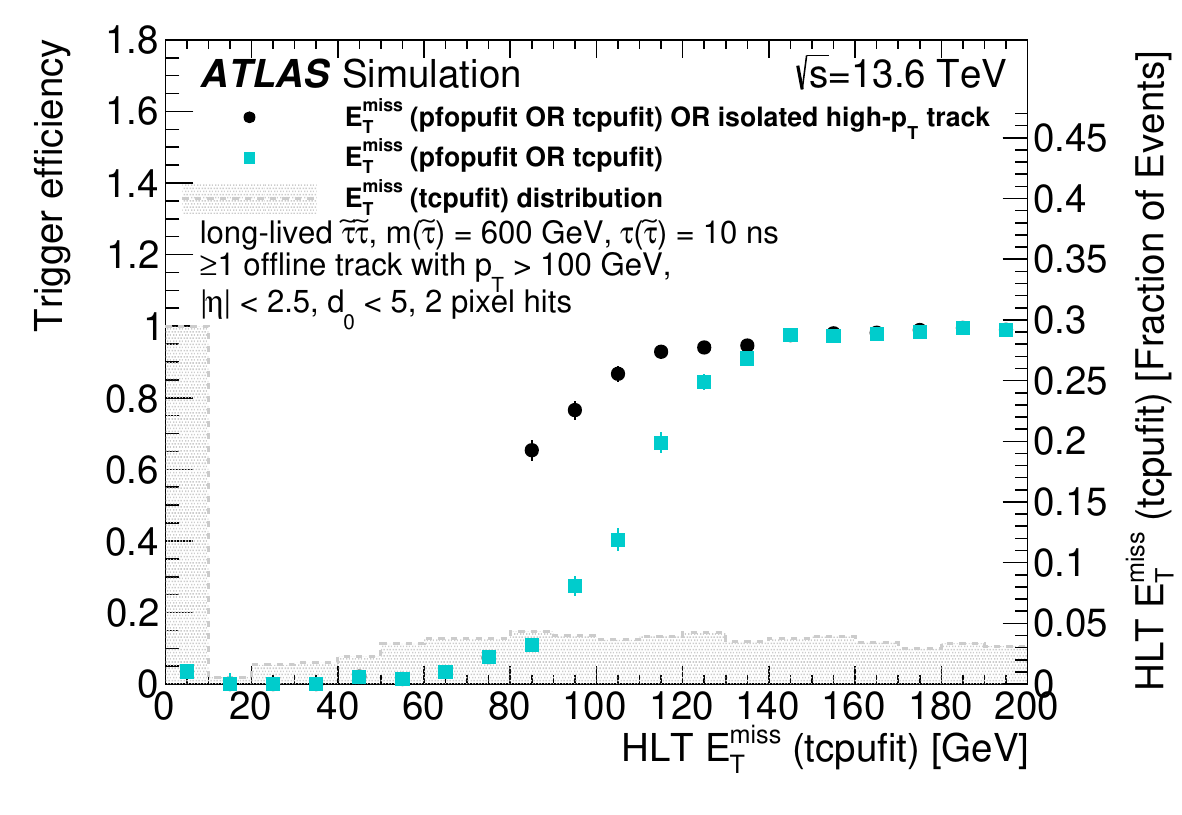}
\caption{Isolated high-\pt\ track trigger expected efficiency using Run-3 MC simulation of SUSY di-stau events with a mass of 600~\GeV\ and a lifetime of 10~ns
vs (left) the \texttt{cell} and (right) the \texttt{tcpufit} algorithm \met.
The efficiency of the new trigger (circles) is compared with that of
a logical ``or'' of the lowest unprescaled \met\ triggers based on \texttt{tcpufit} and \texttt{pfopufit} algorithms described in Section~\ref{sec:met} (squares).
Both of these are overlaid on the signal sample \met\ distribution of fraction of events per bin (gray histogram), and include the L1 trigger efficiency.
Only statistical uncertainties are shown.
}
\label{fig:utt:lrt_isohighpt_sig}
\end{figure}
 
\textbf{Large $dE/dx$ triggers}\newline
New heavy, charged particles, mentioned above, may also leave large energy deposits in the \ac{ID} silicon layers compared to what is expected from a minimally ionizing particle.
The measurement of these large ionisation energy losses per unit pathlength, $dE/dx$, in the Pixel detector is a handle to identify tracks as signal candidates.
A new trigger targeting long-lived, heavy, charged particles~\cite{SUSY-2018-42} 
uses the $dE/dx$ measurement capabilities of the \ac{ID} for trigger decisions.
The same \met\ triggers, as used for the isolated track trigger above, are required before running the full scan tracking.
The $dE/dx$ trigger selects events with at least one track with $\pt>50$~\GeV,
an average $dE/dx>1.7\,$\MeV$/$cm, at least two hits with
$dE/dx>1.7\,$\MeV$/$cm, track $|d_{0}|<2.5$~mm, and $|\eta|<2.5$.
Figure~\ref{fig:utt:lrt_dedx_sig}\, shows
the online $dE/dx$ distribution of tracks with $\pt>10\,$\GeV\ in data collected during 2022 with
\met, single-jet, and multi-jet triggers where full scan tracking was run.
It also shows the correlation of the online $dE/dx$ measurement to the offline measurement.
 
\begin{figure}[htbp]
\centering
\includegraphics[width=0.49\textwidth]{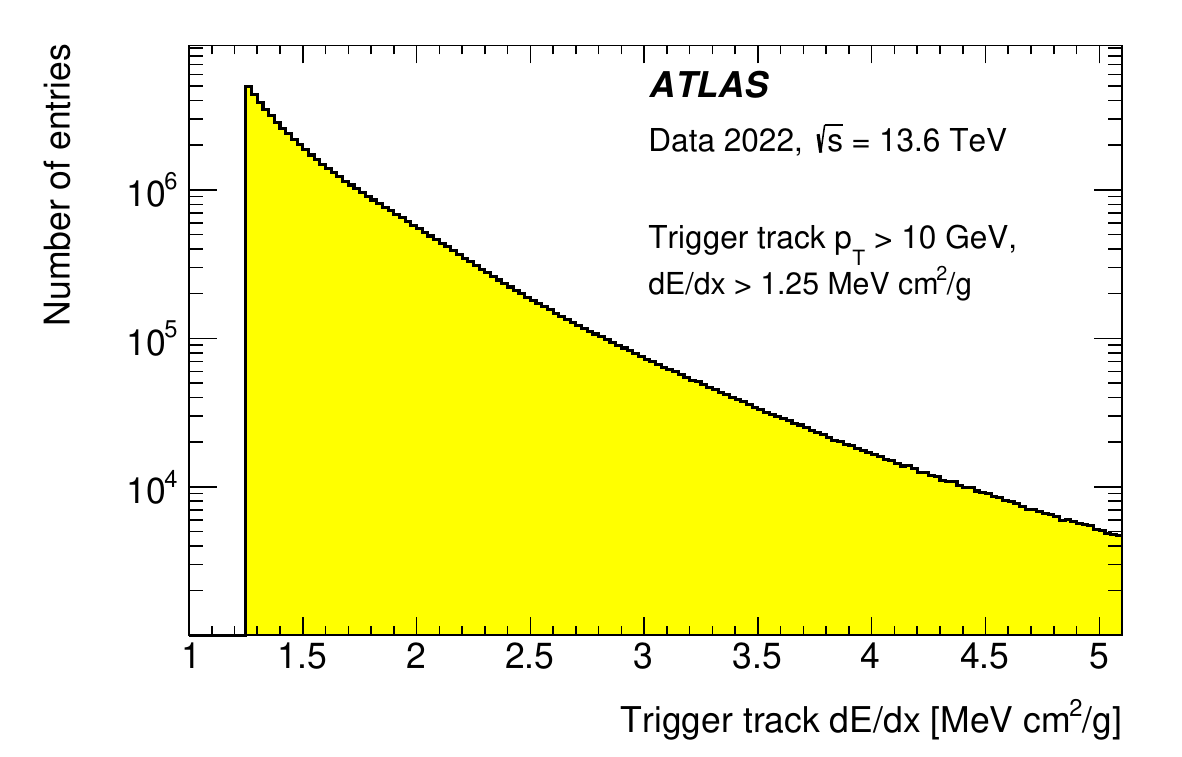}
\includegraphics[width=0.49\textwidth]{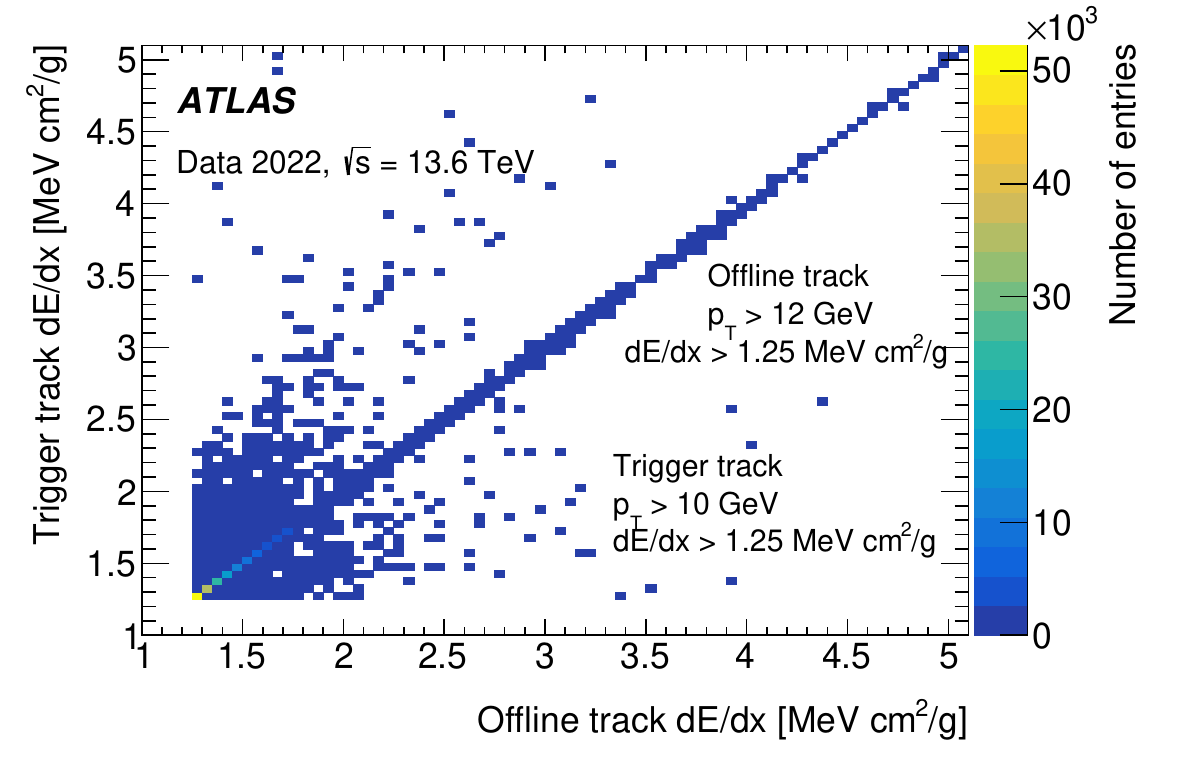}
\caption{(left) Distribution of $dE/dx$ for tracks reconstructed in the trigger used for monitoring the dE/dx-based trigger.
(right) Correlation of online vs offline $dE/dx$ for offline tracks with $\pt>12~\GeV$.
}
\label{fig:utt:lrt_dedx_sig}
\end{figure}
 
\textbf{Disappearing track triggers}\newline
Charged particles with slightly shorter lifetimes than the signatures above can decay part way through the \ac{ID} and leave a short tracklet of a few hits.
These tracklets are referred to as disappearing tracks when the charged particle decays into invisible and low-\pt\ particles that are not reconstructed.
As in the previous two cases, the disappearing track trigger 
makes use of full scan tracking
executed after the same \met\ trigger requirements, which are lower than the lowest unprescaled \met\ trigger.
The fast tracking algorithm is modified to save tracklets with four hits in the inner layers of the \ac{ID} that fail to become tracks.
In order to reduce the large background to this signature,
a \ac{BDT} based on the track parameters, quality of fit, and number
of hits in the Pixel and SCT detectors is used to separate signal-like tracklets from background.
The disappearing track trigger selects events with at least one tracklet with $\pt>20$~\GeV\ that passes a stringent requirement on the \ac{BDT} score.
Figure~\ref{fig:utt:lrt_disptrk_sig} shows the expected performance of the trigger compared to that of the \met\ trigger with
a 110~\GeV\ threshold, which was used in \runii, for a model with long-lived charginos~\cite{SUSY-2018-19}.
The acceptance of the trigger to events with chargino momentum below 150~\GeV\ is greatly improved.
 
\begin{figure}[htbp]
\centering
\includegraphics[width=0.49\textwidth]{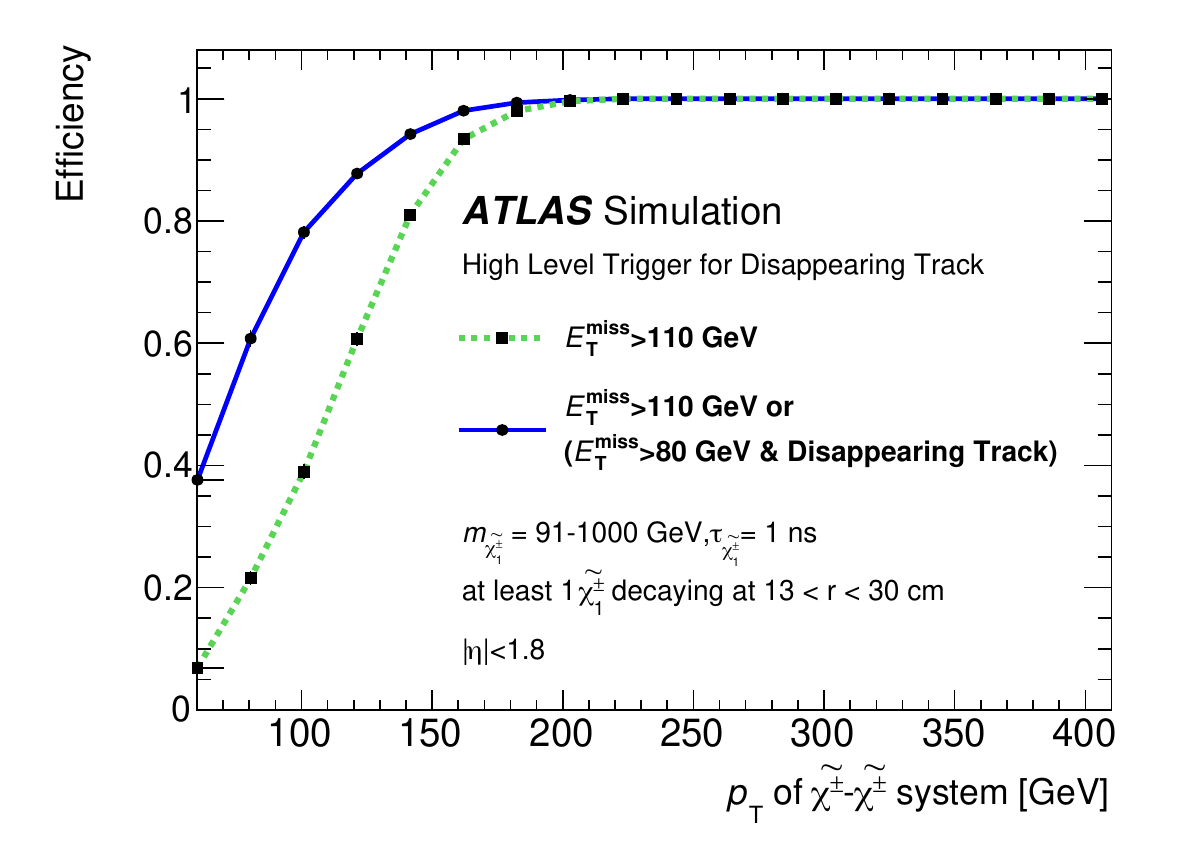}
\caption{Efficiency as a function of the chargino-pair system transverse momentum for the \ac{HLT} \met\ trigger (dotted line)
and a logical OR with the disappearing track trigger (solid line) with respect to the L1 $\met>50$\,\GeV\ trigger.
Both triggers are based on \trig{tcpufit} \met\ algorithm. MC simulation with Run-2 conditions is used.
The events are taken from models with a chargino lifetime of 1\,ns and the
chargino mass of either 91\,\gev\ or in the range 200--1000\,\gev\ with 100\,\gev\
spacing. Only events with $\met>60\,$\gev\ and at least one chargino decaying
between the pixel and SCT detectors (13--30\,cm) in the central region, $|\eta|<1.8$,
are considered.
}
\label{fig:utt:lrt_disptrk_sig}
\end{figure}
 
\subsubsection{Long-lived particle decays into jets}
A second set of three new triggers have been developed targeting long-lived particle decays into jets, which may be displaced themselves or contain displaced tracks.
The triggers make use of full scan tracking run after a \met\ or jet requirement and in some cases use LRT as an additional handle to select events.
 
\textbf{Hit-based displaced vertex triggers}\newline
Neutral \acp{LLP} may travel some distance into the detector before decaying, resulting in a displaced vertex (DV) or jet.
Previous searches relied on a variety of triggers looking for other objects in the final state.
Two hit-based DV triggers 
make use of hits not associated to tracks after the standard full scan tracking is performed in events passing the L1 jet and \met\ triggers, respectively.
Jets with a large number of remaining hits on the outer layers of the \ac{ID}, and few on the inner layers, are indicative of DVs with displaced tracks that are not reconstructed.
Using this as an extra requirement allows for lower \met\ and jet thresholds compared to the lowest unprescaled triggers.
A BDT trained on the fraction of hits in the layers of the \ac{ID} is used to implement this selection on jets.
The first trigger seeded by the same 80\,\GeV\ \met\ threshold as above selects events containing jets with $\pt>200\,$\GeV\ and $|\eta|<1$.
Figure~\ref{fig:utt:lrt_dv_sig} shows its expected performance versus the number of pile-up interactions
using MC simulation of a heavy Higgs boson decaying into two long-lived scalars, each subsequently decaying
into two $b$-jets compared to a background process of \ttbar\ with an all-hadronic final state.
The signal efficiency is around 70\%, compared to a background efficiency of less than 5\%, for high numbers of additional interactions.
The algorithm is tuned such that there is no strong dependence of the signal efficiency on pile-up.
The second trigger seeded by a L1 jet with \pt$>100\,$\gev\ selects events containing  jets with $\pt>260\,$\GeV\ and $|\eta|<1$.
It requires a jet with $\pt>180\,$\GeV\ at the HLT calorimeter preselection step before running the tracking.
 
Another hit-based trigger under development is seeded by a L1 trigger with \met$>50\,$\gev\ and runs only the fast tracking step of \ac{LRT} in a composite \roi around the jets.
Tracks with $\pt>2~\GeV$ are then used to build vertices using a modified version of the offline secondary-vertexing algorithm~\cite{ATL-PHYS-PUB-2019-013}.
The algorithm is optimised to be faster by requiring track pairs to be consistent with an approximate vertex position and
by reducing the combinatorics of clustering by binning the vertex positions.
 
\begin{figure}[htbp]
\centering
\includegraphics[width=0.49\textwidth]{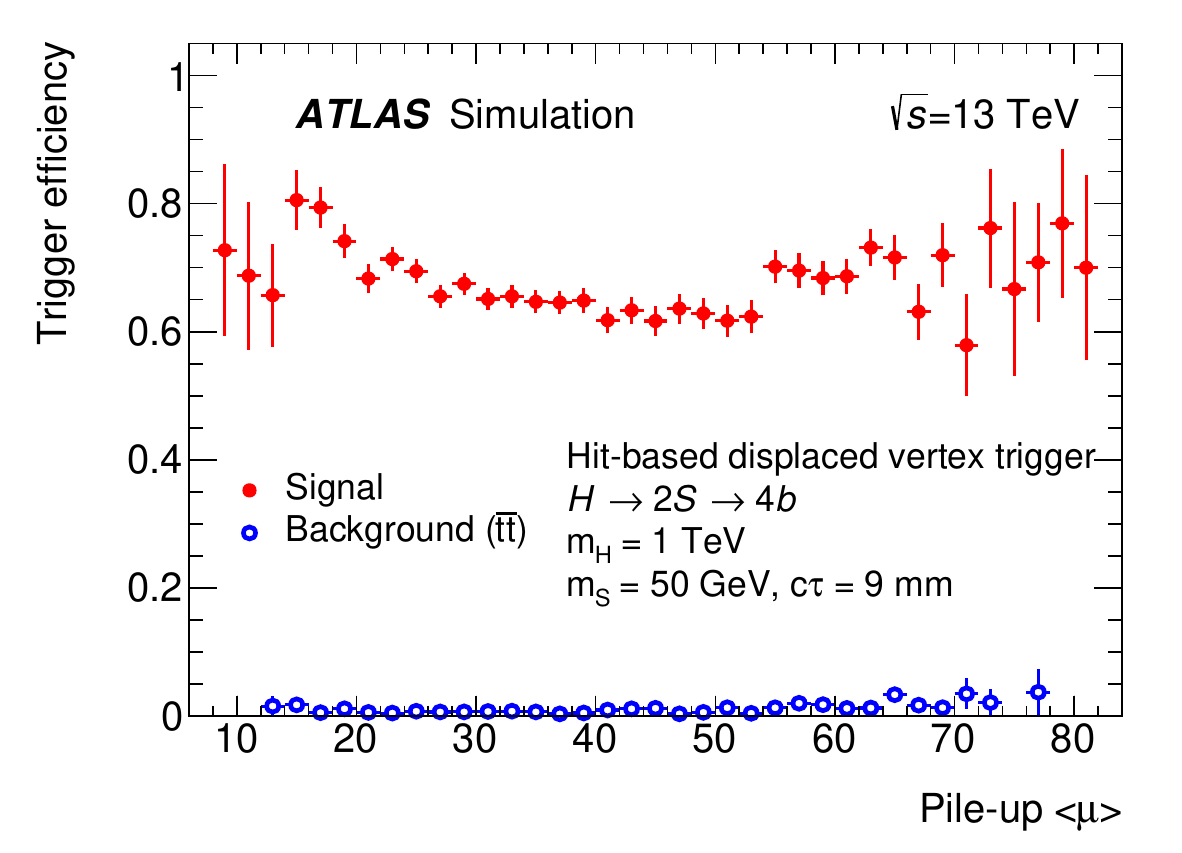}
\caption{Expected performance of the hit-based displaced vertex trigger seeded by a \trig{tcpufit} \met\ threshold of 80\,\gev\ versus the average pile-up.
MC simulation is used for signal and background processes. The signal process is a heavy Higgs boson ($m_H=1\,\TeV$) decaying into two scalars ($m_S=50\,$\GeV) with a proper decay length of 9\,mm,
which subsequently decay into two $b$-jets each.
The background process is \ttbar\ production with an all-hadronic final state.
Only statistical uncertainties are shown.
}
\label{fig:utt:lrt_dv_sig}
\end{figure}
 
\textbf{Emerging jet triggers}\newline
In the models of new phenomena with a dark sector, decays of particles into hadronically interacting particles may result in emerging jets with a large fraction of events possessing a DV and tracks with large $d_{0}$.
The fraction of momentum associated to prompt tracks in the jet relative to the total jet momentum is used to identify
those with a large fraction of displaced tracks, i.e. a low momentum fraction associated to prompt tracks.
This trigger is run in events that 
contain a L1 jet with $\pt>100\,$\gev\ and an HLT jet with $\pt>200\,$\GeV\ in order to reduce the rate of full scan tracking.
The \pt\ of tracks within $\Delta R<1.2$ of the leading large-R jet axis are summed.
Jets are only considered if within $|\eta|<1.8$ and the tracks must have $\pt>2~\GeV$, $|\eta|<2.4$, $|d_0|< 8$~mm, and $|d_{0}|<2.5\cdot \sigma_{d_{0}}$~mm, where the $d_0$ resolution $(\sigma_{d_0})$ is computed from $Z+$jets events in data.
The ratio of the sum of track \pt\ over the jet \pt is computed, and events are selected if this ratio, PTF, is less than 0.08.
This extra requirement allows for a reduction of more than a factor of two of the jet \pt\ threshold as compared to that of the lowest unprescaled trigger (420\,\gev\ in 2022).
 
A second trigger of this type 
is seeded by a 45~\GeV\ photon and selects events with two large-R jets with $|\eta|<2.0$, $\pt>55~\GeV$, and $\text{PTF}<0.1$.
Figure~\ref{fig:utt:lrt_emergjet_lrtdispjet_sig} (left) shows that the emerging jet trigger is efficient down to much lower jet \pt\ than the single large-R jet trigger.
A model of a 1.5 TeV $Z'$ decaying into two 20\,\GeV\ dark pions with a proper decay length of 50\,mm is used~\cite{Linthorne:2021oiz}.
The overall efficiency depends on the acceptance of the PTF requirement.
 
\begin{figure}[htbp]
\centering
\includegraphics[width=0.47\textwidth]{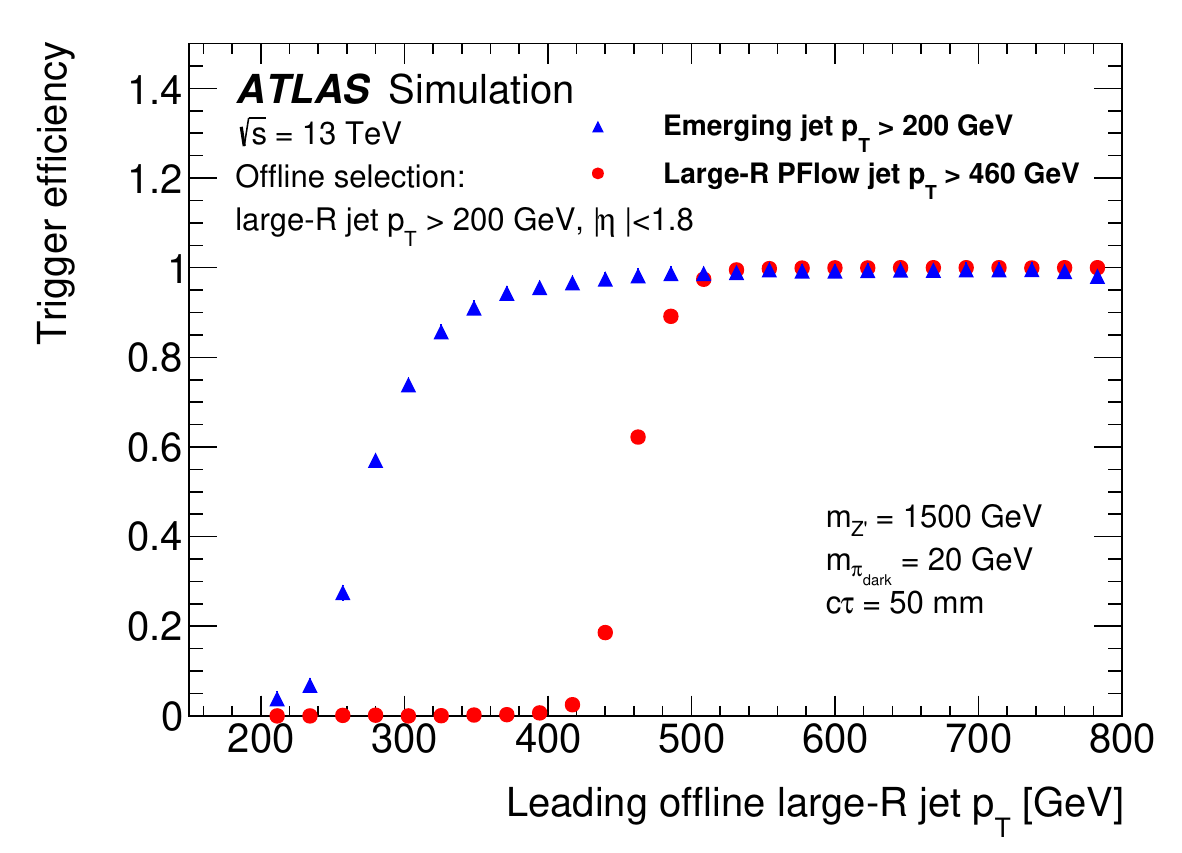}
\includegraphics[width=0.48\textwidth]{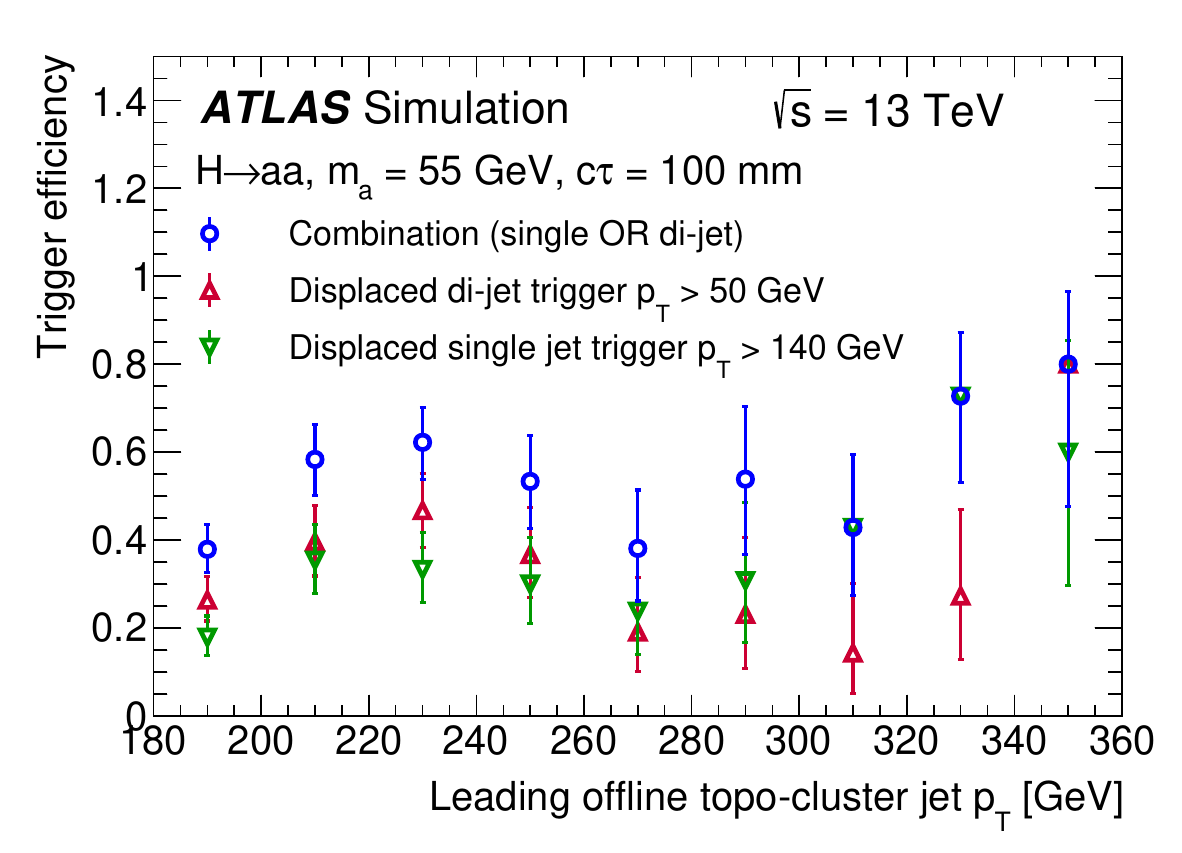}
\caption{(left)
Efficiency of the new emerging jet trigger compared to a single large-radius jet trigger for a $Z'$
decaying into two dark pions with a proper decay length (c$\tau$) of 50\,mm. The efficiency is calculated
as a function of the \pt\ of the leading offline anti-$k_t$ R=1.0 jet, reclustered from R=0.4 topocluster jets.
(right) Efficiency of the \ac{LRT}-based displaced jets triggers plotted against the leading offline anti-$k_t$
R=0.4 jet \pt\ for a model of exotic decays of the Higgs boson into long-lived pseudoscalars $a$ with mass 55~\GeV\ and proper decay length of 100~mm.  
Both studies use MC simulation with Run-2 conditions.
Only statistical uncertainties are shown.
}
\label{fig:utt:lrt_emergjet_lrtdispjet_sig}
\end{figure}
 
\textbf{Displaced jet triggers}\newline
Similar to the previous case, neutral \acp{LLP} decaying into quarks or gluons may result in displaced jets.
Two new \ac{LRT}-based triggers select such events with a jet from initial-state radiation and at least one or two displaced jets,
but with a lower \pt\ threshold than the lowest unprescaled single-jet trigger at 420~\GeV.
Tracks with $\pt>1$~\GeV\ and $|\eta|<2.4$ within $\Delta R<0.4$ of the jet are counted as prompt (\ntrkprompt) or displaced (\ntrkdisp) if they have $|d_{0}|<\text{or}>3~\text{mm}$, respectively.
After the standard tracking is run for the events that contain
a L1 jet with $\pt>100\,$\gev\ and an HLT jet with $\pt>180\,$\GeV, 
jets are preselected by requiring $\ntrkprompt\leq2$.
\ac{LRT} is then run on the remaining hits in \rois around the three leading jets passing the preselection.
In addition to the leading jet $\pt>180~\GeV$ requirement, the displaced single-jet trigger 
selects events with a $\pt>140~\GeV$ jet with $\ntrkdisp\geq3$ and $\ntrkprompt\leq1$.
Similarly, the displaced di-jet trigger 
selects events with two $\pt>50~\GeV$ jets where at least one must satisfy $\ntrkdisp\geq3$ and $\ntrkprompt\leq2$, and the other may satisfy only $\ntrkdisp\geq0$ if $\ntrkprompt\leq1$.
Figure~\ref{fig:utt:lrt_emergjet_lrtdispjet_sig} (right) shows the expected performance of the displaced jets triggers based on \ac{MC} simulation of a model with exotic Higgs decay into long-lived pseudoscalars $a$ in the jet \pt\ region below the primary single jet trigger. Details of the model can be found in Ref.~\cite{EXOT-2018-57}.
 
\subsubsection{Long-lived particle decays into SM leptons}
\acp{LLP} may also decay into \ac{SM} leptons resulting in displaced electrons, muons, and taus.
Three new sets of triggers target these decays using LRT, and also standard tracking in the case of taus.
 
To select such events in \runii, searches generally relied on triggers without tracking information such as photon triggers to select displaced electrons and MS-only triggers to select muons \cite{SUSY-2018-14}.
These triggers had high \pt\ thresholds of 50--120~\GeV, requiring two objects for the 50~\GeV\ threshold, and restrictions in $\eta$ in the case of muons.
New triggers for displaced electrons and muons directly trigger on these signatures allowing for lower thresholds with respect to those used in \runii.
The displaced electron trigger 
runs \ac{LRT} in \rois in events passing the same L1 threshold as the primary prompt electron chain, described in Section~\ref{sec:egamma}.
It selects events that have an electron with $\pt>30\,$\GeV\ and $|d_{0}|>3\,$mm that passes a loose likelihood electron identification~\cite{TRIG-2018-05} without the use of $d_{0}$ or requirements on the number of hits in the Pixel detector.
The displaced muon trigger 
runs \ac{LRT} in \rois in events passing the same L1 threshold as the primary prompt muon chain, described in Section~\ref{sec:muons}.
It selects events that have a muon with $\pt>20$~\GeV\ and $|d_{0}|>2$~mm.
Figures~\ref{fig:utt:lrt_elmu_sig1} and \ref{fig:utt:lrt_elmu_sig} show the expected efficiency of the displaced
electron and muon HLT triggers with respect to their L1 seeds in terms of the offline reconstructed lepton transverse impact parameter ($d_0$)
and the production radius of offline reconstructed electrons and muons, respectively, using MC simulation of LLP di-stau production.
MC simulation of pair production of staus with a 1~ns lifetime is used, and
stau masses of 100--500\,\GeV\ in 100\,\GeV\ steps are merged.
The offline electrons are required to pass the same loose likelihood identification as used in the trigger and have $|d_0|>3$~mm.
The offline muons are required to pass the medium working point described in Ref.~\cite{MUON-2018-03} without a cut on the number of pixel hits and have $|d_0|>2$~mm.
The acceptance of the LRT-based triggers extends out to the first layer of the SCT at 300~mm, where the layout of the detector no longer allows for eight hits on the track in most regions.
The standard tracking runs out to a $|d_0|$ of 5 and 10~mm for electrons and muons respectively, limiting the acceptance of the standard prompt lepton triggers.
The larger value for muon tracking increases the acceptance of $B$-meson decays.
Combining the standard and LRT light lepton triggers provides continuous acceptance from small to large displacements.
 
\begin{figure}[htbp]
\centering
\includegraphics[width=0.49\textwidth]{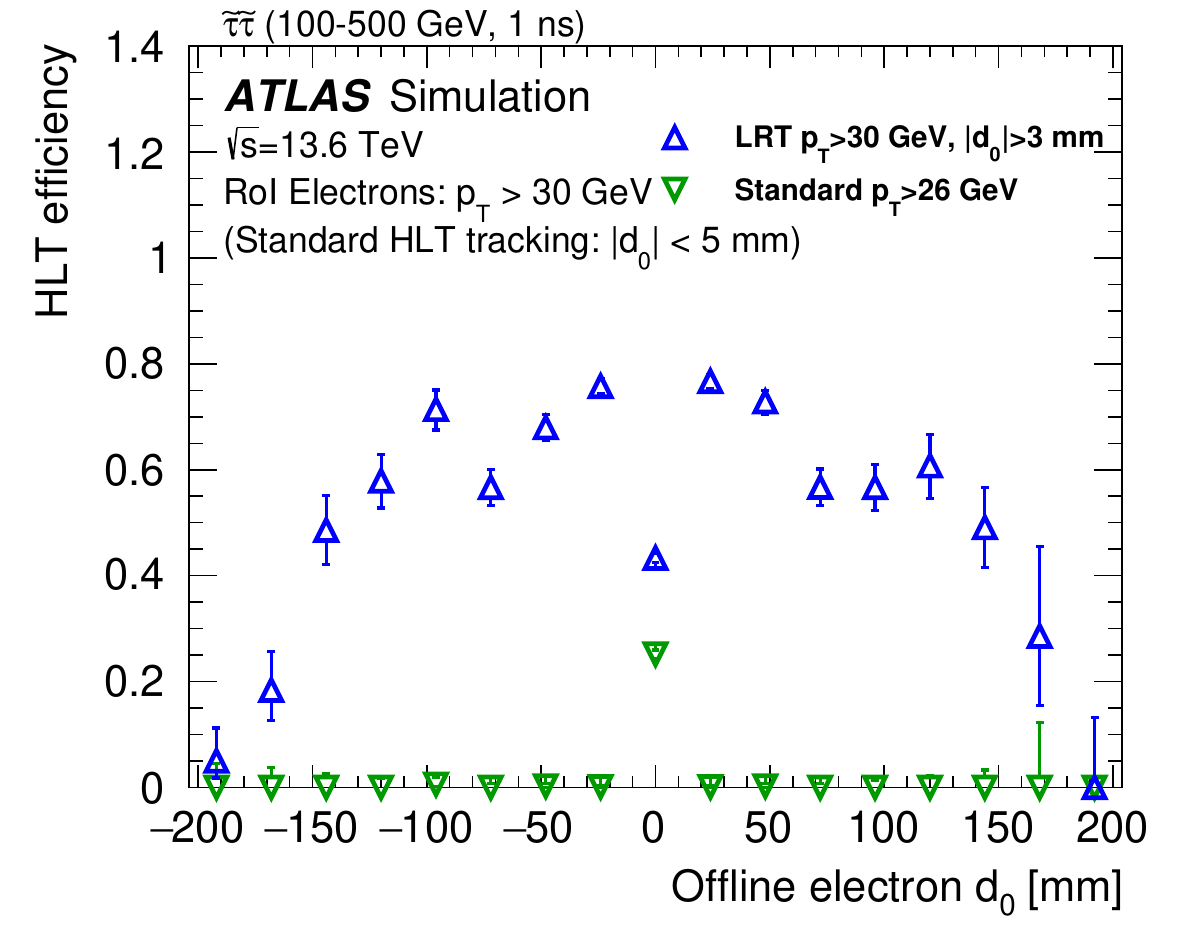}
\includegraphics[width=0.49\textwidth]{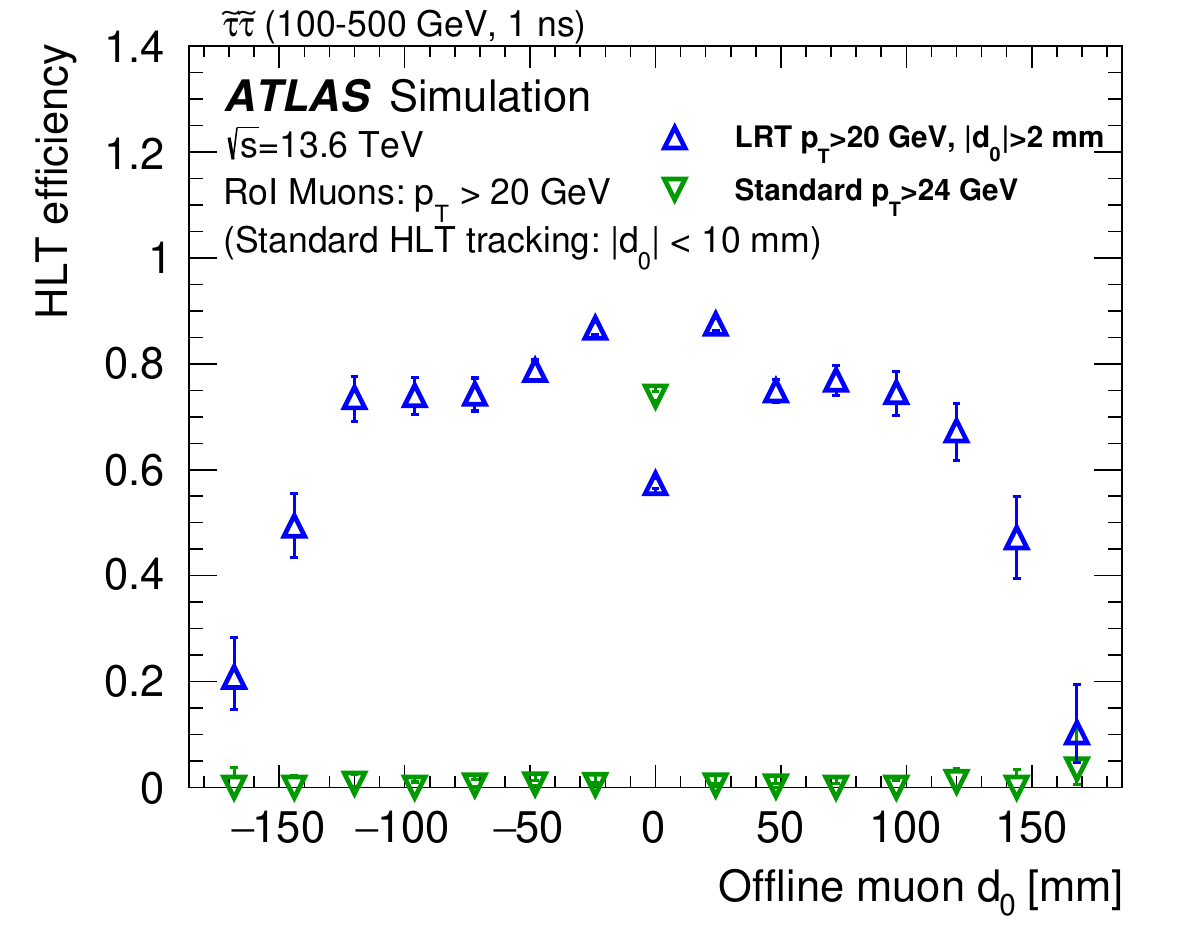}
\caption{
Displaced (left) electron and (right) muon trigger efficiencies with respect to their L1 seeds versus the offline lepton $d_0$ (LRT, open triangles).
The efficiencies for the isolated primary single electron and muon triggers (Standard), described in Sections~\ref{subsec:egammamenu} and~\ref{subsec:muonmenu},
are shown as inverted triangles. MC simulation samples with Run-3 conditions are used.
Only statistical uncertainties are shown.
}
\label{fig:utt:lrt_elmu_sig1}
\end{figure}
 
\begin{figure}[htbp]
\centering
\includegraphics[width=0.49\textwidth]{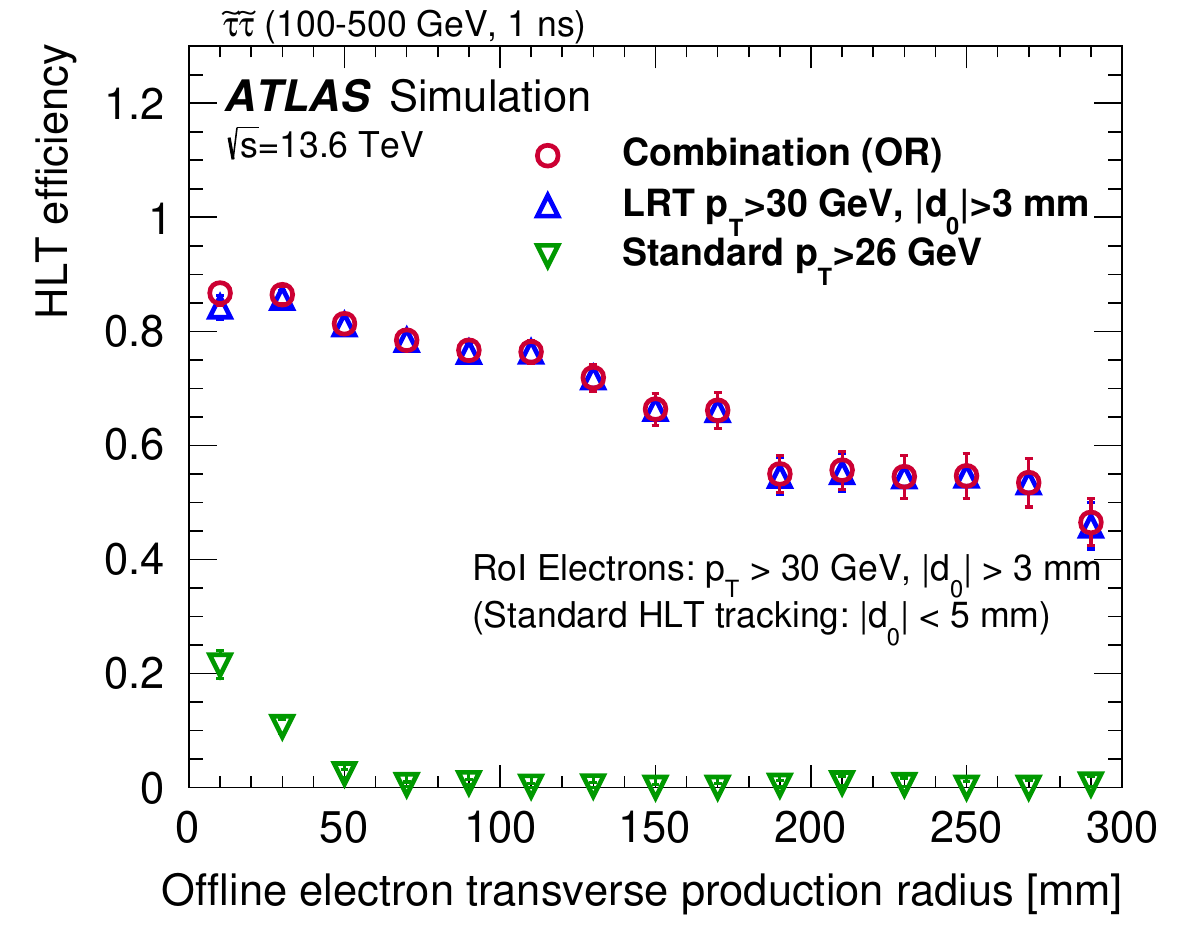}
\includegraphics[width=0.49\textwidth]{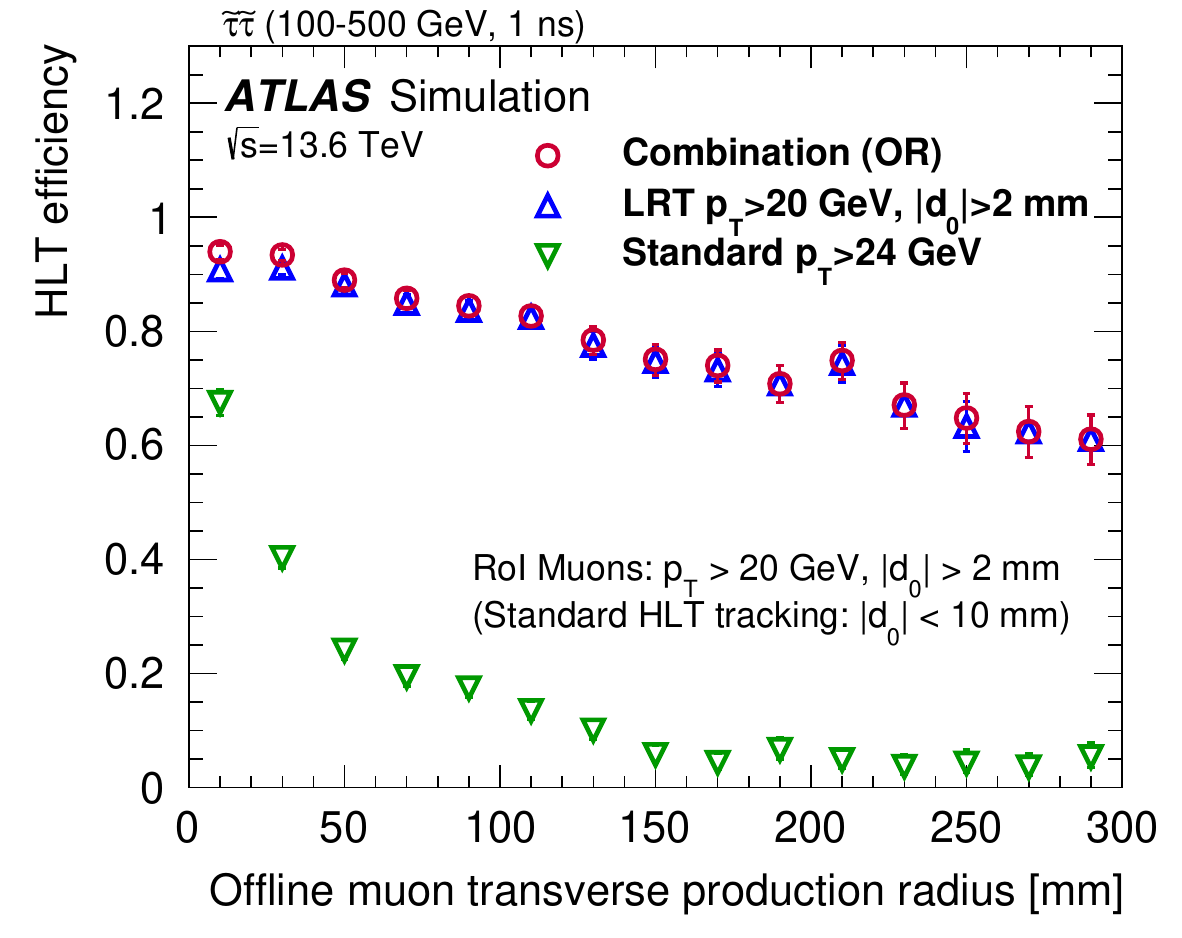}
\caption{Displaced (left) electron and (right) muon trigger efficiencies with respect to their L1 seeds versus
the offline reconstructed lepton production radius (LRT, open triangles).
The efficiencies for the isolated primary single electron and muon triggers (Standard), described in Sections~\ref{subsec:egammamenu} and~\ref{subsec:muonmenu},
are shown as inverted triangles. A logical OR between the LRT and standard triggers (Combination) is marked with open circles.
MC simulation samples with Run-3 conditions are used.
Only statistical uncertainties are shown.
}
\label{fig:utt:lrt_elmu_sig}
\end{figure}
 
Two dedicated triggers for displaced hadronically decaying taus are in development for \runiii. One is based on standard tracking and the other on \ac{LRT}.
A search during \runii\ with this signature was not performed, but would have needed to use jet or \met\ triggers with high thresholds.
The standard tau identification RNN~\cite{ATL-PHYS-PUB-2019-033}
is retrained to use standard tracking to identify displaced taus.
Standard input samples are replaced by representative signal samples
with displaced tau content, which additionally include association of large radius tracks
in the tau reconstruction. MC samples with lifetimes of 0.01--100\,ns are combined for the training.
The first tau LLP trigger 
selects events containing a tau with $\pt>200\,$\GeV\ passing the medium working point of this displaced tau identification.
This identification is also used in multi-object triggers looking for a tau+X, which allows for lower thresholds on the tau object. 
Primary L1 di-tau, single muon, single isolated electron and \met triggers are used to seed them.
The second tau LLP trigger 
runs \ac{LRT} in an \roi around the calorimeter tau seed and uses the same RNN-based tau identification as trained for the previous trigger.
Events are selected if they contain at least one tau passing the displaced tau identification.
The trigger is expected to have a \pt requirement slightly lower than the lowest unprescaled single tau trigger, 160\,\GeV.
Figure~\ref{fig:utt:lrt_tau_sig} shows the expected performance of these triggers compared to the prompt tau trigger based on an MC simulation of 100\,\GeV, 1\,ns staus.
The efficiency is computed with respect to standard and large radius offline tau tracks that are truth-matched to signal tau decays.
Hence, multiple tracks from the same generated tau may be included in the computation.
New LLP and LLP LRT tau triggers significantly improve sensitivity to the displaced tau signals.
 
\begin{figure}[htbp]
\centering
\includegraphics[width=0.49\textwidth]{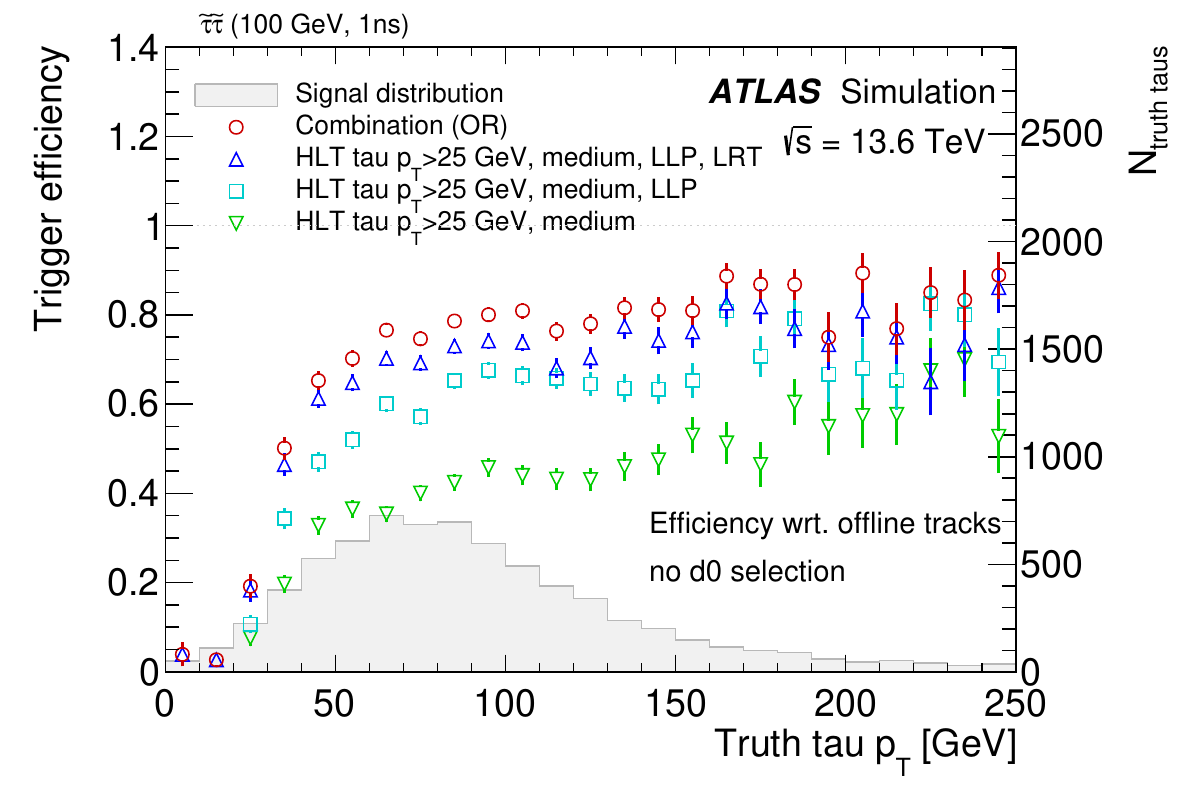}
\includegraphics[width=0.49\textwidth]{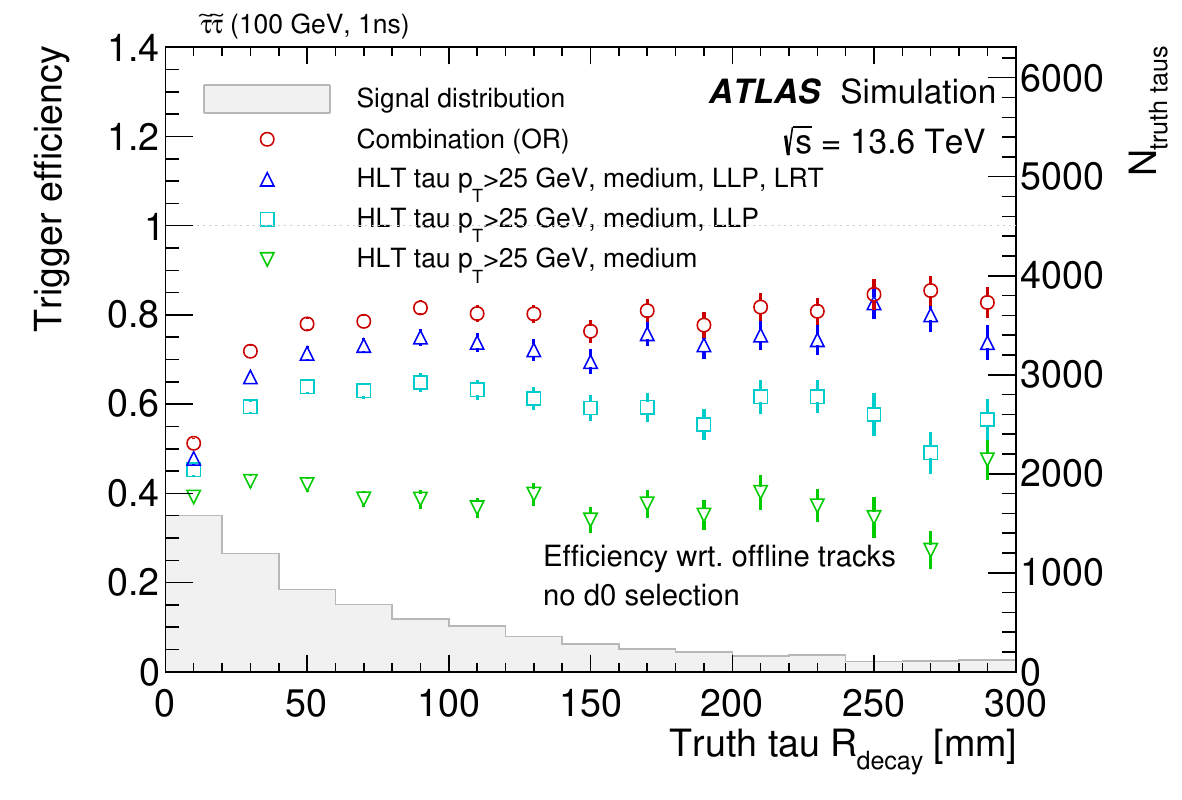}
\caption{Displaced tau trigger efficiency as a function of truth signal tau (left) \pt\ and (right) decay radius of the tau.
MC simulation with Run-3 conditions of staus with a mass of 100\,\GeV\ and a 1\,ns lifetime is used.
The efficiencies for the prompt tau trigger (inverted triangle), LLP tau trigger using standard tracking (square),
and LLP tau trigger using LRT (triangle) are overlaid by the logical OR of all three (circle).
The gray histogram shows the distribution of truth taus for this model.
Only statistical uncertainties are shown.
}
\label{fig:utt:lrt_tau_sig}
\end{figure}


\section{Auxiliary triggers}
\label{sec:special}

\subsection{Non-collision background}
\label{sec:ncb}
 
Non-collision backgrounds comprise detector signals which do not originate from the collisions of paired bunches at the ATLAS interaction point. They are categorised into beam-induced and cosmic-induced backgrounds.
Both beam-induced and cosmic-induced backgrounds are non-negligible background sources in searches for new phenomena targeting delayed or displaced detector signatures, like those discussed in Section~\ref{sec:unconTrack}. A number of dedicated triggers are used to select non-collision backgrounds for study.
 
\subsubsection{Cosmic ray triggers}
\label{sec:cosmic}
 
The cosmic-induced background results from the impact of extremely energetic cosmic muons on the detector,
which induce signals in the muon spectrometers or energy deposits in the calorimeters. 
These events are collected either during LHC data taking (with L1 triggers in the empty bunch crossings which do not have \pp collisions)
or during periods with no LHC data taking (e.g. for detector commissioning as described in Section~\ref{sec:L1Muon}).
The cosmic-muon triggers during data taking do not run any muon reconstruction at the HLT and they only run for dedicated cosmic data taking when there is no beam. Further details on studies of the ATLAS detector performance with cosmic-ray muons can be found in Ref.~\cite{PERF-2010-02}.
 
The \trig{IDCosmic} stream is designed to have a total rate of about 20\,Hz.
It consists of two single L1 muon triggers with \pt thresholds of 3\,\gev\ and 8\,\gev\ at approximate rates of 8\,Hz and 5\,Hz respectively, the latter being unprescaled.
In the same stream, a dedicated TRT-based L1 trigger for cosmics data taking uses a fast read-out path for groups
of channels in the detector~\cite{ATL-INDET-PUB-2009-002}, running at a typical rate of around 10\,Hz.
 
The \trig{CosmicCalo} stream aims for a target rate of 5 Hz and includes legacy L1Calo EM triggers with $\et >3\,$\gev and 7\,\gev,
a tau lepton trigger with \et$>8\,$\gev\ 
and jet triggers with \et$>12\,$\gev\ and 30\,\gev\ 
for $|\eta| < 3.1$ and \et$>30\,$\gev\ 
for $3.1 < |\eta| < 4.9$. 
 
\subsubsection{Beam-induced background triggers}
\label{sec:bib}
 
Beam-induced background (BIB)~\cite{DAPR-2014-01} originates from (1) the inelastic interactions of protons with residual gas molecules upstream and nearby the detector, producing showers of secondary particles (beam-gas background), (2) protons with high transverse amplitude, or (3) from protons deflected in beam-gas scattering hitting the tertiary collimators resulting
in background/secondary particles entering the detector (beam halo).
The online monitoring of BIB is essential to track live information about the beam conditions, which is also provided to the LHC. 
Further studies of BIB are essential to understand its origin and composition, and maintain and develop adequate monitoring.
 
Two single jet triggers with a prescaled L1 threshold of 12\,\gev\ and an unprescaled L1 threshold of 50\,\gev\ are used to record BIB events in bunch crossings where a proton bunch is present in only one or in neither of the beams.
In addition, triggers based on hits in the Beam Conditions Monitor (BCM)~\cite{BCM} are used to record BIB events.
The BCM consists of two stations of detectors located symmetrically around the interaction point at $z = \pm$184\,cm and $r=55\,$mm ($\eta\approx 4.2$).
Each station has four modules of two diamond sensors read out in parallel.
The implementation of the BCM hit-based trigger is based on a coincidence of an early hit on one side (A or C side) and one hit in-time with the bunch crossing on the other side (C or A side)
in unpaired bunch crossings, where only one of the two beams is filled with a proton bunch. 
In order to be more independent of the presence of unpaired bunch crossings and to improve the purity of the selected events, a new logic is implemented
for \runiii which relies on two early hits on the same side (2A or 2C). 
This allows for the triggering on paired bunch crossings while maintaining a similar rate to the AC/CA counterparts,
especially when triggering on the first colliding pair of bunches in a train which helps to mitigate against the impact of afterglow (increased cavern backgrounds following \pp collisions at high $\mu$).
Furthermore, this allows for a measurement of the composition and fraction of BIB in paired bunch crossings, which was found to be different compared to unpaired bunch crossings.
The contribution of ghost collisions\footnote{Ghost collisions occur between protons from a filled bunch, which typically
has $>10^{11}$ protons, and protons in an unfilled (empty) bunch, which has $<10^8$ protons, due to diffusion from filled bunches which takes place at the interaction point.}
is eliminated and thus the purity of the recorded BIB events is increased.
A new bunch group with first paired bunch crossings in a train, containing all paired bunch crossings following a gap of at least 29 empty bunch crossings is defined.
In order to evaluate the performance of the new triggers, they are defined in different flavours, triggering on empty, unpaired, and the aforementioned first paired bunch crossings in a train. Figure \ref{fig:ncb} shows the L1 rate of the old and new 
BCM triggers as a function of a bunch crossing identification (BCID) value.
For each trigger the per-BCID rate is shown for five different bunch groups, paired colliding bunches,
the first colliding bunches in a train, empty bunch crossings, unpaired bunch crossings with either beam 1 or beam 2 filled with a proton bunch.
As expected the rate of the 2A/2C variants is much lower than the AC/CA variants in BCIDs with paired colliding bunches. Further, the rate in unpaired BCIDs is similar for both variations.
 
\begin{figure}[htbp]
\centering
\includegraphics[width=0.49\textwidth]{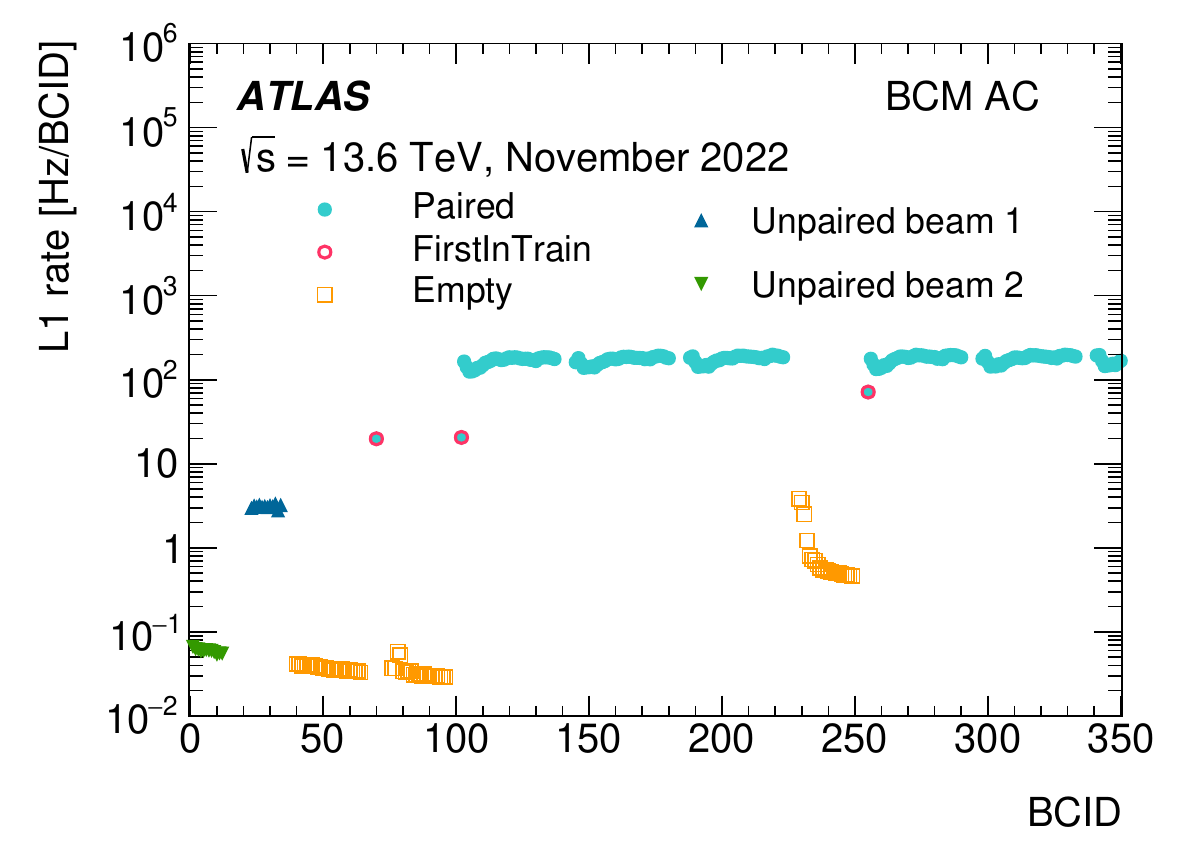}
\includegraphics[width=0.49\textwidth]{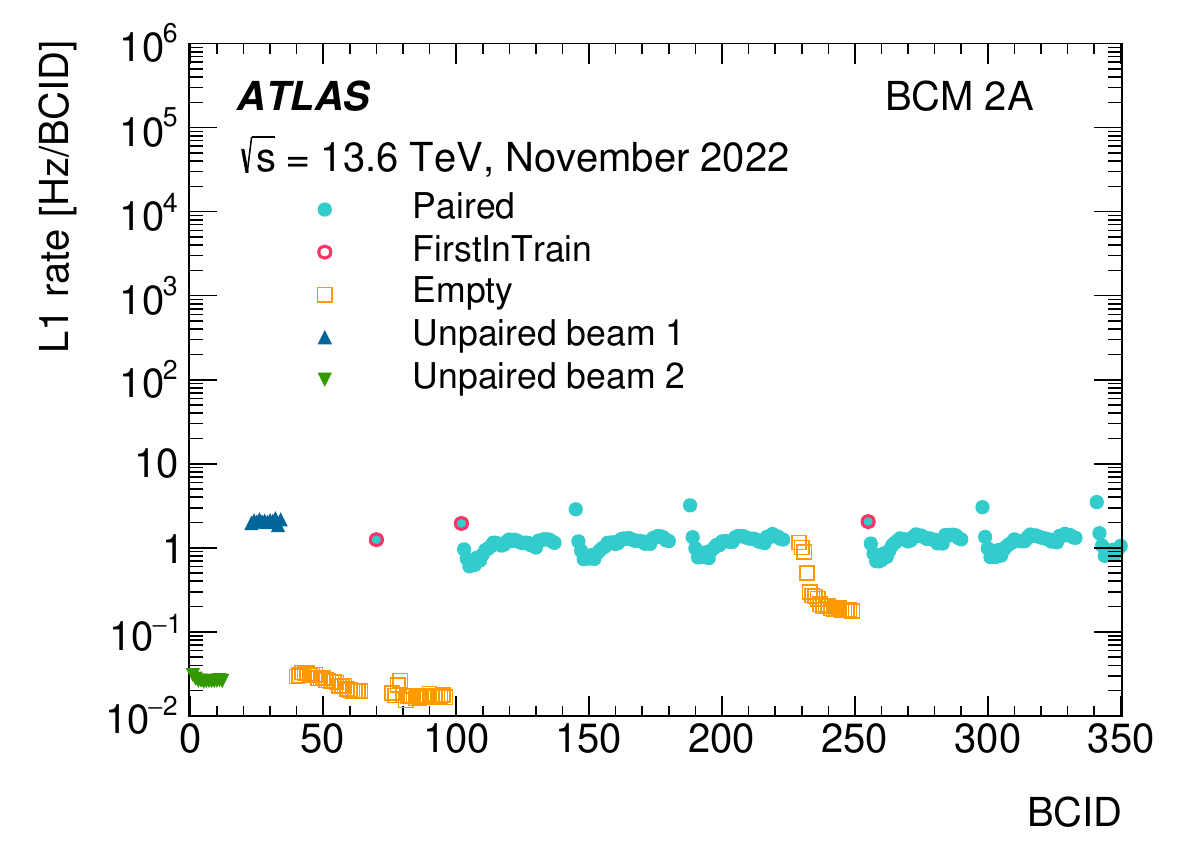}\\
\includegraphics[width=0.49\textwidth]{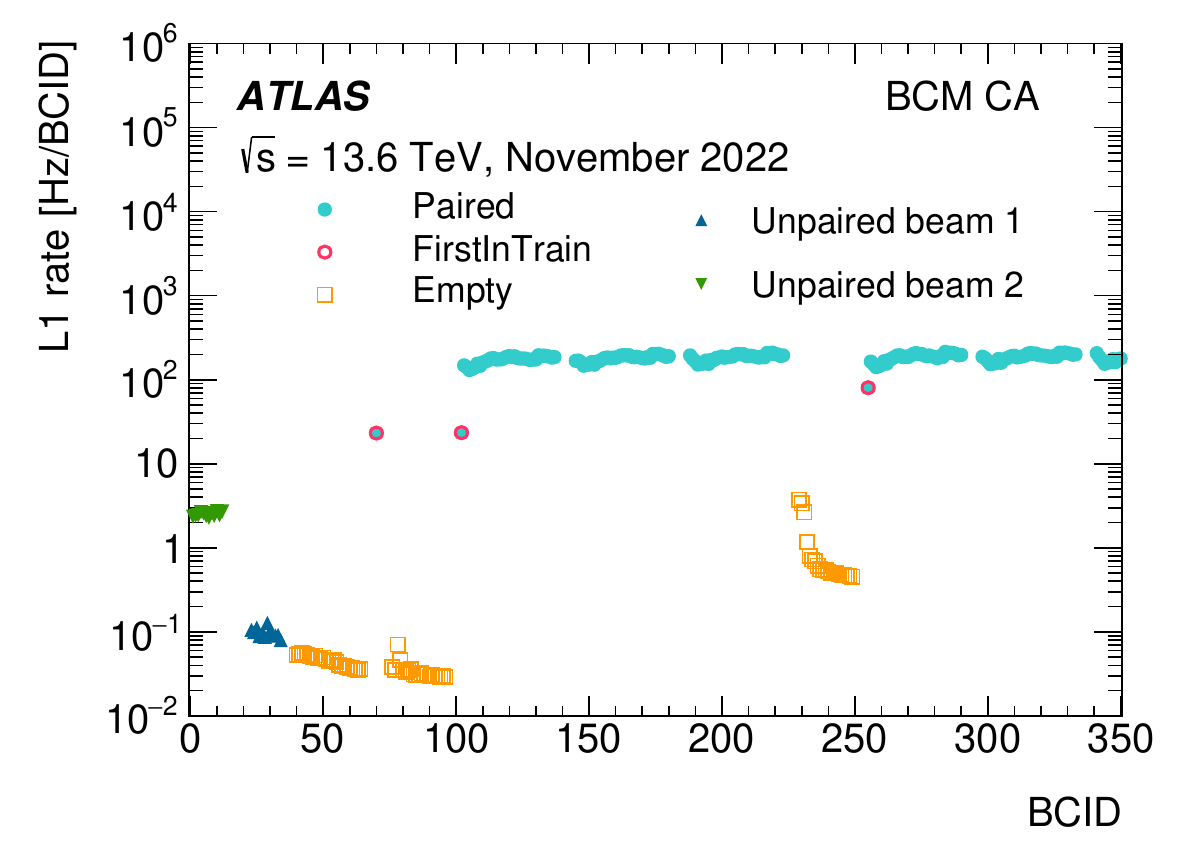}
\includegraphics[width=0.49\textwidth]{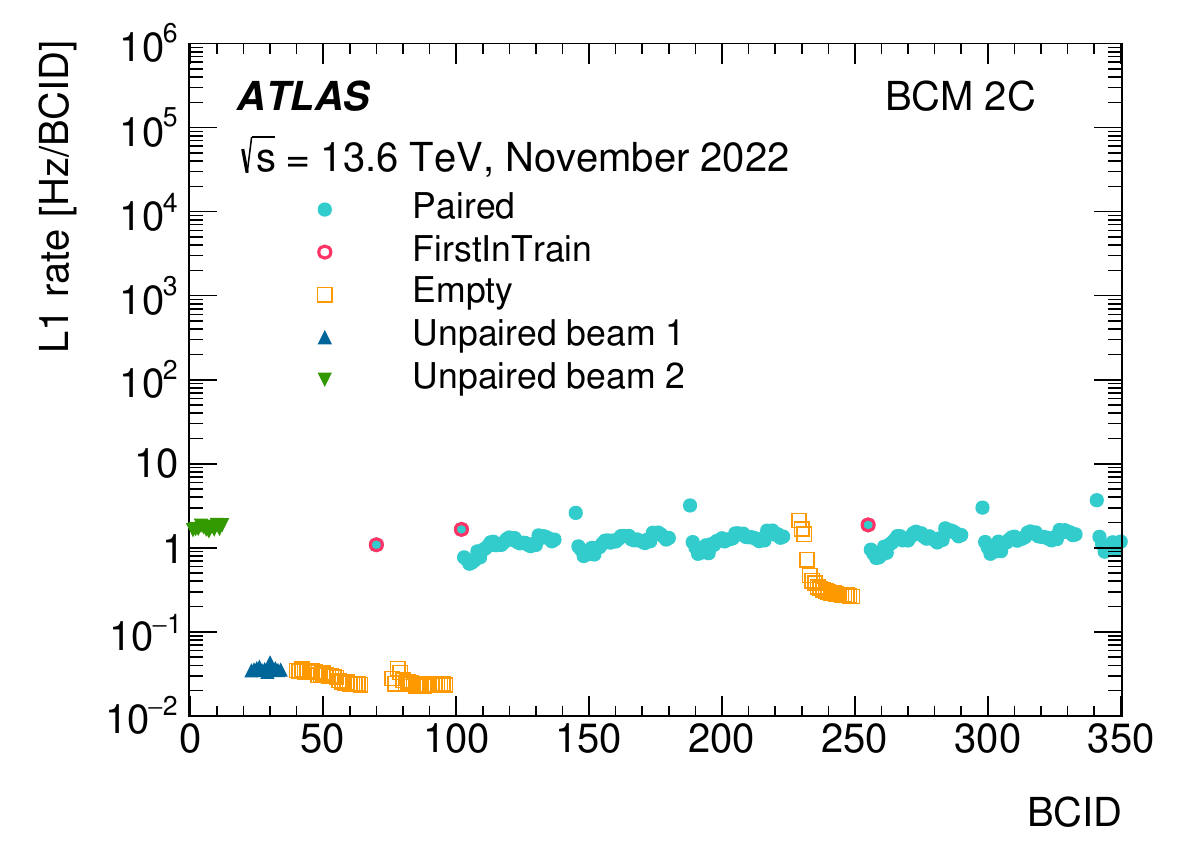}
\caption{
Per-BCID L1 rate of BCM hit-based triggers based on coincidence of (top left) AC 
and (top right) 2A, 
and (bottom left) CA 
and (bottom right) 2C. 
The per-BCID rates are shown for paired colliding bunches, the first colliding bunches in a train, empty bunch crossings (orange), unpaired bunch crossings with either beam 1 or beam 2 filled with a proton bunch. The BCID range shown is restricted to the first 350 BCIDs.
}
\label{fig:ncb}
\end{figure}
 
\subsection{Zero-bias trigger}
\label{sec:zb}
 
The zero-bias data are collected using a dedicated trigger, which fires one LHC turn after a L1 EM trigger with $\et>15\,$\gev\ fires.
This approach allows to collect data which is unbiased with respect to the activity in the event and at the same time proportional to the
luminosity in each bunch crossing, which can not be achieved with random triggers.
Such triggers are used for detector studies as well as for
the MC simulation overlay method~\cite{Haas_2017}, which relies on
the zero-bias data to account for the pile-up background, cavern background, and detector noise.
The zero-bias trigger, prescaled to a constant rate of about 10\,Hz, was based on the legacy L1Calo EM trigger in 2022.
A new Run-3 zero-bias algorithm is implemented in the new L1Topo system to be used following decommissioning of the L1Calo legacy system,
because the data connections between the new L1Calo system and CTP prevent the use of the legacy zero-bias trigger.


\subsection{Triggers for tracking performance studies}
\label{sec:idperf}
 
There are dedicated triggers in the trigger menu to study lepton tracking performance, described in Section~\ref{sec:id}.
These \trig{idperf} chains run the same selection as the corresponding triggers under study but do not apply any requirements on tracking.
This means that they are unbiased with respect to tracking and can be used for efficiency measurements.
For instance, if an electron candidate is formed from a track and a cluster, the idperf electron chain does not make any selection on the electron candidate, as this would include requirements on
track-cluster matching to define the candidate, as described in Section~\ref{sec:egamma}.
 
Two types of \trig{idperf} triggers are employed in the trigger menu: di-lepton triggers and single lepton triggers.
The di-lepton \trig{idperf} triggers target $Z$ boson or $J/\psi$ decays to leptons in order to study the performance with the tag-and-probe method.
These chains are based on primary single or di-object electron and muon triggers, described in Sections~\ref{sec:egamma} and \ref{sec:muons},
and can run unprescaled at about 15-30\,Hz not adding any unique rate.
These \trig{idperf} triggers do not apply any requirements on tracking on the subleading trigger leg which can be an electron, muon or tau candidate.
The single object \trig{idperf} triggers have thresholds of 5, 14, 20, 26 and 30\,\gev\ for electrons, 6, 20, 24 and 40\,\gev\ for muons
and 25, 35, 80 and 160\,\gev for taus. These are heavily prescaled to record events at 0.5 -- 1\,Hz.
 
There is at most one reference object (one track) per \roi from the leptonic triggers. The purity of the muon triggers
is quite high. For the tau trigger, it is very low, and for the electron trigger, lower still.
There are very few real electrons in the sample accumulated by single electron \trig{idperf} triggers,
so the dilepton \trig{idperf} triggers are crucial to evaluate electron tracking performance.
 
\subsection{Triggers for detector performance studies (calibration, noise)}
\label{sec:dettrig}
 
Triggers used for detector performance studies often run at very high rates.
To compensate for this, only partial detector information is recorded through a strategy called Partial
Event Building (PEB), which has the potential to significantly reduce the event size and
thus overall bandwidth. These triggers operate either with a fixed subset of ROBs which are
to be included in events recorded to the PEB stream, or a list of ROBs can be derived dynamically on an event-by-event basis based on a set of \rois.
A combination of both static and dynamic ROB lists is also possible.
These triggers are used for the calibration of muon subdetectors, low-\pt offline muon
calibration and LAr calorimeter performance studies.
 
For example, a dedicated calibration trigger, called \emph{a laser trigger}, is active during collisions runs to monitor the stability of the Tile calorimeter channels.
The LHC abort gap of about 3\,$\mu$s is used to send and register laser pulses that are arbitrated,
timed and controlled with respect to the LHC signals~\cite{laser}.
Laser pulses are sent to the Tile calorimeter at the increased rate of 12\,Hz for \runiii and recorded for an analysis.
Laser events are used to track possible fast gain changes in the photomultiplier tubes and monitor the timing calibration.
 
Random events with PEB information for Pixel and SCT detectors are recorded in the unfilled LHC bunches into dedicated streams to identify noisy channels which need to be masked\footnote{Not used for the track reconstruction.} in these subdetectors.
These triggers typically run at about 10\,Hz. A pixel with noise hit occupancy above $5\times 10^{-4}$ is classified as a noisy channel and masked in DAQ.

\subsubsection{LArNoiseBurst algorithm}
\label{sec:larnoiseburst}
 
The LAr calorimeter has a small but luminosity-dependent probability of generating large noise signals
involving a considerable number of cells, as discussed in detail in Ref.~\cite{LARG-2013-01}.
While this happens with milli-Hertz frequency,
high spurious values of energy are provided by these cells and it is therefore important
to veto affected events from the data quality monitoring and downstream of this from potentially affected physics analyses.
Veto windows occur across a much shorter timescale than that of a single luminosity block. 
The detection and registration of LAr noise bursts is done via a LArNoiseBurst algorithm which runs in the HLT.
 
Noise bursts generate electrical pulses which do not correspond to physics objects and thus these have a bad quality of fit when energy reconstruction is attempted.
Offline tools examine the quality factor of the cell energy reconstruction procedure, counting the number of cells with a bad quality factor in a given event.
Given that many other factors could interfere with the cell quality factor (e.g. intense pile-up in a particular event),
a minimum of two events that are strongly correlated in time is required to declare a veto interval.
 
In the HLT, the offline algorithm is run for every accepted event that is written to certain streams, including the \textit{Main} stream, the TLA, delayed streams and streams recording events with cosmic signatures in the calorimeter.
Events that are considered bad by the offline tool have their time stamp published online via an information
messaging system. A dedicated application then picks the time
stamps, performs a coincidence operation between them (a maximum time difference of 250 $\mu$s is
required) and declares the interval to a database used by the offline reconstruction.
The veto is then applied to reject events in the post-run analysis.
The computing resources taken by this algorithm are negligible due to very low rate of the Noise Bursts and large overlap with other calorimeter triggers.



\subsection{Beam-spot algorithms}
\label{sec:beamspot}
 
A dedicated system allows the HLT to continuously measure the position, size and orientation of the luminous region (also known as beam spot) at the ATLAS interaction point~\cite{Bartoldus:2011tma,Winklmeier:2012nta}.
Beam-spot parameters can change from fill to fill, and some, in particular the transverse position and size, show significant variation over the course of a fill as well.
Knowledge of the current beam-spot parameters is crucial for several HLT algorithms, most notably for the selection of events with $b$-jets, and for HLT tracking itself.
The parameters are also transmitted online to the LHC status display~\cite{LHC:LHCCONFIG}.
 
The beam-spot parameters are continuously monitored and archived, determined bunch-by-bunch as well as fill average, and sampled over different intervals from one to many minutes depending on the required statistics.
One dedicated set that is used by the HLT is only updated whenever significant deviations from the currently used values are detected.
The process of obtaining the parameters, updating them and feeding them to the algorithms as condition parameters is referred to as beam-spot calibration.
 
The calibration and update process is not trivial as it consists of multiple steps involving several cooperating sub-systems to synchronise the beam-spot parameters across the distributed HLT farm at the same time. This requires orchestrating a series of steps, primarily through the CTP:
 
\begin{itemize}
\item HLT algorithms extract tracking and vertexing information and publish their distributions in the form of histograms,
\item histograms from all individual HLT instances are aggregated by a monitoring infrastructure referred to as the Gatherer,
\item on each new luminosity block the merged histograms are processed and new estimates of the beam-spot parameters are calculated by an external application called the BeamspotTool,
\item when the BeamspotTool determines that the new estimate constitutes a significant change with respect to the previous one, it sends these new parameters to the \ac{CTP} process,
\item the \ac{CTP} process writes new beam-spot parameters to the conditions database with a validity interval starting with the next luminosity block and then notifies the HLT processes of the pending update via its event fragment,
\item individual HLT processes read the new beam-spot parameters from the conditions database when they receive events from the next luminosity block, for which
the CoralServer and CoralProxy~\cite{CoralServerOrProxyRef} infrastructure provides scalable access to the conditions database from the HLT farm.
\end{itemize}
 
The online beam-spot calibration received significant improvements in preparation for \runiii.
 
One long-standing issue with the online beam-spot calibration during \runii was a small (5 -- 15\,$\mu m$) systematic difference observed in the transverse beam position with respect to the offline calibration.
During LS2 this was tracked down to an incorrect transverse position at the coordinate system origin being used for track clustering, which caused a systematic bias towards the origin.
The issue was resolved by using the current estimate of the transverse beam location for track clustering.
Reprocessing of the Run-2 data with the fixed vertex finder showed the online calibration matching the offline measurement with excellent precision.
In 2022, there was still an observable difference in transverse position between offline and online due to different alignment constants, which makes a direct comparison difficult.

The original calibration used for \runi and \runii is based on optimised vertex finding and fitting algorithms.
The Run-3 calibration introduces an additional algorithm based only on track information that does not involve vertexing.
The luminous region is narrower than the typical vertexing uncertainty, and an accurate estimation of the transverse widths requires precise knowledge of the vertexing resolution.
The resolution itself depends on detector and trigger conditions, and is therefore evaluated in real time through a split-vertex method that was introduced in \runi.
However, the smaller beam-spot sizes of Run 2 required larger resolution corrections and the vertex method has shown limitations in case of low statistics. To study alternatives, a new method was developed for Run 3.
This method utilises information from reconstructed tracks only and determines beam-spot parameters by a likelihood fit to the observed set of $d_0$ and $\phi$ parameters of tracks.
In the ATLAS online environment the tracking information is local to each of the HLT processing nodes, and it is not possible to collect all track data at one location to perform fitting.
To support fitting of all available tracking data at a single location, the likelihood function is approximated to a sufficiently small set of additive terms.
These terms are calculated from the local set of tracks on each HLT node and merged for the final fit by utilizing the Gatherer infrastructure.
The new method demonstrated reasonable performance with the reprocessed Run-2 data and a shorter ramp-up period between the start of data collection and calibration availability compared to the vertex-based method.
For \runiii both vertex-based and track-based algorithms are in use to further study their performance.
During the initial 2022 data-taking period the track-based algorithm was used preferentially due to its better robustness, lower demand on statistics and earlier availability of the beam spot. For high-luminosity running the vertex algorithm is still the preferred method for its accuracy over a wider range in beam sizes.

The multi-step calibration process outlined above introduces a delay in availability of the calibrated beam-spot position for HLT algorithms, with a typical latency of several luminosity blocks.
This is not an issue for the HLT during a run since the variations are small on that timescale.
However, at the start of each new fill this requires a short bootstrap period, during which certain HLT algorithms, such as those used for $b$-tagging, have to be held off until a first measurement of the beam spot has succeeded.
This results in a loss of a few luminosity blocks of data for the individual triggers: on average 4.5 luminosity blocks were missing beam-spot calibration at the start of each data-taking run in \runii, which corresponds to approximately one percent of a duration of a typical run.
 
Improvements in the Gatherer infrastructure
helped to reduce the calibration bootstrap delay.
The calibration depends on histograms produced by the HLT and merged by the Gatherer.
Improvements in handling of histograms for each of the luminosity blocks in the Gatherer have reduced propagation delays for those types of histograms.
This results in a faster availability of those histograms for the BeamspotTool application and helps in shortening the inherent ramp-up delay, which has reduced to an average of 2.5 luminosity blocks in \runiii.
 
With the higher bunch intensities achieved by the injector upgrades, the LHC had for the first time begun $\beta^*$-levelling in order to limit pile-up in 2022.
This involves a large variation in spot sizes, in turn demanding more frequent updates of the HLT during the levelling phase between $\beta^*$ of 60 cm to 30 cm.
In 2023 the levelling range is extended to be between 120$\,$cm and 30\,cm.
 
Another development begun in 2022 that can be applied in 2023 is to constrain HLT tracking to a 3$\sigma$ window around the longitudinal beam position which is projected to save a significant fraction of CPU time in the HLT.
This will involve yet another bootstrap procedure to settle on the beam spot at the start of each run.
 
Figure~\ref{fig:beamspot} shows the time evolution of the vertical beam-spot position as measured, and subsequently applied, by the HLT over the course of
one LHC fill during November 2022 data taking.
An update of all parameters is performed whenever the position changes
by more than 10\% of the width, or the width changes by more than 10\%, or
the uncertainty on any of the parameters decreases by more than half.
 
\begin{figure}[h]
\centering
\includegraphics[width=.49\textwidth]{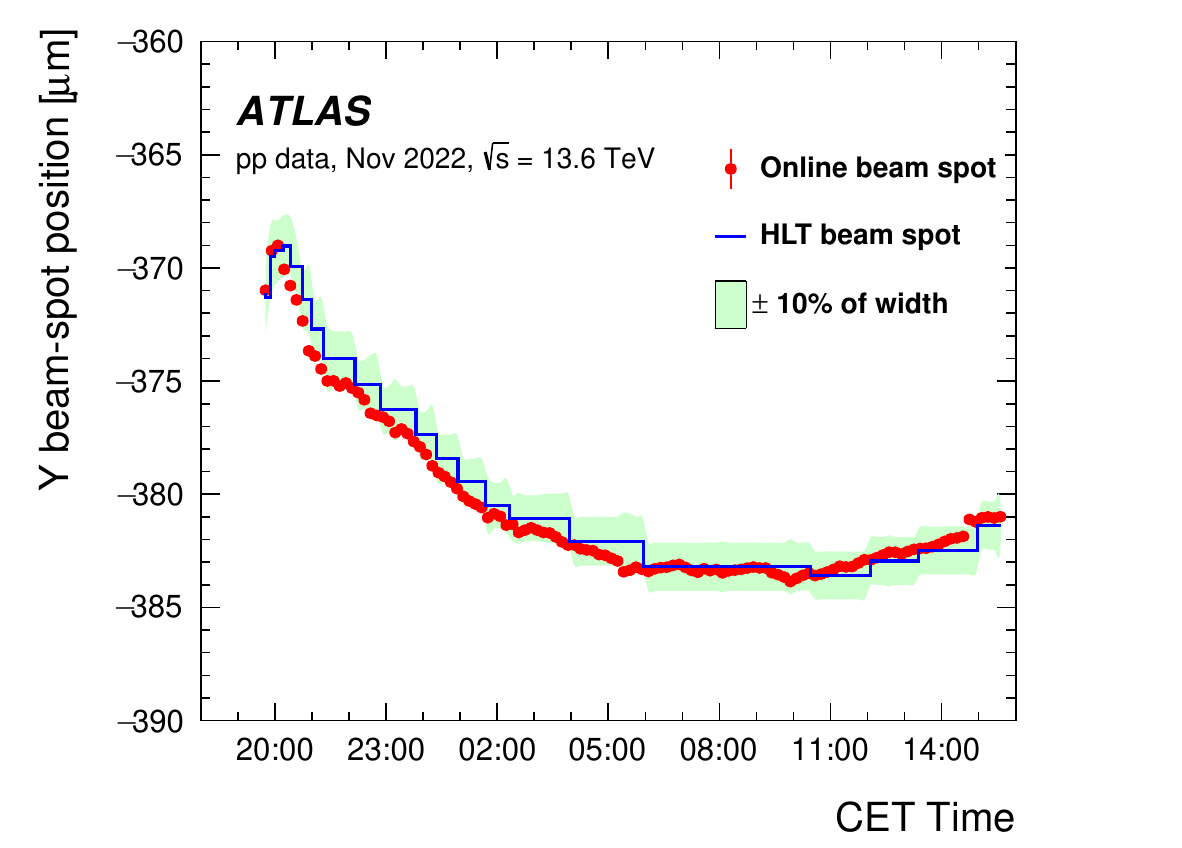}
\caption{
Time evolution of the vertical beam-spot position over the course of one LHC fill.
The dots represent the monitored beam-spot position measured in luminosity blocks. The line represents the position currently
used by the HLT algorithms which is measured as an average over a
sliding window of a series of luminosity blocks. The band indicates a range corresponding
to 10\% of the vertical beam-spot width that was measured at the same time.}
\label{fig:beamspot}
\end{figure}


\section{Trigger software performance}
\label{sec:softwarePerf}

 
\subsection{Trigger software performance scaling}
\label{sec:swperfscaling}
 
For Run 3 the same process forking architecture of the HLTMPPU was kept as in Run 2. However, due to the multithreaded event selection, each worker process can now contain multiple event slots, which are processed in parallel. In Run 2, each worker process handled one event, and the selection progressed sequentially. This allowed the HLTSV to assign new events directly to idle worker processes. In Run 3, the AthenaMT scheduler decides when an event slot is freed and when a new event can be read from the input source. The HLTSV, therefore, assigns events to an event queue for each HLTMPPU from which new events are pulled by the AthenaMT scheduler for each worker process.
 
In the Run-3 system, the number of forked worker processes per HLTMPPU and the number of parallel event slots in each worker process are freely configurable. This allows for an adiabatic transition of a pure multi-process-based event selection like in Run 2 (i.e. many worker processes with one event slot each) to a pure multithreaded configuration with one worker process and many parallel event slots. Event throughput measurements are used to decide what configuration will be used.
 
Figure~\ref{fig:swperfscaling} shows the trigger software performance scaling in terms of application throughput in events/s (left), CPU usage (middle), and memory usage (right) as a function of the number of events processed in parallel for AthenaMT executing trigger selection algorithms.
The measurements were performed in a standalone local environment using a machine identical to those used in the ATLAS HLT computing farm during data taking. It is a dual processor machine with 128 GB RAM using a NUMA memory architecture and two AMD EPYC 7302 CPUs, where each CPU has 16 real cores with two hyper-threads per core, giving the total number of 64 threads.
The data sample contains a mix of events representative of the real HLT input data and trigger selection configuration identical to one used during data taking. Four ways of achieving the parallelism are presented.
Data were taken in 2022 using a pure multi-process configuration with 48 forks as event throughput is the most critical metric for the HLT. Different constraints apply, however, to Grid~\cite{grid} processing of MC simulated events. Here, available memory per core is significantly more restricted than on the dedicated HLT machines, and these tasks execute in \runiii in a pure multi-threaded configuration, the trigger simulation included. A typical Grid site in 2022 ran the multi-threaded configuration with eight event slots. As ATLAS' transition to multi-threading was new for \runiii, a number of components still make use of mutex locking to provide safe shared access to certain common resources. In aggregate, these bottlenecks severely limit event throughput at very high levels of multi-threaded execution, as seen in Figure~\ref{fig:swperfscaling}. Work continues to refactor these components to minimise blocking behaviour in future software releases for both Run 3 and Run 4. In addition, the hybrid modes of operation continue to be studied as significant memory savings are made even in hybrid modes which use a small number of multi-threading event slots, hence minimising losses due to resource contention.
 
\begin{figure}[tbp]
\centering
\includegraphics[width=0.32\textwidth]{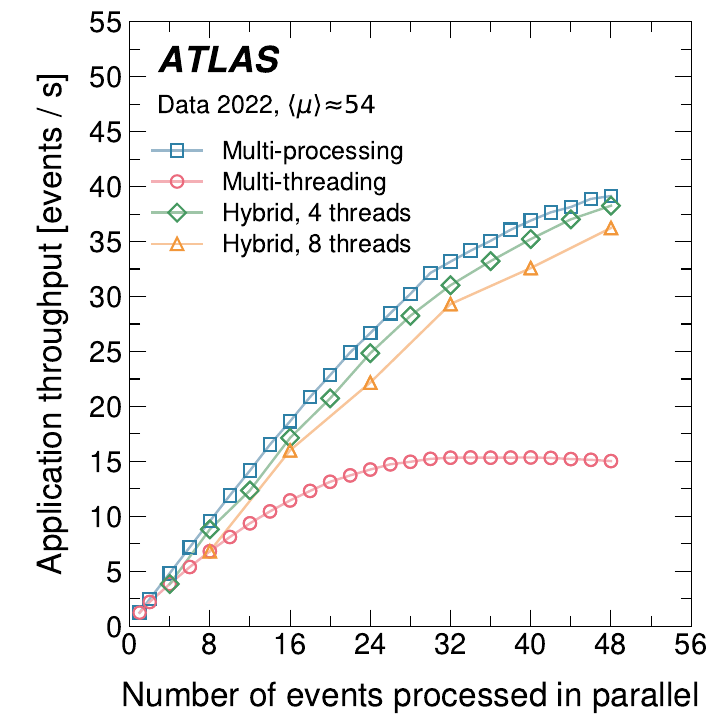}
\includegraphics[width=0.32\textwidth]{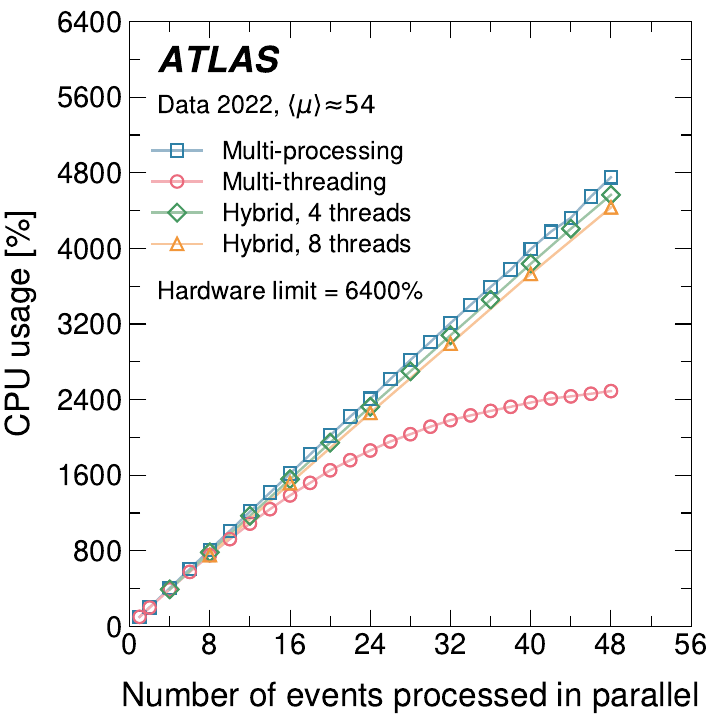}
\includegraphics[width=0.32\textwidth]{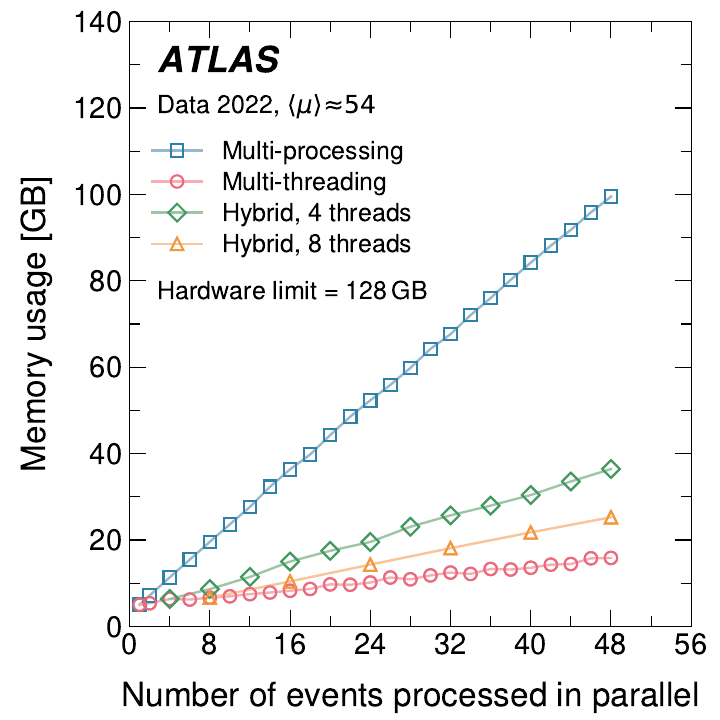}
\caption{(left) Application throughput in events/s, (middle) CPU usage (CPU time divided by wall time) in percent and (right) memory usage in GB as a function of the number of events processed in parallel for AthenaMT executing trigger selection algorithms.  Blue squares represent a multi-processing approach.
Pink circles represent a multi-threading approach with a single process using a number of threads equal to the number of events requested to process in parallel.
Threads are not bound to events; instead, a pool of a number of threads ($N_\text{thread}$) is used to process an equivalent sized pool of events ($N_\text{event}=N_\text{thread}$). Green diamonds and orange triangles represent a hybrid approach where a number of processes, $N_\text{process} = N_\text{event} / N_\text{thread}$, forked after initialisation use a fixed number of threads each ($N_\text{thread} = 4$ and 8, respectively) to process an equivalent number of events in parallel. Differences between these approaches are discussed in Section~\ref{sec:AthenaMT}.}
\label{fig:swperfscaling}
\end{figure}
 
Figure~\ref{fig:calo:app} shows the calorimeter processing time dependency on the number of threads obtained
offline by concurrently processing various numbers of data events with different thread configurations.
In the Run-3 AthenaMT framework, the same block of cells cannot be requested by two different algorithms
(e.g. two overlapping \rois or a full scan request at the same time as any \roi) as it may cause simultaneous reading and writing to a given memory area.
To avoid this, locks are added to the HLTCalo data preparation service, leading to a non-linear dependency between the processing time
and the number of parallel processing threads allocated for intra-event processing
as shown in Figure~\ref{fig:calo:app}. Unpacking of the calorimeter cells and topological clustering algorithms do not show any dependence on the number of
threads, contrary to fast reconstruction, which exhibits turn-on-like dependence on the number of threads
the shape of which depends on the number of concurrently processed events.
Optimisations are planned to either employ a more fine-grained locking or through an updated processing model.
The time of the full scan calorimeter cell unpacking per call is slightly extended (mostly due to locks associated to \rois) with the
increased number of threads, but more linear scaling performance is achieved 
through increasing the number of inter-event parallel processing slots. 
This does not affect the HLTCalo algorithm's functional performance, such as cell and cluster parameters or reconstruction efficiency.
 
\begin{figure}[tph]
\centering
\includegraphics[width=0.49\textwidth]{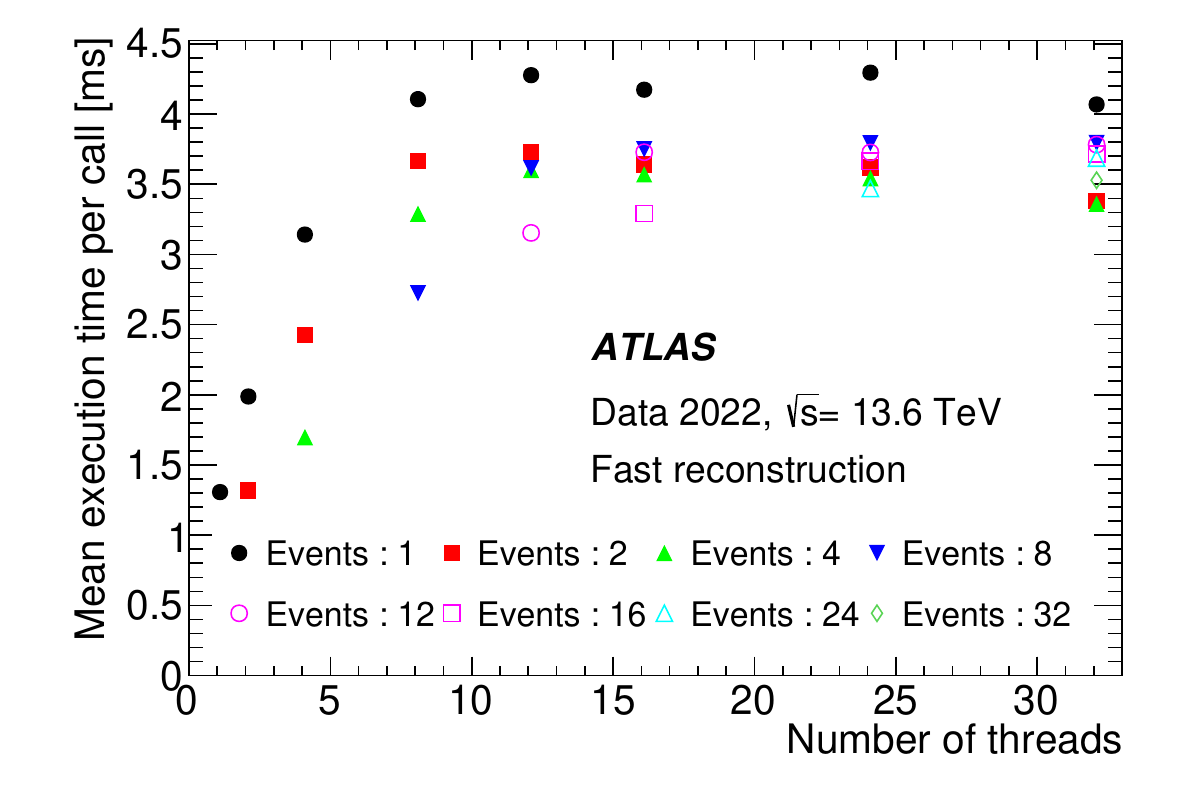}
\includegraphics[width=0.49\textwidth]{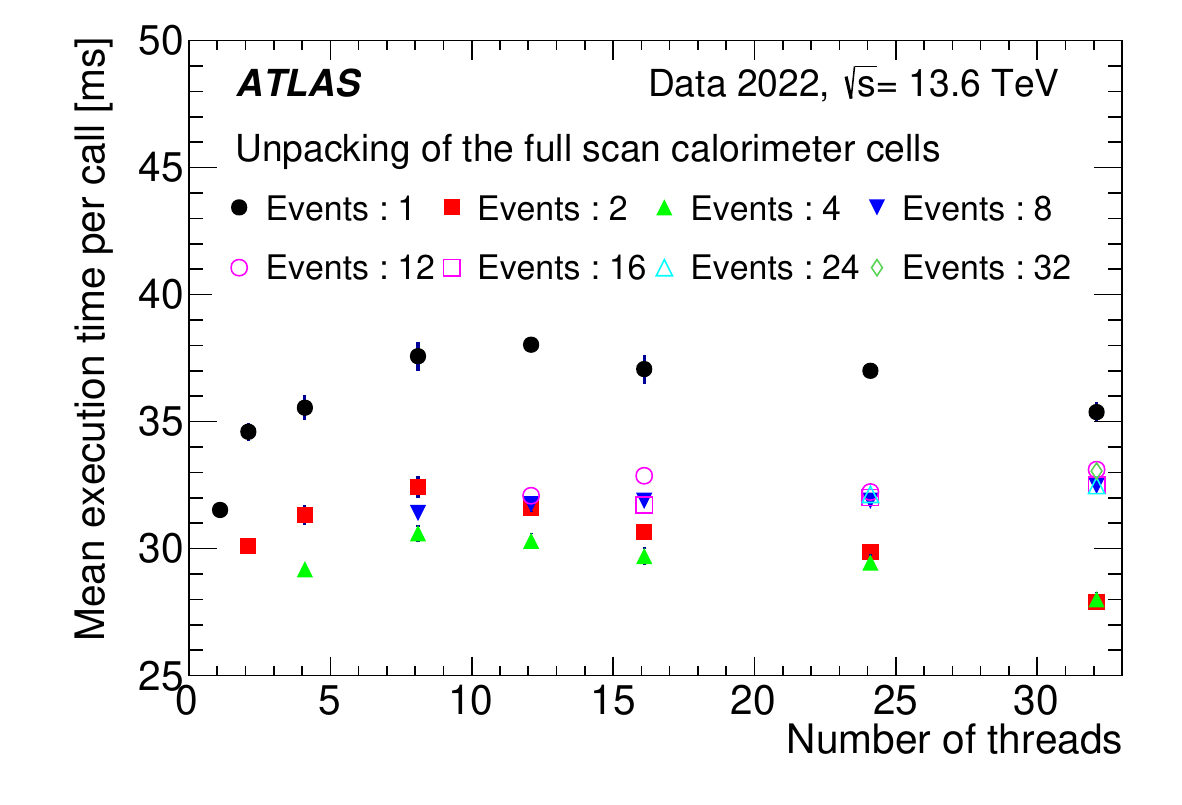}
\includegraphics[width=0.49\textwidth]{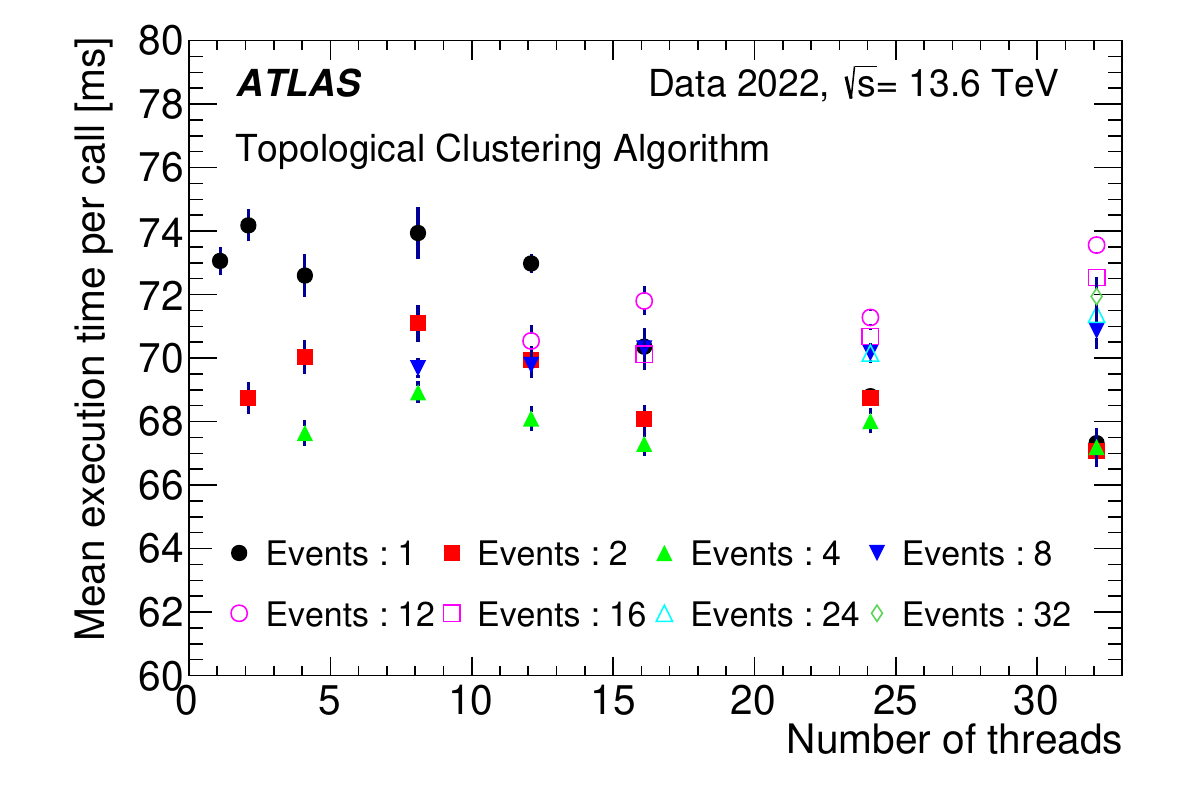}
\caption{Mean execution time per call of
(top left) the fast calorimeter reconstruction,
(top right) the unpacking of the full scan calorimeter cells and
(bottom) the full scan topo-clustering algorithm
as a function of the number of threads for various numbers of concurrently processed events.}
\label{fig:calo:app}
\end{figure}


\subsection{HLT processing time}
\label{sec:menuPerf}
 
The overall utilisation of the HLT is dependent on the mean time taken to process each event and the incoming event rate from L1.
The mean time to process an event is a product of the trigger menu, its associated luminosity-dependent prescale sets,
and of the collisions being supplied by the LHC. Here the processing time is most heavily dependent on the mean number of pile-up interactions.
An example of the mean HLT processing time per event at different
instantaneous luminosity values throughout a run is presented in Figure~\ref{fig:hltprocessing_time}\,(left).
The mean HLT processing time decreases with decreasing average pile-up due to a reduction in
event complexity.
The menu is adjusted based on the current rate of events in order to fully use available resources
while retaining a constant 5000 slot safety margin (about 6\% of the final 2022 farm)
to protect against losing slots due to power glitches, technical problems, etc.
The slope of the distribution in Figure~\ref{fig:hltprocessing_time}\,(left) becomes less steep at lower $\mu$ values where additional trigger selections are enabled.
Additional event processing restrictions came from the delayed delivery of replacement ROS servers.
In 2022 the ROS (like in \runii) was able to supply full detector information at half of the maximum L1 rate (50 kHz). However, it is possible to reach the L1 rate of 100 kHz for a subset of detectors by re-configuring the read-out, for example by increasing the number of ROS servers assigned to it.
 
In addition to the online monitoring of high-level HLT processing statistics, offline tools, such as Cost Monitoring~\cite{ATL-DAQ-PUB-2016-002}, are available to investigate the detailed performance of the
HLT (e.g. the average processing times of individual algorithms or whole trigger chains).
Data required for such studies is saved to a dedicated calibration stream which enables recording of this data for all events, not just those accepted. This stream contains information on algorithms' execution times and on any data requests they make to the ROSes.
By default, only the first 250 LBs are monitored in this way during physics data taking.
After the run has finished, the data from the stream undergo additional
post-processing and the performance details are available on a dedicated website.
 
The Cost Monitoring used during Runs~1 and 2 had to be adapted to the multi-threaded framework for \runiii. While the following figures and tables present data taken using the MP configuration in 2022, some changes and challenges which become more relevant in more MT-like configuration are pointed out in the following.
Upgraded monitoring includes a new MT-compatible algorithm gathering the data, a redesigned post-processing framework, and additional monitors (including thread monitoring).
Given that the new HLT runs under the AthenaMT scheduler, the analysis of the results is no longer as simple as in a sequential framework, therefore multiple metrics were prepared to provide an overview of the event processing time, for example, an event wall time or the sum of the execution times of all algorithms.
The wall time includes the framework operations/delays between scheduled algorithms.
A comparison between those values can be seen in Figure~\ref{fig:hltprocessing_time} (right), which
shows the wall time which includes both algorithm execution time and time spent on framework operations (including algorithms scheduling, data traffic) as well as the total time of just the algorithms. These data were recorded in a MP-mode, where the scheduler was only tasked with processing a single event at a time using a single CPU core.
 
In the figure, three peaks can be identified, representing fast (approximately 30 ms), medium (approximately 300 ms), and
slow (approximately 2s) events. The last type of the event is the rarest due to the early rejection mechanism.
For faster events, when only fast algorithms are executed and most of the trigger chains are rejected,
the impact of steering operations from the underlying framework is observable as an overhead of about 10 ms.
For events with execution times much longer than 100\,ms, when time-consuming algorithms, for example, tracking algorithms, are executed, it is negligible.
Table~\ref{tab:hltprocessing_time_grouped} shows an example of collected Cost Monitoring metrics for a given period of time during a run, broken down into the main types of object reconstruction and the framework.
The Cost Monitoring data are used in an iterative process to identify areas
where optimisation of code and strategy can yield the greatest impact with zero or negligible physics impact.
The difference between the wall time and the total time of all algorithms per event can further diverge in different ways when running in MT mode. For the MP case, the total event wall time is always greater than the algorithm time, but with sufficient intra-event parallelism in an MT configuration, the total event wall time can be smaller than the algorithm time, as some of the event processing may occur simultaneously on multiple cores.

\begin{figure}[htbp]
\centering
\includegraphics[width=0.47\textwidth]{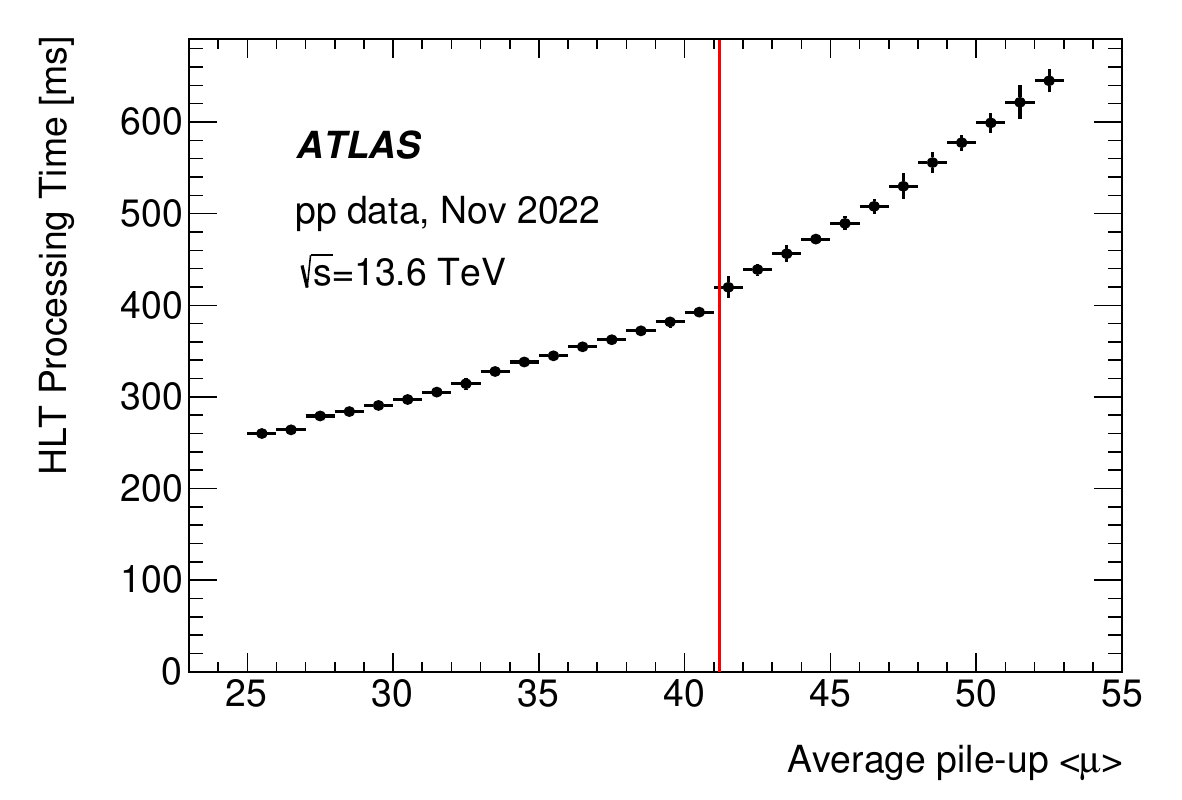}
\qquad
\includegraphics[width=0.47\textwidth]{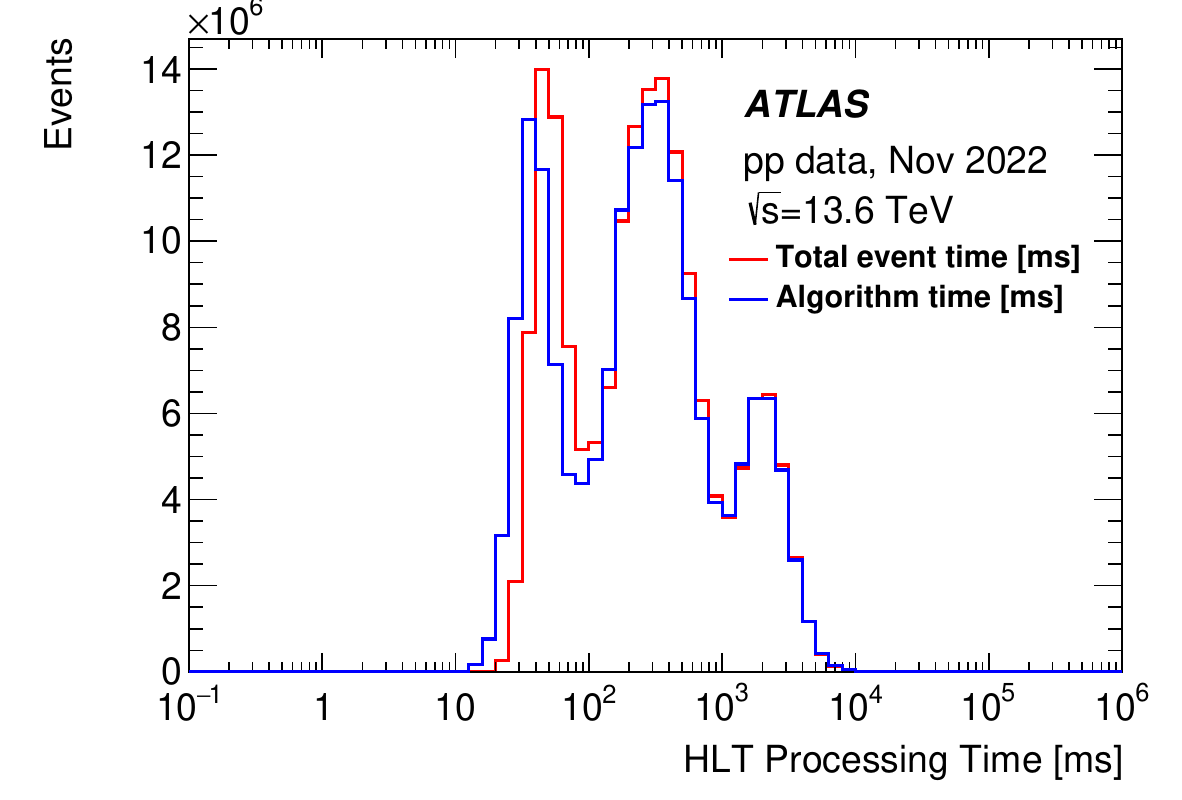}
\caption{(left) Mean HLT wall time as a function of the average pile-up throughout a run.
A vertical line marks the instantaneous luminosity at which additional trigger selections are enabled.
Error bars denote the Gaussian width of the underlying per event measurements.
(right) HLT processing time distribution per event for an instantaneous luminosity of \lumibenchmark at pile-up of 51.
Shown are the wall time spent per event as well as the sum of the algorithms only processing time. The difference between these two distributions is due to the time spent in framework operations.}
\label{fig:hltprocessing_time}
\end{figure}
 
\begin{table}[htp]
\begin{center}
\caption{
Example of collected Cost Monitoring metrics for 50 LBs for an instantaneous luminosity of \lumibenchmark at pile-up of 51,
showing how the total time is distributed between the main types of object reconstruction and the framework software.
The fractional time consumption of algorithms is calculated based on the sum of the execution times of all algorithms.
}
\label{tab:hltprocessing_time_grouped}
\begin{tabular}{lccc}
\toprule
& Total time [\%] \\
\midrule
ID reconstruction & 59 \\
Muon reconstruction & 14 \\
Calorimeter reconstruction & 11 \\
Combined reconstruction and hypothesis algorithms & 8 \\
Trigger infrastructure & 2 \\
Other & 6 \\
\bottomrule
\end{tabular}
\end{center}
\end{table}
 
At the peak instantaneous luminosity during the run, the HLT processing time of one event is approximately 600 ms (590 ms algorithm time), compared to 500 ms achieved during \runii.
Out of this number about 59\% of the total event time is spent on the ID tracking, despite the improvements discussed in Section~\ref{sec:id}.
This is due to the expanded use of the full scan tracking for the hadronic signatures in \runiii.
The HLT farm computing capacity has been increased for Run 3 by routine replacement of old servers, benefiting from the sustained industry trend of increasing processor performance. This has enabled the additional HLT processing for Run 3 that is described in this paper to support the physics goals of the experiment.
During \runii, the HLT farm consisted of processors with a performance of 22.8 HS06/core~\cite{Michelotto_2010} (0.8 MHS06 total farm performance). For \runiii, 60\% of the farm was upgraded to 36.2 HS06/core (1.7 MHS06 total farm performance) for early 2022, rising to 100\% by November 2022 (2.0 MHS06 available for the remainder of \runiii starting from 2023).


\section{Conclusion}
\label{sec:conclusion}

A large number of trigger upgrades and developments for the ATLAS experiment were made during the
second long shutdown of the LHC in preparation for the Run-3 data taking. A summary of the various
updates as well as the first Run-3 performance studies can be found in this paper.
 
Many changes in the L1 trigger system improve both its rejection of background events and
acceptance for interesting physics processes.
Upgrades in the L1 calorimeter trigger increased the granularity of information used by the trigger
and enable it to run more sophisticated algorithms to identify physics objects,
and to calculate missing transverse momentum in the event with higher precision.
The L1 muon system was enhanced through the addition of New Small Wheels, new resistive plate chambers
in the barrel/endcap transition region as well as upgraded electronics.
The topological trigger, installed during \runii, has also undergone a hardware upgrade.
Its refined kinematic measurements of muons and other objects as well as new flexibility to
define multiplicity triggers allow for more sophisticated event selections at L1.
 
The underlying framework of the HLT was completely rewritten in order to execute trigger algorithms
within the multi-threaded software framework AthenaMT, sharing common features between the trigger
software and the software used for offline reconstruction. These changes help to optimise the use of the
HLT computing resources, both in terms of computing power and memory consumption.
New-and-improved selection algorithms and strategies further improve the reconstruction of objects at the trigger level, in the case of some of the hadronic signatures, at the expense of higher CPU time needs due to the expanded use of full scan tracking.
 
While maintaining a level of consistency with the Run-2 trigger menu, the Run-3 trigger menu
sets out to exploit the newly implemented detector features, more performant HLT hardware, and algorithmic advancements.
Trigger thresholds at L1 and HLT were generally kept the same as during \runii, benefiting from improvements to reduce trigger rate.
The additional available HLT rate compared to \runii is dedicated to expanding the physics menu in the physics and TLA streams, lowering trigger thresholds and including new triggers for previously unexplored phase space, which make extensive use of the various inner detector tracking algorithms in the HLT.
 
The ATLAS trigger system was successfully (re-)commissioned with the first data acquired at 13.6\,\TeV,
with some final commissioning steps for L1Calo and L1Muon to be completed during Run 3. First performance studies of the
different trigger signatures and trigger efficiencies with respect to the offline quantities are presented using
the 13.6\,\TeV\ proton-proton collision data collected during 2022.


\section*{Acknowledgements}


We thank CERN for the very successful operation of the LHC and its injectors, as well as the support staff at
CERN and at our institutions worldwide without whom ATLAS could not be operated efficiently.
 
The crucial computing support from all WLCG partners is acknowledged gratefully, in particular from CERN, the ATLAS Tier-1 facilities at TRIUMF/SFU (Canada), NDGF (Denmark, Norway, Sweden), CC-IN2P3 (France), KIT/GridKA (Germany), INFN-CNAF (Italy), NL-T1 (Netherlands), PIC (Spain), RAL (UK) and BNL (USA), the Tier-2 facilities worldwide and large non-WLCG resource providers. Major contributors of computing resources are listed in Ref.~\cite{ATL-SOFT-PUB-2023-001}.
 
We gratefully acknowledge the support of ANPCyT, Argentina; YerPhI, Armenia; ARC, Australia; BMWFW and FWF, Austria; ANAS, Azerbaijan; CNPq and FAPESP, Brazil; NSERC, NRC and CFI, Canada; CERN; ANID, Chile; CAS, MOST and NSFC, China; Minciencias, Colombia; MEYS CR, Czech Republic; DNRF and DNSRC, Denmark; IN2P3-CNRS and CEA-DRF/IRFU, France; SRNSFG, Georgia; BMBF, HGF and MPG, Germany; GSRI, Greece; RGC and Hong Kong SAR, China; ISF and Benoziyo Center, Israel; INFN, Italy; MEXT and JSPS, Japan; CNRST, Morocco; NWO, Netherlands; RCN, Norway; MEiN, Poland; FCT, Portugal; MNE/IFA, Romania; MESTD, Serbia; MSSR, Slovakia; ARRS and MIZ\v{S}, Slovenia; DSI/NRF, South Africa; MICINN, Spain; SRC and Wallenberg Foundation, Sweden; SERI, SNSF and Cantons of Bern and Geneva, Switzerland; MOST, Taipei; TENMAK, T\"urkiye; STFC, United Kingdom; DOE and NSF, United States of America.
 
Individual groups and members have received support from BCKDF, CANARIE, CRC and DRAC, Canada; CERN-CZ, PRIMUS 21/SCI/017 and UNCE SCI/013, Czech Republic; COST, ERC, ERDF, Horizon 2020, ICSC-NextGenerationEU and Marie Sk{\l}odowska-Curie Actions, European Union; Investissements d'Avenir Labex, Investissements d'Avenir Idex and ANR, France; DFG and AvH Foundation, Germany; Herakleitos, Thales and Aristeia programmes co-financed by EU-ESF and the Greek NSRF, Greece; BSF-NSF and MINERVA, Israel; Norwegian Financial Mechanism 2014-2021, Norway; NCN and NAWA, Poland; La Caixa Banking Foundation, CERCA Programme Generalitat de Catalunya and PROMETEO and GenT Programmes Generalitat Valenciana, Spain; G\"{o}ran Gustafssons Stiftelse, Sweden; The Royal Society and Leverhulme Trust, United Kingdom.
 
In addition, individual members wish to acknowledge support from CERN: European Organization for Nuclear Research (CERN PJAS); Chile: Agencia Nacional de Investigaci\'on y Desarrollo (FONDECYT 1190886, FONDECYT 1210400, FONDECYT 1230812, FONDECYT 1230987); China: National Natural Science Foundation of China (NSFC - 12175119, NSFC 12275265, NSFC-12075060); Czech Republic: PRIMUS Research Programme (PRIMUS/21/SCI/017); EU: H2020 European Research Council (ERC - 101002463); European Union: European Research Council (ERC - 948254), Horizon 2020 Framework Programme (MUCCA - CHIST-ERA-19-XAI-00), European Union, Future Artificial Intelligence Research (FAIR-NextGenerationEU PE00000013), Italian Center for High Performance Computing, Big Data and Quantum Computing (ICSC, NextGenerationEU), Marie Sklodowska-Curie Actions (EU H2020 MSC IF GRANT NO 101033496); France: Agence Nationale de la Recherche (ANR-20-CE31-0013, ANR-21-CE31-0013, ANR-21-CE31-0022), Investissements d'Avenir Idex (ANR-11-LABX-0012), Investissements d'Avenir Labex (ANR-11-LABX-0012); Germany: Baden-Württemberg Stiftung (BW Stiftung-Postdoc Eliteprogramme), Deutsche Forschungsgemeinschaft (DFG - 469666862, DFG - CR 312/5-1); Italy: Istituto Nazionale di Fisica Nucleare (FELLINI G.A. n. 754496, ICSC, NextGenerationEU); Japan: Japan Society for the Promotion of Science (JSPS KAKENHI JP21H05085, JSPS KAKENHI JP22H01227, JSPS KAKENHI JP22H04944, JSPS KAKENHI JP22KK0227); Netherlands: Netherlands Organisation for Scientific Research (NWO Veni 2020 - VI.Veni.202.179); Norway: Research Council of Norway (RCN-314472); Poland: Polish National Agency for Academic Exchange (PPN/PPO/2020/1/00002/U/00001), Polish National Science Centre (NCN 2021/42/E/ST2/00350, NCN OPUS nr 2022/47/B/ST2/03059, NCN UMO-2019/34/E/ST2/00393, UMO-2020/37/B/ST2/01043, UMO-2021/40/C/ST2/00187); Slovenia: Slovenian Research Agency (ARIS grant J1-3010); Spain: BBVA Foundation (LEO22-1-603), Generalitat Valenciana (Artemisa, FEDER, IDIFEDER/2018/048), La Caixa Banking Foundation (LCF/BQ/PI20/11760025), Ministry of Science and Innovation (MCIN \& NextGenEU PCI2022-135018-2, MICIN \& FEDER PID2021-125273NB, RYC2019-028510-I, RYC2020-030254-I, RYC2021-031273-I, RYC2022-038164-I), PROMETEO and GenT Programmes Generalitat Valenciana (CIDEGENT/2019/023, CIDEGENT/2019/027); Sweden: Swedish Research Council (VR 2018-00482, VR 2022-03845, VR 2022-04683, VR grant 2021-03651), Knut and Alice Wallenberg Foundation (KAW 2017.0100, KAW 2018.0157, KAW 2018.0458, KAW 2019.0447); Switzerland: Swiss National Science Foundation (SNSF - PCEFP2\_194658); United Kingdom: Leverhulme Trust (Leverhulme Trust RPG-2020-004); United States of America: U.S. Department of Energy (ECA DE-AC02-76SF00515), Neubauer Family Foundation.


\printbibliography
 
\clearpage
 
\begin{flushleft}
\hypersetup{urlcolor=black}
{\Large The ATLAS Collaboration}

\bigskip

\AtlasOrcid[0000-0002-6665-4934]{G.~Aad}$^\textrm{\scriptsize 102}$,
\AtlasOrcid[0000-0001-7616-1554]{E.~Aakvaag}$^\textrm{\scriptsize 16}$,
\AtlasOrcid[0000-0002-5888-2734]{B.~Abbott}$^\textrm{\scriptsize 120}$,
\AtlasOrcid[0000-0002-1002-1652]{K.~Abeling}$^\textrm{\scriptsize 55}$,
\AtlasOrcid[0000-0001-5763-2760]{N.J.~Abicht}$^\textrm{\scriptsize 49}$,
\AtlasOrcid[0000-0002-8496-9294]{S.H.~Abidi}$^\textrm{\scriptsize 29}$,
\AtlasOrcid[0000-0002-9987-2292]{A.~Aboulhorma}$^\textrm{\scriptsize 35e}$,
\AtlasOrcid[0000-0001-5329-6640]{H.~Abramowicz}$^\textrm{\scriptsize 151}$,
\AtlasOrcid[0000-0002-1599-2896]{H.~Abreu}$^\textrm{\scriptsize 150}$,
\AtlasOrcid[0000-0003-0403-3697]{Y.~Abulaiti}$^\textrm{\scriptsize 117}$
% ...
\bigskip
\\

% ...
$^{16}$Department for Physics and Technology, University of Bergen, Bergen; Norway.\\
% ...
$^{29}$Physics Department, Brookhaven National Laboratory, Upton NY; United States of America.\\
% ...
$^{35}$$^{(a)}$Facult\'e des Sciences Ain Chock, R\'eseau Universitaire de Physique des Hautes Energies - Universit\'e Hassan II, Casablanca;$^{(b)}$Facult\'{e} des Sciences, Universit\'{e} Ibn-Tofail, K\'{e}nitra;$^{(c)}$Facult\'e des Sciences Semlalia, Universit\'e Cadi Ayyad, LPHEA-Marrakech;$^{(d)}$LPMR, Facult\'e des Sciences, Universit\'e Mohamed Premier, Oujda;$^{(e)}$Facult\'e des sciences, Universit\'e Mohammed V, Rabat;$^{(f)}$Institute of Applied Physics, Mohammed VI Polytechnic University, Ben Guerir; Morocco.\\
% ...
$^{49}$Fakult\"{a}t Physik , Technische Universit{\"a}t Dortmund, Dortmund; Germany.\\
% ...
$^{55}$II. Physikalisches Institut, Georg-August-Universit\"{a}t G\"ottingen, G\"ottingen; Germany.\\
% ...
$^{102}$CPPM, Aix-Marseille Universit\'e, CNRS/IN2P3, Marseille; France.\\
% ...
$^{117}$Department of Physics, New York University, New York NY; United States of America.\\
% ...
$^{120}$Homer L. Dodge Department of Physics and Astronomy, University of Oklahoma, Norman OK; United States of America.\\
% ...
$^{150}$Department of Physics, Technion, Israel Institute of Technology, Haifa; Israel.\\
$^{151}$Raymond and Beverly Sackler School of Physics and Astronomy, Tel Aviv University, Tel Aviv; Israel.\\
% ...
$^{a}$ Also Affiliated with an institute covered by a cooperation agreement with CERN.\\
$^{b}$ Also at An-Najah National University, Nablus; Palestine.\\
$^{c}$ Also at Borough of Manhattan Community College, City University of New York, New York NY; United States of America.\\
$^{d}$ Also at Center for High Energy Physics, Peking University; China.\\
$^{e}$ Also at Center for Interdisciplinary Research and Innovation (CIRI-AUTH), Thessaloniki; Greece.\\
$^{f}$ Also at Centro Studi e Ricerche Enrico Fermi; Italy.\\
% ...

\end{flushleft}


\end{document}